\documentclass[a4paper,UKenglish,cleveref, autoref, thm-restate]{article}

\newcommand{\proofspath}{./proofs}

\newcommand{\rootpath}{./}
\newcommand{\currentpath}{./}

\usepackage{ifthen}

\newboolean{talk}
\setboolean{talk}{false}
\newboolean{paper}
\setboolean{paper}{false}

\newboolean{IEEEstyle}
\setboolean{IEEEstyle}{false}
\newboolean{lipicsstyle}
\setboolean{lipicsstyle}{false}
\newboolean{eptcsstyle}
\setboolean{eptcsstyle}{false}
\newboolean{entcsstyle}
\setboolean{entcsstyle}{false}
\newboolean{sigplanstyle}
\setboolean{sigplanstyle}{false}
\newboolean{easychairstyle}
\setboolean{easychairstyle}{false}
\newboolean{acmartstyle}
\setboolean{acmartstyle}{false}

\newboolean{lncsstyle}
\setboolean{lncsstyle}{false}

\newboolean{needstheorems}
\setboolean{needstheorems}{false}
\newboolean{withimages}
\setboolean{withimages}{false}
\newboolean{withproofs}
\setboolean{withproofs}{true}

\newboolean{french}
\setboolean{french}{false}

\usepackage{tikz}
\usetikzlibrary{calc}
\usetikzlibrary{matrix,arrows}
\usetikzlibrary{decorations.shapes}
\usetikzlibrary{decorations.text}
\usetikzlibrary{decorations.pathmorphing}
\usetikzlibrary{decorations.markings}
\usetikzlibrary{positioning}

\tikzset{
node distance=1.3cm, auto,
every node/.style={font=\scriptsize },
ocenter/.style={baseline={([yshift=-.5ex, xshift=-.5ex]current bounding box)}},  
labelBeginAbove/.style={postaction={decorate,decoration={markings,mark=at position 0 with {\node[inner sep= 0.6pt, above=1pt]{\tiny #1};}} } },
labelBeginBelow/.style={postaction={decorate,decoration={markings,mark=at position 0 with {\node[inner sep= 0.6pt, below=1pt]{\tiny #1};}}}},
labelEndAbove/.style={postaction={decorate,decoration={markings,mark=at position 1 with {\node[inner sep= 0.6pt, above=1pt]{\tiny #1};}}}},
labelEndBelow/.style={postaction={decorate,decoration={markings,mark=at position 1 with {\node[inner sep= 0.6pt, below=1pt]{\tiny #1};}}}},
labelEndRight/.style={postaction={decorate,decoration={markings,mark=at position 1 with {\node[inner sep= 0.6pt, right=1pt]{\tiny #1};}}}},
labelEndLeft/.style={postaction={decorate,decoration={markings,mark=at position 1 with {\node[inner sep= 0.6pt, left=1pt]{\tiny #1};}}}}
}

\newcommand{\verticaldiagramdistance}{25pt}

\usepackage{amsthm}

\theoremstyle{plain}
    \newtheorem{theorem}{Theorem}[section]
    \newtheorem{lemma}[theorem]{Lemma}

    \newtheorem{proposition}[theorem]{Proposition}
    
\theoremstyle{definition}
    \newtheorem{definition}{Definition}
\theoremstyle{remark}

\usepackage{hyperref}
\usepackage[utf8]{inputenc}
\usepackage{graphicx}
\usepackage{complexity}
\usepackage{cmll}
\usepackage{stmaryrd}
\usepackage{multirow}
\usepackage{fancybox}
\usepackage{xspace}
\usepackage{multicol}
\usepackage{mathtools}
\usepackage{yfonts}
\usepackage{xcolor}
\usepackage[normalem]{ulem}
\usepackage{nameref}
\makeatletter
\newcommand{\@chapapp}{\relax}%
\makeatother
\usepackage{appendix}
\capturecounter{theorem}
\capturecounter{figure}
\usepackage{proof}
\usepackage{ebproof}
\usepackage{rotating}
\usepackage{mathrsfs}

\ifthenelse{\boolean{easychairstyle}}{
\setboolean{needstheorems}{true}}{}

\ifthenelse{\boolean{IEEEstyle}}{
\setboolean{needstheorems}{true}}{}

\ifthenelse{\boolean{eptcsstyle}}{
\setboolean{needstheorems}{true}}{}

\ifthenelse{\boolean{sigplanstyle}}{
\setboolean{needstheorems}{true}}{}

\ifthenelse{\boolean{needstheorems}}{
}{}

\newcommand{\myproof}[1]{
\ifthenelse{\boolean{withproofs}}{#1}{}
}

\newcommand\dom[1]{{\symfont{dom}}({#1})}
\newcommand\eqdef{\eqqcolon}

\newcommand{\onormalpr}[1]{\onormalSym({#1})}
    \newcommand{\onormalentrypr}[2]{\onormalSym \progentry{#1}{#2}}

\newcommand{\inert}{{\symfont{inert}}}
    
\newcommand{\abs}{{\symfont{abs}}}
    
\newcommand{\absPredSym}{\abs}
    
\newcommand{\uabsSym}{{\symfont{uabs}}}

\newcommand{\inerttype}{\symfont{inert}}

\newcommand{\onormalSym}{{\sf{onorm}}\xspace}
    
\newcommand{\ufnormalSym}{{\sf{unorm}}\xspace}

\newcommand{\genVarSetpr}[1]{\genVarPredSym_{\#}(#1)}

\newcommand{\genVarpr}[2]{\genVarPredSym_{#1}(#2)}
\newcommand{\genVarentrypr}[3]{\genVarPredSym_{#1}(#2, #3)}

\newcommand{\inertPredSym}{\inerttype}
    
\newcommand{\inertpr}[1]{\inerttype(#1)}
\newcommand{\inertentrypr}[2]{\inerttype(#1, #2)}
\newcommand{\ufinertPredSym}{{\sf{uinert}}\xspace}
\newcommand{\ufinertpr}[1]{\ufinertPredSym({#1})}
\newcommand{\ufinertentrypr}[2]{\ufinertPredSym(#1, #2)}

\newcommand{\uselessPredSym}{\symfont{ul}\, }
\newcommand{\uselesspr}[1]{\uselessPredSym(#1)}
\newcommand{\uselessentrypr}[2]{\uselessPredSym(#1,#2)\, }

\newcommand{\abspr}[1]{\abs(#1)}
\newcommand{\uabspr}[1]{\uabsSym(#1)}

\newcommand{\ufnormalpr}[1]{\ufnormalSym(#1)}
\newcommand{\ufnormalentrypr}[2]{\ufnormalSym(#1, #2)}

\newcommand\Deribbase[5]{{#3}\ {\pmb\vdash}_{#2}^{#1} {#4} \hastype  {#5}}

\makeatletter

\newcommand{\Deribase}[1]{%
  \def\DeribW[##1]{\Deribbase{##1}{#1}}%
  \def\DeribWO{\Deribbase{}{#1}}%
  \@ifnextchar[\DeribW\DeribWO%
  }

  \newcommand{\Deri}{%
  \def\DeriW_##1{\Deribase{##1}}%
  \def\DeriWO{\Deribase{}}%
  \@ifnextchar_\DeriW\DeriWO%
  }
\makeatother

\makeatletter
\newcommand{\exder}{%
  \def\exderW[##1]{\triangleright_{##1}}%
  \def\exderWO{\triangleright\ }%
  \@ifnextchar[\exderW\exderWO%
  }
\makeatother

 \renewcommand\Deri[4][]{{#2} \vdash^{#1} {#3} : {#4}}

\newcommand{\ndsym}{\symfont{nd}}
\newcommand{\undSym}{\symfont{und}}
\newcommand{\ondSym}{\symfont{ond}}

\newcommand{\sizend}[1]{\size{#1}_{\ndsym}}

\newcommand{\ES}{\text{ES}\xspace}

\newcommand{\la}[1]{\lambda #1.}

\newcommand{\tm}{t}
\newcommand{\tmtwo}{u}
\newcommand{\tmthree}{s}
\newcommand{\tmfour}{m}
\newcommand{\tmfive}{{\tilde{\tm}}}
\newcommand{\tmsix}{{\tilde{\tmtwo}}}
\newcommand{\tmseven}{{\tilde{\tmthree}}}

\newcommand{\var}{x}
\newcommand{\vartwo}{y}
\newcommand{\varthree}{z}
\newcommand{\varfour}{{\tilde{\var}}}

\newcommand{\Id}{{\symfont I}}

\newcommand{\updSym}{\symfont{upd}}
\newcommand{\updp}[3]{\updSym(#1, #2, #3)}

\newcommand{\varset}{{\mathcal{V}}}
    \newcommand{\varsetp}{{{\mathcal{V}}\,'}}
\newcommand{\varsettwo}{\mathcal{W}}
    \newcommand{\varsettwop}{{{\mathcal{W}}\,'}}

\newcommand{\uvarset}{{\mathcal{U}}}
\newcommand{\uvarsettwo}{{\mathcal{V}}}
\newcommand{\uvarsetthree}{{\mathcal{W}}}
\newcommand{\uvarsetfour}{{\mathcal{X}}}
\newcommand{\uvarsetfive}{{\mathcal{\tilde{U}}}}
\newcommand{\uvarsetsix}{{\mathcal{\tilde{V}}}}
\newcommand{\uvarsetseven}{{\mathcal{\tilde{W}}}}
\newcommand{\uvarseteight}{{\mathcal{\tilde{X}}}}

\newcommand{\uvarseteleven}{{\mathcal{\tilde{W}'}}}

\newcommand{\avarset}{{\mathcal{A}}}
\newcommand{\avarsettwo}{{\mathcal{B}}}
\newcommand{\avarsetthree}{{\mathcal{C}}}
\newcommand{\avarsetfour}{{\mathcal{D}}}
\newcommand{\avarsetfive}{{\mathcal{\tilde{A}}}}
\newcommand{\avarsetsix}{{\mathcal{\tilde{B}}}}
\newcommand{\avarsetseven}{{\mathcal{\tilde{C}}}}
\newcommand{\avarseteight}{{\mathcal{\tilde{D}}}}

\newcommand{\avarseteleven}{{\mathcal{\tilde{C}'}}}

\newcommand{\Var}{{\symfont{Var}}}
\newcommand{\genVarSym}{{\symfont{GV}}}
\newcommand{\genVarPredSym}{{\symfont{genVar}}}

 \newcommand{\Rew}[1]{\rightarrow_{#1}}

\makeatletter

\newcommand{\Rewbase}{%
  \def\RewbaseW[##1]##2##3{\ {\xrightarrow{##1}}{}_{##2}^{##3}\xspace }%
  \def\RewbaseWO##1##2{\ {\xrightarrow{}}{}_{##1}^{##2}\xspace }%
  \@ifnextchar[\RewbaseW\RewbaseWO%
  }

\newcommand{\Rewn}[2][*]{%
  \def\RewnW[##1]{\Rewbase[##1]{#2}{#1}}%
  \def\RewnWO{\Rewbase{#2}{#1}}%
  \@ifnextchar[\RewnW\RewnWO%
  }

\makeatother

\newcommand{\tob}{\Rew{\beta}}

\newcommand{\symfont}[1]{\mathsf{#1}}

\newcommand{\esym}{{\symfont e}}

\newcommand{\msym}{{\symfont m}}

\newcommand{\cbv}{CbV\xspace}

\newcommand{\cbn}{CbN\xspace}

\newcommand{\cbneed}{CbNeed\xspace}

\newcommand{\val}{v}
    
\newcommand{\valtwo}{w}

\newcommand{\ctxholep}[1]{\langle #1\rangle}
\newcommand{\ctxhole}{\ctxholep{\cdot}}

\newcommand{\ctx}{C}
\renewcommand{\ctx}{T}
\newcommand{\ctxtwo}{\ctx'}
\newcommand{\ctxthree}{\ctx''}

\newcommand{\ctxp}[1]{\ctx\ctxholep{#1}}

\newcommand{\evctx}{{\mathcal E}}
\newcommand{\evctxp}[1]{\evctx_{#1}}

\newcommand{\evctxApp}{\evctx^{\symfont{@}}}
\newcommand{\evctxAppp}[2]{\evctxApp_{#1, #2}}

\newcommand{\evctxUsep}[2]{\evctx_{#1, #2}}

\newcommand{\defeq}{\coloneqq}
\newcommand{\grameq}{\Coloneqq}

\newcommand{\isub}[2]{\{#1{\leftarrow}#2\}}
\newcommand{\esub}[2]{[#1{\leftarrow}#2]}

\newcommand{\letin}[3]{{\sf let}\ #1=#2\ {\sf in}\ #3}

\renewcommand{\l}{\lambda}

\newcommand{\ie}{{\em i.e.}\xspace}
\newcommand{\eg}{{\em e.g.}\xspace}
\newcommand{\ih}{{\emph{i.h.}}\xspace}
\newcommand{\fv}[1]{{\symfont {fv}}(#1)}

\newcommand{\erase}[1]{}

\newcommand{\ignore}[1]{}

\newcommand{\myinput}[1]{\ifthenelse{\boolean{withproofs}}{\input{#1}}{}}

\newcommand{\reflemma}[1]{Lemma~\ref{l:#1}}
\newcommand{\reflemmap}[2]{Lemma~\ref{l:#1}.\ref{p:#1-#2}}
\newcommand{\reflemmaeq}[1]{{L.\ref{l:#1}}}
\newcommand{\reflemmaeqp}[2]{{L.\ref{l:#1}.\ref{p:#1-#2}}}

\newcommand{\refthm}[1]{Theorem~\ref{thm:#1}}

\newcommand{\refprop}[1]{Proposition~\ref{prop:#1}}

\newcommand{\refsect}[1]{Sect.~\ref{sect:#1}}

\ifthenelse{\boolean{easychairstyle}}{}{
}
\newcommand{\reffig}[1]{Fig.~\ref{fig:#1}}

\newcommand{\refdef}[1]{Definition~\ref{def:#1}}

\newcommand{\refequa}[1]{(\ref{eq:#1})}

\newcommand{\levy}{{L{\'e}vy}\xspace}

\newcommand{\set}[1]{\{#1\}}
\newcommand{\tom}{\Rew{\msym}}
    
\newcommand{\toe}{\Rew{\esym}}

\newcommand{\size}[1]{|#1|}
    
\newcommand{\sizep}[2]{|#1|_{#2}}
\newcommand{\sizem}[1]{\sizep{#1}{\msym}}

\newcommand{\emptyenv}{\epsilon}
\newcommand{\env}{E} 	
\renewcommand{\env}{e} 	
\newcommand{\envtwo}{\env'} 	
\newcommand{\envthree}{\env''} 

\newcommand{\herenv}{G}
\renewcommand{\herenv}{E}
\newcommand{\herenvtwo}{\herenv'}

\newcommand{\deriv}{\ensuremath{d}}

\newcommand{\derivtwo}{\ensuremath{d'}}

\newcommand{\rename}[1]{#1^\alpha}
\renewcommand{\rename}[1]{#1}
\newcommand{\fire}{f}
\newcommand{\firetwo}{g}

\newcommand{\hastype}{\,{:}\,}

\newcommand{\barendregtconv}{{\emph{Barendregt's variable convention}}}

\newcommand{\closedcbneed}{\emph{Closed CbNeed}}

\renewcommand{\closedcbneed}{Closed CbNeed\xspace}

\newcommand{\opencbneed}{{\opencbneednsp}\space}
    \newcommand{\opencbneednsp}{\text{Open CbNeed}}
        \renewcommand{\opencbneed}{\text{Open CbNeed}\xspace}

\newcommand{\opencbv}{{{\opencbvnsp}\,}}
    \newcommand{\ocbv}{\opencbv} 
    \newcommand{\opencbvnsp}{\text{Open CbV}}

\newcommand{\usefulopencbneed}{{\usefulopencbneednsp}\xspace}
    \newcommand{\usefulopencbneednsp}{\text{Useful Open CbNeed}}
\newcommand{\useless}{\emph{useless}\,}

\newcommand{\ufabspr}[1]{\uabspr{#1}}
\newcommand{\ufabsentrypr}[2]{\uabspr{#1,#2}}

\newcommand{\itm}{i}
\newcommand{\itmtwo}{j}

\newcommand{\itmplus}{{\itm}^{+}}
\newcommand{\itmplustwo}{{\itmtwo}^{+}}

\newcommand{\ntm}{n}

\newcommand{\prog}{p}
\newcommand{\progtwo}{q}
\newcommand{\progthree}{r}

\newcommand{\pctx}{P}
\newcommand{\pctxtwo}{Q}
\newcommand{\pctxthree}{R}
\newcommand{\pctxfour}{S}

\newcommand{\pctxp}[1]{\pctx \ctxholep{#1}}
\newcommand{\pctxtwop}[1]{\pctxtwo \ctxholep{#1}}
\newcommand{\pctxthreep}[1]{\pctxthree \ctxholep{#1}}

\newcommand{\progentry}[2]{(#1, #2)}

\newcommand{\nvSym}{\mathsf{nv}}
\newcommand{\anvSym}{\mathsf{a}}

\newcommand{\unvSym}{\mathsf{u}}

\newcommand{\nv}[1]{\nvSym(#1)}
\newcommand{\anv}[1]{\anvSym(#1)}
\newcommand{\unv}[1]{\unvSym(#1)}

\newcommand{\unfold}{\downarrow}
\newcommand{\unfoldpr}[1]{\unfold #1}

\newcommand{\nvprogentry}[2]{\mathsf{nv}(#1, #2)}
\newcommand{\domprogentry}[2]{\mathsf{dom}(#1, #2)}

\newcommand{\nventryp}[2]{\nv{#1, #2}}

\newcommand{\ctxholeentryp}[2]{\ctxholep{#1, #2}}
\newcommand{\pctxentryp}[2]{\pctxp{#1, #2}}
\newcommand{\pctxtwoentryp}[2]{\pctxtwop{#1, #2}}

\newcommand{\anventryp}[2]{\mathsf{a}(#1, #2)}
\newcommand{\unventryp}[2]{\mathsf{u}(#1, #2)}

\newcommand{\appES}[2]{#1 @ #2}

\newcommand{\usefulMultSym}{\symfont{um}}

\newcommand{\toum}{\Rew{\usefulMultSym}}

\newcommand{\usefulExpSym}{\symfont{ue}}

\newcommand{\toue}{\Rew{\usefulExpSym}}

\newcommand{\tound}{\rightarrow_{\undSym}}

\newcommand{\tonm}{\Rew{\symfont{om}}}
    \newcommand{\toom}{\tonm} 
\newcommand{\tone}{\Rew{\symfont{oe}}}
    \newcommand{\tooe}{\tone} 
\newcommand{\tonnd}{\rightarrow_{\ondSym}}
    \newcommand{\toond}{\tonnd} 

\newcommand{\hctx}{{\mathcal{H}}}
\newcommand{\hctxtwo}{{\mathcal{J}}}
\newcommand{\hctxthree}{{\mathcal{I}}}
\renewcommand{\hctx}{H}
\renewcommand{\hctxtwo}{J}
\renewcommand{\hctxthree}{I}

\newcommand{\hctxp}[1]{\hctx\ctxholep{#1}}
\newcommand{\hctxtwop}[1]{\hctxtwo\ctxholep{#1}}

\newcommand{\hctxapp}{{\mathcal{\hctx^{@}}}}
\renewcommand{\hctxapp}{{\hctx^{@}}}
\newcommand{\hctxappsub}[1]{\hctx_{#1}^{@}}
\newcommand{\hctxapptwo}{\mathcal{\hctxtwo^{@}}}
\renewcommand{\hctxapptwo}{\hctxtwo^{@}}
\newcommand{\hctxapptwosub}[1]{{\hctxtwo_{#1}^{@}}}
\newcommand{\hctxappthree}{\mathcal{\hctxthree^{@}}}
\renewcommand{\hctxappthree}{\hctxthree^{@}}

\newcommand{\hctxappp}[1]{\hctxapp \ctxholep{#1}}
\newcommand{\hctxapptwop}[1]{\hctxapptwo \ctxholep{#1}}

\newcommand{\inertSym}{{\symfont{I}}}
\newcommand{\absSym}{{\symfont{A}}}
\newcommand{\axSym}{{\symfont{AX}}}
\newcommand{\varSym}{{\symfont{VAR}}}

\newcommand{\herSym}{{\symfont{HER}}}
\newcommand{\gcSym}{{\symfont{GC}}}

\newcommand{\promSym}{{\symfont{Lift}}} 
\newcommand{\multSym}{{\symfont{M}}}
\newcommand{\expSym}{{\symfont{E}}}

\newcommand{\openCbNeedSym}{{\symfont{O}}}
\newcommand{\ruleEvcAx}{\openCbNeedSym_{\axSym}}
\newcommand{\ruleEvcInert}{\openCbNeedSym_{\inertSym}}
\newcommand{\ruleEvcGc}{\openCbNeedSym_{\gcSym}}
\newcommand{\ruleEvcHer}{\openCbNeedSym_{\herSym}}

\newcommand{\ruleUseAxMult}{\multSym_{\axSym}}
\newcommand{\ruleUseAxOneExp}{\expSym_{\axSym_{1}}}
\newcommand{\ruleUseAxTwoExp}{\expSym_{{\axSym}_{2}}}
\newcommand{\ruleUseVarMult}{\multSym_{\varSym}}
\newcommand{\ruleUseNonVarMult}{\multSym_{\nvarSym}}
\newcommand{\ruleUseVarExp}{\expSym_{\varSym}}
\newcommand{\ruleUseNonVarExp}{\expSym_{\nvarSym}}
\newcommand{\ruleUseGcMult}{\multSym_{\gcSym}}
\newcommand{\ruleUseGcExp}{\expSym_{\gcSym}}
\newcommand{\ruleUseHerMult}{\multSym_{\herSym}}
\newcommand{\ruleUseHerExp}{\expSym_{\NonApplicativeSym}}

\newcommand{\ruleUseUselessMult}{\multSym_{\uselessSym}}
\newcommand{\ruleUseUselessExp}{\expSym_{\uselessSym}}

\newcommand{\NonApplicativeSym}{{\symfont{NA}}}

\newcommand{\absentrypr}[2]{\abs(#1, #2)}

\newcommand{\ruleInertAx}{\inertSym_{\axSym}}

\newcommand{\ruleAbsAx}{\absSym_{\axSym}}
\newcommand{\ruleAbsGenVar}{\absSym_{\genVarSym}}

\newcommand{\ruleInertProm}{\inertSym_{\promSym}}
\newcommand{\ruleAbsProm}{\absSym_{\promSym}}

\newcommand{\ruleInertInert}{\inertSym_{\inertSym}}

\newcommand{\ruleInertGc}{\inertSym_{\gcSym}}
\newcommand{\ruleAbsGc}{\absSym_{\gcSym}}

\newcommand{\nvarSym}{{\symfont{I}}}
\newcommand{\uselessSym}{{\symfont{U}}}

\newcommand{\ruleGenVarAx}{\genVarSym_{\axSym}}
\newcommand{\ruleGenVarHer}{\genVarSym_{\herSym}}
\newcommand{\ruleGenVarGc}{\genVarSym_{\gcSym}}

\newcommand{\ufRuleGenVarHer}{\ruleGenVarHer}
\newcommand{\ufRuleGenVarGc}{\ruleGenVarGc}

\newcommand{\ufRuleNormPrograms}{\ufnormalSym_{\, P}}

\newcommand{\ufRuleInertAx}{\inertSym_{\axSym}}

\newcommand{\ufRuleInertInert}{\inertSym_{\inertSym}}

\newcommand{\ufRuleAbsGc}{\ruleAbsGc}
\newcommand{\ufRuleAbsGenVar}{\ruleAbsGenVar}

\newcommand{\ufRuleInertProm}{\ruleInertProm}
\newcommand{\ufRuleAbsProm}{\ruleAbsProm}

\newcommand{\ufRuleInertGenVar}{\inertSym_{\genVarSym}}
\newcommand{\ufRuleInertUseless}{\inertSym_{\uselessSym}}

\newcommand{\ufRuleInertGc}{\ruleInertGc}

\newcommand{\inputProof}[1]{
\ifthenelse {\boolean{includeProofs}}
	{\input{#1}}
	{}
}

\newcommand{\un}[1]{#1^{\circ}}
\renewcommand{\un}[1]{#1}

\newcommand{\gregoire}{Gr{\'{e}}goire\xspace}

    \newcommand{\tmpsub}[1]{{\tm_{1}}^{'}}

\newcommand{\unfsym}{\rotatebox[origin=c]{-90}{$\rightarrow$}}
\newcommand{\unf}[1]{#1\unfsym\,}

\newcommand{\myparagraph}[1]{\smallskip \myparagraphnoskip{#1}}
\newcommand{\myparagraphnoskip}[1]{\emph{#1}.}

\newcommand\bigo[1]{\mathcal{O}(#1)}

\newcommand{\hhctx}{H^{*}}
\newcommand{\hhctxtwo}{J^{*}}

\newcommand{\hhctxp}[1]{\hhctx\ctxholep{#1}}

\newboolean{includeProofs}
\setboolean{includeProofs}{true}

\begin{document}
\title{Useful Open Call-by-Need}
\author{Beniamino Accattoli}
\author{Beniamino Accattoli \and Maico Leberle}
\date{}
\maketitle

\begin{abstract}
 This paper studies useful sharing, which is a sophisticated optimization for $\l$-calculi, in the context of call-by-need evaluation in presence of open terms. Useful sharing  turns out to be harder in call-by-need than in call-by-name or call-by-value, because call-by-need evaluates inside environments, making it harder to specify when a substitution step is useful. We isolate the key involved concepts and prove the correctness and the completeness of useful sharing in this setting.
\end{abstract}

\bigskip

\begin{center}
This is the version with proofs (in the Appendix, starting on page \pageref{sect:Proofs_of_opencbneed}) of the paper with the same title and authors  in the proceedings of CSL 2022.
\end{center}

\section{Introduction}
Despite decades of research on how to best evaluate $\l$-terms, the topic is still actively studied and recent years have actually seen a surge in new results and sophisticated techniques. This paper is an attempt at harmonizing two of them, namely, \emph{strong call-by-need} and \emph{useful sharing}, under the influence of a third recently identified setting, \emph{open call-by-value}. To describe our results, we have to first outline each of these approaches.

\myparagraph{Call-by-Need} Call-by-need (shortened to \cbneed) is an evaluation scheme for the $\l$-calculus introduced in 1971 by Wadsworth \cite{Wad:SemPra:71}  as an optimization of call-by-name (\cbn), and nowadays lying at the core of the Haskell programming language. In the 
'90s, it was reformulated as operational semantics by Launchbury \cite{DBLP:conf/popl/Launchbury93}, Ariola and Felleisen \cite{DBLP:journals/jfp/AriolaF97}, and Maraist et al. \cite{DBLP:journals/jfp/MaraistOW98}, and implemented by Sestoft \cite{DBLP:journals/jfp/Sestoft97} and further studied by Kutzner and Schmidt-Schau\ss\  \cite{DBLP:conf/icfp/KutznerS98}. 
Despite being decades old, \cbneed is still actively studied, perhaps more than ever before. The last decade indeed saw a number of studies by \eg Ariola et al. \cite{DBLP:conf/tlca/AriolaHS11}, Chang and Felleisen \cite{DBLP:conf/esop/ChangF12}, Danvy and Zerny \cite{DBLP:conf/ppdp/DanvyZ13}, Downen et al. \cite{DBLP:conf/ppdp/DownenMAV14}, Garcia et al. \cite{DBLP:journals/corr/abs-1003-5197}, Hackett
and Hutton \cite{DBLP:journals/pacmpl/HackettH19}, Pédrot and Saurin \cite{DBLP:conf/esop/PedrotS16}, Mizuno and Sumii \cite{DBLP:conf/aplas/MizunoS19}, Herbelin and Miquey \cite{DBLP:conf/lics/HerbelinM20}, and Kesner et al. \cite{DBLP:conf/fossacs/KesnerPV21}, plus those mentioned in the following paragraphs.

 In the untyped, effect-free setting of the $\l$-calculus, \cbneed can be seen as borrowing the best aspects of call-by-value (\cbv), of which it takes efficiency, and of \cbn, of which it retains the better terminating behavior, as stressed in particular by Accattoli et al. \cite{DBLP:conf/esop/AccattoliGL19}. 
In contrast to \cbn and \cbv, however, \cbneed cannot easily be managed at the small-step level of the usual operational semantics of the $\l$-calculus, based on $\beta$-reduction and meta-level substitution. Its fine dynamics, indeed, requires a decomposition of the substitution process acting on single variable occurrences at a time---what we refer to as \emph{micro-step (operational) semantics}---and enriching $\l$-terms with some form of first-class \emph{sharing}. While Wadsworth's original presentation is quite difficult to manage, along the years presentations of \cbneed have improved considerably (\cite{DBLP:conf/popl/Launchbury93,DBLP:journals/jfp/MaraistOW98,DBLP:journals/jfp/AriolaF97,DBLP:conf/esop/ChangF12}), up to obtaining neat definitions, as the one by Accattoli et al. \cite{DBLP:conf/icfp/AccattoliBM14} (2014) in the \emph{linear substitution calculus} (shortened to LSC), which led to elegant proofs of its correctness with respect to \cbn, as done by Kesner \cite{DBLP:conf/fossacs/Kesner16} (2016), and of its relationship with neededness from a rewriting point of view, by Kesner et al. \cite{DBLP:conf/fossacs/KesnerRV18} (2018).

\myparagraph{Strong Call-by-Need} Being motivated by functional languages, \cbneed is usually studied considering two restrictions with respect to the ordinary $\l$-calculus: 1) terms are closed, and 2) abstraction bodies are not evaluated. Let us call this setting \emph{\closedcbneed}. Extensions of \cbneed removing both these restrictions have been considered, obtaining what we shall refer to as \emph{Strong \cbneed}. In his PhD thesis \cite{barras-phd} (1999), Barras designs and implements an abstract machine for Strong \cbneed, which has then been used in the kernel of the Coq proof assistant to decide the convertibility of terms. Balabonski et al. \cite{DBLP:journals/pacmpl/BalabonskiBBK17} (2017) give instead the first formal operational semantics of Strong \cbneed, proving it correct with respect to Strong \cbn---see also Barenbaum et al. \cite{DBLP:conf/ppdp/BarenbaumBM18}, where the semantics of \cite{DBLP:journals/pacmpl/BalabonskiBBK17} is extended towards Barras's work; Biernacka and Charatonik \cite{DBLP:conf/rta/BiernackaC19}, where it is studied via an abstract machine; Balabonski et al. \cite{DBLP:conf/fscd/BalabonskiLM21} where it has  recently been revisited and partially formalized.

\myparagraph{\cbneed and the Strong Barrier} The definition of Strong \cbneed in \cite{DBLP:journals/pacmpl/BalabonskiBBK17} builds over the simple one in the LSC, and yet is very sophisticated and far from obvious. This is an instance of a more general fact concerning implementation techniques: dealing with the strong setting is orders of magnitude more difficult than with the closed setting, it is not just a matter of adapting a few definitions. New complex issues show up, requiring new techniques and concepts---let us refer to this fact as to \emph{the strong barrier}. Another instance is the fact that \levy's optimality \cite{thesislevy} is far more complex in the strong case than in the weak one \cite{DBLP:conf/birthday/BlancLM05,DBLP:conf/icfp/Balabonski13}.

For neededness, the tool to break the strong barrier is a complex notion of \emph{needed evaluation context}, parametrized and defined by mutual induction with their sets of \emph{needed variables}. Specifying the positions in a term where needed redexes take place is very subtle.

\myparagraph{Reasonable Cost Models and the Strong Barrier} Another sophisticated form of sharing for $\l$-calculi arose recently in the study of whether the $\l$-calculus admits  reasonable evaluation strategies, that is, strategies whose number of $\beta$ steps is a reasonable time cost model (\ie measure of time complexity) for $\l$-terms. The number of function calls (that is, $\beta$-steps) is the cost model often used in practice for functional programs---this is done for instance by Chargu{\'{e}}raud and Pottier in \cite{DBLP:conf/itp/ChargueraudP15}. A time cost model is \emph{reasonable} when it is polynomially equivalent to the one of Turing machines, which is the requirement for good time cost models. For the $\l$-calculus, the theory justifying the practice of taking the number of function calls as a time cost model is far from trivial. It is an active research topic, see Accattoli \cite{DBLP:conf/rta/Accattoli19}.

The first result about $\l$-calculus reasonable strategies is due to Blelloch and Greiner \cite{DBLP:conf/fpca/BlellochG95} (1995), and concerns Closed \cbv. In The 2000s have seen similar results for Closed \cbn and Closed \cbneed by Sands, Gustavson, and Moran \cite{DBLP:conf/birthday/SandsGM02} and Dal Lago and Martini \cite{DBLP:journals/corr/abs-1208-0515,DBLP:conf/fopara/LagoM09}. These cases are based on simulating the $\l$-calculus via simple forms of sharing such as those at work in abstract machines. The same kind of sharing can also be represented in the LSC, as shown by Accattoli et al. \cite{DBLP:conf/icfp/AccattoliBM14}. The strong case seemed elusive and was suspected not to be reasonable, because of Asperti and Mairson's result that  \levy's optimal (strong) strategy is not reasonable \cite{DBLP:journals/iandc/AspertiM01}---the elusiveness was just another instance of the strong barrier.

\myparagraph{Useful Sharing} In 2014, Accattoli and Dal Lago managed to break the barrier, proving that Strong \cbn (also known as leftmost-outermost evaluation, or \emph{normal order}) is a reasonable strategy \cite{DBLP:journals/corr/AccattoliL16}. The proof rests on a simulation of Strong \cbn in a refinement of the LSC with a new further level of sharing, deemed \emph{useful sharing}. They also show useful sharing to be \emph{mandatory} for breaking the strong barrier for reasonability. 

Useful sharing amounts to doing minimal \emph{unsharing} work, namely only when it contributes to creating $\beta$-steps, while avoiding to unfold the sharing (\ie to substitute) when it only makes the term grow in size. Similarly to \cbneed, the specification of useful sharing can take place only at the micro-step level. Note that the replacement of a variable $\var$ in $\tm$ with $\tmtwo$ can create a $\beta$ redex only if $\tmtwo$ is (or shall reduce to) an abstraction \emph{and} there is an applied occurrence of $\var$ in $\tm$ (that is, $\tm = \ctxp{\var\tmthree}$ for some context $\ctx$). Therefore, restricting to useful substitutions---that is, useful sharing---amounts to two optimizations of the substitution/unfolding process:
\begin{enumerate}
\item \emph{Never substitute normal applications}: one must avoid substitutions of terms which are not---and shall not reduce to---abstractions, such as, say, $\vartwo \varthree$, because their substitution cannot create $\beta$-redexes. Indeed, $\ctxp{(\vartwo \varthree)\tmthree}$ has a $\beta$ redex if and only if $\ctxp{\var\tmthree}$ does.
\item \emph{Substituting abstractions on-demand}: when the term to substitute is an abstraction, one needs to be sure that the variable occurrence to replace is applied, because, for instance, replacing $\var$ with $\Id$ in $\vartwo\var$ (obtaining $\vartwo\Id$) is useless, as no $\beta$-redexes are created.
\end{enumerate}
 The first optimization is easy to specify, because it concerns the shape of the terms to substitute, that is, \emph{what} to substitute---it has a small-step nature. The second one instead is very delicate, as it also concerns \emph{where} to substitute. It depends on single variable occurrences and thus it is inherently \emph{micro-step}---note that $\var$ has both a useful and a useless occurrence in $\var\var$. 
 Similarly to Strong \cbneed, the difficulty is specifying \emph{useful evaluation contexts}.

\myparagraph{Strong \cbneed and Useful Sharing} Given the similar micro-step traits of \cbneed and useful sharing, and their similar difficulties, it is natural to wonder whether they can be combined. The operational semantics of Strong \cbneed in \cite{DBLP:journals/pacmpl/BalabonskiBBK17} has the easy useful optimization hardcoded, as it substitutes only abstractions. However, it ignores the delicate second optimization, and its number of $\beta$ steps is therefore not a reasonable cost model. Concretely, this means that the practice of counting  function calls does not reflect the cost of  Balabonski et al.'s operational semantics for Strong \cbneed. Since Strong \cbneed is used in the implementation of Coq, this issue has both theoretical and practical relevance. 

The aim of this paper is to start adapting useful sharing to call-by-need, developing reasonable operational semantics for \cbneed beyond the closed setting, and continuing a research line about \cbneed started by Accattoli and Barras \cite{DBLP:conf/ppdp/AccattoliB17,DBLP:conf/aplas/AccattoliB17}. To explain our approach, we first need to overview a recent new perspective on the strong barrier.

\myparagraph{Opening the Strong Barrier} The theory of the $\l$-calculus has mainly been developed in \cbn. Historically, Barendregt stressed the importance of \emph{head} evaluation (which does not evaluate arguments) for a meaningful representation of partial recursive functions---this is the leading theme of his famous book \cite{Barendregt84}. A decade later, Abramsky and Ong stressed the relevance of \emph{weak} head evaluation (which does not evaluate abstraction bodies either) to model functional programming languages \cite{DBLP:journals/iandc/AbramskyO93}. Therefore, the usual incremental way to understand strong evaluation is to start with the \emph{closed} \cbn case (\ie, weak head evaluation and closed terms), then turn to the \emph{head} case (head evaluation and open terms), and finally add evaluation into arguments obtaining the \emph{strong} case (and leftmost-outermost evaluation). This is for instance the progression that has been followed by Accattoli and Dal Lago to obtain a reasonable cost model for Strong \cbn  \cite{DBLP:conf/rta/AccattoliL12,DBLP:journals/corr/AccattoliL16}.

In a line of work by Accattoli and co-authors \cite{DBLP:conf/lics/AccattoliC15,DBLP:conf/aplas/AccattoliG16,DBLP:journals/scp/AccattoliG19,DBLP:conf/ppdp/AccattoliCGC19} aimed at developing a theory of \cbv beyond the usual closed case, it became clear that there is an alternative and better route to the strong setting. The idea is to consider the intermediate \emph{open} setting (rather than the head one) obtained by enabling evaluation in arguments and open terms (as in the strong case), while still forbidding evaluation in abstraction bodies (as in the closed case). One can summarize the situation with the following diagram:
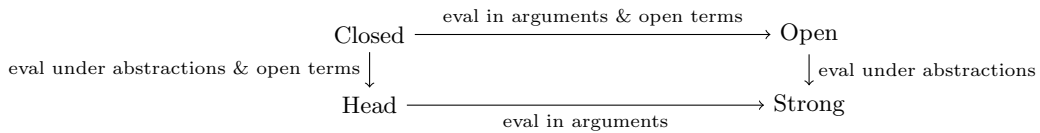
\begin{center}
\begin{tikzpicture}[ocenter]
  \node (s) {\small Closed};
  \node at (s) [right =6*\verticaldiagramdistance] (s1){\small Open};
  \node at (s) [below =0.8*\verticaldiagramdistance] (s2){\small Head};
  \node at (s1|-s2) (t) {\small Strong};
  \draw[->] (s) to node[above] {\scriptsize eval in arguments \& open terms} (s1);
  \draw[->] (s) to node[left] {\scriptsize eval under abstractions \& open terms} (s2);
  \draw[->] (s2) to node[below] {\scriptsize eval in arguments} (t);
  \draw[->] (s1) to node[right] {\scriptsize eval under abstractions} (t);
\end{tikzpicture}
\end{center}
They also show that useful sharing \emph{factors} through the open setting, rather than through the head one: the two useful optimizations are irrelevant in the head case, while they make sense in the open one, where they can be studied without facing the whole of the strong barrier. 

The strong setting can be seen as the iteration of the open one under abstraction, (but not as the iteration of the closed one, because diving into abstractions forces to deal with open terms). This view is adopted by \gregoire and Leroy in the design of the second strong abstract machine at work in  Coq \cite{DBLP:conf/icfp/GregoireL02}. Useful sharing for the strong case then amounts to understanding how open useful sharing and the iteration interact, which is subtle and yet is an orthogonal problem. Studying the open case first is the progression followed recently by Accattoli and co-authors to prove that Strong \cbv is reasonable for time  \cite{DBLP:conf/lics/AccattoliC15,DBLP:journals/scp/AccattoliG19,DBLP:conf/lics/AccattoliCC21}.

\myparagraph{This Paper} According to the decomposition of the strong barrier, here we study, as a first step, useful sharing for \cbneed in the open setting. Let us stress that, because of the barrier, it is not practicable to directly study the strong setting---this is also how the study for \cbn and \cbv, which are simpler than \cbneed, have been carried out in the literature.

An interesting aspect of useful sharing is that, while the underlying principle is the same, its \cbn and \cbv incarnations look very different, as the two strategies provide different invariants, leading to different realizations of the required optimizations. It is then interesting to explore useful sharing in \cbneed, which can be seen as a merge of \cbn and \cbv.

\myparagraph{Difficulties} It turns out that useful sharing is quite more difficult to specify in \cbneed than in \cbn or \cbv. 
Useful sharing requires to know, for every variable replacement, both \emph{what}  is being substituted (is it an abstraction?) and \emph{where} (is the variable to replace applied?). Evaluating only needed arguments, and only once, means that \cbneed evaluation moves deeply into a partially evaluated  environment, making hard to keep track of both the \emph{what} and the \emph{where} of variable replacements. In particular, a variable might not be applied in the environment but at the same time be meant to replace an applied variable---thus being applied \emph{up to sharing}---making the identification of applied variables a major difficulty.

The definition of useful rewriting steps is always involved. In \cbn and \cbv, they can nonetheless be specified compactly via the concept of \emph{unfolding}, that is, iterated meta-level substitutions \cite{DBLP:journals/corr/AccattoliL16,DBLP:conf/lics/AccattoliC15}. These definitions can be called \emph{semantical}, as they define useful \emph{micro} steps via side conditions of a \emph{small}-step nature. They are also somewhat \emph{ineffective}, because they require further work to be made operational. Unfortunately, it is unclear how to give a semantic definition of usefulness in \cbneed. In particular, defining useful \cbneed evaluation contexts seems to require the unfolding of contexts, which is tricky, given that in \cbneed the context hole might be shared, thus risking being duplicated by the unfolding.

\myparagraph{Outcome} Despite these difficulties, we succeed in designing an operational semantics for Open \cbneed with useful sharing, and proving that it validates the expected properties.

We proceed in three incremental steps. First, we provide a new \emph{split} presentation of Closed \cbneed  tuned for the study of useful sharing developed later on. Second, we extend it to the open setting, essentially mimicking Balabonski et al.'s approach \cite{DBLP:journals/pacmpl/BalabonskiBBK17}, but limiting it to the open fragment. The real novelty is the third step, providing the refinement into a \emph{useful} open \cbneed calculus, of which we prove the good properties. The crucial and sophisticated concept is the one of \emph{useful (\cbneed) evaluation contexts}, which isolate where useful needed substitutions can be triggered. They are parametrized and defined by mutual induction with the notions of both \emph{applied} and \emph{unapplied variables}, similarly to how needed evaluation contexts are parametrized and mutually dependent with needed variables. The isolation of these concepts and the proof of their properties are our main contribution.

Our definition of useful step is \emph{operational} rather than semantical, as we give a direct---and unfortunately involved---definition of useful evaluation contexts, being unclear how to give a semantic definition based on unfoldings in \cbneed. On the positive side, ours is the first fully operational definition of usefulness in the literature. Previous work (in \cbn and \cbv) has either adopted semantical ones \cite{DBLP:journals/corr/AccattoliL16,DBLP:conf/lics/AccattoliC15}, or has given abstract machines realizing the useful optimizations, but avoiding defining a useful calculus on purpose \cite{DBLP:conf/wollic/Accattoli16,DBLP:journals/scp/AccattoliG19,DBLP:conf/lics/AccattoliCC21}.

Among the properties that we prove, two can be seen as capturing the correctness and the completeness of useful sharing with respect to Open \cbneed:
\begin{itemize}
	\item \emph{Correctness}: every useful substitution step is eventually followed by a $\beta$ step, the one that it contributes to create. Essentially, our  definition of useful step correctly captures the intended semantics, as no useless steps are mistakenly considered as useful.
	\item \emph{Completeness}: normal forms in Useful Open \cbneed unfold to normal forms in Open \cbneed (the unfolding of normal forms is easy to deal with). Essentially, useful substitutions do not stop \emph{too soon}: no useful steps are mistakenly considered as useless.
\end{itemize}

\myparagraphnoskip{Sketched Complexity Analysis} The third essential property for useful sharing, and its reason to be, is \emph{reasonability}: the useful calculus can be implemented within a polynomial (or even linear) overhead in the number of $\beta$-steps. We sketch the complexity analysis at the end of the paper. A formal proof requires introducing an abstract machine implementing the calculus. We have developed the machine, but left it to a forthcoming paper for lack of space.

\myparagraph{Intersection Types in the Background} Because of the inherent difficulties mentioned above, our calculus is involved, even very involved. To remove the suspicion that it is an ad-hoc calculus, we paired it with a characterization of its key properties via intersection types, used as a validation tool with a denotational flavor, refining the type-based studies in \cite{DBLP:conf/fossacs/Kesner16,DBLP:journals/pacmpl/BalabonskiBBK17,DBLP:conf/esop/AccattoliGL19}. In such typing system, the delicate notions of useful evaluation contexts, and applied and unapplied variables have natural counterparts, and type derivations can be used to measure both evaluation lengths and the size of normal forms \emph{exactly}. 
 Such a companion study---omitted for lack of space---is in Leberle's PhD thesis \cite{leberlePhD}.

\myparagraph{Proofs} We adopt a meticulous  approach, developing proofs in full details, almost at the level of a formalization in a proof assistant. The many technical details, mostly of a tedious nature, are in the Appendix, while the body of the paper explains the relevant concepts.

\section{The Need For Useful Sharing}
Here we show a paradigmatic case of size exploding family---which is a family of terms whose size grows exponentially with the number of $\beta$-steps---motivating the key optimization of useful sharing for open and strong evaluation. Actually, there are \emph{two} paradigmatic cases of size explosion and, accordingly, \emph{two} optimizations characterizing useful sharing. The first optimization amounts to forbid the substitution of normal applications, and it is hardcoded into \cbneed evaluation, which by definition substitutes only values. Therefore, we omit discussing the first case of size explosion---more details can be found in \cite{DBLP:journals/corr/AccattoliL16,DBLP:conf/lics/AccattoliCC21}.

\myparagraph{Size Explosion} The example of size-explosion we are concerned with is due to Accattoli \cite{DBLP:journals/corr/Accattoli17} and based on the following families of terms, the $\tm_{i}$, and results, the $\tmtwo_{i}$ (where $\Id \defeq \la\varthree\varthree$):
\begin{center}
$\begin{array}{ccccc|ccccc}
  \tm_1   \defeq   \la{\var}\la{\vartwo}\vartwo \var \var
   &&&\tm_{n+1}  \defeq  \la{\var}\tm_n (\la{\vartwo}\vartwo \var \var)
   &&&
\tmtwo_0   \defeq  \Id
  &&&
  \tmtwo_{n+1}  \defeq  \la{\vartwo}\vartwo \tmtwo_{n} \tmtwo_{n}
\end{array}$  
\end{center}

\begin{proposition}[Closed and strategy-independent size explosion, \cite{DBLP:journals/corr/Accattoli17}]
\label{prop:abs-size-explosion}
  Let 
  $n \!>\! 0$. Then $\tm_n \Id \tob^n  \tmtwo_n$. Moreover, $\size{\tm_n \Id} = \bigo{n}$, $\size{\tmtwo_n} = \Omega(2^n)$, 
$\tm_n \Id$ is closed, and $\tmtwo_n$ is normal.
\end{proposition}

\myparagraphnoskip{The Useful Optimization} It is easily seen that all the terms substituted along the evaluation of the family are abstractions, namely the identity $\Id$ and instances of $\tmtwo_{i}$, and that none of these abstractions ever becomes the abstraction (on the left) of a $\beta$-redex---that is, their substitution does not create, or it is not \emph{useful} for, $\beta$-redexes. These abstractions are however duplicated and nested inside each other, being responsible for the exponential growth of the term size. Useful sharing is about avoiding such useless duplications.

If evaluation is weak, and substitution is \emph{micro-step} (\ie  one variable occurrence at a time, when in evaluation position, in a formalism with sharing), then the family does not cause an explosion. The replaced variables indeed are all instances of $\var$ in some $\tm_{i}$ which are under abstraction, and which are then never replaced in micro-step \emph{weak} evaluation. With micro-step \emph{strong} evaluation, however, these replacement do happen, and the size explodes. When evaluated with Balabonski et al. Strong \cbneed \cite{DBLP:journals/pacmpl/BalabonskiBBK17}, this family takes a number of micro-steps exponential in the number of $\beta$ steps, showing that---for as efficient as Strong \cbneed may be---the number of $\beta$ steps does not reasonably measure its evaluation time.

To tame this problem, one needs to avoid useless substitutions, resting on an optimization sometimes called \emph{substituting abstractions on-demand}, which is tricky. It requires abstractions to be substituted \emph{only} on applied variable occurrences: note that the explosion is caused by replacements of variables (namely the instances of $\var$) which are \emph{not} applied, and that thus do not create $\beta$-redexes. For instance, the optimization should allow us substituting $\Id$ on $\vartwo$ in $\vartwo\var$, because \emph{it is useful}, that is, it creates a $\beta$ redex, while it should forbid substituting it on $\var$ because it is \emph{useless} for $\beta$-redexes. Note that this optimization makes sense only when one switches to micro-step evaluation, that is, at the level of machines, because in $\var\var$ there are both a useful and a useless occurrence of $\var$. The implementation of \emph{substituting abstractions on-demand} is very subtle, also because by not performing useless substitutions, it breaks invariants of the usual open/strong evaluation process.

As shown by Accattoli and Guerrieri \cite{DBLP:journals/scp/AccattoliG19}, in an open (but not strong) setting, substituting abstractions on-demand is not mandatory for reasonability. They also show, however, that it makes nonetheless sense to study it because it is mandatory for obtaining efficient implementations, as it reduces the complexity of the overhead from \emph{quadratic} to \emph{linear} with respect to the size of the initial term. On the other hand, the optimization is mandatory in strong settings, and it is easier to first study it in the open setting, because the iteration under abstraction (required to handle the strong case) introduces new complex subtleties.
\section{The Split Presentation of Closed Call-by-Need} 
In this section we give an unusual \emph{split} presentation of \closedcbneed that shall be the starting point for our study of the open and the useful open cases of the next sections.

\myparagraph{The Need to Split} The linear substitution calculus (LSC) provides a simple and elegant setting for studying \cbneed, as shown repeatedly by Accattoli, Kesner and co-authors \cite{DBLP:conf/icfp/AccattoliBM14,DBLP:conf/fossacs/Kesner16,DBLP:journals/iandc/AccattoliC17,DBLP:journals/pacmpl/BalabonskiBBK17,DBLP:conf/fossacs/KesnerRV18,DBLP:conf/esop/AccattoliGL19,DBLP:conf/fossacs/KesnerPV21}. The LSC extends the $\l$-calculus with \emph{explicit substitutions} (shortened to ES), noted $\tm\esub\var\tmtwo$, which are a compact notation for $\letin\var\tmtwo\tm$. Capture-avoiding meta-level substitution is  noted $\tm\isub\var\tmtwo$. To model the useful optimization explained above, we shall need to substitute abstractions only on applied variables. Now, in the LSC, ES can appear everywhere in the term, for instance there are terms such as $\tm \defeq \var\esub\var\Id\tmtwo$. Note that in $\tm$ it is hard to say whether the replacement of $\var$ with $\Id$ is useful by looking only at the scope of the ES (which is the left of the $\esub{\cdot}{\cdot}$ construct): the subtlety being that the replacement is indeed useful, because the variable is applied and $\Id$ is an abstraction, but the application it is involved in is outside the scope of the ES. To avoid this complication, we give a presentation of \cbneed where ES are separated from the term they act upon, and cannot be nested into each other, similarly to what happens in abstract machines. The split presentation is not mandatory to study useful sharing, but it is quite convenient.

\myparagraph{Split Grammars}
In the split syntax, a \emph{term} is an ordinary $\l$-term (without ESs), and a \emph{program} is a term together with---in a separate place---a list of ESs, called \emph{environment}. 

Given a countable set of variables $\Var$, the syntax of \closedcbneed is given by:
\[\begin{array}{rrcl@{\hspace{.5cm}}@{\hspace{.5cm}} rrcl}
    	\textsc{Values} & \val, \valtwo & \grameq & \la{\var}{\tm} 
	&
	\textsc{Environments} & \env, \envtwo & \grameq & \emptyenv \mid \env\esub\var\tm\\ 		
	\textsc{Terms} & \tm, \tmtwo, \tmthree & \grameq & \var\in \Var \mid \val \mid \tm \tmtwo
	&
	\textsc{Programs} & \prog, \progtwo & \grameq & \progentry{\tm}{\env} \\
\end{array}\]
Note that the body of a $\l$-abstraction is a term and not a program. Of course, extending the framework to strong evaluation---which is left for future work---requires to allow programs under $\l$-abstractions. Note also that variables are not values. This is standard in works dealing with implementations or efficiency, as excluding them brings a speed-up, as shown by Accattoli and Sacerdoti Coen \cite{DBLP:journals/iandc/AccattoliC17}. In $\env\esub\var\tm$ and $\progentry\tmtwo{\env\esub\var\tm}$ the variable $\var$ is bound in $\env$ and $\tmtwo$. Terms and programs are identified modulo $\alpha$-renaming. Environments are concatenated by simple juxtaposition. We also define the environment look-up operation as follows: set $\env(\var) \defeq \tm$ if $\env = \envtwo\esub\var\tm\envthree$ and $\var$ is not bound in $\envtwo$, and $\env(\var) \defeq \bot$ otherwise.

\myparagraph{Split Contexts and Plugging} Micro-step \cbn and \cbv evaluation have easy \emph{split} presentations, because their evaluation contexts may be seen as term contexts, using the environment only for look up, see for instance \cite{DBLP:conf/aplas/AccattoliB17}. \cbneed evaluation contexts, instead, need to enter into the environment. Typically in a program such as $(\var\vartwo,\esub\var\varthree\esub\varthree\tm)$, whose head variable $\var$ has been found, \cbneed evaluation has to enter inside $\esub\var\cdot$, finding another (hereditary) head variable $\varthree$, and in turn enter inside $\esub\varthree\cdot$ and evaluate $\tm$. The subtlety is that the evaluation of $\tm$ can create new ESs, which should be added to the program without breaking its structure, that is, outside the ES which is being evaluated ($\esub\varthree\cdot$ in the example). The trick to make it work, is using an unusual notion of context plugging. Before defining evaluation contexts we simply discuss split contexts, which are used throughout the paper.
\[\small\begin{array}{rrcl@{\hspace{.5cm}} rrcl}
	\textsc{Term ctxs} &\ctx, \ctxtwo & \grameq & \ctxhole \mid \ctx\tm \mid \tm \ctx 
&	\textsc{Env. ctxs} & \herenv, \herenvtwo & \grameq & \emptyenv \mid \env \, \esub{\var}{\ctx} \mid \herenv \, \esub{\var}{\tmtwo} \\	
  	\textsc{Prog. ctxs} &\pctx,\pctxtwo & \grameq & \progentry{\ctx}{\env} \mid \progentry{\tm}{\herenv}
\end{array}\]
Plugging of a term into a context is defined as expected. Plugging of a program into a context, the tricky bit, requires an auxiliary notation $	\appES\prog{\esub\var\tm}$ for the appending of an ES $\esub\var\tm$ to the end of the environment of a program $\prog$:
\[\begin{array}{rcl|rcl}
		\multicolumn{3}{c}{\textsc{Appending ES}} 
		&
		\multicolumn{3}{c}{\textsc{Plugging of programs}}
		\\
	\appES{\progentry{\tm}{\env}}{\esub{\var}{\tmtwo}}
	& \defeq  &
	\progentry{\tm}{\env \esub{\var}{\tmtwo}} 
	&
	\progentry{\ctx}{\env} \ctxholeentryp{\tm}{\envtwo} 
	& \defeq  &
	\progentry{\ctxp{\tm}}{\envtwo \env}
	\\
	\appES{\progentry{\ctx}{\env}}{\esub{\var}{\tmtwo}}
	& \defeq &
	\progentry{\ctx }{\env \esub{\var}{\tmtwo}} 
	&
	\progentry{\tmtwo}{\env \esub{\var}{\ctx}} \ctxholeentryp{\tm}{\envtwo} 
	& \defeq  &
	\progentry{\tmtwo}{\env \esub{\var}{\ctxp{\tm}} \envtwo} 
	\\
	\appES{\progentry{\tm}{\herenv}}{\esub{\var}{\tmtwo}} 
	& \defeq &
	\progentry{\tm}{\herenv \esub{\var}{\tmtwo}}	
	&
	\progentry{\tmtwo}{\herenv \esub{\var}{\tmthree}} \ctxholeentryp{\tm}{\env} 
	& \defeq  &
	\appES{\progentry{\tmtwo}{\herenv} \ctxholeentryp{\tm}{\env}}{\esub{\var}{\tmthree}} \\
\end{array}\]
For instance, $\progentry{\var\vartwo}{\esub\var\tm\esub\vartwo\ctxhole\esub\varthree\tmtwo}\ctxholeentryp{\tmthree}{\esub{\var'}{\tm'}} = \progentry{\var\vartwo}{\esub\var\tm\esub\vartwo\tmthree\esub{\var'}{\tm'}\esub\varthree\tmtwo}$. The look-up operation is extended to environment and program contexts as expected.

Next, we define the \cbneed evaluation contexts in the split approach.
\[\begin{array}{rrclll}
	\textsc{Head contexts}    &\hctx, \hctxtwo & \grameq & \ctxhole \mid \hctx \tm\\
	\textsc{Hereditary head contexts} &\hhctx,\hhctxtwo & \grameq & \progentry{\hctx}{\env} \mid \appES{\hhctx}{\esub{\var}{\tm}} \mid \appES{\hhctxp{\var}}{\esub{\var}{\hctx}}
\end{array}\]
The third production for $\hhctx$ is what allows evaluation to be iterated inside ES, seeing for instance $(\var\vartwo,\esub\var\varthree\esub\varthree\ctxhole)$ as a hereditary head context of $(\var\vartwo,\esub\var\varthree\esub\varthree\tm)$ (by applying the production twice, the first time obtaining $(\var\vartwo,\esub\var\ctxhole)$).

\myparagraph{Split Evaluation Rules} In contrast to most $\l$-calculi, we do not define the root cases of the rules and then extend them by a closure by evaluation contexts. We rather define them directly at the global level. Adopting global rules is not mandatory, and yet it  shall be convenient for dealing with the useful calculus---we use them here too for uniformity.
\begin{center}
	$\begin{array}{rrcl@{\hspace{.5cm}} l}
		\multicolumn{5}{c}{\textsc{\closedcbneed evaluation rules}}
		\\
		\textsc{Multiplicative} 
		&
		\hhctxp{(\la{\var}{\tm}) \tmtwo}
		& \tom &
		\hhctxp{\tm,\esub{\var}{\tmtwo}}
		& \\
		\textsc{Exponential} 
		&
		\hhctxp{\var}
		& \toe &
		\hhctxp{\rename{\val}}
		&
		\text{ if $\hhctx(\var) = \val$}
	\end{array}$
\end{center}
The names of the rules are due to the link between the LSC and  linear logic, see Accattoli \cite{DBLP:conf/ictac/Accattoli18}.
Note that we use both plugging of terms and programs, to ease up notations. An example of how rule $\tom$ exploits the unusual notion of plugging is $\progentry{\var\tm}{\esub\var{(\la\vartwo \tmtwo) \tmthree \tm'}\esub\varthree {\tmtwo'}} \tom \progentry{\var\tm}{\esub\var{\tmtwo\tm'} \esub\vartwo\tmthree \esub\varthree {\tmtwo'}}$. As it is standard in the study of \cbneed, garbage collection is simply ignored, because it is postponable at the micro-step level.
Normal forms of \closedcbneed are programs of the form $(\val,\env)$, which are sometimes called \emph{answers}.

%
%
%
%



\section{Open Call-by-Need}
\label{sect:open-cbneed}

We now shift to \opencbneed, an evaluation strategy extending \closedcbneed and allowing reduction to act on possibly \emph{open} programs. Roughly, the strategy iterates \cbneed evaluation on the arguments of the head variable, when the normal form of ordinary \cbneed evaluation is not an abstraction, which can happen when terms are not necessarily closed. Various aspects of \closedcbneed become subtler in Open \cbneed, namely the definition of evaluation contexts and the structure of normal forms, together with the new notion of needed variables. Essentially, we are giving an alternative presentation of the open fragment of Balabonski et al.'s Strong Call-by-Need, with which we compare at the end of the section.

\myparagraph{Some Motivating Examples} We show a few examples of the rewriting relation that we aim at defining, as to guide the reader through the technical aspects. We want reduction to take place in arguments, after a (hereditary) head variable has been found, having \eg:
\begin{center}$
			\progentry{\var((\la{\varthree}{\varthree}) \Id)}{\esub{\vartwo}{\tm}} \tom \progentry{\var\varthree}{\esub{\varthree}{\Id}\esub{\vartwo}{\tm}}
			\ \ \ \mbox{ and }\ \ \ 
			\progentry{\var \varthree}{\esub{\varthree}{\Id}\esub{\vartwo}{\tm}} \toe \progentry{\var \Id}{\esub{\varthree}{\Id}\esub{\vartwo}{\tm}}
			$\end{center}
For appropriate generalizations of $\tom$ and $\toe$. Of course, we retain and extend to arguments the hereditary character of the reduction rules, therefore having also steps such as:
			\begin{center}\begin{tabular}{llllll}$
			\progentry{\vartwo\var }{\esub{\var}{\vartwo((\la{\varthree}{\varthree}) \Id)}} \tom \progentry{\vartwo\var}{\esub{\var}{\vartwo \varthree}\esub{\varthree}{\Id}}$,\   and \  
			$\progentry{\vartwo\var}{\esub{\var}{\vartwo \varthree}\esub{\varthree}{\Id}} \toe \progentry{\vartwo\var}{\esub{\var}{\vartwo \rename{\Id}}\esub{\varthree}{\Id}}$
			\end{tabular}
		\end{center}
While the intended behavior is---we hope---clear, specifying these steps via evaluation contexts requires some care and a few definitions. Essentially, we need to understand when evaluation can pass to the next argument, and thus characterize when terms are normal. This is easy for terms but becomes tricky for programs.

\myparagraph{Evaluation Places and Needed Variables} The grammars of the language are the same as for split \closedcbneed, but defining the open evaluation contexts is quite subtler. In \closedcbneed there is only one place of the term where evaluation can take place, the hereditary head context $\hhctx$. In the open setting the situation is more general: there is one \emph{active} evaluation place \emph{plus} potentially many \emph{passive} ones, which are those places where evaluation already passed and ended. On some of these passive places, evaluation ended on a free variable (occurrence). We refer to these free variables as \emph{needed} (definition below\footnote{Needed variables are intended to be considered only for normal terms (or normal programs, or normal parts of a context), and yet  the definition is given here for every term (in particular every applications, instead of  only inert applications $\itm\fire$). The reason for our lax definition is that the technical development requires at times to consider the needed variables of a term that is not yet known to be normal. The lax definition goes against the \emph{needed} intuition, as one of the reviewers understandably complained about, suggesting to call these variables \emph{frozen}, following Balabonski et al. \cite{DBLP:journals/pacmpl/BalabonskiBBK17}. We preferred to keep \emph{needed} because they are similar but different from the frozen variables in \cite{DBLP:journals/pacmpl/BalabonskiBBK17}, see the end of this section.}), as they shall end up in the normal form, given that at least one of their occurrences has already been evaluated and cannot be erased. For instance in $\prog \defeq (\var(\vartwo\Id), \esub\varthree\var\esub\vartwo\Id)$ the active place is $\vartwo$, the first occurrence of $\var$ is a needed occurrence, while the second one is not. \begin{center}
\begin{tabular}{cc}
$\arraycolsep=2.5pt\begin{array}{rcl}
	\multicolumn{3}{c}{\textsc{Needed vars for terms}}
	\\
	\nv{\var} & \defeq & \set{\var} 
	\\
	\nv{\la{\var}{\tm}} & \defeq & \emptyset 
	\\
	\nv{\tm \tmtwo} & \defeq & \nv{\tm} \cup \nv{\tmtwo} 
\end{array}$
&
	 $\arraycolsep=2.5pt\begin{array}{rcl}
\multicolumn{3}{c}{\textsc{Needed variables for programs}} \\
	\nvprogentry{\tm}{\emptyenv} & \defeq & \nv{\tm}  \\

	\nvprogentry{\tm}{\env \esub{\var}{\tmtwo}} & \defeq & 
		\begin{cases}
			\nvprogentry{\tm}{\env} & \var \notin \nvprogentry{\tm}{\env} 
			\\
			(\nvprogentry{\tm}{\env} \setminus \set{\var}) \cup \nv{\tmtwo} & \var \in \nvprogentry{\tm}{\env} 
		\end{cases}
		\end{array}$
\end{tabular}
\end{center}
The difficulty in defining \opencbneed is in the inductive definition of both normal forms and evaluation contexts. The problem is that extending a term or a context with a new ES may re-activate a passive evaluation place, if the ES binds a needed variable occurrence. For instance, appending $\esub\var\delta$ to $\prog$ above would reactivate the needed occurrence of $\var$.
 
\myparagraph{Normal Terms} In \opencbneed normal forms are not simply answers (\ie abstractions together with an environment), as free variables induce a richer structure. We shall later characterize the subtle inductive structure of normal programs. For now, we need predicates (that shall be later shown) characterizing normal \emph{terms}, as they are used to define evaluation contexts. The definition and the terminology are borrowed from \ocbv \cite{DBLP:conf/lics/AccattoliC15,DBLP:conf/aplas/AccattoliG16}, where normal terms are called \emph{fireballs}  and are defined by mutual induction with \emph{inert terms}:
\begin{center}$\begin{array}{rrcl @{\hspace{.5cm}} @{\hspace{.5cm}} rrcl}
    	\textsc{Values} & \val, \valtwo & \grameq & \la{\var}{\tm} 
	&
  	\textsc{Inert terms} & \itm, \itmtwo & \grameq & \var\in\Var \mid \itm \fire \\
	 	\textsc{Fireballs} & \fire, \firetwo & \grameq & \val \mid \itm
		&
	\textsc{Non-var inert terms} & \itmplus  & \grameq & \itm \fire
\end{array}$\end{center}
Later on, we shall often need to refer to inert terms that are not variables, which is why we introduce now a dedicated notation. We shall sometimes write $\inertpr{\tm}$ (resp., $\abspr{\tm}$) to express that $\tm$ is an inert term  (resp., an abstraction). 
\begin{figure}[t]
	\begin{center}\begin{tabular}{c|c}
	$\arraycolsep=2.5pt\begin{array}{rcl}
		\multicolumn{3}{c}{\textsc{Needed vars for term ctxs}}\\
			\nv{\ctxhole} & \defeq & \emptyset \\
			\nv{\hctx \tm} & \defeq & \nv{\hctx} \\
			\nv{\itm \hctx} & \defeq & \nv{\itm} \cup \nv{\hctx} \\
		\end{array}$
		&
		$\begin{array}{c@{\hspace{.5cm}} c}
		\multicolumn{2}{c}{\textsc{Open evaluation contexts and their needed vars}}\\
			\infer
				[\ruleEvcAx]
				{\progentry{\hctx}{\emptyenv} \in \evctxp{\nv{\hctx}}}
				{}
		
			&
		
			\infer
				[\ruleEvcInert]
				{\appES{\pctx}{\esub{\var}{\itm}} \in \evctxp{(\varset \setminus \set{\var}) \cup \nv{\itm}}}
				{\pctx \in \evctxp{\varset}
				\quad
				\var \in \varset
				}
			
			\\\\
		
			\infer
				[\ruleEvcGc]
				{\appES{\pctx}{\esub{\var}{\tm}} \in \evctxp{\varset}}
				{\pctx \in \evctxp{\varset}
				\quad
				\var \notin \varset}
			
			&
			
			\infer
				[\ruleEvcHer]
				{\appES{\pctxp{\var}}{\esub{\var}{\hctx}} \in \evctxp{\varset \cup \nv{\hctx}}}
				{\pctx \in \evctxp{\varset}
				\quad			
				\var \notin \varset}
		
		\end{array}$
		\end{tabular}
	\end{center}
	\vspace*{-3mm}
		\caption{Needed variables for term contexts and the derivation rules for open evaluation contexts.}
	\label{fig:derivation-rules-for-open-cbneed-evaluation-contexts}
\end{figure}

\myparagraph{Evaluation Contexts} Open evaluation contexts cannot be defined with a grammar, as for the closed case, because they are defined by mutual induction with their own set of needed variables, see the right part of \reffig{derivation-rules-for-open-cbneed-evaluation-contexts}. The notation  $\pctx \in \evctxp{\varset}$ means that $\pctx$ is an open evaluation context of needed variables $\varset$. We assume that $\var \notin \dom{\pctx}$ in rules $\ruleEvcGc$, $\ruleEvcInert$ and $\ruleEvcHer$, in accordance with \barendregtconv. The base case $\ruleEvcAx$ requires the notion of needed variables for term contexts, which is on the left side of \reffig{derivation-rules-for-open-cbneed-evaluation-contexts}.

Rule $\ruleEvcAx$ simply coerces term contexts to program contexts. The production $\appES{\pctx}{\esub{\var}{\tm}}$ for the closed case here splits into the two rules $\ruleEvcGc$ and $\ruleEvcInert$. This is relative to needed variables: one can append the ES $\esub{\var}{\tm}$ only if $\var$ is not needed ($\ruleEvcGc$) or, when $\var$ is needed, if the content $\tm$ of the ES is inert ($\ruleEvcInert$), as to avoid re-activation of a passive evaluation place on $\var$. Rule $\ruleEvcHer$ is the open version of the production $\appES{\pctxp{\var}}{\esub{\var}{\hctx}}$, with the needed variables constraint to prevent re-activations.
 Examples: $(\var\vartwo,\esub\vartwo\ctxhole)$ and $(\var\vartwo,\esub\vartwo\varthree\esub\varthree\ctxhole)$ for $\ruleEvcHer$, $(\var\vartwo,\esub\vartwo\ctxhole\esub\var{\varthree\varthree})$ for $\ruleEvcInert$, $(\var\vartwo,\esub\vartwo\ctxhole\esub\varthree{\varthree\varthree})$ for $\ruleEvcInert$.

\begin{toappendix}
\begin{lemma}[Unique parameterization of open evaluation contexts]
\hypertarget{targ-st:Unique_derivation_parameterization_of_open_evaluation_contexts}{}
\hfill

Let $\pctx \in \evctxp{\varset}$ and $\pctx \in \evctxp{\varsettwo}$. Then $\varset = \varsettwo$.

\label{l:Unique_derivation_parameterization_of_open_evaluation_contexts}
\end{lemma}
\end{toappendix}


\myparagraphnoskip{Open Evaluation Rules} The definition of the evaluation rules mimics exactly the one for the split closed case. Given an \opencbneed evaluation context $\pctx \in \evctx_{\varset}$, we have:

\begin{center}
	$\begin{array}{rrcl@{\hspace{.5cm}} l}
		\multicolumn{5}{c}{\textsc{\opencbneed evaluation rules}}
		\\
		\textsc{Open multiplicative} &
		\pctxp{(\la{\var}{\tm}) \tmtwo}
		& \tonm &
		\pctxentryp{\tm}{\esub{\var}{\tmtwo}}
		& \\
	
		\textsc{Open exponential} &
		\pctxp{\var}
		& \tone &
		\pctxp{\rename{\val}}
		&
		\text{ if $\pctx(\var) = \val$ \erase{and $\var\notin\varset$}}

		\\
	\end{array}$
	
\end{center}

We shall say that $\prog$ reduces to $\progtwo$ in the \opencbneed evaluation strategy, and write $\prog \tonnd \progtwo$, whenever $\prog \tonm \progtwo$ or $\prog \tone \progtwo$. 

\begin{toappendix}
\begin{proposition}[Determinism of {\opencbneednsp}]
\hypertarget{targ-st:Determinism_of_opencbneed}{}
Reduction $\toond$ is deterministic.


\label{c:Determinism_of_opencbneed}
\end{proposition}
\end{toappendix}

\myparagraphnoskip{Normal Programs} Normal programs mimic normal terms and are of two kinds, inert or abstractions. The definition however now depends on needed variables and cannot be given as a simple grammar. The two predicates $\inert$ and $\abs$ are defined in \reffig{useful-predicates-for-terms-and-programs}. Finally, predicate $\onormalSym$ is defined as the union of $\inertPredSym$ and $\absPredSym$, that is, $\onormalpr{\prog}$ if $\inertpr{\prog}$ or $\abspr{\prog}$. The intended meaning is that it characterizes programs in {\opencbneednsp}-normal form. 
\begin{figure}[t]
	\begin{center}
		\begin{tabular}{c}
		$\begin{array}{r @{\hspace{1.5cm}} c @{\hspace{1.5cm}} l}
			
			\infer
				[\ruleInertAx]
				{\inertentrypr{\itm}{\emptyenv}}
				{}
			&
			\infer
				[\ruleInertInert]
				{\inertpr{\appES{\prog}{\esub{\var}{\itm}}}}
				{\inertpr{\prog}
				\quad
				\var \in \nv{\prog}}
			&
			\infer
				[\ruleInertGc]
				{\inertpr{\appES{\prog}{\esub{\var}{\tm}}}}
				{\inertpr{\prog}
				\quad
				\var \notin \nv{\prog}} 
			\\[5pt]
			\hline
			\\[-5pt]
			\multicolumn{3}{c}{
				\begin{array}{c @{\hspace{1.5cm}} c}
					\infer
						[\ruleAbsAx]
						{\absentrypr{\val}{\emptyenv}}
						{}
					&
					\infer
						[\ruleAbsGc]
						{\abspr{\appES{\prog}{\esub{\var}{\tm}}}}
						{\abspr{\prog}}
				\end{array}
			}
		\end{array}$
		\end{tabular}
	\end{center}
	\vspace*{-5mm}
	\caption{Predicates for \opencbneed normal programs.}

	\label{fig:useful-predicates-for-terms-and-programs}

\end{figure}

\begin{toappendix}
\begin{proposition}[Syntactic characterization of {\opencbneednsp}-normal forms]
\hypertarget{targ-st:Syntactic_characterization_of_opencbneed-normal_forms}{}

Let $\prog$ be a program. Then $\prog$ is in $\toond$-normal form if and only if $\onormalpr{\prog}$.

\label{prop:Syntactic_characterization_of_opencbneed-normal_forms}
\end{proposition}
\end{toappendix}

The proofs of Prop. \ref{c:Determinism_of_opencbneed} and \ref{prop:Syntactic_characterization_of_opencbneed-normal_forms} (in the Appendix) are  subtler and longer than one might expect, because of the fact that evaluation contexts and needed variables are mutually defined. 


\myparagraph{Relationship with Balabonski et al} With respect to the definition of Strong \cbneed in \cite{DBLP:journals/pacmpl/BalabonskiBBK17}, we follow essentially the same approach up to two differences, not counting the  obvious fact that we are open and not strong. First, we use a split calculus, while they do not, because they do not study useful sharing. 

Second, they have a similar but different parametrization of evaluation contexts. They are more liberal, as their sets of \emph{frozen variables} used as parameters are supersets of our needed variables, but they also parametrize reduction steps, which we avoid. Our 'tighter' choice is related to the fine study of intersection types for \opencbneed, which can be found in the Leberle's PhD thesis \cite{leberlePhD}, and it is also essential for the refinement required by the useful extension of \refsect{useful-open-cbneed}. In \cite{DBLP:conf/fscd/BalabonskiLM21}, a reformulation of \cite{DBLP:journals/pacmpl/BalabonskiBBK17} using a deductive system (parametrized by frozen variables) rather than evaluation contexts is used---it could also be used here.




\section{Useful Open Call-by-Need}
\label{sect:useful-open-cbneed}
Roughly, useful sharing is an optimization of micro-step substitutions, that is, of exponential steps. The idea is that there are substitution steps that are useful to create $\beta$/multiplicative redexes and steps that are useless. For instance (the underline stresses the created $\beta$-redex):
\[\begin{array}{c@{\hspace{.5cm}} @{\hspace{.5cm}} @{\hspace{.5cm}} ccc}
\textsc{Example of useful step}
&
\textsc{Example of useless step}
\\
(\var\vartwo,\esub\var\Id) \tone (\underline{\Id\vartwo},\esub\var\Id)
&
(\var\vartwo,\esub\vartwo\Id) \tone (\var\Id,\esub\vartwo\Id)
\end{array}\]
The main idea is that useful steps replace applied variable occurrences, while useless steps replace unapplied occurrences. The definition of the useful calculus then shall refine the open one by replacing the set of needed variables with \emph{two} sets, one for applied and one for unapplied variable occurrences. Note a subtlety: variables can have both applied and unapplied needed occurrences, as $\var$ in $\var\var$. Therefore, usefulness is a concept that can be properly expressed only when considering replacements of single variable occurrences.

Usefulness unfortunately is not so simple. Consider the following  step replacing $\varthree$ with $\Id$:
\begin{equation}
(\var\vartwo,\esub\var\varthree \esub\varthree\Id) \tone (\var\vartwo,\esub\var\Id \esub\varthree\Id)
\label{eq:subtle-useful-step}
\end{equation}
Is it useful or useless? It does not create a multiplicative redex---therefore it looks useless---but without it we cannot perform the next step $(\var\vartwo,\esub\var\Id \esub\varthree\Id) \tone (\Id\vartwo,\esub\var\Id \esub\varthree\Id)$ replacing $\var$ with $\Id$ which is certainly useful---thus step \refequa{subtle-useful-step} \emph{has to be useful}.

We then have to refine the defining principle for usefulness: useful steps replace \emph{hereditarily applied} variable occurrences, that is, occurrences that are applied, or that are by themselves (\ie not in an application) and that are meant to replace a hereditarily applied occurrence. 

Handling hereditarily applied variables is specific to \cbneed, and makes defining Useful \opencbneed quite painful. The key point is the \emph{global} character of the hereditary notion, that requires checking the evaluation context leading to the variable occurrence and it is then not of a local nature. We believe that hereditarily applied variables, nonetheless, are an unavoidable ingredient of usefulness in a \cbneed scenario, and not an ad-hoc point of our study. This opinion is backed by the fact that such a convoluted mechanism is modeled very naturally at the level of intersection types, as it is shown in Leberle's PhD Thesis \cite{leberlePhD}.
\textbf{Important}: from now on, we ease the language saying \emph{applied} to mean \emph{hereditarily applied}.

\myparagraph{Applied and Unapplied Variables} We now define, for terms, programs, and term contexts, the sets of applied  and unapplied variables $\anv\cdot$ and $\unv\cdot$, that are subsets of needed variables $\nv\cdot$. We shall prove that $\nv\tm = \anv\tm \cup \unv\tm$ (\ie, the two sets cover $\nv\tm$ exactly). As already pointed out, applied and unapplied variables, however, are \emph{not} a partition of needed variables, that is, in general $\anv\tm \cap \unv\tm \neq \emptyset$ as a variable can have both applied and unapplied (needed) occurrences, as $\var$ in $\var\var$. The same holds also for programs and term contexts.



\begin{figure}[t]
\begin{center}
\begin{tabular}{c}
\textsc{Applied variables for terms and programs}
\\
$\begin{array}{l @{\hspace{.5cm}}@{\hspace{.5cm}} c @{\hspace{.5cm}}@{\hspace{.5cm}} r@{\hspace{.5cm}}c}
	\anv{\la{\var}{\tm}} \defeq \emptyset  &
	\anv{\var} \defeq \emptyset &
	\anv{\tm \tmtwo} \defeq
		\begin{cases}
			\set{\var} \cup \anv{\tmtwo} & \tm = \var \in \Var\\
			\anv{\tm} \cup \anv{\tmtwo} & \tm \notin \Var \\
		\end{cases}
		&
		\anventryp{\tm}{\emptyenv} \defeq \anv{\tm}
\end{array}$
\\
{\small$\anventryp{\tm}{\env \esub{\var}{\tmtwo}} \defeq 
		\begin{cases}
			\anventryp{\tm}{\env} & \var \notin \nventryp{\tm}{\env}, \\
			(\anventryp{\tm}{\env} {\setminus} \set{\var}) \cup \anv{\tmtwo} & \var \in \nventryp{\tm}{\env} \, \land (\var \notin \anventryp{\tm}{\env} \lor \tmtwo \notin \Var),\\
			(\anventryp{\tm}{\env} {\setminus} \set{\var}) \cup \set{\vartwo} & \var \in \nventryp{\tm}{\env} \, \land \var \in \anventryp{\tm}{\env} \, \land \tmtwo = \vartwo \in \Var \\
			
		\end{cases}$}
\\\hline
\textsc{Unapplied variables for terms and programs}\\

$\begin{array}{l @{\hspace{.5cm}}@{\hspace{.5cm}} c @{\hspace{.5cm}}@{\hspace{.5cm}} r@{\hspace{.5cm}}c}
	\unv{\la{\var}{\tm}} \defeq \emptyset &
 	\unv{\var} \defeq \set{\var} &
	\unv{\tm \tmtwo} \defeq 
		\begin{cases}
			\unv{\tmtwo} & \tm \in \Var \\
			\unv{\tm} \cup \unv{\tmtwo} & \tm \notin \Var \\
		\end{cases} 
		&
		\unventryp{\tm}{\emptyenv} \defeq \unv{\tm}
\end{array}$
\\
{\small$\unventryp{\tm}{\env \esub{\var}{\tmtwo}} \defeq 
		\begin{cases}
			\unventryp{\tm}{\env} & \var \notin \unventryp{\tm}{\env} \, \land (\var \notin \nventryp{\tm}{\env} \, \lor  \tmtwo = \vartwo \in \Var) \\
			(\unventryp{\tm}{\env} {\setminus} \set{\var}) \cup \unv{\tmtwo} & \var \in \unventryp{\tm}{\env} \, \lor (\var \in \nventryp{\tm}{\env} \, \land \tmtwo \notin \Var) \\
		\end{cases}$
		}
		\\\hline
		\begin{tabular}{c@{\hspace{.5cm}} | @{\hspace{.5cm}} c}
	\textsc{Applied vars of term contexts}
	&
	\textsc{Unapplied vars of term contexts}
	\\	
		$\begin{array}{rcl}
				\anv{\ctxhole} & \defeq & \emptyset \\
			\anv{\hctx \tm} & \defeq & \anv{\hctx} \\
			\anv{\itm \hctx} & \defeq & \anv{\itm} \cup \anv{\hctx} \\
		\end{array}$
		&
				$\begin{array}{rcl}
						
			\unv{\ctxhole} & \defeq & \emptyset \\
			\unv{\hctx \tm} & \defeq & \unv{\hctx} \\
			\unv{\itm \hctx} & \defeq & \unv{\itm} \cup \unv{\hctx} \\
		\end{array}$
		\end{tabular}

\end{tabular}
\end{center}
\vspace*{-4mm}
\caption{Applied and unapplied variables for terms, programs, and term contexts.}
\label{fig:ap-and-unapp-vars}
\end{figure}
The set of applied variables of terms, programs, and term contexts are defined in \reffig{ap-and-unapp-vars}---explanations follow. Having in mind that we want to define $\anv{\prog}$ in such a way that it satisfies $\anv{\prog} \subseteq \nv{\prog}$, note that condition $\var \in \nventryp{\tm}{\env} \land \var \in \anventryp{\tm}{\env} \land \tmtwo = \vartwo \in \Var$ in the definition of $\anventryp{\tm}{\env \esub{\var}{\tmtwo}}$ would more simply be $\var \in \anventryp{\tm}{\env} \land \tmtwo = \vartwo \in \Var$. However, we have not proved yet that $\anv\prog \subseteq \nv\prog$, which is why the definition is given in this more general form.

We give some examples. As expected, $\vartwo$ is an applied variable of $\vartwo\, \varthree$ and $\varthree\, (\vartwo\, \varthree)$. It is also applied in $\prog \defeq \progentry{\varthree \, \var}{\esub{\var}{\vartwo \, \varthree}}$, even if $\var$ is not applied in $\progentry{\varthree \, \var}{\emptyenv}$. Thus,  {\usefulopencbneednsp} evaluation  shall be defined as to include exponential steps such as  $\progentry{\varthree \, \var}{\esub{\var}{\vartwo \, \varthree} \esub{\varthree}{\val}} \tone \progentry{\varthree \, \var}{\esub{\var}{\val \, \varthree} \esub{\vartwo}{\val}}$, which are useful. Note that $\vartwo$ is not applied in $\progentry{\var}{\esub\varthree{\vartwo\var}}$, because applied variables have to be needed variables, and $\vartwo$ is not needed.
Another example:
if $\prog \defeq \progentry{\var \tm}{\esub{\var}{\vartwo}}$, then $\vartwo \in \anv{\prog}$ (and also $\varthree \in \anv{\progentry{\var \tm}{\esub{\var}{\vartwo}\esub{\vartwo}{\varthree}}}$). \usefulopencbneed, then, shall retain the following two exponential steps of the open case, since the sequence is supposed to continue with a $\tom$ step, contracting the redex given by $\val \tm$:
			\begin{center}$
			\progentry{\var \tm}{\esub{\var}{\vartwo} \esub{\vartwo}{\val}} \tone \progentry{\var \tm}{\esub{\var}{\val} \esub{\vartwo}{\val}} \tone \progentry{\val \tm}{\esub{\var}{\val} \esub{\vartwo}{\val}}
			$\end{center}
		
The set of unapplied variables of terms, programs, and term contexts are defined in \reffig{ap-and-unapp-vars}.
Once again, in the second clause defining $\unventryp{\tm}{\env \esub{\var}{\tmtwo}}$ the side condition $\var \in \nventryp{\tm}{\env}$ can be replaced  by $\var \in \anventryp{\tm}{\env}$, after \reflemma{Unapplied_applied_and_needed_variables} below is proved. 

We give some examples. A consequence of the definition is that, as for applied variables, $\vartwo$ is not unapplied in $\progentry{\var\var}{\esub\varthree{\var\vartwo}}$ because it is not needed. As it is probably expected, $\vartwo$ is unapplied in $\progentry{\varthree\var}{\esub\varthree{\var\vartwo}}$, even if ${\var\vartwo}$ is meant to replace $\varthree$ which is applied in $\varthree\var$.
Perhaps counter-intuitively, instead, our definitions imply  both $\vartwo \in \anv\prog$ and $\vartwo \in \unv\prog$ for $\prog \defeq \progentry{\var\var}{\esub{\var}{\vartwo}}$, that is,  the unique occurrence of $\vartwo$ is both applied and unapplied in $\prog$\footnote{This fact is in accordance with the companion intersection type study in Leberle's PhD thesis \cite{leberlePhD} mentioned in the introduction: in spite of $\vartwo$ having only one syntactic occurrence in $\prog$, it is needed twice, and so intersection types derivations of $\prog$ do type $\vartwo$ twice.}.

	\begin{toappendix}
\begin{lemma}[Unapplied and applied cover needed variables]
\hypertarget{targ-st:Unapplied_applied_and_needed_variables}{}
\hfill

\begin{enumerate}
	\item
	\label{p:Unapplied_applied_and_needed_variables-one}
	\emph{Terms}: $\nv{\tm} = \unv{\tm} \cup \anv{\tm}$ for every term $\tm$.
	
	\item 
	\label{p:Unapplied_applied_and_needed_variables-two}
	\emph{Programs}: $\nv{\prog} = \unv{\prog} \cup \anv{\prog}$ for every program $\prog$.
	\item \label{l:Term_contexts_Unapplied_applied_and_needed_variables}
			\emph{Term contexts}: 	$\nv{\hctx} = \unv{\hctx} \cup \anv{\hctx}$, for every term context $\hctx$.
\end{enumerate}
\label{l:Unapplied_applied_and_needed_variables}
\end{lemma}
\end{toappendix}

Finally, the derived concept of useless variable shall also be used.
\begin{definition}[Useless variables]
\label{def:useless-variables}

Given a term $\tm$, we define the set of \useless variables as  $\uselesspr{\tm} \defeq \unv{\tm} {\setminus} \anv{\tm}$. The set of \useless variables of a program $\prog$ is defined analogously.
\end{definition}

Useless variables are crucial in differentiating Useful Open \cbneed from \opencbneed. We shall prove that if $\prog$ is a useful open normal form and $\var \in \uselesspr{\prog}$, then $\appES{\prog}{\esub{\var}{\val}}$ is also a useful open normal form (while it is not a open normal form). The notion of useless variables is intuitively simple but technically complex. Some examples. First, note that $\uselessentrypr{\var \, \var}{\emptyenv} = \emptyset$. The example can be extended to a hereditary setting, noting that $\uselessentrypr{\vartwo}{\esub{\vartwo}{\var \, \var}} = \emptyset$. However, the reasoning takes into account only needed occurrences, that is, note that $\var \in \uselessentrypr{\varthree \, \var}{\esub{\vartwo}{\var \, \var}}$, as the occurrence of $\var$ that is applied to an argument is not needed.

	\begin{figure}[t]
	\small
		\begin{center}

			$\begin{array}{c@{\hspace{.8cm}} c}
				\infer
					[\ruleUseAxMult]
					{\progentry{\hctx}{\emptyenv} \in \evctxUsep{\unv{\hctx}}{\anv{\hctx}}}
					{}
				
				&
				
				\infer
					[\ruleUseVarMult]
					{\appES{\pctx}{\esub{\var}{\vartwo}} \in \evctxUsep{\updp{\uvarset}{\var}{\vartwo}}{\updp{\avarset}{\var}{\vartwo}}}
					{\pctx \in \evctxUsep{\uvarset}{\avarset}
					\quad
					\var \in (\uvarset \cup \avarset)}
					
				\\\\

				\infer
					[\ruleUseGcMult]
					{\appES{\pctx}{\esub{\var}{\tm}} \in \evctxUsep{\uvarset}{\avarset}}
					{\pctx \in \evctxUsep{\uvarset}{\avarset}
					\quad
					\var \notin (\uvarset \cup \avarset)}
					
					&
					
					\infer
					[\ruleUseNonVarMult]
					{\appES{\pctx}{\esub{\var}{\itmplus}} \in \evctxUsep{(\uvarset \setminus \set{\var}) \cup 
		\unv{\itmplus}}{(\avarset  \setminus \set{\var}) \cup \anv{\itmplus}}}
					{\pctx \in \evctxUsep{\uvarset}{\avarset}
					\quad
					\var \in (\uvarset \cup \avarset)}
					
				\\\\
					
					\infer
					[\ruleUseUselessMult]
					{\appES{\pctx}{\esub{\var}{\val}} \in \evctxUsep{\uvarset \setminus \set{\var}}{\avarset}}
					{\pctx \in \evctxUsep{\uvarset}{\avarset}
					\quad
					\var \in (\uvarset \setminus \avarset)}
				
				&
				
				\infer
					[\ruleUseHerMult]
					{\appES{\pctxp{\var}}{\esub{\var}{\hctx}} \in \evctxUsep{\uvarset \cup 
		\unv{\hctx}}{\avarset  \cup \anv{\hctx}}}
					{\pctx \in \evctxUsep{\uvarset}{\avarset}
					\quad
					\var \notin (\uvarset \cup \avarset)}
					
				\\
				
			\end{array}$
		\end{center}
		\vspace*{-4mm}
		\caption{Derivation rules for multiplicative evaluation contexts.}
		\label{fig:Derivation_rules_for_multiplicative_evaluation_contexts}
	\end{figure}

\myparagraph{Evaluation Contexts} The definition of evaluation contexts is particularly subtle in the useful case. First of all, their set $\evctxUsep{\uvarset}{\avarset}$ is indexed by \emph{two} sets of variables (rather than one as in the open case), the applied $\avarset$ and the unapplied $\uvarset$ variables of the context, defined by mutual induction with the contexts themselves. The second key point is that there are two different kinds of evaluation contexts, a permissive one for multiplicative redexes, whose set is  noted $\evctxUsep{\uvarset}{\avarset}$, and a restrictive one for exponential redexes, noted $\evctxAppp{\uvarset}{\avarset}$ and implementing the fact that the variable occurrence to be replaced has to be in an applied position. The asymmetry is unavoidable, because useful sharing concerns only exponential steps.

\myparagraph{Multiplicative Contexts} They are a refinement of the open contexts defined in \reffig{Derivation_rules_for_multiplicative_evaluation_contexts}. Their set is noted $\evctxUsep{\uvarset}{\avarset}$. The refinement is needed even if useful sharing concerns only exponential steps: a multiplicative context such as $\progentry{(\vartwo \var) \ctxhole}{\esub\var\val} \in \evctxUsep{\emptyset}{\set\vartwo}$ indeed is not an open context, because it contains a useless substitution step that in Open \cbneed would be fired before evaluating the hole. The definition of multiplicative contexts follows the one for  {\opencbneednsp} contexts ($\ruleUseAxMult$, $\ruleUseGcMult$, and $\ruleUseHerMult$ are essentially as before) \emph{except for} rule:
	\begin{center}$
	\infer
		[\ruleEvcInert]
		{\appES{\pctx}{\esub{\var}{\itm}} \in \evctxp{(\varset \setminus \set{\var}) \cup \nv{\itm}}}
		{\pctx \in \evctxp{\varset}
		\quad
		\var \in \varset}
	$\end{center}
which is now generalized into 3 rules, depending on the kind of term contained in the ES. That is, given $\pctx \in \evctxUsep{\uvarset}{\avarset}$ and $\var \in (\uvarset \cup \avarset)$, the constraints to extend $\pctx$ with an ES $\esub{\var}{\tm}$ are:
\begin{itemize}
	\item \emph{Rule $\ruleUseNonVarMult$}: there are no constraints if $\tm$ is a non-variable inert term $\itmplus$. Note that $\ruleUseNonVarMult$ and $\ruleUseGcMult$ together imply that we can \emph{always} append ESs containing inert terms to multiplicative contexts, without altering the {\usefulopencbneednsp} order of reduction.

	\item \emph{Rule $\ruleUseVarMult$}: this rule covers the case where $\tm$ is a variable $\vartwo$. It is used to handle the global applicative constraint, as in such a case, if the 
 evaluation context is $\appES\pctx{\esub\var\vartwo}$, then $\vartwo$ has to be added to the 
applied and/or unapplied variables of the context, according to the role played by $\var$ in $\pctx$, which is realized via the function $\updSym$ defined as follows:
	\begin{center}$
	\updp{S}{\var}{\vartwo} \defeq
		\begin{cases}
			S & \var \notin S \\
			(S \setminus \set{\var}) \cup \set{\vartwo} & \var \in S \\
		\end{cases}
	$\end{center}

	\item \emph{Rule $\ruleUseUselessMult$}: it covers the case where $\tm$ is a value $\val$, requiring that $\var$ is not applied, that is, $\notin \avarset$. Such an extension would have re-activated $\var$ in the plain open case, and created a (useless) exponential redex, but here it shall not be the case. Note that it means that $\appES{\pctx}{\esub{\var}{\tm}}$ is a multiplicative context \emph{only} if $\var \in (\uvarset \setminus \avarset)$, \ie if $\var$ is a \emph{useless} variable of $\pctx$.
\end{itemize}

\myparagraphnoskip{Exponential Contexts} Exponential contexts are even more involved, because they have to select only \emph{applicative} variable occurrences and the applicative constraint is of a global nature. First, we need a notion of applicative term context, where the hole is applied.

\begin{definition}[Applicative term contexts]
A term context $\hctx$ shall be called an \emph{applicative term context} if it is derived using the grammar
$\hctxapp, \hctxapptwo, \hctxappthree \grameq \ctxhole \tm \mid \hctxapp \tm \mid \itm \hctxapp$.
\end{definition}

\begin{definition}[Exponential evaluation contexts]
\label{def:exponential-evaluation-contexts}

We shall say that an evaluation context $\pctx$ is a \emph{exponential evaluation context} if it is derived with the rules in \reffig{Derivation_rules_for_exponential_evaluation_contexts}.
\end{definition}
\begin{figure}[t]
\small
		\begin{center}
			$\begin{array}{c @{\hspace{.6cm}} c}
				\infer
					[\ruleUseAxOneExp]
					{\progentry{\hctxapp}{\emptyenv} \in \evctxAppp{\unv{\hctxapp}}{\anv{\hctxapp}}}
					{}

				&
				
				\infer
					[\ruleUseAxTwoExp]
					{\appES{\pctxp{\var}}{\esub{\var}{\hctxapp}} \in \evctxAppp{(\uvarset \setminus \set{\var}) \cup \unv{\hctxapp}}{\avarset \cup \anv{\hctxapp}}}
					{\pctx \in \evctxUsep{\uvarset}{\avarset}
					\quad
					\var \notin (\uvarset \cup \avarset)}
			
				\\\\
			
				\infer
					[\ruleUseVarExp]
					{\appES{\pctx}{\esub{\var}{\vartwo}} \in \evctxAppp{\updp{\uvarset}{\var}{\vartwo}}{
	\updp{\avarset}{\var}{\vartwo}}}
					{\pctx \in \evctxAppp{\uvarset}{\avarset}
					\quad
					\var \in (\uvarset \cup \avarset)}
				
				&
			
				\infer
					[\ruleUseNonVarExp]
					{\appES{\pctx}{\esub{\var}{\itmplus}} \in \evctxAppp{(\uvarset \setminus \set{\var}) \cup 
	\unv{\itmplus}}{(\avarset \setminus \set{\var}) \cup \anv{\itmplus}}}
					{\pctx \in \evctxAppp{\uvarset}{\avarset}
					\quad
					\var \in (\uvarset \cup \avarset)}
						\end{array}$\bigskip
			
			$\begin{array}{c @{\hspace{.3cm}} c @{\hspace{.3cm}} c}
			
				\infer
					[\ruleUseGcExp]
					{\appES{\pctx}{\esub{\var}{\tm}} \in \evctxAppp{\uvarset}{\avarset}}
					{\pctx \in \evctxAppp{\uvarset}{\avarset}
					\quad
					\var \notin (\uvarset \cup \avarset)}
				
				&
				
					\infer
					[\ruleUseUselessExp]
					{\appES{\pctx}{\esub{\var}{\val}} \in \evctxAppp{\uvarset \setminus \set{\var}}{\avarset}}
					{\pctx \in \evctxAppp{\uvarset}{\avarset}
					\quad
					\var \in (\uvarset \setminus \avarset)}

&
			
				\infer
					[\ruleUseHerExp]
					{\appES{\pctxp{\var}}{\esub{\var}{\ctxhole}} \in \evctxAppp{\uvarset \setminus \set{\var}}{\avarset}}
					{\pctx \in \evctxAppp{\uvarset}{\avarset}			
					\quad
					\var \notin \avarset}

				\\
			
			\end{array}$
		\end{center}
				\vspace*{-4mm}
		\caption{Derivation rules for exponential evaluation contexts.}
		\label{fig:Derivation_rules_for_exponential_evaluation_contexts}
	\end{figure}

Applicative term contexts serve as the base case of exponential evaluation contexts, now given by two refinements of the multiplicative case:
\begin{enumerate}
	\item the base case $\ruleUseAxOneExp$ is akin to the base case $\ruleUseAxMult$ for multiplicative contexts, except that it requires the term context to be applicative. 
	\item the plugging-based rule $\ruleUseHerMult$ splits in two. A first rule $\ruleUseAxTwoExp$ which simply plugs an applicative context $\hctxapp$ into a \emph{multiplicative} evaluation context---note that this rule gives another base case for exponential evaluation contexts. A second rule $\ruleUseHerExp$ that handles the special case of the global applicative constraint. \end{enumerate}

Let us see the differences between rules $\ruleUseHerExp$ and $\ruleUseAxTwoExp$ with two examples. Their side conditions ($\var \notin (\uvarset \cup \avarset)$ and $\var \notin \avarset$) shall be explained after the examples.
	\begin{itemize}
		\item $\ruleUseHerExp$: consider the program $\prog \defeq \progentry{\var \, \tm}{\esub{\var}{\varthree}}$, where $\varthree$ is in applied position due to the global applicative constraint, as it substitutes $\var$ which is applied to $\tm$. We may derive an exponential evaluation context $\pctx$ that isolates $\varthree$, that is, such that $\pctxp{\varthree} = \prog$, as follows:
			\begin{center}$
			\infer
				[\ruleUseHerExp]
				{\pctx \defeq \appES{\left(\progentry{\ctxhole \, \tm}{\emptyenv} \ctxholep{\var}\right)}{\esub{\var}{\ctxhole}} \in \evctxAppp{\emptyset}{\emptyset}}
				{\infer
					[\ruleUseAxOneExp]
					{\progentry{\ctxhole \, \tm}{\emptyenv} \in \evctxAppp{\emptyset}{\emptyset}}
					{}
				\quad
				\var \notin \emptyset
				\quad
				\quad}
			$\end{center}
		noting that $\pctx = \appES{\left(\progentry{\ctxhole \, \tm}{\emptyenv} \ctxholep{\var}\right)}{\esub{\var}{\ctxhole}} = \progentry{\var \, \tm}{\esub{\var}{\ctxhole}}$, and so $\prog = \pctxp{\varthree}$ as expected. In this case, we are extending an exponential context, which is  already applied.
		
		\item $\ruleUseAxTwoExp$: consider $\prog \defeq \progentry{\var}{\esub{\var}{\varthree \, \tm}}$, for which $\varthree$ is an applied variable because it is itself applied, while its ES binds the needed but unapplied variable $\var$. Let us derive an exponential evaluation context $\pctx$ focusing on $\varthree$ in such a way that $\pctxp{\varthree} = \prog$ as follows:
			\begin{center}$
			\infer
				[\ruleUseAxTwoExp]
				{\pctx \defeq \appES{\left(\progentry{\ctxhole}{\emptyenv} \right) \ctxholep{\var}}{\esub{\var}{\ctxhole \, \tm}} \in \evctxAppp{\emptyset}{\emptyset}}
				{\infer
					[\ruleEvcAx]
					{\progentry{\ctxhole}{\emptyenv} \in \evctxUsep{\emptyset}{\emptyset}}
					{}
				\quad
				\var \notin (\emptyset \cup \emptyset)
				\quad
				\quad
				\quad}
			$\end{center}
		noting that $\pctx = \appES{\left(\progentry{\ctxhole}{\emptyenv} \right) \ctxholep{\var}}{\esub{\var}{\ctxhole \, \tm}} = \progentry{\var}{\esub{\var}{\ctxhole \, \tm}}$, and so $\prog = \pctxp{\varthree}$ as expected. Here the context $\progentry{\ctxhole}{\emptyenv} $ in the hypothesis is \emph{multiplicative} and it becomes \emph{exponential} once extended with an ES containing an applicative term context.
		
	\end{itemize}
Last, we explain the side conditions $\var \notin (\uvarset \cup \avarset)$ in rule $\ruleUseAxTwoExp$  and $\var \notin \avarset$ in rule $\ruleUseHerExp$. As a design choice, we want evaluation to be deterministic and the derivation of an evaluation context to be unique (the choice is not mandatory, but it seemed the natural way to proceed when we started our study). Essentially, our definitions force a total order over redexes and evaluation contexts akin to the leftmost-outermost order in \cbn, and yet different. The order is not made explicit because it does not admit a description as simple and concise as \emph{leftmost-outermost}. The difficulty is given by the jumping into the environment typical of \cbneed, which breaks the left-to-right order. Now, the side conditions in rules $\ruleUseAxTwoExp$ and $\ruleUseHerExp$ are there to force the unique derivability of the evaluation context. Roughly, they force the context $\pctx$ in the hypotheses of the rules to isolate the \emph{first} needed occurrence of $\var$ for $\ruleUseAxTwoExp$ and the \emph{first} applied occurrence of $\var$ in $\ruleUseHerExp$, where \emph{first} is relative to the (implicit) total order on contexts.

The next proposition guarantees that exponential contexts are a restriction of multiplicative contexts, that is, that the introduced variations over the deduction rules do not add contexts that were not already available before.
\begin{toappendix}
\begin{proposition}[Exponential contexts are multiplicative]
\label{prop:exp-ctx-are-mult} 
	Let $\pctx \in \evctxAppp{\uvarset}{\avarset}$. Then $\pctx \in \evctxUsep{\uvarsettwo}{\avarsettwo}$, for some $\uvarsettwo \subseteq \uvarset$ and $\avarsettwo \subseteq \avarset$.
\end{proposition}
\end{toappendix}
Let us repeat that, instead, multiplicative contexts are not in general exponential contexts, because they are not required to be applicative, for instance $\pctx \defeq \progentry{\var\ctxhole}{\esub\var{\vartwo\vartwo}}\in \evctxUsep{\set\vartwo}{\set\vartwo}$ is a multiplicative context but not an exponential one.

\myparagraph{Evaluation Rules} The reduction rules for the \usefulopencbneed strategy are:
\begin{center}
	$\begin{array}{rc l}
		\multicolumn{3}{c}{\textsc{\usefulopencbneed evaluation rules}}
		\\
		\textsc{Useful multiplicative} 
		&
		\pctxp{(\la{\var}{\tm}) \tmtwo}
		\toum
		\pctxentryp{\tm}{\esub{\var}{\tmtwo}}
		&
		\text{if $\pctx \in \evctxUsep{\uvarset}{\avarset}$}
		\\
		\textsc{Useful exponential} 
		&
		\pctxp{\var}
		\toue
		\pctxp{\rename{\val}}
		&
		\text{if $\pctx \in \evctxAppp{\uvarset}{\avarset}$ and $\pctx(\var) = \val$}

		\\
	\end{array}$
\end{center}
Moreover, we shall say that $\prog$ reduces in the \usefulopencbneed strategy to $\progtwo$, and write $\prog \tound \, \progtwo$, if $\prog \toum \progtwo$ or $\prog \toue \progtwo$.

\myparagraph{Determinism} The first property of useful evaluation that we consider is determinism, that is proved similarly for the open case, but for some further technicalities due to the existence of two sets of variables parametrizing evaluation contexts.

\begin{toappendix}
\begin{proposition}[Determinism of {\usefulopencbneednsp}]
$\tound$ is deterministic.
\label{prop:Determinism_of_usefulopencbneed}
\end{proposition}
\end{toappendix}

\emph{Usefulness.} We prove two properties ensuring that the defined reduction captures useful sharing. The first one is a \emph{correctness} property, stating that useful exponential steps are eventually followed by a multiplicative step---\emph{no useless exponential steps are possible}.

\begin{toappendix}
\begin{proposition}[Usefulness of exponential steps]
\label{prop:Usefulness-is-correct}{}
\hypertarget{targ-st:Usefulness_of_exponential_steps}{}
Let $\prog= \pctxp{\var} \toue \pctxp{\rename\val} = \progtwo$ with $\pctx \in \evctxAppp{\uvarset}{\avarset}$ and $\pctx(\var) = \val$. Then there exists a program $\progthree$ and a reduction sequence $\deriv:\progtwo \toue^k \toum \progthree$ s.t.:
	\begin{enumerate}
		\item the evaluation context of each $\toue$ steps in $\deriv$ is in $\evctxAppp{\uvarset}{\avarset}$, and the one of $\toum$ is in $\evctxUsep{\uvarset}{\avarset}$.

		\item $k\geq 0$ is the number of $\ruleUseHerExp$ rules in the derivation of $\pctx \in \evctxAppp{\uvarset}{\avarset}$.
		
	\end{enumerate}

\label{prop:Usefulness_of_exponential_steps}
\end{proposition}
\end{toappendix}

\emph{Completeness} amounts to proving that useful normal forms, when unshared, give a \opencbneed normal term. The point is that useful substitutions, if erroneously designed, might stop too soon, on programs that still contain---up to unsharing---some redexes. Completeness is developed in the following paragraph about useful normal forms.

\myparagraph{Useful Normal Forms} We are now going to develop an inductive description of useful normal forms, that is, programs that are $\tound$-normal. 
The key property guiding the characterization of a useful normal program $\prog$ is that if the sharing in $\prog$ is unfolded (by turning ES into meta-level substitutions and obtaining a term) it produces a normal term of the open system, where the unfolding operation is defined as follows:
\begin{center}
\textsc{Unfolding of programs}\ \ \ 
$\begin{array}{c@{\hspace{.5cm}} @{\hspace{.5cm}} c}
\unf{\progentry{\tm}{\emptyenv}} \defeq \tm
&
\unf{\progentry{\tm}{\env\esub\var\tmtwo}} \defeq \unf{\progentry{\tm}{\env}}\isub\var\tmtwo
\end{array}$
\end{center}
 The characterization rests on 3 predicates, defined in \reffig{Predicates_meant_to_characterize_usefulopencbneed-normal_forms}, for programs unfolding to variables ($\genVarpr{\var}{\prog}$), values ($\ufabspr{\prog}$), and non-variable inert terms ($\ufinertpr{\prog}$). Programs satisfying $\genVarpr{\var}{\prog}$ are called \emph{generalized variable} of \emph{(hereditary) head} variable $\var$---we also write $\genVarSetpr{\prog}$ to state that there exists $\var \in \Var$ such that $\genVarpr{\var}{\prog}$. Programs satisfying  $\ufabspr{\prog}$ (resp. $\ufinertpr{\prog}$), instead, are \emph{useful abstractions} (resp. \emph{useful inerts}). The predicate $\ufnormalpr{\prog}$ holds for programs satisfying either of the three described predicates, which we shall show being exactly programs that are normal in \usefulopencbneed.
 
 Generalized variables play a special role, because they can be extended to unfold to values or non-variable inert terms, by appending an appropriate ES to their environment with rule $\ruleAbsGenVar$ or $\ufRuleInertGenVar$. For instance, a useful normal program such as $(\var,\esub\var\vartwo)$ unfolds to a variable but its useful normal extension $(\var,\esub\var\vartwo\esub\vartwo\Id)$ unfolds to the value $\Id$, while $(\var,\esub\var\vartwo\esub\vartwo{\varthree\varthree})$ unfolds to the non-variable inert term $\varthree\varthree$. 


\begin{toappendix}
\begin{proposition}[Disjointness and unfolding of useful predicates]
\label{l:unf-usef-preds}
For every program $\prog$, \emph{at most one} of the following holds: $\genVarSetpr{\prog}$, $\ufabspr{\prog}$, or
$\ufinertpr{\prog}$. Moreover,
\begin{enumerate}
\item If $\genVarpr{\var}{\prog}$ then $\unf{\prog} = \var$.
\item If $\ufabspr{\prog}$ then $\unf{\prog}$ is a value.
\item If $\ufinertpr{\prog}$ then $\unf{\prog}$ is a non-variable inert term.
\end{enumerate}
\end{proposition}
\end{toappendix}
\begin{figure}[t]
\small
		\begin{center}
			$\begin{array}{c @{\hspace{.8cm}} c @{\hspace{.8cm}} c}
				\infer
					[\ruleGenVarAx]
					{\genVarentrypr{\var}{\var}{\emptyenv}}
					{}
					
				&

				\infer
					[\ruleGenVarHer]
					{\genVarpr{\vartwo}{\appES{\prog}{\esub{\var}{\vartwo}}}}
					{\genVarpr{\var}{\prog}}
					
				&
				
				\infer
					[\ruleGenVarGc]
					{\genVarpr{\var}{\appES{\prog}{\esub{\varthree}{\tm}}}}
					{\genVarpr{\var}{\prog}
					\quad
					\varthree \neq \var}
					
				\\[5pt]
			\hline
			\\[-5pt]
				
				\infer
					[\ruleAbsProm]
					{\ufabsentrypr{\val}{\emptyenv}}
					{}
				
				&
				
				\infer
					[\ruleAbsGenVar]
					{\ufabspr{\appES{\prog}{\esub{\var}{\val}}}}
					{\genVarpr{\var}{\prog}}
				
				&
			
				\infer
					[\ruleAbsGc]
					{\ufabspr{\appES{\prog}{\esub{\var}{\tm}}}}
					{\ufabspr{\prog}}
				
				\\[5pt]
			\hline
			\\[-5pt]
				
				\infer
					[\ufRuleInertProm]
					{\ufinertentrypr{\itmplus}{\emptyenv}}
					{}
				
				&
			
				\infer
					[\ufRuleInertGenVar]
					{\ufinertpr{\appES{\prog}{\esub{\var}{\itmplus}}}}
					{\genVarpr{\var}{\prog}}

				&
				
				\infer
					[\ufRuleInertInert]
					{\ufinertpr{\appES{\prog}{\esub{\var}{\itm}}}}
					{\ufinertpr{\prog}
					\quad
					\var \in \nv{\prog}}
				
				\\\\
				
				\multicolumn{2}{c}{
					
						\infer
							[\ufRuleInertUseless]
							{\ufinertpr{\appES{\prog}{\esub{\var}{\val}}}}
							{\ufinertpr{\prog}
							\quad
							\var \in \unv{\prog}
							\quad
							\var \notin \anv{\prog}}
			}
						&
			
						\infer
							[\ufRuleInertGc]
							{\ufinertpr{\appES{\prog}{\esub{\var}{\tm}}}}
							{\ufinertpr{\prog}
							\quad
							\var \notin \nv{\prog}}

				\\[5pt]
			\hline
			\\[-5pt]
				
				\multicolumn{3}{c}{
					\infer
						[\ufRuleNormPrograms]
						{\ufnormalpr{\prog}}
						{\ufinertpr{\prog} 
						\, \lor \,
						\ufabspr{\prog}
						\, \lor \,
						\genVarSetpr{\prog}}
				}	
					
			\end{array}$
		\end{center}
				\vspace*{-4mm}
		\caption{Predicates characterizing {\usefulopencbneednsp} normal forms.}
		\label{fig:Predicates_meant_to_characterize_usefulopencbneed-normal_forms}
	\end{figure}
	
While the concepts in the characterization of useful normal programs are relatively simple and natural, the proof of the next proposition is long and tedious, because of the complex shape of useful evaluation contexts and of their parametrization, see the Appendix.

\begin{toappendix}
\begin{proposition}[Syntactic characterization of {\usefulopencbneednsp}-normal forms]
\hypertarget{targ-st:Syntactic_characterization_of_usefulopencbneed-normal_forms}{}

Let $\prog$ be a program. Then $\prog$ is in $\tound$-normal form if and only if $\ufnormalpr{\prog}$.
\label{prop:Syntactic_characterization_of_usefulopencbneed-normal_forms}
\end{proposition}
\end{toappendix}

The characterization of useful normal forms together with the fact that they unfold to normal terms (\reflemma{unf-usef-preds}) express the \emph{completeness} of useful sharing: our useful evaluation does compute---up to unfolding---representations of \opencbneed normal terms.

\myparagraph{Complexity} A precise complexity analysis requires an abstract machine implementing the search for redexes specified by evaluation contexts. The machine---which we have developed---is left to a forthcoming paper, for lack of space. Crucially, it avoids tracing sets of applied and unapplied variables by simply using a boolean that indicates---when evaluation moves into the environment---whether the current evaluation position is hereditarily applied.

We provide a sketch of the complexity analysis. The $k$ in point 2 of \refprop{Usefulness-is-correct} allows us to bound any sequence of consecutive $\toue$ steps with the length of the environment, which---via the same reasoning used for the \cbn case by Accattoli and Dal Lago \cite{DBLP:conf/rta/AccattoliL12}---gives a quadratic bound to the whole number of $\toue$ steps in terms of $\toum$ steps. A finer amortized analysis, following Accattoli and Sacerdoti Coen \cite{DBLP:journals/iandc/AccattoliC17}, gives a linear bound. The cost of duplications in exponential steps $\toue$ is bound by the size of the initial program, because the calculus evidently has the \emph{subterm property} (\ie only subterms of the initial programs are duplicated): it duplicates values but it does not substitute nor evaluate into them, therefore the initial ones are preserved. Then, the cost of implementing a reduction sequence $\deriv: \prog \tound^{k} \progtwo$, omitting the cost of searching for redexes (itself usually realized linearly in the size $\size\prog$ of $\prog$ by abstract machines \cite{DBLP:conf/icfp/AccattoliBM14,DBLP:conf/lics/AccattoliC15}), is linear in $\size\prog$ and in the number $\sizem\deriv$ of multiplicative/$\beta$ steps in $\deriv$. 

Therefore, the number of multiplicative/$\beta$ steps in our Useful Open \cbneed calculus is a reasonable time cost model, even realizable within an efficient, bilinear overhead.




\newpage

\renewcommand{\currentpath}{\rootpath/98_Appendix}


\appendix


\renewcommand{\proofspath}{\rootpath/98_Appendix/07_opencbneed/Proofs}

\section{Proofs of \refsect{open-cbneed} (\opencbneednsp)}
\label{sect:Proofs_of_opencbneed}

\gettoappendix{l:Unique_derivation_parameterization_of_open_evaluation_contexts}

\begin{proof}
By induction on the derivation of $\pctx \in \evctxp{\varset}$, proceeding by case analysis on the last applied derivation rule. The case where $\pctx = \progentry{\hctx}{\emptyenv}$ for some term context $\hctx$ is trivial, since $\nv{\cdot}$ is a \emph{function}. The statement follows in the cases of derivation rules $\ruleEvcInert$ and $\ruleEvcGc$ from the \ih, and by definition of $\nv{\cdot}$ in the case of $\ruleEvcInert$. As an example, let us prove the case of $\ruleEvcHer$:
let $\pctx \in \evctxp{\varset}$ be derived as follows:
	$$
	\infer
		[\ruleEvcHer]
		{\appES{\pctxtwop{\var}}{\esub{\var}{\hctx}} \in \evctxp{\varsettwo \cup \nv{\hctx}}}
		{\pctxtwo \in \evctxp{\varsettwo}
		\quad
		\var \notin \varsettwo}
	$$
where $\pctx = \appES{\pctxtwop{\var}}{\esub{\var}{\hctx}}$ and $\varset = \varsettwo \cup \nv{\hctx}$. Let us assume that $\pctx = \appES{\pctxtwop{\var}}{\esub{\var}{\hctx}} \in \evctxp{\varsetp}$ for some set of variables $\varsetp$. Given the shape of $\pctx$, it could only be derived as follows
	$$
	\infer
		[\ruleEvcHer]
		{\appES{\pctxtwop{\var}}{\esub{\var}{\hctx}} \in \evctxp{\varsettwop \cup \nv{\hctx}}}
		{\pctxtwo \in \evctxp{\varsettwop}
		\quad
		\var \notin \varsettwop}
	$$
for some $\varsettwop$ such that $\varsetp = \varsettwop \cup \nv{\hctx}$. The statement then follows by application of \ih on $\pctxtwo \in \evctxp{\varsettwop}$, yielding that $\varsettwop = \varsettwo$.\qedhere

\end{proof}

\subsection{Determinism}

To prove determinism of {\opencbneednsp}, we need to prove that for every program $\prog$, if $\prog = \pctx_{1} \ctxholep{\tm_{1}} = \pctx_{2} \ctxholep{\tm_{2}}$ for some $\pctx_{1} \in \evctxp{\varset_{1}}$, $\pctx_{2} \in \evctxp{\varset_{2}}$, terms $\tm_{1}, \tm_{2} $, and $\prog$ $\toond$-reduces rewriting $\tm_{1}$ and $\tm_{2}$, then $\pctx_{1} = \pctx_{2}$ and $\tm_{1} = \tm_{2}$.

While multiplicative redexes in {\opencbneednsp} are simply given by the $\beta$-redex in the 	 plugged into the open evaluation context, note that exponential redexes are instead defined in terms of the variable occurrence to be substituted \emph{and} the \ES in the environment of the open evaluation context that binds that variable. Take, for example, $\progentry{\var}{\esub{\var}{\Id}} = \appES{(\progentry{\ctxhole}{\emptyenv} \ctxholep{\var})}{\esub{\var}{\Id}}$, which is an exponential redex, but whose base case $\progentry{\var}{\emptyenv} = \progentry{\ctxhole}{\emptyenv} \ctxholep{\var}$ is not.

Therefore, the first thing to do to prove determinism of {\opencbneednsp} consists in generalizing the notion of multiplicative and exponential redexes, via what are called \emph{reduction places}, thus devising the kind of induction required in the proof of determinism.

\begin{definition}[Reduction places in {\opencbneednsp}]
\label{def:Reduction_places_in_opencbneed}
\hfill

Let $\tm$ be a term, $\hctx$ be a term context, and let $S \supseteq \nv{\hctx}$. We say that $\tm$ is a \emph{$S$-reduction place} of $\hctxp{\tm}$ if one of the following conditions hold:
	\begin{itemize}
		\item \emph{Multiplicative redex}: $\tm = (\la\var\tmtwo) \tmthree$;

		\item \emph{New needed variable}: $\tm = \var$, $\var\notin S$.

	\end{itemize}

Let $\tm $ be a term, $\pctx \in \evctxp{\varset}$ be an open evaluation context, and let $S \supseteq \varset$. We say that $\tm$ is a \emph{$S$-reduction place} of $\pctxp{\tm}$ if one of the following conditions hold:
	\begin{itemize}
		\item \emph{Multiplicative redex}: $\tm = (\la{\var}{\tmtwo}) \tmthree$;

		\item \emph{Exponential redex}: $\tm = \var$, $\var \in \dom{\pctx}$ and $\pctx(\var) = \val$;

		\item \emph{New needed variable}: $\tm = \var$, $\var \notin S$ and $\var \notin \dom\pctx$.

	\end{itemize}

\end{definition} 

The notion of reduction place is enough to prove that given a term $\tm$, we can single out the ``\emph{first}'' multiplicative redex---if there is any---relative to the {\opencbneednsp} evaluation strategy. We believe that this vague notion of a \emph{first} multiplicative redex in the strategy---meant to serve as a guiding \emph{intuition} to understanding determinism of {\opencbneednsp}---becomes clearer when considering terms with a total order in their multiplicative redexes. For instance, if we consider a term $(...(\var \tm_{1})... \tm_{n-1}) \tm_{n}$, then we note that {\opencbneednsp} proceeds to reduce $\progentry{(...(\var \tm_{1})... \tm_{n-1}) \tm_{n}}{\emptyenv}$ by reducing every $\tm_{i}$ in a left-to-right fashion, first reducing $\tm_{1}$ to a normal term (if necessary), then reducing $\tm_{2}$ to a normal term (again, only if necessary), and so on until finally reducing $\tm_{n}$ to a normal term. Thus, we can say that the ``first'' multiplicative redex in $\progentry{(...(\var \tm_{1})... \tm_{n-1}) \tm_{n}}{\emptyenv}$ is the smallest $i$ such that $\tm_{i}$ is not a normal term.

In this sense, the following Lemma implies that for every term context $\hctx$ and term of the shape $(\la{\var}{\tm}) \tmtwo$, the {\opencbneednsp} evaluation strategy reduces $\progentry{\hctxp{(\la{\var}{\tm}) \tmtwo}}{\emptyenv}$ by contracting $(\la{\var}{\tm}) \tmtwo$.

\begin{lemma}[Unique decomposition of terms]
\hypertarget{targ-st:Unique_decomposition_of_Lambda-terms-opencbneed}{}
\hfill

Let $\hctx_{1} \ctxholep{\tm_{1}} = \hctx_{2} \ctxholep{\tm_{2}}$, with $\hctx_{1}, \hctx_{2}$ term contexts, let $S \supseteq (\nv{\hctx_{1}} \cup \nv{\hctx_{2}})$, and let $\tm_{i}$ be a $S$-reduction place of $\hctx_{i} \ctxholep{\tm_{i}}$, for $i = 1, 2$.
Then $\tm_{1} = \tm_{2}$ and $\hctx_{1} = \hctx_{2}$.

\label{l:Unique_decomposition_of_Lambda-terms-opencbneed}
\end{lemma}


\begin{proof}
\hypertarget{targ-pr:Unique_decomposition_of_Lambda-terms-opencbneed}{}
By induction on $\hctx_{1}$. Cases:
\begin{itemize}
	\item \emph{Empty}: $\hctx_{1} = \ctxhole$. If $\tm_{1}$ is a multiplicative redex then $\hctx_{2} = \ctxhole$ and $\tm_{2}=\tm_{1}$. The same is true if $\tm_{1}$ is a variable not in S.
	
	\item \emph{Left of an application}: $\hctx_{1} = \hctxtwo_{1} \tmtwo$. Cases of $\hctx_{2}$:
	\begin{itemize}
		\item \emph{Empty}: $\hctx_{2} = \ctxhole$, then $\tm_{2}$ is a multiplicative redex, implying that $\hctxtwo_{1}$ is empty and $\tm_{1}$ is a value, which is absurd.	Therefore, this case is impossible.
		
		\item \emph{Left of an application}: $\hctx_{2} = \hctxtwo_{2} \tmtwo$. Then $\hctxtwo_{1}\ctxholep{\tm_{1}} = \hctxtwo_{2}\ctxholep{\tm_{2}}$ and $\tm_{i}$ is a S-reduction places of $\hctx_{i} \ctxholep{\tm_{i}}$ for $i =1,2$. By \ih, $\tm_{1} = \tm_{2}$ and $\hctxtwo_{1} = \hctxtwo_{2}$, and then $\hctx_{1} = \hctx_{2}$.
		
		\item \emph{Right of an application}: $\hctx_{2} = \itm \hctxtwo_{2}$. Then $\hctxtwo_{1}\ctxholep{\tm_{1}} = \itm$, and---by \reflemma{Redex_in_non-normal_terms-opencbneed} (Redex in non-normal terms)---$\tm_{1}$ can only be a free variable $\var$ such that $\var \notin S$. Note however, that $\var\in\fv\itm \subseteq \nv{\hctx_{2}} \subseteq S$, which is absurd. Therefore, this case is impossible.
	\end{itemize}
	
	\item \emph{Right of an application}: $\hctx_{1} = \itm \hctxtwo_{1}$. Cases of $\hctx_{2}$:
		\begin{itemize}
		\item \emph{Empty}: $\hctx_{2} = \ctxhole$, then $\tm_{2}$ is an application that is not a multiplicative redex, absurd.	Therefore, this case is impossible.
		
		\item \emph{Left of an application}: $\hctx_{2} = \hctxtwo_{2} \tmtwo$. This case is identical to the case where the hole of $\hctx_{1}$ is on the left of the application while the one of $\hctx_{2}$ is on the right, treated above.	
		
		\item \emph{Right of an application}: $\hctx_{2} = \itm \hctxtwo_{2}$. Then $\hctxtwo_{1}\ctxholep{\tm_{1}} = \hctxtwo_{2}\ctxholep{\tm_{2}}$ and $\tm_{i}$ is a S-reduction places of $\hctx_{i} \ctxholep{\tm_{i}}$ for $i =1,2$. By \ih, $\tm_{1} = \tm_{2}$ and $\hctxtwo_{1} = \hctxtwo_{2}$, and then $\hctx_{1} = \hctx_{2}$.\qedhere
		
	\end{itemize}

\end{itemize}

\end{proof}

With \reflemma{Unique_decomposition_of_Lambda-terms-opencbneed} serving as the base case, we can now prove that there exists a ``\emph{first}'' multiplicative or exponential redex in a program (provided it $\toond$-reduces), in the sense that it is the only reduction step defined in the {\opencbneednsp} reduction relation:

\begin{theorem}[Unique decomposition of programs]
\hypertarget{targ-st:Unique_decomposition_of_programs-opencbneed}{}
\hfill

Let $\pctx_{1} \ctxholep{\tm_{1}} = \pctx_{2} \ctxholep{\tm_{2}}$, with $\pctx_{1} \in \evctxp{\varset_{1}}$, $\pctx_{2} \in \evctxp{\varset_{2}}$, $S \supseteq (\varset_{1} \cup \varset_{2})$, and $\tm_{i}$ be a S-reduction place of $\pctx_{i} \ctxholep{\tm_i}$ for $i =1,2$. 
Then $\tm_{1} = \tm_{2}$ and $\pctx_{1} = \pctx_{2}$.

\label{thm:Unique_decomposition_of_programs-opencbneed}
\end{theorem}


\begin{proof}
By induction on $\pctx_{1}$. Cases:
	\begin{itemize}
		\item Rule $\ruleEvcAx$: then $\pctx_{1} = \progentry{\hctx_{1}}{\emptyenv}$. Then $\pctx_{2} = \progentry{\hctx_{2}}{\emptyenv}$. By \reflemma{Unique_decomposition_of_Lambda-terms-opencbneed} (Unique decomposition of terms), we obtain $\tm_{1} = \tm_{2}$ and $\hctx_{1} = \hctx_{2}$, and then also $\pctx_{1} = \pctx_{2}$.
	
		\item Rule $\ruleEvcInert$: Let $\pctx_{1} \in \evctxp{\varset_{1}}$ be derived as
			$$
			\infer
				[\ruleEvcInert]
				{\appES{\pctxtwo_{1}}{\esub{\var}{\itm}} \in \evctxp{(\varsettwo_{1} \setminus \set{\var}) \cup \nv{\itm}}}
				{\pctxtwo_{1} \in \evctxp{\varsettwo_{1}}
				\quad
				\var \in \varsettwo_{1}}
			$$
		with $\pctx_{1} = \appES{\pctxtwo_{1}}{\esub{\var}{\itm}}$ with $\varset_{1} = (\varsettwo_{1} \setminus \set{\var}) \cup \nv{\itm}$. Cases of $\pctx_{2} \in \evctxp{\varset_{2}}$:
			\begin{itemize}
				\item Rule $\ruleEvcAx$: impossible.
				
				\item Rule $\ruleEvcInert$: Let $\pctx_{2} \in \evctxp{\varset_{2}}$ be derived as
					$$
					\infer
						[\ruleEvcInert]
						{\appES{\pctxtwo_{2}}{\esub{\var}{\itm}} \in \evctxp{(\varsettwo_{2} \setminus \set{\var}) \cup \nv{\itm}}}
						{\pctxtwo_{2} \in \evctxp{\varsettwo_{2}}
						\quad
						\var \in \varsettwo_{2}}
					$$
				with $\pctx_{2} = \appES{\pctxtwo_{2}}{\esub{\var}{\itm}}$ with $\varset_{2} = (\varsettwo_{2} \setminus \set{\var}) \cup \nv{\itm}$.
	
				Then $\pctxtwo_{1} \ctxholep{\tm_{1}} = \pctxtwo_{2} \ctxholep{\tm_{2}}$ and $\tm_{i}$ is a S-reduction place of $\pctxtwo_{i} \ctxholep{\tm_{i}}$ for $i = 1,2$. By \ih, $\tm_{1} = \tm_{2}$ and $\pctxtwo_{1} = \pctxtwo_{2}$, and then $\pctx_{1} = \pctx_{2}$.
		
				\item Rule $\ruleEvcGc$: Let $\pctx_{2} \in \evctxp{\varset_{2}}$ be derived as
					$$
					\infer
						[\ruleEvcGc]
						{\appES{\pctxtwo_{2}}{\esub{\var}{\itm}} \in \evctxp{\varsettwo_{2}}}
						{\pctxtwo_{2} \in \evctxp{\varsettwo_{2}}
						\quad
						\var \notin \varsettwo_{2}}
					$$
				with $\pctx_{2} = \appES{\pctxtwo_{2}}{\esub{\var}{\itm}}$ and $\varset_{2} = (\varsettwo_{2} \setminus \set{\var}) \cup \nv{\itm}$. Note that $\tm_{i} \neq \var$ for $i = 1,2$ because $\var$ is bound by $\pctx_{1}$ and $\pctx_{2}$ but not associated to a value. Note that $\pctxtwo_{1} \ctxholep{\tm_{1}} = \pctxtwo_{2} \ctxholep{\tm_{2}}$ and $\tm_{i}$ is a S-reduction place of  $\pctxtwo_{i} \ctxholep{\tm_{i}}$ for $i =1,2$. By \ih, $\tm_{1} = \tm_{2}$ and $\pctxtwo_{1} = \pctxtwo_{2}$. This is absurd, because by \reflemma{Unique_derivation_parameterization_of_open_evaluation_contexts} (Unique derivation parameterization of open evaluation contexts) we would have that $\varsettwo_{1} = \varsettwo_{2}$, although we know that $\var \in \varsettwo_{1}$ and $\var \notin \varsettwo_{2}$. Therefore, this case is impossible.
		
				\item Rule $\ruleEvcHer$: Let $\pctx_{2} \in \evctxp{\varset_{2}}$ be derived as
					$$		
					\infer
						[\ruleEvcHer]
						{\appES{\pctxtwo_{2} \ctxholep{\var}}{\esub{\var}{\hctx}} \in \evctxp{\varsettwo_{2} \cup \nv{\hctx}}}
						{\pctxtwo_{2} \in \evctxp{\varsettwo_{2}}
						\quad			
						\var \notin \varsettwo_{2}}
					$$
				with $\pctx_{2} = \appES{\pctxtwo_{2} \ctxholep{\var}}{\esub{\var}{\hctx}}$ and $\varset_{2} = \varsettwo_{2} \cup \nv{\hctx}$.
		
				Note that $\tm_{2}$ cannot be a bound variable, because it occurs in the rightmost ES of the program, nor a $\beta$-redex, otherwise by \reflemma{Redex_in_non-normal_terms-opencbneed}  (Redex in non-normal terms) $\hctxp{\tm_{2}} = \itm$ would not be an inert term. Thus, it can only be that $\tm_{2} = \vartwo \in \Var$ and $\vartwo \in \nv{\itm}$. But then $\vartwo \in \nv{\itm} \subseteq \varset_{1} \subseteq \varsettwo$, implying that $\tm_{2}$ is not a S-reduction place of $\pctx_{2} \ctxholep{\tm_{2}}$; absurd. Therefore, this case is impossible.
	
			\end{itemize}

		\item Rule $\ruleEvcGc$: Let $\pctx \in \evctxp{\varset_{1}}$ be derived as
			$$		
			\infer
				[\ruleEvcGc]
				{\appES{\pctxtwo_{1}}{\esub{\var}{\tmtwo}} \in \evctxp{\varset_{1}}}
				{\pctxtwo_{1} \in \evctxp{\varset_{1}}
				\quad
				\var \notin \varset_{1}}
			$$
		with $\pctx_{1} = \appES{\pctxtwo_{1}}{\esub{\var}{\tm}}$. Cases of $\pctx_{2} \in \evctxp{\varset_{2}}$:
			\begin{itemize}
				\item Rule $\ruleEvcAx$: impossible.
		
				\item Rule $\ruleEvcInert$: this case follows an identical but inversed analysis to that of the case of rule $\ruleEvcInert$ for $\pctx_{1} \in \evctxp{\varset_{1}}$ and $\ruleEvcGc$ for $\pctx_{2} \in \evctxp{\varset_{2}}$, treated above.
			
				\item Rule $\ruleEvcGc$: Let $\pctx_{2} \in \evctxp{\varset_{2}}$ be derived as  
					$$
					\infer
						[\ruleEvcGc]
						{\appES{\pctxtwo_{2}}{\esub{\var}{\tmtwo}} \in \evctxp{\varset_{2}}}
						{\pctxtwo_{2} \in \evctxp{\varset_{2}}
						\quad
						\var \notin \varset_{2}}
					$$
				where $\pctx_{2} = \appES{\pctxtwo_{2}}{\esub{\var}{\tm}}$.
				
				Then $\pctxtwo_{1}\ctxholep{\tm_{1}} = \pctxtwo_{2}\ctxholep{\tm_{2}}$ and $\tm_{i}$ is a S-reduction place of $\pctxtwo_{i}\ctxholep{\tm_{i}}$ for $i =1,2$. By \ih, $\tm_{1} = \tm_{2}$ and $\pctxtwo_{1} = \pctxtwo_{2}$, and then $\pctx_{1} = \pctx_{2}$.
		
				\item Rule $\ruleEvcHer$: Let $\pctx_{2} \in \evctxp{\varset_{2}}$ be derived as
					$$
					\infer
						[\ruleEvcHer]
						{\appES{\pctxtwo_{2} \ctxholep{\var}}{\esub{\var}{\hctx}} \in \evctxp{\varsettwo_{2} \cup \nv{\hctx}}}
						{\pctxtwo_{2} \in \evctxp{\varsettwo_{2}}
						\quad			
						\var \notin \varsettwo_{2}}
					$$
				where $\pctx_{2} = \appES{\pctxtwo_{2} \ctxholep{\var}}{\esub{\var}{\hctx}}$ and $\varset_{2} = \varsettwo_{2} \cup \nv{\hctx}$.					
						
				Then $\pctxtwo_{1} \ctxholep{\tm_{1}} = \pctxtwo_{2} \ctxholep{\var}$. Note we can assume that $\var\notin \varsettwo$, since $\var$ was a bound name. Note also that $\pctxtwo_{1}\ctxholep{\tm_{1}}$ is a $\varsettwo$-reduction place and that  $\pctxtwo_{2}\ctxholep\var$ is also a $\varsettwo$-reduction place, because $\var\notin \varsettwo$. By \ih, $\tm_{1} = \var$, $\pctxtwo_{1} = \pctxtwo_{2}$, and $\varset_{1}' = \varset'_{2}$. 

				Now, let us look at the reduction places $\tm_{1} = \var$ and $\tm_{2}$. The fact that $\tm_{1} = \var$ is a reduction place in $\pctx_{1}$ implies that $\tmtwo$ is a value. Then $\hctx$ is the empty context $\ctxhole$ and $\tm_{2}$ is a value, which is absurd, since values cannot be reduction places. Therefore, this case is impossible.
			
			\end{itemize}
	
		\item Rule $\ruleEvcHer$: Let $\pctx_{1} \in \evctxp{\varset_{1}}$ be derived as 
			$$
			\infer
				[\ruleEvcHer]
				{\appES{\pctxtwo_{1} \ctxholep{\var}}{\esub{\var}{\hctx_{1}}} \in \evctxp{\varsettwo_{1} \cup \nv{\hctx_{1}}}}
				{\pctxtwo_{1} \in \evctxp{\varsettwo_{1}}
				\quad			
				\var \notin \varsettwo_{1}}
			$$
		where $\pctx_{1} = \appES{\pctxtwo_{1} \ctxholep{\var}}{\esub{\var}{\hctx_{1}}}$ and $\varset_{1} = \varsettwo_{1} \cup \nv{\hctx_{1}}$. Cases of $\pctx_{2} \in \evctxp{\varset_{2}}$:
			\begin{itemize}
				\item Rule $\ruleEvcAx$: impossible.
		
				\item Rule $\ruleEvcInert$: this case follows an identical but inversed analysis to that of the case of rule $\ruleEvcInert$ for $\pctx_{1} \in \evctxp{\varset_{1}}$ and rule $\ruleEvcHer$ for $\pctx_{2} \in \evctxp{\varset_{2}}$, treated above.		
		
				\item Rule $\ruleEvcGc$: this case follows an identical but inversed analysis to that of the case of rule $\ruleEvcGc$ for $\pctx_{1} \in \evctxp{\varset_{1}}$ and rule $\ruleEvcHer$ for $\pctx_{2} \in \evctxp{\varset_{2}}$, treated above.				
		
				\item Rule $\ruleEvcHer$: Let $\pctx_{2} \in \evctxp{\varset_{2}}$ be derived as
					$$		
					\infer
						[\ruleEvcHer]
						{\appES{\pctxtwo_{2} \ctxholep{\var}}{\esub{\var}{\hctx_{2}}} \in \evctxp{\varsettwo_{2} \cup \nv{\hctx_{2}}}}
						{\pctxtwo_{2} \in \evctxp{\varsettwo_{2}}
						\quad			
						\var \notin \varsettwo_{2}}
					$$
				where $\pctx_{2} = \appES{\pctxtwo_{2} \ctxholep{\var}}{\esub{\var}{\hctx_{2}}}$ and $\varset_{2} = \varsettwo_{2} \cup \nv{\hctx_{2}}$.
		
				Then $\pctxtwo_{1} \ctxholep{\var} = \pctxtwo_{2} \ctxholep{\var}$ and $\tm_{i}$ is a S-reduction place of $\pctxtwo_{i} \ctxholep{\tm_{i}}$ for $i =1,2$. By \ih, $\tm_{1} = \tm_{2}$ and $\pctxtwo_{1} = \pctxtwo_{2}$. 
				
				Moreover, note that $\hctx_{1} \ctxholep{\tm_{1}} = \hctx_{2} \ctxholep{\tm_{2}}$ and that $\tm_{i}$ is a S-reduction place of $\hctx_{i} \ctxholep{\tm_{i}}$. Hence, by \reflemma{Unique_decomposition_of_Lambda-terms-opencbneed} (Unique decomposition of terms), we have that $\tm_{1} = \tm_{2}$ and $\hctx = \hctxtwo$. The latter allows us to finally conclude that $\pctx_{1} = \pctx_{2}$.\qedhere
			
			\end{itemize}
	
	\end{itemize}

\end{proof}

Finally,

\gettoappendix{c:Determinism_of_opencbneed}

\begin{proof}
Let $\prog, \progtwo$, and $\progthree$ be programs. We have to prove that if $\prog \toond \progtwo$ and $\prog \toond \progthree$, then $\progtwo = \progthree$.
 Let $\tm_{1}, \tm_{2}$ be terms, $\pctx_{1} \in \evctxp{\varset_{1}}$ and $\pctx_{2} \in \evctxp{\varset_{2}}$ be such that $\prog = \pctx_{1} \ctxholep{\tm_{1}} = \pctx_{2} \ctxholep{\tm_{2}}$. Let moreover $S \defeq \varset_{1} \, \cup \, \varset_{2}$, noting that $\tm_{1}$ is a $S$-reduction place of $\pctx_{1} \ctxholep{\tm_{1}}$ and $\tm_{2}$ is a reduction place of $S$-reduction place of $\pctx_{2} \ctxholep{\tm_{2}}$. By \refthm{Unique_decomposition_of_programs-opencbneed} (Unique decomposition of programs), $\pctx_{1} = \pctx_{2}$ and $\tm_{1} = \tm_{2}$. Therefore, $\progtwo = \progthree$.\qedhere

\end{proof}


\subsection{Characterizing \opencbneednsp-normal forms.}
\label{subsubsect:Characterizing_opencbneed-normal_forms-Proofs_of_opencbneed}

\begin{lemma}[Redex in non-normal terms]
$\tm$ is not a normal term if and only if there exist term context $\hctx$, and terms $\la{\var}{\tmtwo}$ and $\tmthree$ such that $\tm = \hctxp{(\la{\var}{\tmtwo})\tmthree}$.

\label{l:Redex_in_non-normal_terms-opencbneed}
\end{lemma}

\begin{proof}
\hypertarget{targ-pr:Redex_in_non-normal_terms-opencbneed}{}
\begin{itemize}
	\item[$\Rightarrow$] Let $\tm$ be a non-normal term. We proceed by induction on the structure of $\tm$, noting that $\tm$ is neither a variable nor a value. 

	Let $\tm = \tm_{1} \tm_{2}$ such that $\tm_{1}$ is not inert or $\tm_{2}$ is not a normal term. 

	If $\tm_{1}$ is not a normal term, then by \ih $\tm_{1} = \hctx_{1} \ctxholep{(\la{\var_{1}}{\tmtwo_{1}})\tmthree_{1}}$ and so by defining $\hctx \defeq \hctx_{1}\tm_{2}$ we have that $\tm = \hctx \ctxholep{(\la{\var_{1}}{\tmtwo_{1}})\tmthree_{1}}$.

	Let $\tm_{1}$ be a normal term. If $\tm_{1} = \val_{1}$, then by defining $\hctx \defeq \ctxhole$ we have that $\tm = \hctx \ctxholep{\val_{1} \tm_{2}}$. Let $\tm_{1} = \itm_{1}$ and $\tm_{2}$ be a non-normal term. Then by \ih $\tm_{2} = \hctx_{2} \ctxholep{(\la{\var_{2}}{\tmtwo_{2}}) \tmthree_{2}}$ and so by defining $\hctx \defeq \itm_{1} \hctx_{2}$ we have that $\tm = \hctx \ctxholep{(\la{\var_{2}}{\tmtwo_{2}}) \tmthree_{2}}$.
	
	\item[$\Leftarrow$] By induction on the structure of $\hctx$.
		\begin{itemize}
			\item If $\hctx = \ctxhole$ then $\tm = (\la{\var}{\tmtwo})\tmthree$, which does not correspond to any term derivable from the grammar of inert terms.
			
			\item If $\hctx = \hctxtwo \tmfour$, then $\tm = \hctxtwo \ctxholep{(\la{\var}{\tmtwo})\tmthree} \tmfour$. By \ih $\hctxtwo \ctxholep{(\la{\var}{\tmtwo})\tmthree}$ is not a normal term, therefore nor is $\hctxtwo \ctxholep{(\la{\var}{\tmtwo})\tmthree} \tmfour = \tm$.
			
			\item If $\hctx = \itm \hctxtwo$, then $\tm = \itm \hctxtwo \ctxholep{(\la{\var}{\tmtwo})\tmthree}$. By \ih $\hctxtwo \ctxholep{(\la{\var}{\tmtwo})\tmthree}$ is not a normal term, therefore nor is $\itm \hctxtwo \ctxholep{(\la{\var}{\tmtwo})\tmthree} = \tm$.\qedhere
			
		\end{itemize}
	
\end{itemize}

\end{proof}

Before proving the characterization of \opencbneednsp-normal forms by means of predicate $\onormalpr{.}$, we first need to give a series of properties. The first set concerns term contexts, and revolves around the notion of needed variables:
\begin{lemma}[Rewriting: term contexts]
\hfill

	\begin{enumerate}		
		\item
		\label{p:Rewriting_term_contexts-opencbneed-one} 
		\emph{Term contexts give needed variables}: For every $\var \in \Var$ and term context $\hctx$, $\var \in \nv{\hctxp{\var}}$.
		
		\item
		\label{p:Rewriting_term_contexts-opencbneed-two}
		\emph{Focusing inert terms on needed variables}: Let $\itm$ be an inert term and $\var \in \nv{\itm}$. Then there exists a term context $\hctx_{\var}$ such that $\var \notin \nv{\hctx_{\var}} \subset \nv{\itm}$ and that $\hctx_{\var} \ctxholep{\var} = \itm$.
		
		\item
		\label{p:Rewriting_term_contexts-opencbneed-three}
		\emph{Focusing term contexts on needed variabless}: Let $\var \in \nv{\hctx}$. Then for every term $\tm$ there exists a term context $\hctx_{\tm}$ such that $\var \notin \nv{\hctx_{\tm}} \subset \nv{\hctx}$ and that $\hctx_{\tm} \ctxholep{\var} = \hctxp{\tm}$.
				
	\end{enumerate}

\label{l:Rewriting_term_contexts-opencbneed}
\end{lemma}


\begin{proof}
\hfill

\begin{enumerate}
	\item 
	\emph{Term contexts give needed variables}: By structural induction on $\hctx$:
		\begin{itemize}
			\item 
			\emph{Context hole}: If $\hctx = \ctxhole$, then it holds trivially.
			
			\item 
			\emph{Left of an application}: Let $\hctx = \hctxtwo \tm$. By \ih, $\var \in \nv{\hctxtwop{\var}}$. Hence,
				$$
					\var \in \nv{\hctxtwop{\var}} = \nv{\hctxtwop{\var} \tm} = \nv{\hctxp{\var}}
				$$
			
			\item
			\emph{Right of an application}: Let $\hctx = \itm \hctxtwo$. By \ih, $\var \in \nv{\hctxtwop{\var}}$. Hence,
				$$
					\var \in \nv{\hctxtwop{\var}} \subseteq \nv{\itm} \cup \nv{\hctxtwop{\var}} = \nv{\hctxp{\var}}
				$$
			
		\end{itemize}

	\item
	\emph{Focusing inert terms on needed variables}: By structural induction on $\itm$:
		\begin{itemize}				
			\item If $\itm = \var$ then the statement holds by defining $\hctx \defeq \ctxhole$. 
		
			\item Let $\itm = \itmtwo \fire$. If $\var \in \nv{\itmtwo}$, then applying the \ih on $\itmtwo$ yields term context $\hctx_{\itmtwo}$ such that $\var \notin \nv{\hctx_{\itmtwo}} \subset \nv{\itmtwo}$ and such that $\hctx_{\itmtwo} \ctxholep{\var} = \itmtwo$. Thus, the statement holds by defining $\hctx \defeq \hctx_{\itmtwo} \fire$.
		
			If $\var \notin \nv{\itmtwo}$ then $\var \in \nv{\fire}$, implying in turn that $\fire$ is an inert term---since values have no needed variable. Then, applying the \ih on $\fire$ yields term context $\hctx_{\fire}$ such that $\var \notin \nv{\hctx_{\fire}} \subset \nv{\fire}$ and $\hctx_{\fire} \ctxholep{\var} = \fire$. Thus, the statement holds by defining $\hctx \defeq \itmtwo \hctx_{\fire}$.
		
		\end{itemize}

	\item
	\emph{Focusing term contexts on needed variables}: Let $\tm$ be a term. We proceed by structural induction on $\hctx$:
			\begin{itemize}
				\item 
				\emph{Context hole}: This case is impossible, because $\nv{\ctxhole} = \emptyset$.
					
				\item
				\emph{Left of an application}: Let $\hctx = \hctxtwo \tmthree$. Then $\var \in \nv{\hctxtwo} = \nv{\hctx}$. By \ih, there exists term context $\hctxtwo_{\tm}$ such that $\var \notin \nv{\hctxtwo_{\tm}} \subset \nv{\hctxtwo}$ and $\hctxtwo_{\tm} \ctxholep{\var} = \hctxtwop{\tm}$. Therefore, the statement holds by defining $\hctx_{\tm} \defeq \hctxtwo_{\tm} \tmthree$.
				
				\item 
				\emph{Right of an application}: Let $\hctx = \itm \hctxtwo$. Case analysis on whether $\var \in \nv{\itm}$:
					\begin{itemize}
						\item Let $\var \in \nv{\itm}$. By \reflemmap{Rewriting_term_contexts-opencbneed}{two} (Focusing inert terms on needed variables), there exists term context $\hctxtwo_{\var}$ such that $\var \notin \nv{\hctxtwo_{\var}} \subset \nv{\itm}$ and $\itm = \hctxtwo_{\var} \ctxholep{\var}$. The statement follows by defining $\hctx_{\tm} \defeq \hctxtwo_{\var} \hctxtwop{\tm}$, verifying that $\var \notin \nv{\hctxtwo_{\var}} = \nv{\hctxtwo_{\var} \hctxtwop{\tm}} = \nv{\hctx_{\tm}}$, that $\nv{\hctx_{\tm}} =\nv{\hctxtwo_{\var} \hctxtwop{\tm}} = \nv{\hctxtwo_{\var}} \subset \nv{\itm} \subseteq \nv{\itm} \cup \nv{\hctxtwo} = \nv{\hctx}$ and that $\hctx_{\tm} \ctxholep{\var} = \hctxtwo_{\var} \ctxholep{\var} \hctxtwop{\tm} = \itm \hctxtwop{\tm} = \tm$.
	
						\item Let $\var \notin \nv{\itm}$. Then $\var \in \nv{\hctxtwo}$. By \ih, there exists $\hctxtwo_{\tm}$ such that $\var \notin \nv{\hctxtwo_{\tm}} \subset \nv{\hctxtwo}$ and that $\hctxtwo_{\tm} \ctxholep{\var} = \hctxtwop{\tm}$. The statement follows by defining $\hctx_{\tm} \defeq \itm \hctxtwo_{\tm}$, verifying that $\var \notin \nv{\itm} \cup \nv{\hctxtwo_{\tm}} = \nv{\hctx_{\tm}}$, that $\nv{\hctx_{\tm}} = \nv{\itm} \cup \nv{\hctxtwo_{\tm}} \subset \nv{\itm} \cup \nv{\hctxtwo} = \nv{\hctx}$, and that $\hctx_{\tm} \ctxholep{\var} = \itm \hctxtwo_{\tm} \ctxholep{\var} = \itm \hctxtwop{\tm} = \hctxp{\tm}$.\qedhere
						
					\end{itemize}
				
			\end{itemize}

\end{enumerate}

\end{proof}

Next, we proceed to lift \reflemmap{Rewriting_term_contexts-opencbneed}{one} and \reflemmap{Rewriting_term_contexts-opencbneed}{three} to open evaluation contexts, as follows
\begin{lemma}[Rewriting open evaluation contexts]
\hfill
	
	\begin{enumerate}
		\item 
		\label{p:Rewriting_open_evaluation_contexts-one} 
		\emph{Open evaluation contexts give needed variables}: Let $\pctx \in \evctxp{\varset}$ and $\var \notin \dom{\pctx}$. Then $\var \in \nv{\pctxp{\var}}$.
				
		\item 
		\label{p:Rewriting_open_evaluation_contexts-two} 
		\emph{Focusing open evaluation contexts on needed variables}: Let $\pctx \in \evctxp{\varset}$ and $\var \in \varset$. Then for every term $\tm$ there exists an open evaluation context $\pctx_{\tm} \in \evctxp{\varset_{\tm}}$ such that $\pctx_{\tm} \ctxholep{\var} = \pctxp{\tm}$ and $\var \notin \varset_{\tm} \subset \varset$.

	\end{enumerate}

\label{l:Rewriting_open_evaluation_contexts}
\end{lemma}


\begin{proof}
\hfill

\begin{enumerate}		
	\item By induction on the derivation of $\pctx \in \evctxp{\varset}$:
		\begin{itemize}
			\item
			\emph{Rule $\ruleEvcAx$}: Let $\pctx = \progentry{\hctx}{\emptyenv} \in \evctxp{\nv{\hctx}}$. By \reflemmap{Rewriting_term_contexts-opencbneed}{one} (Term contexts give needed variables), $\var \in \nv{\hctxp{\var}}$. Hence, $\var \in \nv{\pctxp{\var}} = \nv{\hctxp{\var}}$.
			
			\item 
			\emph{Rule $\ruleEvcGc$}: Let $\pctx \in \evctxp{\varset}$ be derived as follows
				$$
				\infer
					[\ruleEvcGc]
					{\appES{\pctxtwo}{\esub{\vartwo}{\tm}} \in \evctxp{\varset}}
					{\pctxtwo \in \evctxp{\varset}
					\quad
					\vartwo \notin \varset}
				$$
			where $\pctx = \appES{\pctxtwo}{\esub{\vartwo}{\tm}}$. By hypothesis, $\var \neq \vartwo$ and $\var \notin \dom{\pctxtwop{\var}}$. By \ih, $\var \in \nv{\pctxtwop{\var}}$. Case analysis on whether $\vartwo \in \nv{\pctxtwop{\var}}$:
				\begin{itemize}
					\item Let $\vartwo \notin \nv{\pctxtwop{\var}}$. The statement follows because then $\nv{\pctxp{\var}} = \nv{\pctxtwop{\var}}$.
					
					\item Let $\vartwo \in \nv{\pctxtwop{\var}}$. Since $\var \neq \vartwo$, then
						$$
						\var \in (\nv{\pctxtwop{\var}} \setminus \set{\vartwo}) \subseteq (\nv{\pctxtwop{\var}} \setminus \set{\vartwo}) \cup \nv{\tm} = \nv{\pctxp{\var}}
						$$

				\end{itemize}
			
			\item
			\emph{Rule $\ruleEvcInert$}: Let $\pctx \in \evctxp{\varset}$ be derived as follows
				$$
				\infer
					[\ruleEvcInert]
					{\appES{\pctxtwo}{\esub{\vartwo}{\itm}} \in \evctxp{(\varsettwo \setminus \set{\vartwo}) \cup \nv{\itm}}}
					{\pctxtwo \in \evctxp{\varsettwo}
					\quad
					\vartwo \in \varsettwo}
				$$
			where $\pctx = \appES{\pctxtwo}{\esub{\vartwo}{\itm}}$ and $\varset = (\varsettwo \setminus \set{\vartwo}) \cup \nv{\itm}$. By \ih, we have that $\var \in \nv{\pctxtwop{\var}}$. Case analysis on whether $\vartwo \in \nv{\pctxtwop{\var}}$:
				\begin{itemize}
					\item Let $\vartwo \notin \nv{\pctxtwop{\var}}$. The statement follows because then $\var \in \nv{\pctxtwop{\var}} = \nv{\pctxp{\var}}$.
					
					\item Let $\vartwo \in \nv{\pctxtwop{\var}}$. Since $\var \neq \vartwo$, then
						$$
							\var \in \nv{\pctxtwop{\var}} \setminus \set{\vartwo} \subseteq (\nv{\pctxtwop{\var}} \setminus \set{\vartwo}) \cup \nv{\itm} = \nv{\pctxp{\var}}
						$$
					
				\end{itemize}
			
			\item
			\emph{Rule $\ruleEvcHer$}: Let $\pctx \in \evctxp{\varset}$ be derived as follows:
				$$
				\infer
					[\ruleEvcHer]
					{\appES{\pctxtwop{\vartwo}}{\esub{\vartwo}{\hctx}} \in \evctxp{\varsettwo \cup \nv{\hctx}}}
					{\pctxtwo \in \evctxp{\varsettwo}
					\quad
					\vartwo \notin \varsettwo}
				$$
			where $\pctx = \appES{\pctxtwop{\vartwo}}{\esub{\vartwo}{\hctx}}$ and $\varset = \varsettwo \cup \nv{\hctx}$. By $\alpha$-conversion, we may safely assume that $\vartwo \notin \dom{\pctxtwo}$. By \ih, we have that $\vartwo \in \nv{\pctxtwop{\vartwo}}$. Moreover, \reflemmap{Rewriting_term_contexts-opencbneed}{one} (Term contexts give needed variables) gives that $\var \in \nv{\hctxp{\var}}$. Hence,
				\begin{center}
					$
					\begin{array}{rcl}
						\var \in \nv{\hctxp{\var}} 
						& \subseteq & (\nv{\pctxtwop{\vartwo}} \setminus \set{\vartwo}) \cup \nv{\hctxp{\var}} \\
						& = & \nv{(\appES{\pctxtwop{\vartwo}}{\esub{\vartwo}{\hctx}}) \ctxholep{\var}} \\
						& = & \nv{\pctxp{\var}}
					\end{array}
					$
				\end{center}
				
		\end{itemize}
		
		\item Let $\tm$ be a term. We proceed by induction on the derivation of $\pctx \in \evctxp{\varset}$:
		\begin{itemize}
			\item
			\emph{Rule $\ruleEvcAx$}: Let $\pctx = \progentry{\hctx}{\emptyenv} \in \evctxp{\nv{\hctx}}$, with $\var \in \nv{\hctx}$. By \reflemmap{Rewriting_term_contexts-opencbneed}{three} (Focusing term contexts on needed variables), there exists a term context $\hctx_{\tm}$ such that $\var \notin \nv{\hctx_{\tm}} \subset \nv{\hctx}$ and that $\hctx_{\tm} \ctxholep{\var} = \hctxp{\tm}$. Thus, the statement holds by defining $\pctx_{\tm} \defeq \progentry{\hctx_{\tm}}{\emptyenv}$.
			
			\item 
			\emph{Rule $\ruleEvcGc$}: Let $\pctx \in \evctxp{\varset}$ be derived as follows
				$$
				\infer
					[\ruleEvcGc]
					{\appES{\pctxtwo}{\esub{\vartwo}{\tmtwo}} \in \evctxp{\varset}}
					{\pctxtwo \in \evctxp{\varset}
					\quad
					\vartwo \notin \varset}
				$$
			where $\pctx = \appES{\pctxtwo}{\esub{\vartwo}{\tmtwo}}$. By \ih with respect to $\var$ and $\pctxtwo$, there exists $\pctxtwo_{\tm} \in \evctxp{\varset_{\tm}}$ such that $\var \notin \varset_{\tm} \subset \varset$ and $\pctxtwo_{\tm} \ctxholep{\var} = \pctxtwop{\tm}$. We may then derive $\pctx_{\tm} \in \evctxp{\varset_{\tm}}$ as follows
				$$
				\infer
					[\ruleEvcGc]
					{\appES{\pctxtwo_{\tm}}{\esub{\vartwo}{\tmtwo}} \in \evctxp{\varset_{\tm}}}
					{\pctxtwo_{\tm} \in \evctxp{\varset_{\tm}}
					\quad
					\vartwo \notin \varset_{\tm}}
				$$
			where $\pctx_{\tm} = \appES{\pctxtwo_{\tm}}{\esub{\vartwo}{\tmtwo}}$, noting that $\pctx_{\tm} \ctxholep{\var} =  \appES{(\pctxtwo_{\tm} \ctxholep{\var})}{\esub{\vartwo}{\tmtwo}}  = \appES{(\pctxtwop{\tm})}{\esub{\vartwo}{\tmtwo}} = \pctxp{\tm}$.
								
			\item
			\emph{Rule $\ruleEvcInert$}: Let $\pctx \in \evctxp{\varset}$ be derived as
				$$
				\infer
					[\ruleEvcInert]
					{\appES{\pctxtwo}{\esub{\vartwo}{\itm}} \in \evctxp{(\varsettwo \setminus \set{\vartwo}) \cup \nv{\itm}}}
					{\pctxtwo \in \evctxp{\varsettwo}
					\quad
					\vartwo \in \varsettwo}
				$$
			where $\pctx = \appES{\pctxtwo}{\esub{\vartwo}{\itm}}$ and $\varset = (\varsettwo \setminus \set{\vartwo}) \cup \nv{\itm} $. Case analysis on whether $\var \in (\varsettwo \setminus \set{\vartwo})$:
				\begin{itemize}
					\item Let $\var \in (\varsettwo \setminus \set{\vartwo})$. Note that then $\var \notin \vartwo$. By application of the \ih with respect to $\var$ and $\pctxtwo$, there exists $\pctxtwo_{\tm} \in \evctxp{\varsettwo_{\tm}}$ such that $\var \notin \varsettwo_{\tm} \subset \varsettwo$ and $\pctxtwo_{\tm} \ctxholep{\var} = \pctxtwop{\tm}$. Now, if $\vartwo \notin \varsettwo_{\tm}$, then the statement holds by deriving $\pctx_{\tm}$ as follows
						$$
						\infer
							[\ruleEvcGc]
							{\appES{\pctxtwo_{\tm}}{\esub{\vartwo}{\itm}} \in \evctxp{\varsettwo_{\tm}}}
							{\pctxtwo_{\tm} \in \evctxp{\varsettwo_{\tm}}
							\quad
							\vartwo \notin \varsettwo_{\tm}}
						$$
					where $\pctx_{\tm} = \appES{\pctxtwo_{\tm}}{\esub{\vartwo}{\itm}}$. If $\vartwo \in \varsettwo_{\tm}$ instead, we proceed by case analysis on whether $\var \in \nv{\itm}$:
						\begin{enumerate}
							\item Let $\var \notin \nv{\itm}$. The statement then holds by deriving $\pctx_{\tm}$ as follows
								$$
								\infer
									[\ruleEvcInert]
									{\appES{\pctxtwo_{\tm}}{\esub{\vartwo}{\itm}} \in \evctxp{(\varsettwo_{\tm} \setminus \set{\var}) \cup \nv{\itm}}}
									{\pctxtwo_{\tm} \in \evctxp{\varsettwo_{\tm}}
									\quad
									\vartwo \in \varsettwo_{\tm}}
								$$
							where $\pctx_{\tm} = \appES{\pctxtwo_{\tm}}{\esub{\vartwo}{\itm}}$. Thus, and since $\var \neq \vartwo$ and $\var \notin \varsettwo_{\tm}$, we have that 
								$$
								\var \notin \varsettwo_{\tm} \setminus \set{\vartwo} \subset \varsettwo \setminus \set{\vartwo} \subseteq (\varsettwo \setminus \set{\vartwo}) \cup \nv{\itm} = \varset
								$$

							\item Let $\var \in \nv{\itm}$. By application of the \ih with respect to $\vartwo$ and $\pctxtwo_{\tm}$, there exists $\pctxtwo_{\var} \in \evctxp{\varsettwo_{\var}}$ such that $\vartwo \notin \varsettwo_{\var} \subset \varsettwo_{\tm} \subset \varsettwo$, and that $\pctxtwo_{\var} \ctxholep{\vartwo} = \pctxtwo_{\tm} \ctxholep{\var} = \pctxtwop{\tm}$.
							
							Moreover, by \reflemmap{Rewriting_term_contexts-opencbneed}{two} (Focusing inert terms on needed variables), there exists a term context $\hctx_{\var}$ such that $\var \notin \nv{\hctx_{\var}} \subset \itm$ and that $\hctx_{\var} \ctxholep{\var} = \itm$. Thus, we may derive $\pctx_{\tm} \in \evctxp{\varset_{\tm}}$ 
								$$
								\infer
									[\ruleEvcHer]
									{\appES{\pctxtwo_{\var} \ctxholep{\vartwo}}{\esub{\vartwo}{\hctx_{\var}}} \in \evctxp{\varsettwo_{\var} \cup \nv{\hctx_{\var}}}}
									{\pctxtwo_{\var} \in \evctxp{\varsettwo_{\var}}
									\quad
									\vartwo \notin \varsettwo_{\var}}
								$$
							where $\pctx_{\tm} = \appES{\pctxtwo_{\var} \ctxholep{\vartwo}}{\esub{\vartwo}{\hctx_{\var}}}$ and $\varset_{\tm} = \varsettwo_{\var} \cup \nv{\hctx_{\var}}$.

						\end{enumerate}
					
					\item Let $\var \notin (\varsettwo \setminus \set{\vartwo})$. Since $\var \in \varset = (\varsettwo \setminus \set{\vartwo}) \cup \nv{\itm}$, then it must be that $\var \in \nv{\itm}$. By application of the \ih with respect to $\vartwo$ and $\pctxtwo$, there exists $\pctxtwo_{\tm} \in \evctxp{\varsettwo_{\tm}}$ such that $\vartwo \notin \varsettwo_{\tm} \subset \varsettwo$ and $\pctxtwo_{\tm} \ctxholep{\vartwo} = \pctxtwop{\tm}$. Moreover, by \reflemmap{Rewriting_term_contexts-opencbneed}{two} (Focusing inert terms on needed variables), there exists term context $\hctx_{\var}$ such that $\var \notin \nv{\hctx_{\var}} \subset \nv{\itm}$ and $\hctx_{\var} \ctxholep{\var} = \itm$. Thus, we may derive $\pctx_{\tm} \in \evctxp{\varset_{\tm}}$ as follows
						$$
						\infer
							[\ruleEvcHer]
							{\appES{\pctxtwo_{\vartwo} \ctxholep{\vartwo}}{\esub{\vartwo}{\hctx_{\var}}} \in \evctxp{\varsettwo_{\vartwo} \cup \nv{\hctx_{\var}}}}
							{\pctxtwo_{\vartwo} \in \evctxp{\varsettwo_{\vartwo}}
							\quad
							\vartwo \notin \varsettwo_{\vartwo}}
						$$
					where $\pctx_{\tm} = \appES{\pctxtwo_{\vartwo} \ctxholep{\vartwo}}{\esub{\vartwo}{\hctx_{\var}}}$, verifying in particular that
						$$
							\pctx_{\var} \ctxholep{\var} = \appES{(\pctxtwo_{\vartwo} \ctxholep{\vartwo})}{\esub{\vartwo}{\hctx_{\var} \ctxholep{\var}}} = \appES{\pctxtwop{\tm}}{\esub{\vartwo}{\itm}} = \pctxp{\tm}
						$$							
				\end{itemize}
				
			\item 
			\emph{Rule $\ruleEvcHer$}: Let $\pctx \in \evctxp{\varset}$ be derived as
				$$
				\infer
					[\ruleEvcHer]
					{\appES{\pctxtwop{\vartwo}}{\esub{\vartwo}{\hctx}} \in \evctxp{\varsettwo \cup \nv{\hctx}}}
					{\pctxtwo \in \evctxp{\varsettwo}
					\quad
					\vartwo \notin \varsettwo}
				$$
			where $\pctx = \appES{\pctxtwop{\vartwo}}{\esub{\vartwo}{\hctx}}$ and $\varset = \varsettwo \cup \nv{\hctx}$. We do case analysis on whether $\var \in \varsettwo$:
				\begin{itemize}
					\item Let $\var \in \varsettwo$. By application of the \ih with respect to $\var$ and $\pctxtwo$, there exists $\pctxtwo_{\vartwo} \in \evctxp{\varsettwo_{\vartwo}}$ such that $\var \notin \varsettwo_{\vartwo} \subset \varsettwo$ and $\pctxtwo_{\vartwo} \ctxholep{\var} = \pctxtwop{\vartwo}$. Since $\vartwo \notin \varsettwo \supset \varsettwo_{\var}$, we can then derive $\pctx_{\tm} \in \evctxp{\varset_{\tm}}$ as follows
						$$
						\infer
							[\ruleEvcGc]
							{\appES{\pctxtwo_{\vartwo}}{\esub{\vartwo}{\hctxp{\tm}}} \in \evctxp{\varsettwo_{\vartwo}}}
							{\pctxtwo_{\vartwo} \in \evctxp{\varsettwo_{\vartwo}}
							\quad
							\vartwo \notin \varsettwo_{\vartwo}}
						$$
					where $\pctx_{\tm} = \appES{\pctxtwo_{\vartwo}}{\esub{\vartwo}{\hctxp{\tm}}}$, verifying in particular that
						$$
							\pctx_{\tm} \ctxholep{\var} = \appES{(\pctxtwo_{\vartwo} \ctxholep{\var})}{\esub{\vartwo}{\hctxp{\tm}}} = \appES{(\pctxtwo \ctxholep{\vartwo})}{\esub{\vartwo}{\hctxp{\tm}}} = \pctxp{\tm}
						$$
				
					\item Let $\var \notin \varsettwo$. Since $\var \in \varset = \varsettwo \cup \nv{\hctx}$, then it must be that $\var \in \nv{\hctx}$. By \reflemmap{Rewriting_term_contexts-opencbneed}{three} (Focusing term contexts on needed variables), there exists term context $\hctx_{\tm}$ such that $\var \notin \nv{\hctx_{\tm}} \subset \nv{\hctx}$ and $\hctx_{\tm} \ctxholep{\var} = \hctxp{\tm}$. Thus, we can derive $\pctx_{\tm} \in \evctxp{\varset_{\tm}}$ as follows
						$$
						\infer
							[\ruleEvcHer]
							{\appES{\pctxtwop{\vartwo}}{\esub{\vartwo}{\hctx_{\tm}}} \in \evctxp{\varsettwo \cup \nv{\hctx_{\tm}}}}
							{\pctxtwo \in \evctxp{\varsettwo}
							\quad
							\vartwo \notin \varsettwo}
						$$
					where $\pctx_{\tm} = \appES{\pctxtwop{\vartwo}}{\esub{\vartwo}{\hctx_{\tm}}}$.
											
				\end{itemize}
			
		\end{itemize}

\end{enumerate}

\end{proof}

Given that programs are structured inductively with respect to the length of their environments, we shall later see how the proof that the $\onormalpr{\cdot}$ predicate characterizes the \opencbneednsp-normal forms requires appending or removing ESs while preserving the $\toond$-normality or the $\toond$-reducibility of programs. More concretely, we need the following
\begin{lemma}[Properties of \opencbneednsp-normal forms and ESs]
\hfill

	\begin{enumerate}
		\item
		\label{p:Properties_of_opencbneed-normal_forms_and_ESs-one}
		\emph{Removing ESs does not create $\toond$-redexes:} if $\progentry{\tm}{\env \esub{\var}{\tmtwo}}$ is $\tonnd$-normal, then $\progentry\tm\env$ is $\tonnd$-normal.
		
		\item 
		\label{p:Properties_of_opencbneed-normal_forms_and_ESs-two}
		\emph{Appending ESs that do not create $\toond$-redexes:} let $\progentry{\tm}{\env}$ be a $\tonnd$-normal form such that if $\var \in \nventryp{\tm}{\env}$ then $\tmtwo$ is inert. Then $\progentry{\tm}{\env \esub{\var}{\tmtwo}}$ is also in $\tonnd$-normal form.

	\end{enumerate}

\label{l:Properties_of_opencbneed-normal_forms_and_ESs}
\end{lemma}


\begin{proof}
\hfill

\begin{enumerate}
	\item
	\emph{Removing ESs does not create $\toond$-redexes}: We prove the contrapositive statement, that is, 
		\begin{center}
			If $\progentry{\tm}{\env}$ is not in $\tonnd$-normal form 
			
			then 
			
			$\progentry{\tm}{\env \esub{\var}{\tmtwo}}$ is not in $\tonnd$-normal form.
		\end{center}
	We proceed by case analysis on the kind of $\toond$-reduction step in $\progentry{\tm}{\env}$; namely, whether it is a multiplicative or exponential step:
		\begin{itemize}
			\item
			\emph{Multiplicative step}: Let $\progentry{\tm}{\env} = \pctxp{(\la{\vartwo}{\tmthree}) \tmfour} \toom \pctxentryp{\tmthree}{\esub{\vartwo}{\tmfour}}$, with $\pctx \in \evctxp{\varset}$. Case analysis on whether $\var \in \varset$:
				\begin{itemize}
					\item Let $\var \notin \varset$. Then we may derive $\appES{\pctx}{\esub{\var}{\tmtwo}} \in \evctxp{\varset}$ via the application of the $\ruleEvcGc$ rule, getting that
						$$
							\progentry{\tm}{\env \esub{\var}{\tmtwo}} = (\appES{\pctx}{\esub{\var}{\tmtwo}}) \ctxholep{(\la{\vartwo}{\tmthree}) \tmfour} \toom (\appES{\pctx}{\esub{\var}{\tmtwo}}) \ctxholeentryp{\tmthree}{\esub{\vartwo}{\tmfour}}
						$$ 
					
					\item Let $\var \in \varset$. Case analysis on the shape of $\tmtwo$:
						\begin{itemize}
							\item Let $\tmtwo$ be inert. Then we can derive 
								$$
								\infer
									[\ruleEvcInert]
									{\appES{\pctx}{\esub{\var}{\tmtwo}} \in \evctxp{(\varset \setminus \set{\var}) \cup \nv{\tmtwo}}}
									{\pctx \in \evctxp{\varset}
									\quad
									\var \in \varset}
								$$
							to obtain that
								$$
									\progentry{\tm}{\env \esub{\var}{\tmtwo}} = (\appES{\pctx}{\esub{\var}{\tmtwo}}) \ctxholep{(\la{\vartwo}{\tmthree}) \tmfour} \toom (\appES{\pctx}{\esub{\var}{\tmtwo}}) \ctxholeentryp{\tmthree}{\esub{\vartwo}{\tmfour}}
								$$
							
							\item Let $\tmtwo$ be a non-inert term. First, let $\tmseven \defeq (\la{\vartwo}{\tmthree}) \tmfour$, and note that by an application of \reflemmap{Rewriting_open_evaluation_contexts}{two} (Focusing open evaluation contexts on needed variables) with respect to $\var$ and $\pctx$, there exists $\pctx_{\tmseven} \in \evctxp{\varset_{\tmseven}}$ such that $\var \notin \varset_{\tmseven} \subset \varset$, and that $\pctx_{\tmseven} \ctxholep{\var} = \pctxp{\tmseven}$. Two sub-cases:
							\begin{enumerate}
								\item Let $\tmtwo$ be a value. Then we can derive 
									$$
									\infer
										[\ruleEvcGc]
										{\appES{\pctx_{\tmseven}}{\esub{\var}{\tmtwo}} \in \evctxp{\varset_{\tmseven}}}
										{\pctx_{\tmseven} \in \evctxp{\varset_{\tmseven}}
										\quad
										\var \notin \varset_{\tmseven}}
									$$
								to obtain that
								
								\[\begin{array}{rclcl}
									\progentry{\tm}{\env \esub{\var}{\tmtwo}} 
									& = & \appES{\pctxp{(\la{\vartwo}{\tmthree}) \tmfour}}{\esub{\var}{\tmtwo}} \\
									& = & \appES{\pctx_{\tmseven}\ctxholep{\var}}{\esub{\var}{\tmtwo}} \\
									& = & (\appES{\pctx_{\tmseven}}{\esub{\var}{\tmtwo}}) \ctxholep{\var} \\
									& \tooe & (\appES{\pctx_{\tmseven}}{\esub{\var}{\tmtwo}})\ctxholep{\rename\tmtwo} \\
								\end{array}\]
							
								\item Suppose that  $\tmtwo$ is not a value. Since $\tmtwo$ is also not an inert term, then we may conclude that $\tmtwo$ is a non-normal term. By \reflemma{Redex_in_non-normal_terms-opencbneed} (Redex in non-normal terms), there exists a term context $\hctx_{\tmtwo}$ such that $\tmtwo = \hctx_{\tmtwo} \ctxholep{(\la{\varthree}{\tmfive}) \tmsix}$.  Then we can derive 
									$$
									\infer
										[\ruleEvcHer]
										{\appES{\pctx_{\tmseven} \ctxholep{\var}}{\esub{\var}{\hctx_{\tmtwo}}} \in \evctxp{\varset_{\tmseven} \cup \nv{\hctx_{\tmtwo}}}}
										{\pctx_{\tmseven} \in \varset_{\tmseven}
										\quad
										\var \notin \varset_{\tmseven}}
									$$
								to obtain that
									\[\begin{array}{rclcl}
										\progentry{\tm}{\env \esub{\var}{\tmtwo}} 
										& = & \appES{\pctxp{(\la{\vartwo}{\tmthree}) \tmfour}}{\esub{\var}{\tmtwo}} \\
										& = & \appES{\pctx_{\tmseven}\ctxholep{\var}}{\esub{\var}{\tmtwo}} \\
										& = & \appES{\pctx_{\tmseven}\ctxholep{\var}}{\esub{\var}{\hctx_{\tmtwo} \ctxholep{(\la{\varthree}{\tmfive}) \tmsix}}} \\
										& = & (\appES{\pctx_{\tmseven}\ctxholep{\var}}{\esub{\var}{\hctx_{\tmtwo} }}) \ctxholep{(\la{\varthree}{\tmfive}) \tmsix} \\
										& \toom & (\appES{\pctx_{\tmseven} \ctxholep{\var}}{\esub{\var}{\hctx_{\tmtwo}}}) \ctxholeentryp{\tmfive}{\esub{\varthree}{\tmsix}} \\
									\end{array}\]
									
							\end{enumerate}
							
						\end{itemize}

				\end{itemize}
			
			\item 
			\emph{Exponential step}: Let $\progentry{\tm}{\env} = \pctxp{\vartwo} \tooe \pctxp{\rename{\val}}$, with $\pctx \in \evctxp{\varset}$, $\vartwo \in \dom{\pctx}$, $\pctx(\vartwo) = \val$. By $\alpha$-conversion, we may safely assume that $\vartwo \neq \var$. We now consider two sub-cases, depending on whether $\var \in \varset$:
				\begin{itemize}
					\item If $\var \notin \varset$, then we can derive 
						$$
						\infer
							[\ruleEvcGc]
							{\appES{\pctx}{\esub{\var}{\tmtwo}} \in \evctxp{\varset}}
							{\pctx \in \evctxp{\varset}
							\quad
							\var \notin \varset}
						$$
					to obtain that $\progentry{\tm}{\env \esub{\var}{\tmtwo}} = (\appES{\pctx}{\esub{\var}{\tmtwo}}) \ctxholep{\vartwo} \tooe (\appES{\pctx}{\esub{\var}{\tmtwo}}) \ctxholep{\rename{\val}}$.
					
					\item Let $\var \in \varset$. Case analysis on the shape of $\tmtwo$:
						\begin{itemize}
							\item Let $\tmtwo$ be inert. Then we can derive
								$$
								\infer
									[\ruleEvcInert]
									{\appES{\pctx}{\esub{\var}{\tmtwo}} \in \evctxp{(\varset \setminus \set{\var}) \cup \nv{\tmtwo}}}
									{\pctx \in \evctxp{\varset}
									\quad
									\var \in \varset}
								$$
							to obtain that $\progentry{\tm}{\env \esub{\var}{\tmtwo}} = (\appES{\pctx}{\esub{\var}{\tmtwo}}) \ctxholep{\vartwo} \tooe (\appES{\pctx}{\esub{\var}{\tmtwo}}) \ctxholep{\rename{\val}}$.
							
							\item Let $\tmtwo$ be a non-inert term. First, let $\tmseven \defeq (\la{\vartwo}{\tmthree}) \tmfour$, and note that by an application of \reflemmap{Rewriting_open_evaluation_contexts}{two} (Focusing open evaluation contexts on needed variables) with respect to $\var$ and $\pctx$, there exists $\pctx_{\tmseven} \in \evctxp{\varset_{\tmseven}}$ such that $\var \notin \varset_{\tmseven} \subset \varset$, and that $\pctx_{\tmseven} \ctxholep{\var} = \pctxp{\tmseven}$. There are two sub-cases, depending on the shape of $\tmtwo$:
								\begin{enumerate}
									\item If $\tmtwo$ is a value, then we can derive
										$$
										\infer
											[\ruleEvcGc]
											{\appES{\pctx_{\tmseven}}{\esub{\var}{\tmtwo}} \in \evctxp{\varset_{\tmseven}}}
											{\pctx_{\tmseven} \in \evctxp{\varset_{\tmseven}}
											\quad
											\var \notin \varset_{\tmseven}}
										$$
									to obtain that 
										\[\begin{array}{rclcl}
											\progentry{\tm}{\env \esub{\var}{\tmtwo}}
											& = & \appES{\pctxp{\vartwo}}{\esub{\var}{\tmtwo}} \\
											& = & \appES{\pctx_{\tmseven}\ctxholep{\var}}{\esub{\var}{\tmtwo}} \\
											& = & (\appES{\pctx_{\tmseven}}{\esub{\var}{\tmtwo}})\ctxholep{\var} \\
											& \tooe & (\appES{\pctx_{\tmseven}}{\esub{\var}{\tmtwo}})\ctxholep{\rename{\tmtwo}} \\
										\end{array}\]

								\item Let $\tmtwo$ be a value. Since $\tmtwo$ is also not an inert term, then we may conclude that $\tmtwo$ is a non-normal term. By \reflemma{Redex_in_non-normal_terms-opencbneed} (Redex in non-normal terms), there exists a term context $\hctx_{\tmtwo}$  such that $\hctx_{\tmtwo} \ctxholep{(\la{\varthree}{\tmthree}) \tmfour}$. Then we can derive 
									$$
									\infer
										[\ruleEvcHer]
										{\appES{\pctx_{\tmseven} \ctxholep{\var}}{\esub{\var}{\hctx_{\tmtwo}}} \in \evctxp{\varset_{\tmseven} \cup \nv{\hctx_{\tmtwo}}}}
										{\pctx_{\tmseven} \in \evctxp{\varset_{\tmseven}}
										\quad
										\var \notin \varset_{\tmseven}}
									$$
								to obtain that 
									\[\begin{array}{rclcl}
										\progentry{\tm}{\env \esub{\var}{\tmtwo}} 
										& = & \appES{\pctxp{\vartwo}}{\esub{\var}{\tmtwo}} 	\\
										& = & \appES{\pctx_{\tmseven}\ctxholep{\var}}{\esub{\var}{\tmtwo}} \\
										& = & \appES{\pctx_{\tmseven}\ctxholep{\var}}{\esub{\var}{\hctx_{\tmtwo} \ctxholep{(\la{\varthree}{\tmthree}) \tmfour}}} \\
										& = & (\appES{\pctx_{\tmseven}\ctxholep{\var}}{\esub{\var}{\hctx_{\tmtwo} }}) \ctxholep{(\la{\varthree}{\tmthree}) \tmfour} \\
										& \toom & (\appES{\pctx_{\tmseven} \ctxholep{\var}}{\esub{\var}{\hctx_{\tmtwo}}}) \ctxholeentryp{\tmthree}{\esub{\varthree}{\tmfour}} \\
									\end{array}\]
										
								\end{enumerate}
						\end{itemize}
				
				\end{itemize}
			
		\end{itemize}

	\item
	\emph{Adding ESs that do not create $\toond$-redexes}: We prove the contrapositive statement; that is, that for all $\var \in \Var$ and term $\tmtwo$ satisfying that either $\var \notin \nventryp{\tm}{\env}$ or that $\tmtwo$ is inert, the following holds
			\begin{center}
				If $\progentry{\tm}{\env \esub{\var}{\tmtwo}}$ is not in $\tonnd$-normal form
					
				then
				
				$\progentry{\tm}{\env}$ is not in $\tonnd$-normal form
			\end{center}
			
	Let us assume that $\var \notin \nventryp{\tm}{\env}$ or that $\tmtwo$ is inert. We proceed by case analysis on the kind of $\toond$-reduction step in $\progentry{\tm}{\env \esub{\var}{\tmtwo}}$; namely, whether it is a multiplicative or exponential step:
		\begin{itemize}
			\item Let $\progentry{\tm}{\env \esub{\var}{\tmtwo}} = \pctxp{(\la{\var}{\tmthree})\tmfour} \toom \pctxentryp{\tmthree}{\esub{\var}{\tmfour}}$, with $\pctx \in \evctxp{\varset}$. We proceed by induction on the derivation of $\pctx \in \evctxp{\varset}$: 
				\begin{itemize}
					\item 
					\emph{Rule $\ruleEvcAx$}: This case is clearly not possible.
					
					\item 
					\emph{Rule $\ruleEvcGc$}: Let $\pctx \in \evctxp{\varset}$ be derived as
						$$
						\infer
							[\ruleEvcGc]
							{\appES{\pctxtwo}{\esub{\var}{\tmtwo}} \in \evctxp{\varset}}
							{\pctxtwo \in \evctxp{\varset}
							\quad
							\var \notin \varset}
						$$
					with $\pctx = \appES{\pctxtwo}{\esub{\var}{\tmtwo}}$. Then $\progentry{\tm}{\env} = \pctxtwop{(\la{\var}{\tmthree}) \tmfour} \toom \pctxtwoentryp{\tmthree}{\esub{\var}{\tmfour}}$.
					
					\item 
					\emph{Rule $\ruleEvcInert$}: Let $\pctx \in \evctxp{\varset}$ be derived as
						$$
						\infer
							[\ruleEvcInert]
							{\appES{\pctxtwo}{\esub{\var}{\tmtwo}} \in \evctxp{(\varsettwo \setminus \set{\var}) \cup \nv{\tmtwo}}}
							{\pctxtwo \in \evctxp{\varset}
							\quad
							\var \in \varset}
						$$
					with $\pctx = \appES{\pctxtwo}{\esub{\var}{\tmtwo}}$, $\varset = (\varsettwo \, \setminus \, \set{\var}) \, \cup \, \nv{\tmtwo}$ and $\tmtwo$ inert. Then $\progentry{\tm}{\env} = \pctxtwop{(\la{\var}{\tmthree}) \tmfour} \\ \toom \pctxtwoentryp{\tmthree}{\esub{\var}{\tmfour}}$.
					
					\item 
					\emph{Rule $\ruleEvcHer$}: Suppose $\pctx \in \evctxp{\varset}$ were derived as
						$$
						\infer
							[\ruleEvcHer]
							{\appES{\pctxtwop{\var}}{\esub{\var}{\hctx}} \in \evctxp{\varsettwo \cup \nv{\hctx}}}
							{\pctxtwo \in \evctxp{\varsettwo}
							\quad
							\var \notin \varsettwo}
						$$
					with $\pctx = \appES{\pctxtwop{\var}}{\esub{\var}{\hctx}}$, $\varset = \varsettwo \cup \nv{\hctx}$ and $\tmtwo = \hctxp{(\la{\var}{\tmthree}) \tmfour}$. By \reflemma{Redex_in_non-normal_terms-opencbneed} (Redex in non-normal terms), $\tmtwo$ could not be inert, thus implying by hypothesis that $\var \notin \nventryp{\tm}{\env}$. However, by \reflemmap{Rewriting_open_evaluation_contexts}{one} (Open evaluation contexts give needed variables), we have that $\var \in \nv{\pctxtwop{\var}} = \nventryp{\tm}{\env}$. Hence, this case is absurd.
					
				\end{itemize}
			
			\item Let $\progentry{\tm}{\env \esub{\var}{\tmtwo}} = \pctxp{\vartwo} \tooe \pctxp{\rename{\val}}$, with $\pctx \in \evctxp{\varset}$, $\var \in \dom{\pctx}$ and $\pctx(\var) = \val$. We proceed by case analysis on the last rule applied in $\pctx \in \evctxp{\varset}$:
				\begin{itemize}
					\item 
					\emph{Rule $\ruleEvcAx$}: This case is clearly not possible.
					
					\item
					\emph{Rule $\ruleEvcGc$}: Let $\pctx \in \evctxp{\varset}$ be derived as
						$$
						\infer
							[\ruleEvcGc]
							{\appES{\pctxtwo}{\esub{\var}{\tmtwo}} \in \evctxp{\varset}}
							{\pctxtwo \in \evctxp{\varset}
							\quad
							\var \notin \varset}
						$$
					with $\pctx = \appES{\pctxtwo}{\esub{\var}{\tmtwo}}$. 
					
					Suppose $\vartwo = \var$, and so $\tmtwo = \val$. That is, suppose $\tmtwo$ is not inert, which would imply that $\var \notin \nventryp{\tm}{\env}$. We would then have that $\progentry{\tm}{\env \esub{\var}{\tmtwo}} = \appES{\progentry{\tm}{\env}}{\esub{\var}{\tmtwo}} = \appES{\pctxtwop{\var}}{\esub{\var}{\tmtwo}}$; that is, $\progentry{\tm}{\env} = \pctxtwop{\var}$. However, this would imply by \reflemmap{Rewriting_open_evaluation_contexts}{one} (Open evaluation contexts give needed variables) that $\var \in \nventryp{\tm}{\env}$, which would contradict the hypothesis.
					
					Therefore, it must be that $\vartwo \neq \var$, which implies that $\vartwo \in \dom{\pctxtwo}$ and $\pctxtwo(\vartwo) = \val$, and so $\progentry{\tm}{\env} = \pctxtwop{\vartwo} \tooe \pctxtwop{\rename{\val}}$.
					
					\item 
					\emph{Rule $\ruleEvcInert$}: Let $\pctx \in \evctxp{\varset}$ be derived as
						$$
						\infer
							[\ruleEvcInert]
							{\appES{\pctxtwo}{\esub{\var}{\tmtwo}} \in \evctxp{(\varset \setminus \set{\var}) \cup \nv{\tmtwo}}}
							{\pctxtwo \in \evctxp{\varset}
							\quad
							\var \in \varset}
						$$
					with $\pctx = \appES{\pctxtwo}{\esub{\var}{\tmtwo}}$ and $\tmtwo$ inert. This means that $\tmtwo$ is not a value, and so we have that $\vartwo \in \dom{\pctxtwo}$ and $\pctxtwo(\vartwo) = \val$. That is, $\progentry{\tm}{\env} = \pctxtwop{\vartwo} \tooe \pctxtwop{\rename{\val}}$.
					
					\item 
					\emph{Rule $\ruleEvcHer$}: Suppose $\pctx \in \evctxp{\varset}$ were derived as
						$$
						\infer
							[\ruleEvcHer]
							{\appES{\pctxtwop{\var}}{\esub{\var}{\hctx}} \in \evctxp{\varset \cup \nv{\hctx}}}
							{\pctxtwo \in \evctxp{\varset}
							\quad
							\var \in \varset}
						$$
					with $\pctx = \appES{\pctxtwop{\var}}{\esub{\var}{\hctx}}$ and $\tmtwo = \hctxp{\vartwo}$. But then $\vartwo \notin \dom{\pctx}$, and so this case is absurd.\qedhere
				
				\end{itemize}
			
		\end{itemize}

\end{enumerate}

\end{proof}

In particular, note that the contrapositive of {\reflemmap{Properties_of_opencbneed-normal_forms_and_ESs}{one}} and {\reflemmap{Properties_of_opencbneed-normal_forms_and_ESs}{two}} show how $\toond$-reducibility---\ie, the property of not being a $\toond$-normal form---is preserved when appending or removing an ES.

With \reflemma{Properties_of_opencbneed-normal_forms_and_ESs} (Properties of \opencbneednsp-normal forms), we can now lift \reflemmap{Rewriting_term_contexts-opencbneed}{two} (Focusing inert terms on needed variables) to program contexts as follows
\begin{lemma}[\opencbneednsp-normal forms and needed variables]
\hfill

Let $\prog$ be a program in $\tonnd$-normal form and let $\var \in \nv{\prog}$. Then there exists $\pctx \in \evctxp{\varset}$ such that $\pctx \ctxholep{\var} = \prog$, $\var \notin \varset \subset \nv{\prog}$ and $\var \notin \dom{\pctx}$.

\label{l:opencbneed-normal_forms_and_needed_variables}
\end{lemma}


\begin{proof}
\hfill

Let $\prog = \progentry{\tm}{\env}$. We proceed by induction on the length of $\env$.
	\begin{itemize}
		\item Let $\env = \emptyenv$. That is, $\prog = \progentry{\tm}{\emptyenv}$. First, let us suppose that $\tm$ were a non-normal term. By \reflemma{Redex_in_non-normal_terms-opencbneed} (Redex in non-normal terms), there would exist a term context $\hctx$ such that $\tm = \hctxp{(\la{\vartwo}{\tmthree}) \tmfour}$. But then $\progentry{\hctx}{\emptyenv} \in \evctxp{\nv{\hctx}}$ and so $\prog = \progentry{\hctx}{\emptyenv} \ctxholep{(\la{\vartwo}{\tmthree}) \tmfour} \tonm \progentry{\hctx}{\emptyenv} \ctxholeentryp{\tmthree}{\esub{\vartwo}{\tmfour}}$, which is absurd. 
		
		Thus, $\tm$ must be a normal term. Moreover, if $\tm$ is a value, then $\nv{\tm}$ and the statement holds trivially.

		Finally, let $\tm$ be an inert term and $\var \in \nv{\tm}$. By \reflemmap{Rewriting_term_contexts-opencbneed}{two} (Rewriting term contexts), there exists a term context $\hctx_{\var}$ such that $\var \notin \nv{\hctx_{\var}} \subset \nv{\tm}$ and $\hctx_{\var} \ctxholep{\var} = \tm$. The statement follows by taking $\pctx_{\var} \defeq \progentry{\hctx_{\var}}{\emptyenv} \in \evctxp{\nv{\hctx_{\var}}}$ verifies the statement, since $\var \notin \nv{\hctx_{\var}} \subset \nv{\tm} = \nv{\prog}$ and $\progentry{\hctx_{\var}}{\emptyenv} \ctxholep{\var} = \progentry{\hctx_{\var}\ctxholep{\var}}{\emptyenv} = \progentry{\tm}{\emptyenv} = \prog$.
		
		\item Let $\env = \envtwo \esub{\vartwo}{\tmtwo}$. Note that $\var \neq \vartwo$, because $\var \in \nv{\prog}$ and $\nv{\prog} \cap \dom{\prog} = \emptyset$---easily provable for every program $\prog$.

		We proceed by case analysis on whether $\var \in \nv{\prog}$:
			\begin{itemize}
				\item Let $\var \in \nventryp{\tm}{\envtwo}$. By \reflemmap{Properties_of_opencbneed-normal_forms_and_ESs}{one} (Properties of \opencbneednsp-normal forms and ESs), $\progentry{\tm}{\envtwo}$ is in $\tonnd$-normal form, and so we can apply the \ih on it to obtain $\pctxtwo \in \evctxp{\varsettwo}$ such that $\pctxtwo \ctxholep{\var} = \progentry{\tm}{\envtwo}$, $\var \notin \varsettwo \subset \nventryp{\tm}{\envtwo}$, and $\var \notin \dom{\pctxtwo}$. 
						
				Note that if $\vartwo \notin \varsettwo$ then we may derive $\pctx \in \evctxp{\varset}$ as
					$$
					\infer
						[\ruleEvcGc]
						{\appES{\pctxtwo}{\esub{\vartwo}{\tmtwo}} \in \evctxp{\varsettwo}}
						{\pctxtwo \in \evctxp{\varsettwo}
						\quad
						\vartwo \notin \varsettwo}
					$$
				where $\varset = \varsettwo \subset \nventryp{\tm}{\envtwo} = \nventryp{\tm}{\envtwo \esub{\vartwo}{\tmtwo}}$.

				Let us now consider the case where $\vartwo \in \varsettwo$. By \reflemmap{Rewriting_open_evaluation_contexts}{two} (Focusing open evaluation contexts on needed variables), there exists $\pctxtwo_{\var} \in \evctxp{\varsettwo_{\var}}$ such that $\vartwo \notin \varsettwo_{\var} \subset \varsettwo$ and $\pctxtwo_{\var} \ctxholep{\vartwo} = \pctxtwo \ctxholep{\var} = \progentry{\tm}{\envtwo}$. We proceed by case analysis on the shape of $\tmtwo$:
					\begin{itemize}
						\item Let $\tmtwo$ be inert. Case analysis on whether $\var \in \nv{\tmtwo}$:
							\begin{enumerate}
								\item $\var \in \nv{\tmtwo}$: by \reflemmap{Rewriting_term_contexts-opencbneed}{two} (Rewriting term contexts), there exists a term context $\hctx_{\var}$ such that $\var \notin \nv{\hctx_{\var}} \subset \nv{\tmtwo}$ and $\hctx_{\var} \ctxholep{\var} = \tmtwo$. We may then derive $\pctx \in \evctxp{\varset}$ as follows:
										$$
										\infer
											[\ruleEvcHer]
											{\appES{\pctxtwo_{\var} \ctxholep{\vartwo}}{\esub{\vartwo}{\hctx_{\var}}} \in \evctxp{\varsettwo_{\var} \cup \nv{\hctx_{\var}}}}
											{\pctxtwo_{\var} \in \evctxp{\varsettwo_{\var}}
											\quad
											\vartwo \notin \varsettwo_{\var}}
										$$
								where $\varset = \varsettwo_{\var} \cup \nv{\hctx_{\var}}$. We thus verify  that 
									\[\begin{array}{rcl}
										\varset
										& = & \varsettwo_{\var} \cup \nv{\hctx_{\var}} \\
										& \subset & \nventryp{\tm}{\envtwo} \cup \nv{\tmtwo} \\
										& = & \nventryp{\tm}{\envtwo \esub{\vartwo}{\tmtwo}}
									\end{array}\]
								Moreover, since $\var \notin \varsettwo_{\var}$ and $\var \notin \nv{\hctx_{\var}}$, then $\var \notin \varset$.

								\item Let $\var\notin\nv\tmtwo$. Then, we may derive $\pctx \in \evctxp{\varset}$ as follows
									$$
									\infer
										[\ruleEvcInert]
										{\appES{\pctxtwo}{\esub{\vartwo}{\tmtwo}} \in \evctxp{(\varsettwo \setminus \set{\vartwo}) \cup \nv{\tmtwo}}}
										{\pctxtwo \in \evctxp{\varsettwo}
										\quad
										\vartwo \in \varsettwo}
									$$				
								where $\varset = (\varsettwo \setminus \set{\vartwo}) \cup \nv{\tmtwo}$, and verifying that 
									\[\begin{array}{rcl}
										\varset
										& = & (\varsettwo \setminus \set{\vartwo}) \cup \nv{\tmtwo} \\
										& \subset &  (\nventryp{\tm}{\envtwo} \setminus \set{\vartwo}) \cup \nv{\tmtwo} \\
										& \subseteq & \nventryp{\tm}{\envtwo \esub{\vartwo}{\tmtwo}} \\
									\end{array}\]
									
								Note that the last inequality holds regardlessly of whether $\vartwo \in \nventryp{\tm}{\envtwo}$ or not. Moreover, since $\var \notin \varsettwo$ and $\var \notin \nv{\tmtwo}$, then $\var \notin \varset$.	
								
								\end{enumerate}
								
							\item Suppose that $\tmtwo$ is a value. Then we would be able to derive
								$$
								\infer
									[\ruleEvcGc]
									{\appES{\pctxtwo_{\vartwo}}{\esub{\vartwo}{\tmtwo}} \in \evctxp{\varsettwo_{\vartwo}}}
									{\pctxtwo_{\vartwo} \in \evctxp{\varsettwo_{\vartwo}}
									\quad
									\vartwo \notin \varsettwo_{\vartwo}}
								$$
							and thus obtain that $\prog = (\appES{\pctxtwo}{\esub{\vartwo}{\tmtwo}}) \ctxholep{\vartwo} \tone (\appES{\pctxtwo}{\esub{\vartwo}{\tmtwo}}) \ctxholep{\rename{\tmtwo}}$, which is absurd.			

							\item Suppose $\tmtwo$ were not a normal term. Then, by \reflemma{Redex_in_non-normal_terms-opencbneed} (Redex in non-normal terms), there would exist a term context $\hctx$ such that $\hctx \ctxholep{(\la{\varthree}{\tmthree}) \tmfour} = \tmtwo$. But then we would be able to derive
								$$
								\infer
									[\ruleEvcHer]
									{\appES{\pctxtwo \ctxholep{\vartwo}}{\esub{\vartwo}{\hctx_{\tmtwo}}} \in \evctxp{\varsettwo \cup \nv{\hctx_{\tmtwo}}}}
									{\pctxtwo \in \evctxp{\varsettwo}
									\quad
									\vartwo \notin \varsettwo}
								$$
							and thus obtain that
								$$
									\prog = (\appES{\pctxtwo \ctxholep{\vartwo}}{\esub{\vartwo}{\hctx}}) \ctxholep{(\la{\varthree}{\tmthree}) \tmfour} \tonm (\appES{\pctxtwo \ctxholep{\vartwo}}{\esub{\vartwo}{\hctx}}) \ctxholeentryp{\tmthree}{\esub{\varthree}{\tmfour}}
								$$
							which is absurd.
								
						\end{itemize}
						
				\item Let $\var \notin \nventryp{\tm}{\envtwo}$. Note that since $\var \in \nventryp{\tm}{\envtwo \esub{\vartwo}{\tmtwo}}$ by hypothesis, then it must be that $\vartwo \in \nventryp{\tm}{\envtwo}$, thus having that $\nventryp{\tm}{\envtwo \esub{\vartwo}{\tmtwo}} = (\nventryp{\tm}{\envtwo} \setminus \set{\vartwo}) \cup \nv{\tmtwo}$ and so $\var \in \nv{\tmtwo}$.
				
				By application of the \ih on $\vartwo$ and $\progentry{\tm}{\envtwo}$, there exists $\pctxtwo \in \evctxp{\varsettwo}$ such that $\pctxtwo \ctxholep{\vartwo} = \progentry{\tm}{\envtwo}$, $\vartwo \notin \varsettwo \subset \nventryp{\tm}{\envtwo}$, and $\vartwo \notin \dom{\pctxtwo}$. We proceed by case analysis on the shape of $\tmtwo$:
					\begin{itemize}
						\item Let $\tmtwo$ be inert. By \reflemmap{Rewriting_term_contexts-opencbneed}{two} (Term contexts and needed variables), there exists term context $\hctx_{\var}$ such that $\var \notin \nv{\hctx_{\var}} \subset \nv{\tmtwo}$ and $\hctx_{\var} \ctxholep{\var} = \tmtwo$. We may then derive $\pctx \in \evctxp{\varset}$ as follows
							$$
							\infer
								[\ruleEvcHer]
								{\appES{\pctxtwo \ctxholep{\vartwo}}{\esub{\vartwo}{\hctx_{\var}}} \in \evctxp{\varsettwo \cup \nv{\hctx_{\var}}}}
								{\pctxtwo \in \evctxp{\varsettwo}
								\quad
								\vartwo \notin \varsettwo}
							$$
						
						where $\pctx = \appES{\pctxtwo \ctxholep{\vartwo}}{\esub{\vartwo}{\hctx_{\var}}}$ and $\varset = \varsettwo \cup \nv{\hctx_{\var}}$, and verifying that
							\[\begin{array}{rcl}
								\varset
								& = & \varsettwo \cup \nv{\hctx_{\var}} \\
								& = & (\varsettwo \setminus \set{\vartwo}) \cup \nv{\hctx_{\var}} \\
								& \subset & (\nventryp{\tm}{\envtwo} \setminus \set{\vartwo}) \cup \nv{\tmtwo} \\
								& = & \nventryp{\tm}{\envtwo \esub{\vartwo}{\tmtwo}} \\
							\end{array}\]
						
						Moreover, since $\var \notin \varsettwo$---because $\varsettwo \subset \nventryp{\tm}{\envtwo}$ and $\var \notin \nv{\hctx_{\var}}$ either, then $\var\notin \varset$.
									
						\item Suppose that $\tmtwo$ is a value. We would then be able to derive
							$$
							\infer
								[\ruleEvcGc]
								{\appES{\pctxtwo}{\esub{\vartwo}{\tmtwo}} \in \evctxp{\varsettwo}}
								{\pctxtwo \in \evctxp{\varsettwo}
								\quad
								\vartwo \notin \varsettwo}
							$$
						and thus obtain that
							$$
								\prog = (\appES{\pctxtwo}{\esub{\vartwo}{\tmtwo}}) \ctxholep{\vartwo} \tone (\appES{\pctxtwo}{\esub{\vartwo}{\tmtwo}}) \ctxholep{\rename{\tmtwo}}
							$$
						which is absurd.
										
						\item Suppose $\tmtwo$ were not a normal term. By \reflemma{Redex_in_non-normal_terms-opencbneed} (Redex in non-normal terms), there would exist a term context $\hctx$ such that $\hctxp{(\la{\varthree}{\tmthree}) \tmfour} = \tmtwo$. But then we would be able to derive
							$$
							\infer
								[\ruleEvcHer]
								{\appES{\pctxtwo \ctxholep{\vartwo}}{\esub{\vartwo}{\hctx}} \in \evctxp{\varsettwo \cup \nv{\hctx}}}
								{\pctxtwo \in \evctxp{\varsettwo}
								\quad
								\vartwo \notin \varsettwo}
							$$
						and thus obtain that
							$$
								\prog = (\appES{\pctxtwo \ctxholep{\vartwo}}{\esub{\vartwo}{\hctx}}) \ctxholep{(\la{\varthree}{\tmthree}) \tmfour} \tonm (\appES{\pctxtwo \ctxholep{\vartwo}}{\esub{\vartwo}{\hctx}}) \ctxholeentryp{\tmthree}{\esub{\varthree}{\tmfour}}
							$$
						which is absurd.\qedhere
								
					\end{itemize}
				
			\end{itemize}
		
	\end{itemize}

\end{proof}

\gettoappendix{prop:Syntactic_characterization_of_opencbneed-normal_forms}

\begin{proof}
\item[$\Rightarrow$]: Let $\prog = \progentry{\tm}{\env}$ be in $\tonnd$-normal form. We prove that $\onormalpr{\prog}$ proceeding by induction on the length of $\env$:
	\begin{itemize}
	\item Let $\env = \emptyenv$. That is, $\prog = \progentry{\tm}{\env} = \progentry{\tm}{\emptyenv}$. First, note that if we suppose that $\tm$ is non-normal term, then an application of \reflemma{Redex_in_non-normal_terms-opencbneed} (Redex in non-normal terms) would yield the existence of a term context $\hctx$ such that $\tm = \hctxp{(\la{\var}{\tmtwo})\tmthree}$ and so obtain that
		\begin{center}
			$
			\begin{array}{rcl}
				\prog
				& = & \progentry{\tm}{\env} \\
				& = & \progentry{\hctx \ctxholep{(\la{\var}{\tmtwo})\tmthree}}{\emptyenv} \\
				& = & \progentry{\hctx}{\emptyenv} \ctxholep{(\la{\var}{\tmtwo})\tmthree} \\
				& \toond & \progentry{\hctx}{\emptyenv} \ctxholeentryp{\tmtwo}{\esub{\var}{\tmthree}}
			\end{array}
			$
		\end{center}
	which is absurd. Hence, $\tm$ must be a normal term. Thus, if $\tm$ is an inert term we may derive
		$$
		\infer
			[\ruleInertAx]
			{\inertentrypr{\tm}{\emptyenv}}
			{}
		$$
	and if $\tm$ is a value we may derive
		$$
		\infer
			[\ruleAbsAx]
			{\absentrypr{\tm}{\emptyenv}}
			{}
		$$
	In either case, we conclude that $\onormalentrypr{\tm}{\emptyenv}$.
		
	\item Let $\env = \envtwo \esub{\vartwo}{\tmtwo}$. That is, $\prog = \progentry{\tm}{\env} = \progentry{\tm}{\envtwo \esub{\vartwo}{\tmtwo}}$. Note that $\progentry{\tm}{\envtwo}$ is itself in $\tonnd$-normal form---by \reflemmap{Properties_of_opencbneed-normal_forms_and_ESs}{one} (Properties of \opencbneednsp-normal forms and ESs). Hence, there exists a derivation of $\onormalentrypr{\tm}{\envtwo}$---by \ih 

	First, note that if $\absentrypr{\tm}{\envtwo}$, then we may derive
		$$
		\infer
			[\ruleAbsGc]
			{\absentrypr{\tm}{\envtwo \esub{\vartwo}{\tmtwo}}}
			{\absentrypr{\tm}{\envtwo}}
		$$
	and conclude that $\onormalentrypr{\tm}{\envtwo}$.

	Let us now consider the case of $\inertentrypr{\tm}{\envtwo}$, and prove the statement by case analysis on whether $\vartwo \in \nventryp{\tm}{\envtwo}$:
		\begin{itemize}
			 \item Let $\vartwo \notin \nventryp{\tm}{\envtwo}$. Thus, we may derive $\inertentrypr{\tm}{\envtwo \esub{\vartwo}{\tmtwo}}$ as follows
				$$
				\infer
					[\ruleInertGc]
					{\inertentrypr{\tm}{\envtwo \esub{\vartwo}{\tmtwo}}}
					{\inertentrypr{\tm}{\envtwo}
					\quad
					\vartwo \notin \nventryp{\tm}{\envtwo}}
				$$
			concluding that $\onormalentrypr{\tm}{\env}$.

		   	\item Let $\vartwo \in \nventryp{\tm}{\envtwo}$. By \reflemma{opencbneed-normal_forms_and_needed_variables} (\opencbneednsp-normal forms and needed variables), there exists an $\pctx \in \evctxp{\varset}$ such that $\pctx \ctxholep{\vartwo} = \progentry{\tm}{\envtwo} $, $\vartwo \notin \varset \subset \varsettwo$, and $\vartwo \notin \dom{\pctx}$. We now proceed by case analysis on the shape of $\tmtwo$:
				\begin{itemize}
					\item If $\tmtwo$ is an inert term, then we may derive $\inertentrypr{\tm}{\envtwo \esub{\vartwo}{\tmtwo}}$ as follows
						$$
						\infer
							[\ruleInertInert]
							{\inertentrypr{\tm}{\envtwo \esub{\vartwo}{\tmtwo}}}
							{\inertentrypr{\tm}{\envtwo}
							\quad
							\vartwo \in \nventryp{\tm}{\envtwo}}
						$$
					concluding that $\onormalentrypr{\tm}{\envtwo \esub{\vartwo}{\tmtwo}}$.
						
					\item Suppose $\tmtwo$ were a value. We would then be able to derive
						$$
						\infer
							[\ruleEvcGc]
							{\appES{\pctx}{\esub{\vartwo}{\tmtwo}} \in \evctxp{\varset}}
							{\pctx \in \evctxp{\varset}
							\quad
							\vartwo \notin \varset}
						$$
					to obtain that
						$$
							 \progentry{\tm}{\envtwo \esub{\vartwo}{\tmtwo}} = \appES{\pctx}{\esub{\vartwo}{\tmtwo}} \ctxholep{\vartwo} \tone \appES{\pctx}{\esub{\vartwo}{\tmtwo}} \ctxholep{\rename{\tmtwo}}
						$$
					which is absurd.
					
					\item Suppose $\tmtwo$ were not a normal term. Then there would exist a term context $\hctx$ such that $\hctxp{(\la{\varthree}{\tmthree}) \tmfour}$---by \reflemma{Redex_in_non-normal_terms-opencbneed} (Redex in non-normal terms). But then we would be able to derive
						$$
						\infer
							[\ruleEvcHer]
							{\appES{\pctx \ctxholep{\vartwo}}{\esub{\vartwo}{\hctx}} \in \evctxp{\varset \cup \nv{\hctx}}}
							{\pctx \in \evctxp{\varset}
							\quad
							\vartwo \notin \varset}
						$$
					and obtain that
						$$
							\progentry{\tm}{\envtwo \esub{\vartwo}{\tmtwo}} = (\appES{\pctx \ctxholep{\vartwo}}{\esub{\vartwo}{\hctx}}) \ctxholep{(\la{\varthree}{\tmthree}) \tmfour} \tonm (\appES{\pctxp{\vartwo}}{\esub{\vartwo}{\hctx}}) \ctxholeentryp{\tmthree}{\esub{\varthree}{\tmfour}}
						$$
					which is absurd.
					
				\end{itemize}
			
		\end{itemize}
		
\end{itemize}
	
\item[$\Leftarrow$]: Let $\onormalpr{\prog}$. Case analysis on the predicate from which we derive $\onormalpr{\prog}$:
	\begin{itemize}
		\item 
		\emph{Inert predicate}: We proceed by induction on the derivation of $\inertpr{\prog}$, proceeding by case analysis on the last applied derivation rule:
			\begin{itemize}
				\item
				\emph{Rule $\ruleInertAx$}: Let $\inertentrypr{\tm}{\env}$ be derived as follows
					$$
					\infer
						[\ruleInertAx]
						{\inertentrypr{\itm}{\emptyenv}}
						{}
					$$
				where $\prog = \progentry{\itm}{\emptyenv}$. Note that $\prog$ is in $\tooe$-normal form because the environment is empty. By \reflemma{Redex_in_non-normal_terms-opencbneed} (Redex in non-normal terms), we also have that $\prog$ is in $\toom$-normal form. Therefore, $\prog$ is in $\toond$-normal form.

				\item
				\emph{Rule $\ruleInertInert$}: Let $\inertentrypr{\tm}{\env}$ be derived as follows
					$$
					\infer
						[\ruleInertInert]
						{\inertentrypr{\tm}{\envtwo \esub{\vartwo}{\tmtwo}}}
						{\inertentrypr{\tm}{\envtwo}
						\quad
						\vartwo \in \nventryp{\tm}{\envtwo}}
					$$
				where $\prog = \progentry{\tm}{\envtwo \esub{\vartwo}{\tmtwo}}$ and $\tmtwo$ is an inert term. By \ih, $\progentry{\tm}{\envtwo}$ is in $\toond$-normal form.

				Firstly, since $\tmtwo$ is not a value, note that there cannot exist $\appES{\pctx}{\esub{\vartwo}{\tmtwo}} \in \evctxp{\varset}$---for some $\varset$---with which we would have that
					$$
					\prog = (\appES{\pctx}{\esub{\vartwo}{\tmtwo}}) \ctxholep{\vartwo} \tooe (\appES{\pctx}{\esub{\vartwo}{\tmtwo}}) \ctxholep{\rename{\tmtwo}}
					$$
				Said differently, $\prog$ is in $\tooe$-normal form.

				Secondly, by \reflemma{Redex_in_non-normal_terms-opencbneed} (Redex in non-normal terms), there cannot exist $\appES{\pctxp{\vartwo}}{\esub{\vartwo}{\hctx}} \in \evctxp{\varset}$---for some $\varset$---such that $\tmtwo = \hctxp{(\la{\varthree}{\tmthree}) \tmfour}$---for some $((\la{\varthree}{\tmthree}) \tmfour)$---which would give that 
					$$
					\begin{array}{rcl}
						\prog & = & \appES{\pctxp{\vartwo}}{\esub{\vartwo}{\hctxp{(\la{\varthree}{\tmthree}) \tmfour}}} \\
						& = & (\appES{\pctxp{\vartwo}}{\esub{\vartwo}{\hctx}}) \ctxholep{(\la{\varthree}{\tmthree}) \tmfour} \\
						& \toom & (\appES{\pctxp{\vartwo}}{\esub{\vartwo}{\hctx}}) \ctxholeentryp{\tmthree}{\esub{\varthree}{\tmfour}}
					\end{array}
					$$
				Said differently, $\prog$ is also in $\toom$-normal form. We may thus conclude that $\prog$ is in $\toond$-normal form.

				\item
				\emph{Rule $\ruleInertGc$}: Let $\inertentrypr{\tm}{\env}$ be derived as follows
					$$
					\infer
						[\ruleInertGc]
						{\inertentrypr{\tm}{\envtwo \esub{\vartwo}{\tmtwo}}}
						{\inertentrypr{\tm}{\envtwo}
						\quad
						\vartwo \notin \nventryp{\tm}{\envtwo}}
					$$
				where $\prog = \progentry{\tm}{\envtwo \esub{\vartwo}{\tmtwo}}$. By \ih, $\progentry{\tm}{\envtwo}$ is in $\toond$-normal form. Let us separately consider the $\toond$-reducibility of $\prog$:
					\begin{itemize}
						\item Suppose there existed $\appES{\pctx}{\esub{\vartwo}{\tmtwo}} \in \evctxp{\varset}$---for some $\varset$---such that $\progentry{\tm}{\envtwo \esub{\vartwo}{\tmtwo}} = \appES{\pctxp{\vartwo}}{\esub{\vartwo}{\tmtwo}}$ which would gives us that
							$$
							\begin{array}{rcl}
								\prog & = & \appES{\pctxp{\vartwo}}{\esub{\vartwo}{\tmtwo}} \\
								& = & (\appES{\pctx}{\esub{\vartwo}{\tmtwo}}) \ctxholep{\vartwo} \\
								& \tooe & (\appES{\pctx}{\esub{\vartwo}{\tmtwo}}) \ctxholep{\rename{\tmtwo}}
							\end{array}
							$$
						with $\tmtwo$ a value. However, \reflemmap{Rewriting_open_evaluation_contexts}{two} (Open evaluation contexts give needed variables) would then give that  $\vartwo \in \nventryp{\tm}{\envtwo} = \pctxp{\vartwo}$, which contradicts the hypothesis that $\vartwo \notin \nventryp{\tm}{\envtwo}$. Therefore, this case is impossible.

						\item Suppose there existed $\appES{\pctxp{\vartwo}}{\esub{\vartwo}{\hctx}} \in \evctxp{\varset}$---for some $\varset$---such that $\progentry{\tm}{\envtwo \esub{\vartwo}{\tmtwo}} = \appES{\pctxp{\vartwo}}{\esub{\vartwo}{\hctxp{(\la{\varthree}{\tmthree}) \tmfour}}}$---\ie, with $\tmtwo = \hctxp{(\la{\varthree}{\tmthree}) \tmfour}$---which would give us that
							$$
							\begin{array}{rcl}
								\prog & = & \appES{\pctxp{\vartwo}}{\esub{\vartwo}{\tmtwo}} \\
								& = & \appES{\pctx}{\esub{\vartwo}{\hctxp{(\la{\varthree}{\tmthree}) \tmfour}}} \\
								& = & (\appES{\pctx}{\esub{\vartwo}{\hctx}}) \ctxholep{(\la{\varthree}{\tmthree}) \tmfour} \\
								& \tooe & (\appES{\pctx}{\esub{\vartwo}{\hctx}}) \ctxholeentryp{\tmthree}{\esub{\varthree}{\tmfour}} \\
							\end{array}
							$$
						However, \reflemmap{Rewriting_open_evaluation_contexts}{two} (Open evaluation contexts give needed variables) would then give that  $\vartwo \in \nventryp{\tm}{\envtwo} = \pctxp{\vartwo}$, which contradicts the hypothesis that $\vartwo \notin \nventryp{\tm}{\envtwo}$. Therefore, this case is also impossible. 

					\end{itemize}

				We may thus conclude that $\progentry{\tm}{\envtwo \esub{\vartwo}{\tmtwo}} = \prog$ is in $\toond$-normal form.

			\end{itemize}

		\item
		\emph{Abstraction predicate}: We proceed by induction on the derivation of $\abspr{\prog}$, proceeding by case analysis on the last applied derivation rule:
			\begin{itemize}
				\item
				\emph{Rule $\ruleAbsAx$}: Let $\absentrypr{\tm}{\env}$ be derived as follows
					$$
					\infer
						[\ruleAbsAx]
						{\absentrypr{\val}{\emptyenv}}
						{}
					$$
				where $\prog = \progentry{\val}{\emptyenv}$. Note that $\prog$ is in $\tooe$-normal form because the environment is empty. By \reflemma{Redex_in_non-normal_terms-opencbneed} (Redex in non-normal terms), we also have that $\prog$ is in $\toom$-normal form. Therefore, $\progentry{\val}{\emptyenv} = \prog$ is in $\toond$-normal form.

				\item
				\emph{Rule $\ruleAbsGc$}: Let $\absentrypr{\tm}{\env}$ be derived as follows
					$$
					\infer
						[\ruleAbsGc]
						{\absentrypr{\tm}{\envtwo \esub{\vartwo}{\tmtwo}}}
						{\absentrypr{\tm}{\envtwo}}
					$$
				where $\prog = \progentry{\tm}{\envtwo \esub{\vartwo}{\tmtwo}}$. By \ih, $\progentry{\tm}{\envtwo}$ is in $\toond$-normal form. Let us separately consider the $\toond$-reducibility of $\prog$:
					\begin{itemize}
						\item Suppose there existed $\appES{\pctx}{\esub{\vartwo}{\tmtwo}} \in \evctxp{\varset}$---for some $\varset$---such that $\progentry{\tm}{\envtwo \esub{\vartwo}{\tmtwo}} = \appES{\pctxp{\vartwo}}{\esub{\vartwo}{\tmtwo}}$ which would gives us that
							$$
							\begin{array}{rcl}
								\prog & = & \appES{\pctxp{\vartwo}}{\esub{\vartwo}{\tmtwo}} \\
								& = & (\appES{\pctx}{\esub{\vartwo}{\tmtwo}}) \ctxholep{\vartwo} \\
								& \tooe & (\appES{\pctx}{\esub{\vartwo}{\tmtwo}}) \ctxholep{\rename{\tmtwo}}
							\end{array}
							$$
						with $\tmtwo$ a value. However, \reflemmap{Rewriting_open_evaluation_contexts}{two} (Open evaluation contexts give needed variables) would then give that  $\vartwo \in \nventryp{\tm}{\envtwo} = \pctxp{\vartwo}$, which is absurd because abstraction programs do not have needed variables. Therefore, this case is impossible.

						\item Suppose there existed $\appES{\pctxp{\vartwo}}{\esub{\vartwo}{\hctx}} \in \evctxp{\varset}$---for some $\varset$---such that $\progentry{\tm}{\envtwo \esub{\vartwo}{\tmtwo}} = \appES{\pctxp{\vartwo}}{\esub{\vartwo}{\hctxp{(\la{\varthree}{\tmthree}) \tmfour}}}$---\ie, with $\tmtwo = \hctxp{(\la{\varthree}{\tmthree}) \tmfour}$---which would give us that
							$$
							\begin{array}{rcl}
								\prog & = & \appES{\pctxp{\vartwo}}{\esub{\vartwo}{\tmtwo}} \\
								& = & \appES{\pctx}{\esub{\vartwo}{\hctxp{(\la{\varthree}{\tmthree}) \tmfour}}} \\
								& = & (\appES{\pctx}{\esub{\vartwo}{\hctx}}) \ctxholep{(\la{\varthree}{\tmthree}) \tmfour} \\
								& \tooe & (\appES{\pctx}{\esub{\vartwo}{\hctx}}) \ctxholeentryp{\tmthree}{\esub{\varthree}{\tmfour}} \\
							\end{array}
							$$
						However, \reflemmap{Rewriting_open_evaluation_contexts}{two} (Open evaluation contexts give needed variables) would then give that  $\vartwo \in \nventryp{\tm}{\envtwo} = \pctxp{\vartwo}$, which is absurd because abstraction programs do not have needed variables. Therefore, this case is also impossible.\qedhere

					\end{itemize}

			\end{itemize}

	\end{itemize}

\end{proof}



\newpage


\renewcommand{\proofspath}{\rootpath/98_Appendix/09_usefulopencbneed/Proofs}

\section{Proofs of \refsect{useful-open-cbneed} (\usefulopencbneednsp)}
\label{sect:Proofs_of_usefulopencbneed}

\gettoappendix{l:Unapplied_applied_and_needed_variables}

\begin{proof}	
\hfill
\begin{enumerate}
	\item \emph{Terms}: By induction on the shape of $\tm$:
		\begin{itemize}
			\item Let $\tm = \var$. Then $\nv{\tm} = \set{\var} = \set{\var} \cup \emptyset = \unv{\tm} \cup \anv{\tm}$.
			
			\item Let $\tm = \la{\var}{\tmtwo}$. Then $\nv{\tm} = \emptyset = \unv{\tm} \cup \anv{\tm}$.
			
			\item Let $\tm = \tmthree \tmfour$. We do case analysis on the shape of $\tmthree$:
				\begin{itemize}
					\item Let $\tmthree = \var$. Then $\nv{\tm} = \nv{\tmthree} \cup \nv{\tmfour} = \set{\var} \cup \nv{\tmfour} =^{\ih} \set{\var} \cup (\unv{\tmfour} \cup \anv{\tmfour}) = \unv{\tmfour} \cup (\set{\var} \cup \anv{\tmfour}) = \unv{\tm} \cup \anv{\tm}$.
					
					\item Let $\tmthree \neq \var$. Then $\nv{\tm} = \nv{\tmthree} \cup \nv{\tmfour} =^{\ih} (\unv{\tmthree} \cup \anv{\tmthree}) \cup (\unv{\tmfour} \cup \anv{\tmfour}) = (\unv{\tmthree} \cup \unv{\tmfour}) \cup (\anv{\tmthree} \cup \anv{\tmfour}) = \unv{\tm} \cup \anv{\tm}$.
					
				\end{itemize}
			
		\end{itemize}
	
	\item \emph{Programs}: Let $\prog = \progentry{\tm}{\env}$. We proceed by induction on the lenght of $\env$:
		\begin{itemize}
			\item Let $\env = \emptyenv$. By \reflemmap{Unapplied_applied_and_needed_variables}{one} (Unapplied, applied and needed variabless), we have that $\nv{\tm} = \unv{\tm} \cup \anv{\tm}$, and so $\nventryp{\tm}{\emptyenv} = \nv{\tm} 
= \unv{\tm} \cup \anv{\tm} = 
\unventryp{\tm}{\emptyenv} \cup \anventryp{\tm}{\emptyenv}$.
			
			\item Let $\env = \envtwo \esub{\var}{\tmtwo}$. We proceed by case analysis on $\var \in \nventryp{\tm}{\envtwo}$:
				\begin{itemize}
					\item Let $\var \in \nventryp{\tm}{\envtwo}$. Then, $\nventryp{\tm}{\envtwo \esub{\var}{\tmtwo}} = (\nventryp{\tm}{\envtwo} \setminus \set{\var}) \cup \nv{\tmtwo}$, $\nventryp{\tm}{\envtwo} = \unventryp{\tm}{\envtwo} \cup \anventryp{\tm}{\envtwo}$---by \ih---and $\nv{\tmtwo} = \unv{\tmtwo} \cup \anv{\tmtwo}$---by \reflemmap{Unapplied_applied_and_needed_variables}{one} (Unapplied, applied and needed variables). That is,
						$$
							\nventryp{\tm}{\envtwo \esub{\var}{\tmtwo}} = ((\unventryp{\tm}{\envtwo} \cup \anventryp{\tm}{\envtwo}) \setminus \set{\var}) \cup (\unv{\tmtwo} \cup \anv{\tmtwo})
						$$
						
					We proceed by case analysis on the shape of $\tmtwo$ and on how $\var \in \nventryp{\tm}{\envtwo} = \unventryp{\tm}{\envtwo} \cup \anventryp{\tm}{\envtwo}$:
						\begin{itemize}
							\item Let $\tmtwo = \vartwo$ and $\var \in (\unventryp{\tm}{\envtwo} \setminus 
\anventryp{\tm}{\envtwo})$. Then $\var \notin \anventryp{\tm}{\envtwo}$ and $\anv{\tmtwo} = 
\emptyset$, so that
								$$
								\begin{array}{rcl}
									&    & ((\unventryp{\tm}{\envtwo} \cup \anventryp{\tm}{\envtwo}) \setminus \set{\var}) \cup (\unv{\tmtwo} \cup \anv{\tmtwo}) \\
									& = & ((\unventryp{\tm}{\envtwo} \setminus \set{\var}) \cup \unv{\tmtwo}) \cup \anventryp{\tm}{\envtwo} \\
									& = & \unventryp{\tm}{\envtwo \esub{\var}{\tmtwo}} \cup \anventryp{\tm}{\envtwo \esub{\var}{\tmtwo}} \\
								\end{array}
								$$
							
							\item Let $\tmtwo = \vartwo$ and $\var \in (\anventryp{\tm}{\envtwo} \setminus 
\unventryp{\tm}{\envtwo})$. Then $\var \notin \unventryp{\tm}{\envtwo}$, $\unv{\tmtwo} = \set \vartwo$, and $\anv{\tmtwo} = \emptyset$, so that
								$$
								\begin{array}{rcl}
									&    & ((\unventryp{\tm}{\envtwo} \cup \anventryp{\tm}{\envtwo}) \setminus \set{\var}) \cup (\unv{\tmtwo} \cup \anv{\tmtwo}) \\
									& = & \unventryp{\tm}{\envtwo} \cup ((\anventryp{\tm}{\envtwo} \setminus \set{\var}) \cup \set{\vartwo}) \\
									& = & \unventryp{\tm}{\envtwo \esub{\var}{\tmtwo}} \cup \anventryp{\tm}{\envtwo \esub{\var}{\tmtwo}} \\
								\end{array}
								$$
							
							\item Let $\tmtwo = \vartwo$, $\var \in \unventryp{\tm}{\envtwo}$ and $\var \in \anventryp{\tm}{\envtwo}$. Then,
								$$
								\begin{array}{rcl}
									&    & ((\unventryp{\tm}{\envtwo} \cup \anventryp{\tm}{\envtwo}) \setminus \set{\var}) \cup (\unv{\tmtwo} \cup \anv{\tmtwo}) \\
									& = & (\unventryp{\tm}{\envtwo} \setminus \set{\var}) \cup (\anventryp{\tm}{\envtwo} \setminus \set{\var}) \cup \set{\vartwo} \\
									& = & ((\unventryp{\tm}{\envtwo} \setminus \set{\var}) \cup \set{\vartwo}) \cup ((\anventryp{\tm}{\envtwo} \setminus \set{\var}) \cup \set{\vartwo}) \\
									& = & \unventryp{\tm}{\envtwo \esub{\var}{\tmtwo}} \cup \anventryp{\tm}{\envtwo \esub{\var}{\tmtwo}} \\
								\end{array}								
								$$
							
							\item Let $\tmtwo \neq \vartwo$. Then by 
definition $\unventryp{\tm}{\envtwo \esub{\var}{\tmtwo}} = (\unventryp{\tm}{\envtwo} \setminus 
\set{\var}) \cup \unv{\tmtwo}$ and $\anventryp{\tm}{\envtwo \esub{\var}{\tmtwo}} = 
(\anventryp{\tm}{\envtwo} \setminus \set{\var}) \cup \anv{\tmtwo}$. Thus,
								$$
								\begin{array}{rcl}
									&    & ((\unventryp{\tm}{\envtwo} \cup \anventryp{\tm}{\envtwo}) \setminus \set{\var}) \cup (\unv{\tmtwo} \cup \anv{\tmtwo}) \\
									& = & ((\unventryp{\tm}{\envtwo} \setminus \set{\var}) \cup \unv{\tmtwo}) \cup ((\anventryp{\tm}{\envtwo} \setminus \set{\var}) \cup \anv{\tmtwo}) \\
									& = & \unventryp{\tm}{\envtwo \esub{\var}{\tmtwo}} \cup \anventryp{\tm}{\envtwo \esub{\var}{\tmtwo}} \\
								\end{array}
								$$
														
						\end{itemize}
					
					\item Let $\var \notin \nventryp{\tm}{\envtwo}$. The \ih gives $\nventryp{\tm}{\envtwo} = \unventryp{\tm}{\envtwo} \cup \anventryp{\tm}{\envtwo}$ and by definition $\anventryp{\tm}{\envtwo \esub{\var}{\tmtwo}} = \anventryp{\tm}{\envtwo}$ and $\unventryp{\tm}{\envtwo \esub{\var}{\tmtwo}} = \unventryp{\tm}{\envtwo}$. Hence,
						$$
						\begin{array}{rll}
							\nventryp{\tm}{\envtwo \esub{\var}{\tmtwo}} 
							& = & \nventryp{\tm}{\envtwo} \\
							& =_{\ih} & \unventryp{\tm}{\envtwo} \cup \anventryp{\tm}{\envtwo} \\
							& = & \unventryp{\tm}{\envtwo \esub{\var}{\tmtwo}} \cup \anventryp{\tm}{\envtwo \esub{\var}{\tmtwo}} \\
						\end{array}
						$$
					
				\end{itemize}
			
		\end{itemize}
	\item \emph{Term contexts}: by structural induction on $\hctx$: the base case is trivial, while the inductive ones easily follow from the \ih and \reflemmap{Unapplied_applied_and_needed_variables}{one} (Unapplied, applied and needed variables).\qedhere

\end{enumerate}

\end{proof}


 Let us recall the following lemma, used for \opencbneed, which shall be used repeatedly also for \usefulopencbneed:
\begin{lemma}[Redex in non-normal terms]

$\tm$ is not a normal term if and only if there exist term context $\hctx$, and terms $\la{\var}{\tmtwo}$ and $\tmthree$ such that $\tm = \hctxp{(\la{\var}{\tmtwo})\tmthree}$.

\label{l:Redex_in_non-normal_terms-usefulopencbneed}
\end{lemma}

\subsection{Determinism.}
Another important property {\usefulopencbneednsp} should satisfy is determinism. This is simply proven in a way analogous to that of {\opencbneednsp}, following the same proof schema.
\newline

First, we define what a reduction place in the {\usefulopencbneednsp} case is:
\begin{definition}[Reduction places in {\usefulopencbneednsp}]
\hfill

Let $\tm$ be a term, $\hctx$ be a term context, and let $S \subseteq \Var$. We say that $\tm$ is a \emph{$S$-reduction place} of $\hctxp{\tm}$ if one of the following conditions holds:
	\begin{itemize}
		\item \emph{Multiplicative redex}: $\tm = (\la\var\tmtwo) \tmthree$;
		
		\item \emph{New hereditary jump}: $\tm = \var$ and $\var \notin S \supseteq \nv{\hctx}$.

		\item \emph{New applied variable}: $\tm = \var$, $\hctx$ is applicative, and $\var \notin S \supseteq \anv{\hctx}$.

	\end{itemize}

Let $\tm$ be a term and $\pctx$ be a program context, and let $S \subseteq \Var$. We say that $\tm$ is a \emph{$S$-reduction place} of $\pctxp{\tm}$ if one of the following conditions hold:
	\begin{itemize}
		\item \emph{Multiplicative redex}: $\pctx \in \evctxUsep{\uvarset}{\avarset}$\erase{, $S \supseteq (\uvarset \cup \avarset)$,} and $\tm = (\la{\var}{\tmtwo}) \tmthree$;
		
		\item \emph{Exponential redex}: $\tm = \var$, $\pctx \in \evctxAppp{\uvarset}{\avarset}$, $\var \in \dom{\pctx}$ and $\pctx(\var) = \val$;
		
		\item \emph{New hereditary jump}: $\tm = \var$, $\pctx \in \evctxUsep{\uvarset}{\avarset}$, $\var \notin S \supseteq (\uvarset \cup \avarset)$, and $\var \notin \dom{\pctx}$.

		\item \emph{New applied variable}: $\tm = \var$, $\pctx \in \evctxAppp{\uvarset}{\avarset}$, $\var \notin S \supseteq \avarset$, and $\var \notin \dom\pctx$.

	\end{itemize}

\label{def:Reduction_places_in_usefulopencbneed}
\end{definition}

Note that the conditions in the \emph{New applied variable} item for term contexts above are included in the conditions of \emph{New hereditary jump} for term contexts. In spite of this, we state it as a separate category of reduction places, since the former serves as base case for \emph{New applied variable}---for exponential evaluation contexts)---while the latter serves as base case for \emph{New hereditary jump}---for multiplicative evaluation contexts.

The following is a direct consequence of \reflemma{Unique_decomposition_of_Lambda-terms-opencbneed} (Unique decomposition of terms) in the Determinism subsection of the proofs for {\opencbneednsp} and \reflemma{Term_contexts_Unapplied_applied_and_needed_variables} (Term contexts: Unapplied, applied and needed variables).
\begin{toappendix}
\begin{lemma}[Unique decomposition of terms]

Let $\hctx_{1} \ctxholep{\tm_{1}} = \hctx_{2} \ctxholep{\tm_{2}}$, with $\hctx_{1}, \hctx_{2}$ term contexts, let $S \supseteq (\anv{\hctx_{1}} \cup \anv{\hctx_{2}})$, and let $\tm_{i}$ be an $S$-reduction place of $\hctx_i \ctxholep{\tm_i}$, for $i = 1, 2$.

Then $\tm_{1} = \tm_{2}$ and $\hctx_{1} = \hctx_{2}$.

\label{l:Unique_decomposition_of_Lambda-terms-usefulopencbneed}
\end{lemma}
\end{toappendix}


\begin{proof}
By induction on $\hctx_{1}$. Cases:
\begin{itemize}
	\item \emph{Empty}: $\hctx_{1} = \ctxhole$. If $\tm_{1}$ is a multiplicative redex then it must be that $\hctx_{2} = \ctxhole$ and $\tm_{2} = \tm_{1}$. The same is true if $\tm_{1}$ is a variable not in $S$.
	
	\item \emph{Left of an application}: $\hctx_{1} = \hctxtwo_{1} \tmtwo$. Cases of $\hctx_{2}$:
	\begin{itemize}
		\item \emph{Empty}: If $\hctx_{2} = \ctxhole$, then $\tm_{2}$ is a multiplicative redex, implying that $\hctxtwo_{1}$ is empty and $\tm_{1}$ is a value, which is absurd. Therefore, this case is impossible.
		
		\item \emph{Left of an application}: Let $\hctx_{2} = \hctxtwo_{2} \tmtwo$. Then $\hctxtwo_{1}\ctxholep{\tm_{1}} = \hctxtwo_{2}\ctxholep{\tm_{2}}$ and $\tm_{i}$ is a S-reduction places of $\hctx_{i} \ctxholep{\tm_{i}}$ for $i =1,2$. By \ih, $\tm_{1} = \tm_{2}$ and $\hctxtwo_{1} = \hctxtwo_{2}$, and then $\hctx_{1} = \hctx_{2}$.
		
		\item \emph{Right of an application}: If $\hctx_{2} = \itm \hctxtwo_{2}$, then $\itm = \hctxtwo_{1} \ctxholep{\tm_{1}}$. Two cases:
			\begin{itemize}
				\item If $\itm \in \Var$, then it should be that $\hctxtwo_{1} = \ctxhole$ and $\tm_{1} \in \Var$. But since $\hctxtwo_{1}$ is not applicative, then $\tm_{1}$ cannot be an $S$-reduction place of $\hctx_{1} \ctxholep{\tm_{1}}$, which contradicts the hypothesis.

				\item Suppose $\itm = \itmplustwo$. Then it should be that $\tm_{1} \in \Var$---by \reflemma{Redex_in_non-normal_terms-usefulopencbneed} (Redex in non-normal terms)---and so that $\hctxtwo_{1}$ is applicative. However, this would imply that $\tm_{1} \in \anv{\itm}$---by \reflemmap{Rewriting_applicative_term_contexts}{one} (Applicative term contexts give applied variables)--- which gives that $\tm \in \anv{\hctx_{2}} \subset S$; absurd.

			\end{itemize} 

		Therefore, this case is impossible.
	\end{itemize}
	
	\item \emph{Right of an application}: $\hctx_{1} = \itm \hctxtwo_{1}$. Cases of $\hctx_{2}$:
		\begin{itemize}
		\item \emph{Empty}: If $\hctx_{2} = \ctxhole$, then $\tm_{2}$ is an application that is not a multiplicative redex; absurd.	Therefore, this case is impossible.
		
		\item \emph{Left of an application}: $\hctx_{2} = \hctxtwo_{2} \tmtwo$. This case is identical to the case where the hole of $\hctx_{1}$ is on the left of the application while the one of $\hctx_{2}$ is on the right, treated above.	
		
		\item \emph{Right of an application}: $\hctx_{2} = \itm \hctxtwo_{2}$. Then $\hctxtwo_{1}\ctxholep{\tm_{1}} = \hctxtwo_{2}\ctxholep{\tm_{2}}$ and $\tm_{i}$ is a S-reduction places of $\hctx_{i} \ctxholep{\tm_{i}}$ for $i =1,2$. By \ih, $\tm_{1} = \tm_{2}$ and $\hctxtwo_{1} = \hctxtwo_{2}$, and then $\hctx_{1} = \hctx_{2}$.\qedhere
		
	\end{itemize}
	
\end{itemize}

\end{proof}

\begin{lemma}[Unique decomposition of programs]

Let $\pctx_{1} \ctxholep{\tm_{1}} = \pctx_{2} \ctxholep{\tm_{2}}$, with $\pctx_{1} \in (\evctxUsep{\uvarset_{1}}{\avarset_{1}} \cup \evctxAppp{\uvarset_{1}}{\avarset_{1}})$ and $\pctx_{2} \in (\evctxUsep{\uvarset_{2}}{\avarset_{2}} \cup \evctxAppp{\uvarset_{2}}{\avarset_{2}})$ such that $S \supseteq (\avarset_{1} \cup \avarset_{2})$, and $\tm_{i}$ be an $S$-reduction place of $\pctx_{i} \ctxholep{\tm_{i}}$ for $i = 1,2$. 

Then $\tm_{1} = \tm_{2}$ and $\pctx_{1} = \pctx_{2}$.
\label{l:Unique_decomposition_of_programs-usefulopencbneed}
\end{lemma}


\begin{proof}
We split the analysis in two, namely when $\pctx_{1} \in \evctxUsep{\uvarset_{1}}{\avarset_{1}}$ and $\pctx_{1} \in \evctxAppp{\uvarset_{1}}{\avarset_{1}}$:
	\begin{itemize}
		\item Let $\pctx_{1} \in \evctxUsep{\uvarset_{1}}{\avarset_{1}}$. We further split the analysis in whether $\pctx_{2} \in \evctxUsep{\uvarset_{2}}{\avarset_{2}}$ or $\pctx_{2} \in \evctxAppp{\uvarset_{2}}{\avarset_{2}}$:
			\begin{itemize}
				\item Let $\pctx_{2} \in \evctxUsep{\uvarset_{2}}{\avarset_{2}}$. We proceed by induction on the derivation of $\pctx_{1} \in \evctxUsep{\uvarset_{1}}{\avarset_{1}}$. Case analysis on the last derivation rule:
					\begin{itemize}
						\item \emph{Rule $\ruleUseAxMult$}: Let $\pctx_{1}$ be derived via an application of rule $\ruleUseAxMult$. Then $\pctx_{1} \defeq \progentry{\hctx_{1}}{\emptyenv}$ for some term context $\hctx_{1}$ and so, in addition, it must be that $\pctx_{2} = \progentry{\hctx_{2}}{\emptyenv}$ for some term context $\hctx_{2}$---\ie, $\pctx_{2} \in \evctxUsep{\uvarset_{2}}{\avarset_{2}}$ is derived via an application of $\ruleUseAxMult$. The statement follows by \reflemma{Unique_decomposition_of_Lambda-terms-usefulopencbneed} (Unique decomposition of terms).

						\item \emph{Rule $\ruleUseHerMult$}: Let $\pctx_{1}$ be derived as follows
							$$
							\infer
								[\ruleUseHerMult]
								{\appES{\pctxtwo_{1} \ctxholep{\var}}{\esub{\var}{\hctx_{1}}} \in \evctxUsep{\uvarsettwo_{1} \cup \unv{\hctx_{1}}}{\avarsettwo_{1} \cup \anv{\hctx_{1}}}}
								{\pctxtwo_{1} \in \evctxUsep{\uvarsettwo_{1}}{\avarsettwo_{1}}
								\quad
								\var \notin (\uvarsettwo_{1} \cup \avarsettwo_{1})}
							$$
						where $\pctx_{1} = \appES{\pctxtwo_{1} \ctxholep{\var}}{\esub{\var}{\hctx_{1}}}$, $\uvarset_{1} = \uvarsettwo_{1} \cup \unv{\hctx_{1}}$ and $\avarset_{1} = \avarsettwo_{1} \cup \anv{\hctx_{1}}$. There are two possibilities: either $\tm_{1} \in \Var$ and $\tm_{1} \notin (\uvarsettwo_{1} \cup \avarsettwo_{1})$ or $\tm_{1} = (\la{\var_{1}}{\tmtwo_{1}}) \tmthree_{1}$---by \refdef{Reduction_places_in_usefulopencbneed} (Reduction places in \usefulopencbneed). We shall only cover the case of $\tm_{1} = (\la{\var_{1}}{\tmtwo_{1}}) \tmthree_{1}$, while leaving the other case for the reader.
						
						As a starting point, note that $\var$ is a $S$-reduction place of $\pctxtwo_{1} \ctxholep{\var}$, since $\uvarsettwo_{1} \cup \avarsettwo_{1} \subseteq S$. We now proceed by case analysis on the last derivation rule applied in $\pctx_{2} \in \evctxUsep{\uvarset_{2}}{\avarset_{2}}$. 
							\begin{itemize}
								\item The case of rule $\ruleUseAxMult$ is impossible.

								\item The statement in the case of rule $\ruleUseHerMult$ follows by \ih.

								\item Suppose $\pctx_{2} \defeq \appES{\pctxtwo_{2}}{\esub{\var}{\vartwo}} \in \evctxUsep{\uvarset_{2}}{\avarset_{2}}$ is derived via an application of rule $\ruleUseVarMult$. But then we would have that $\hctx_{1} \ctxholep{\tm_{1}} = \hctx_{1} \ctxholep{(\la{\var_{1}}{\tmtwo_{1}}) \tmthree_{1}} = \vartwo$; absurd by \reflemma{Redex_in_non-normal_terms-usefulopencbneed} (Redex in non-normal terms).

								\item The cases where $\pctx_{2}$ is derived via an application of rule $\ruleUseNonVarMult$ or an application of rule $\ruleUseUselessMult$ are ruled out analogously to the previous case, given that non-variable inerts and values are normal terms---just like $\vartwo$ is in the previous case.

								\item Suppose $\pctx_{2} \in \evctxUsep{\uvarset_{2}}{\avarset_{2}}$ is derived as follows
									$$
									\infer
										[\ruleUseGcMult]
										{\appES{\pctxtwo_{2}}{\esub{\var}{\tmseven}} \in \evctxUsep{\uvarset_{2}}{\avarset_{2}}}
										{\pctxtwo_{2} \in \evctxUsep{\uvarset_{2}}{\avarset_{2}}
										\quad
										\var \notin (\uvarset_{2} \cup \avarset_{2})}
									$$ 
								with $\pctx_{2} = \appES{\pctxtwo_{2}}{\esub{\var}{\tmseven}}$. Then there are two possibilities: either $\tm_{2} = \vartwo \neq \var$, or $\tm_{2} = (\la{\var_{2}}{\tmtwo_{2}}) \tmthree_{2}$---by \refdef{Reduction_places_in_usefulopencbneed} (Reduction places in \usefulopencbneed). On the one hand, if $\tm_{2} = \vartwo$, then the \ih on $\pctxtwo_{1} \ctxholep{\var} = \pctxtwo_{2} \ctxholep{\vartwo}$ gives that $\var = \tmtwo = \vartwo$; absurd. On the other hand, if $\tm_{2} = (\la{\var_{2}}{\tmtwo_{2}}) \tmthree_{2}$, then the \ih on $\pctxtwo_{1} \ctxholep{\var}$ and $\pctxtwo_{2} \ctxholep{(\la{\var_{2}}{\tmtwo_{2}}) \tmthree_{2}}$ with respect to $S \defeq (\uvarsettwo_{1} \cup \uvarset_{2} \cup \avarsettwo_{1} \cup \avarset_{2})$ to get that $\var = \tm_{1} = \tm_{2} = (\la{\var_{2}}{\tmtwo_{2}}) \tmthree_{2}$; absurd.

								\item In the case where $\pctx_{2}$ is derived via an application of rule $\ruleUseHerMult$, the statement follows by \reflemma{Unique_decomposition_of_Lambda-terms-usefulopencbneed} (Unique decomposition of terms).

							\end{itemize}

							\item \emph{Rule $\ruleUseGcMult$}: Let $\pctx_{1}$ be derived as follows
							$$
							\infer
								[\ruleUseGcMult]
								{\appES{\pctxtwo_{1}}{\esub{\var}{\tmseven_{1}}} \in \evctxUsep{\uvarset_{1}}{\avarset_{1}}}
								{\pctxtwo_{1} \in \evctxUsep{\uvarset_{1}}{\avarset_{1}}
								\quad
								\var \notin (\uvarset_{1} \cup \avarset_{1})}
							$$
							where $\pctx_{1} = \appES{\pctxtwo_{1}}{\esub{\var}{\tmseven_{1}}}$. Case analysis on the last derivation rule in $\pctx_{2} \in \evctxUsep{\uvarset_{2}}{\avarset_{2}}$:
								\begin{itemize}
									\item The case of rule $\ruleUseAxMult$ is impossible.

									\item The cases of rules $\ruleUseGcMult$, $\ruleUseVarMult$, $\ruleUseNonVarMult$ and $\ruleUseUselessMult$ are given by the \ih.

									\item Suppose $\pctx_{2} \in \evctxUsep{\uvarset_{2}}{\avarset_{2}}$ is derived as follows:
										$$
										\infer
											[\ruleUseHerMult]
											{\appES{\pctxtwo_{2} \ctxholep{\var}}{\esub{\var}{\hctx_{2}}} \in \evctxUsep{\uvarsettwo_{2} \cup \unv{\hctx_{2}}}{\avarsettwo_{2} \cup \anv{\hctx_{2}}}}
											{\pctxtwo_{2} \in \evctxUsep{\uvarsettwo_{2}}{\avarsettwo_{2}}
											\quad
											\var \notin (\uvarsettwo_{2} \cup \avarsettwo_{2})}
										$$
									with $\pctx_{2} = \appES{\pctxtwo_{2} \ctxholep{\var}}{\esub{\var}{\hctx_{2}}}$, $\uvarset_{2} = \uvarsettwo_{2} \cup \unv{\hctx_{2}}$ and $\avarset_{2} = \avarsettwo_{2} \cup \anv{\hctx_{2}}$. But since $\tm_{1}$ is a $S$-reduction place of $\pctxtwo_{1} \ctxholep{\tm_{1}}$ and $\var$ is a $S$-reduction place of $\pctxtwo_{2} \ctxholep{\var}$, we would then be able to apply the \ih on $\pctxtwo_{1} \ctxholep{\tm_{1}}$ and $\pctxtwo_{2} \ctxholep{\var}$ to get that that $\tm_{1} = \var$, which is absurd by \refdef{Reduction_places_in_usefulopencbneed} (Reduction places in \usefulopencbneed).

								\end{itemize}

							\item \emph{Rule $\ruleUseVarMult$}: Let $\pctx_{1}$ be derived as follows
								$$
								\infer
									[\ruleUseVarMult]
									{\appES{\pctxtwo_{1}}{\esub{\var}{\varthree}} \in \evctxUsep{\updp{\uvarsettwo_{1}}{\var}{\varthree}}{\updp{\avarsettwo_{1}}{\var}{\varthree}}}
									{\pctxtwo_{1} \in \evctxUsep{\uvarsettwo_{1}}{\avarsettwo_{1}}
									\quad
									\var \in (\uvarsettwo_{1} \cup \avarsettwo_{1})}
								$$
							where $\pctx_{1} = \appES{\pctxtwo_{1}}{\esub{\var}{\varthree}}$, $\uvarset_{1} = \updp{\uvarsettwo_{1}}{\var}{\varthree}$ and $\avarset_{1} = \updp{\avarsettwo_{1}}{\var}{\varthree}$. Case analysis on the last derivation rule in $\pctx_{2} \in \evctxUsep{\uvarset_{2}}{\avarset_{2}}$:
								\begin{itemize}
									\item The case of rule $\ruleUseAxMult$ is impossible.

									\item The case of rule $\ruleUseGcMult$ was covered above, when $\pctx_{1}$ was derived by an application of rule $\ruleUseGcMult$ and $\pctx_{2}$ was derived by an application of rule $\ruleUseVarMult$.

									\item The case of rule $\ruleUseHerMult$ was covered above, when $\pctx_{1}$ was derived by an application of rule $\ruleUseHerMult$ and $\pctx_{2}$ was derived by an application of rule $\ruleUseVarMult$.

									\item The case of rule $\ruleUseVarMult$ is given by a simple application of the \ih.

									\item The case of rules $\ruleUseNonVarMult$ and $\ruleUseUselessMult$ are impossible, because the set of variables is pairwise disjoint with the set of values and with the set of non-variale inert terms.

								\end{itemize}

							\item \emph{Rule $\ruleUseNonVarMult$}: This case can be analyzed analogously to the previous case.

							\item \emph{Rule $\ruleUseUselessMult$}: Same as previous item.

					\end{itemize}

				\item Let $\pctx_{2} \in \evctxAppp{\uvarset_{2}}{\avarset_{2}}$. We now proceed by induction on the derivation of $\pctx_{1} \in \evctxUsep{\uvarset_{1}}{\avarset_{1}}$. Case analysis on the last derivation rule:
					\begin{itemize}
						\item \emph{Rule $\ruleUseAxMult$}: Let $\pctx_{1}$ be derived via an application of rule $\ruleUseAxMult$. Then $\pctx_{1} \defeq \progentry{\hctx_{1}}{\emptyenv}$ for some term context $\hctx_{1}$ and so, in addition, it must be that $\pctx_{2} = \progentry{\hctxappsub{2}}{\emptyenv}$ for some term context $\hctx_{2}$---\ie, $\pctx_{2} \in \evctxAppp{\uvarset_{2}}{\avarset_{2}}$ is derived via an application of rule $\ruleUseAxOneExp$. The statement follows by \reflemma{Unique_decomposition_of_Lambda-terms-usefulopencbneed} (Unique decomposition of terms).

						\item \emph{Rule $\ruleUseHerMult$}: Let $\pctx_{1}$ be derived as follows
							$$
							\infer
								[\ruleUseHerMult]
								{\appES{\pctxtwo_{1} \ctxholep{\var}}{\esub{\var}{\hctx_{1}}} \in \evctxUsep{\uvarsettwo_{1} \cup \unv{\hctx_{1}}}{\avarsettwo_{1} \cup \anv{\hctx_{1}}}}
								{\pctxtwo_{1} \in \evctxUsep{\uvarsettwo_{1}}{\avarsettwo_{1}}
								\quad
								\var \notin (\uvarsettwo_{1} \cup \avarsettwo_{1})}
							$$
						where $\pctx_{1} = \appES{\pctxtwo_{1} \ctxholep{\var}}{\esub{\var}{\hctx_{1}}}$, $\uvarset_{1} = \uvarsettwo_{1} \cup \unv{\hctx_{1}}$ and $\avarset_{1} = \avarsettwo_{1} \cup \anv{\hctx_{1}}$. Note that $\var$ is a $S$-reduction place of $\pctxtwo_{1} \ctxholep{\var}$. There are two possibilities: either $\tm_{1} = (\la{\var_{1}}{\tmtwo_{1}}) \tmthree_{1}$ or $\tm_{1} \in \Var$---by \refdef{Reduction_places_in_usefulopencbneed} (Reduction places in \usefulopencbneed). We consider the former case, and leave the latter for the reader.

						We now proceed by case analysis on the last derivation rule applied in $\pctx_{2} \in \evctxAppp{\uvarset_{2}}{\avarset_{2}}$.
							\begin{itemize}
								\item The case of rule $\ruleUseAxOneExp$ is impossible.

								\item Let $\pctx_{2} \in \evctxAppp{\uvarset_{2}}{\avarset_{2}}$ be derived as follows
									$$
									\infer
										[\ruleUseAxTwoExp]
										{\appES{\pctxtwo_{2} \ctxholep{\var}}{\esub{\var}{\hctxappsub{2}}} \in \evctxAppp{(\uvarsettwo_{2} \setminus \set{\var}) \cup \unv{\hctxappsub{2}}}{(\avarsettwo_{2} \setminus \set{\var}) \cup \anv{\hctxappsub{2}}}}
										{\pctxtwo_{2} \in \evctxUsep{\uvarsettwo_{2}}{\avarsettwo_{2}}
										\quad
										\var \notin (\uvarsettwo_{2} \cup \avarsettwo_{2})}
									$$
								where $\pctx_{2} = \appES{\pctxtwo_{2} \ctxholep{\var}}{\esub{\var}{\hctxappsub{2}}}$, $\uvarset_{2} = (\uvarsettwo_{2} \setminus \set{\var}) \cup \unv{\hctxappsub{2}}$ and $\avarset_{2} = (\avarsettwo_{2} \setminus \set{\var}) \cup \anv{\hctxappsub{2}}$.

								The statement follows trivially by application of the \ih on $\pctxtwo_{1} \ctxholep{\var}$ and $\pctxtwo_{2} \ctxholep{\var}$ and by \reflemma{Unique_decomposition_of_Lambda-terms-usefulopencbneed} (Unique decomposition of terms) on $\hctx_{1} \ctxholep{\tm_{1}}$ and $\hctxappsub{2} \ctxholep{\tm_{2}}$.

								\item The case of rules $\ruleUseVarExp$, $\ruleUseNonVarExp$ and $\ruleUseUselessExp$ are all ruled out by \reflemma{Redex_in_non-normal_terms-usefulopencbneed} (Redex in non-normal terms).

								\item Suppose $\pctx_{2} \in \evctxAppp{\uvarset_{2}}{\avarset_{2}}$ is derived as follows
									$$
									\infer
										[\ruleUseGcExp]
										{\appES{\pctxtwo_{2}}{\esub{\var}{\tmseven_{2}}} \in \evctxAppp{\uvarset_{2}}{\avarset_{2}}}
										{\pctxtwo_{2} \in \evctxAppp{\uvarset_{2}}{\avarset_{2}}
										\quad
										\var \notin (\uvarset_{2} \cup \avarset_{2})}
									$$
								with $\pctx_{2} = \appES{\pctxtwo_{2}}{\esub{\var}{\tmseven_{2}}}$. Given \refdef{Reduction_places_in_usefulopencbneed} (Reduction places in Useful Open CbNeed), if $\tm_{2} = \vartwo \neq \var$ then by \ih we would get that $\pctxtwo_{1} = \pctxtwo_{2}$ and that $\var = \tm_{1} = \tm_{2} = \vartwo$; absurd. Moreover, if $\tm_{2} = \var$ then $\tmseven_{2}$ is a value but then also $\hctx_{1} \ctxholep{\tm_{1}} = \tmseven_{2} $ is a value; absurd by \reflemma{Redex_in_non-normal_terms-usefulopencbneed} (Redex in non-normal terms). Therefore, it could only be that $\tm_{2} = (\la{\var_{2}}{\tmtwo_{2}}) \tmthree_{2} $---again, by \refdef{Reduction_places_in_usefulopencbneed} (Reduction places in Useful Ppen CbNeed). However, applying the \ih on $\pctxtwo_{1} \ctxholep{\var}$ and $\pctxtwo_{2} \ctxholep{\tm_{2}}$ gives that $\var = \tm_{2}$; absurd. Therefore, this case is impossible.

								\item Let $\pctx_{2} \in \evctxAppp{\uvarset_{2}}{\avarset_{2}}$ be derived as follows
									$$
									\infer
										[\ruleUseHerExp]
										{\appES{\pctxtwo_{2} \ctxholep{\var}}{\esub{\var}{\ctxhole}} \in \evctxAppp{\uvarsettwo_{2} \setminus \set{\var}}{\avarset_{2}}}
										{\pctxtwo_{2} \in \evctxAppp{\uvarsettwo_{2}}{\avarset_{2}}
										\quad
										\var \notin \avarsettwo_{2}}
									$$
								where $\pctx_{2} = \appES{\pctxtwo_{2} \ctxholep{\var}}{\esub{\var}{\ctxhole}}$ and $\uvarset_{2} = \uvarsettwo_{2} \setminus \set{\var}$. We can apply the \ih on $\pctxtwo_{1} \ctxholep{\var}$ and $\pctxtwo_{2} \ctxholep{\var}$ taking $S \defeq \avarsettwo_{1} \cup \avarsettwo_{2}$ to obtain that $\pctxtwo_{1} = \pctxtwo_{2}$. Finally, we can apply \reflemma{Unique_decomposition_of_Lambda-terms-usefulopencbneed} (Unique decomposition of terms) to obtain that $\hctx = \ctxhole$---and so $\pctx_{1} = \pctx_{2}$---and $\tm_{1} = \tm_{2}$.

							\end{itemize}

						\item \emph{Rule $\ruleUseVarMult$}: Let $\pctx_{1} \in \evctxUsep{\uvarset_{1}}{\avarset_{1}}$ be derived as follows
							$$
							\infer
								[\ruleUseVarMult]
								{\appES{\pctxtwo_{1}}{\esub{\var}{\varthree}} \in \evctxUsep{\updp{\uvarsettwo_{1}}{\var}{\varthree}}{\updp{\avarsettwo_{1}}{\var}{\varthree}}}
								{\pctxtwo_{1} \in \evctxUsep{\uvarsettwo_{1}}{\avarsettwo_{1}}
								\quad
								\var \in (\uvarsettwo_{1} \cup \avarsettwo_{1})}
							$$
						where $\pctx_{1} = \appES{\pctxtwo_{1}}{\esub{\var}{\varthree}}$, $\uvarset_{1} = \updp{\uvarsettwo_{1}}{\var}{\varthree}$ and $\avarset_{1} = \updp{\avarsettwo_{1}}{\var}{\varthree}$. Case analysis on the last derivation rule in $\pctx_{2} \in \evctxAppp{\uvarset_{2}}{\avarset_{2}}$:
							\begin{itemize}
								\item The case of rule $\ruleUseAxOneExp$ is impossible.

								\item The case of rule $\ruleUseAxTwoExp$ is impossible because $\varthree$ cannot be rewritten as an applicative term context with $\tm_{2}$ plugged into it.

								\item The case of rule $\ruleUseVarExp$ is given simply by \ih.

								\item The cases of rules $\ruleUseNonVarExp$ and $\ruleUseUselessExp$ are ruled out by the fact that non-variable inert terms and values cannot be rewritten as $\varthree$.

								\item Let $\pctx_{2} \in \evctxAppp{\uvarset_{2}}{\avarset_{2}}$ be derived as follows:
									$$
									\infer
										[\ruleUseGcExp]
										{\appES{\pctxtwo_{2}}{\esub{\var}{\varthree}} \in \evctxAppp{\uvarset_{2}}{\avarset_{2}}}
										{\pctxtwo_{2} \in \evctxAppp{\uvarset_{2}}{\avarset_{2}}
										\quad
										\var \notin (\uvarset_{2} \cup \avarset_{2})}
									$$
								with $\pctx_{2} = \appES{\pctxtwo_{2}}{\esub{\var}{\varthree}}$. By application of the \ih on $\pctxtwo_{1} \ctxholep{\tm_{1}}$ and $\pctxtwo_{2} \ctxholep{\tm_{2}}$, we have that $\pctxtwo_{1} = \pctxtwo_{2}$---and so $\pctx_{1} = \pctx_{2}$---and $\tm_{1} = \tm_{2}$.

								\item Suppose $\pctx_{2}$ is derived as follows
									$$
									\infer
										[\ruleUseHerExp]
										{\appES{\pctxtwo_{2} \ctxholep{\var}}{\esub{\var}{\ctxhole}} \in \evctxAppp{\uvarsettwo_{2} \setminus \set{\var}}{\avarset_{2}}}
										{\pctxtwo_{2} \in \evctxAppp{\uvarsettwo_{2}}{\avarset_{2}}
										\quad
										\var \notin \avarsettwo_{2}}
									$$
								with $\pctx_{2} = \appES{\pctxtwo_{2} \ctxholep{\var}}{\esub{\var}{\ctxhole}}$ and $\uvarset_{2} = \uvarsettwo_{2} \setminus \set{\var}$. But then application of the \ih gives that $\pctxtwo_{1} = \pctxtwo_{2}$ and $\tm_{1} = \var$, and so by \refdef{Reduction_places_in_usefulopencbneed} (Reduction places in \usefulopencbneed) it must be that $\varthree$ is a value; absurd.

							\end{itemize}

						\item \emph{Rule $\ruleUseNonVarMult$}: Let $\pctx_{1} \in \evctxUsep{\uvarset_{1}}{\avarset_{1}}$ be derived as follows:
							$$
							\infer
								[\ruleUseNonVarMult]
								{\appES{\pctxtwo_{1}}{\esub{\var}{\itmplus_{1}}} \in \evctxUsep{(\uvarsettwo_{1} \setminus \set{\var}) \cup \unv{\itmplus}}{(\avarsettwo_{1} \setminus \set{\var}) \cup \anv{\itmplus}}}
								{\pctxtwo_{1} \in \evctxUsep{\uvarsettwo_{1}}{\avarsettwo_{1}}
								\quad
								\var \in (\uvarsettwo_{1} \cup \avarsettwo_{1})}
							$$
						where $\pctx_{1} = \appES{\pctxtwo_{1}}{\esub{\var}{\itmplus_{1}}}$, $\uvarset_{1} = (\uvarsettwo_{1} \setminus \set{\var}) \cup \unv{\itmplus}$, $\avarset_{1} = (\avarset_{1} \setminus \set{\var}) \cup \anv{\itmplus}$. Case analysis on the last derivation rule in $\pctx_{2} \in \evctxAppp{\uvarset_{2}}{\avarset_{2}}$:
							\begin{itemize}
								\item The case of rule $\ruleUseAxOneExp$ is impossible.

								\item Suppose $\pctx_{2} \in \evctxAppp{\uvarset_{2}}{\avarset_{2}}$ is derived as follows
									$$
									\infer
										[\ruleUseAxTwoExp]
										{\appES{\pctxtwo_{2} \ctxholep{\var}}{\esub{\var}{\hctxappsub{2}}} \in \evctxAppp{(\uvarsettwo_{2} \setminus \set{\var}) \cup \unv{\hctxappsub{2}}}{(\avarsettwo_{2} \setminus \set{\var}) \cup \anv{\hctxappsub{2}}}}
										{\pctxtwo_{2} \in \evctxUsep{\uvarsettwo_{2}}{\avarsettwo_{2}}
										\quad
										\var \notin (\uvarsettwo_{2} \cup \avarsettwo_{2})}
									$$
								with $\pctx_{2} = \appES{\pctxtwo_{2} \ctxholep{\var}}{\esub{\var}{\hctxappsub{2}}}$, $\uvarset_{2} = (\uvarsettwo_{2} \setminus \set{\var}) \cup \unv{\hctxappsub{2}}$ and $\avarset_{2} = (\avarsettwo_{2} \setminus \set{\var}) \cup \anv{\hctxappsub{2}}$. Note that if $\tm_{1} = \var$ then it should be that $\itmplus_{1} $ is a value---by \refdef{Reduction_places_in_usefulopencbneed} (Reduction places in \usefulopencbneed); absurd. Moreover, if $\tm_{1} = \vartwo \neq \var$ or if $\tm_{1} = (\la{\var_{1}}{\tmtwo_{1}}) \tmthree_{1}$, then application of the \ih on $\pctxtwo_{1} \ctxholep{\tm_{1}}$ and $\pctxtwo_{2} \ctxholep{\var}$ gives that $\tm_{1} = \var$; absurd in either case. Therefore, this case is impossible.

								\item The cases of rules $\ruleUseVarExp$ and $\ruleUseUselessExp$ are impossible because $\itmplus$ cannot be rewritten as a variable or a value.

								\item The case of $\ruleUseNonVarExp$ follows by application of the \ih.

								\item  Let $\pctx_{2} \in \evctxAppp{\uvarset_{2}}{\avarset_{2}}$ be derived as follows
									$$
									\infer
										[\ruleUseGcExp]
										{\appES{\pctxtwo_{2}}{\esub{\var}{\itmplus_{1}}} \in \evctxAppp{\uvarset_{2}}{\avarset_{2}}}
										{\pctxtwo_{2} \in \evctxAppp{\uvarset_{2}}{\avarset_{2}}
										\quad
										\var \notin (\uvarset_{2} \cup \avarset_{2})}
									$$
								where $\pctx_{2} = \appES{\pctxtwo_{2}}{\esub{\var}{\itmplus_{1}}}$. Note that $\tm_{1} \neq \var$---otherwise $\itmplus$ is a value; absurd. Moreover, $\tm_{2} \neq \var$---otherwise application of the \ih on $\pctxtwo_{1} \ctxholep{\tm_{1}}$ and $\pctxtwo_{2} \ctxholep{\tm_{2}}$ would give that $\tm_{1} = \tm_{2} = \var$; absurd. If $\tm_{1}, \tm_{2} \neq \var$, we can apply the \ih on $\pctxtwo_{1} \ctxholep{\tm_{1}}$ and $\pctxtwo_{2} \ctxholep{\tm_{2}}$ to get that $\pctxtwo_{1} = \pctxtwo_{2}$---and so $\pctx_{1} = \pctx_{2}$---and $\tm_{1} = \tm_{2}$.

								\item Suppose $\pctx_{2} \in \evctxAppp{\uvarset_{2}}{\avarset_{2}}$ is derived as follows
									$$
									\infer
										[\ruleUseHerExp]
										{\appES{\pctxtwo_{2} \ctxholep{\var}}{\esub{\var}{\ctxhole}} \in \evctxAppp{\uvarsettwo_{2} \setminus \set{\var}}{\avarset_{2}}}
										{\pctxtwo_{2} \in \evctxAppp{\uvarsettwo_{2}}{\avarset_{2}}
										\quad
										\var \notin \avarset_{2}}
									$$
								with $\pctx_{2} = \appES{\pctxtwo_{2} \ctxholep{\var}}{\esub{\var}{\ctxhole}}$ and $\uvarset_{2} = \uvarsettwo_{2} \setminus \set{\var}$. However, application of the \ih on $\pctxtwo_{1} \ctxholep{\tm_{1}}$ and $\pctxtwo_{2} \ctxholep{\var}$ would give that $\tm_{1} = \var$, which would imply that $\itmplus_{1} $ is a value---by \refdef{Reduction_places_in_usefulopencbneed} (Reduction places in \usefulopencbneed); absurd.

							\end{itemize}

						\item \emph{Rule $\ruleUseUselessMult$}: Let $\pctx_{1} \in \evctxUsep{\uvarset_{1}}{\avarset_{1}}$ be derived as follows:
							$$
							\infer
								[\ruleUseUselessMult]
								{\appES{\pctxtwo_{1}}{\esub{\var}{\val_{1}}} \in \evctxUsep{\uvarsettwo_{1} \setminus \set{\var}}{\avarset_{1}}}
								{\pctxtwo_{1} \in \evctxUsep{\uvarsettwo_{1}}{\avarset_{1}}
								\quad
								\var \in (\uvarsettwo_{1} \setminus \avarset_{1})}
							$$
						where $\pctx_{1} = \appES{\pctxtwo_{1}}{\esub{\var}{\val_{1}}}$ and $\uvarset_{1} = \uvarsettwo_{1} \setminus \set{\var}$. Case analysis on the last derivation rule in $\pctx_{2}  \in \evctxAppp{\uvarset_{2}}{\avarset_{2}}$:
							\begin{itemize}
								\item The case of rule $\ruleUseAxOneExp$ is impossible.

								\item Suppose $\pctx_{2} \in \evctxAppp{\uvarset_{2}}{\avarset_{2}}$ is derived as follows
									$$
									\infer
										[\ruleUseAxTwoExp]
										{\appES{\pctxtwo_{2} \ctxholep{\var}}{\esub{\var}{\hctxappsub{2}}} \in \evctxAppp{(\uvarsettwo_{2} \setminus \set{\var}) \cup \unv{\hctxappsub{2}}}{(\avarsettwo_{2} \setminus \set{\var}) \cup \anv{\hctxappsub{2}}}}
										{\pctxtwo_{2} \in \evctxUsep{\uvarsettwo_{2}}{\avarsettwo_{2}}
										\quad
										\var \notin (\uvarsettwo_{2} \cup \avarsettwo_{2})}
									$$
								with $\pctx_{2} = \appES{\pctxtwo_{2} \ctxholep{\var}}{\esub{\var}{\hctxappsub{2}}}$, $\uvarset_{2} = (\uvarsettwo_{2} \setminus \set{\var}) \cup \unv{\hctxappsub{2}}$ and $\avarset_{2} = (\avarsettwo_{2} \setminus \set{\var}) \cup \anv{\hctxappsub{2}}$. Note that by \refdef{Reduction_places_in_usefulopencbneed} (Reduction places in \usefulopencbneed) it would be that $\tm_{2} = (\la{\var_{2}}{\tmtwo_{2}}) \tmthree_{2}$. This is absurd, because $\val_{1}$ cannot be rewritten as $\hctxappsub{2} \ctxholep{(\la{\var_{2}}{\tmtwo_{2}}) \tmthree_{2}}$---\reflemma{Redex_in_non-normal_terms-usefulopencbneed} (Redex in non-normal terms).

								\item The cases of rules $\ruleUseVarExp$ and $\ruleUseNonVarExp$ are impossible because $\val_{1}$ cannot be rewritten as a variable or a value.

								\item The case of $\ruleUseNonVarExp$ follows by application of the \ih.

								\item  Let $\pctx_{2} \in \evctxAppp{\uvarset_{2}}{\avarset_{2}}$ be derived as follows
									$$
									\infer
										[\ruleUseGcExp]
										{\appES{\pctxtwo_{2}}{\esub{\var}{\itmplus_{2}}} \in \evctxAppp{\uvarset_{2}}{\avarset_{2}}}
										{\pctxtwo_{2} \in \evctxAppp{\uvarset_{2}}{\avarset_{2}}
										\quad
										\var \notin (\uvarset_{2} \cup \avarset_{2})}
									$$
								where $\pctx_{2} = \appES{\pctxtwo_{2}}{\esub{\var}{\itmplus}}$. Note that $\tm_{1} \neq \var$---otherwise $\itmplus$ is a value; absurd. By application of the \ih on $\pctxtwo_{1} \ctxholep{\tm_{1}}$ and $\pctxtwo_{2} \ctxholep{\tm_{2}}$ gives that $\pctxtwo_{1} = \pctxtwo_{2}$---and so $\pctx_{1} = \pctx_{2}$---and $\tm_{1} = \tm_{2}$.

								\item In the case where $\pctx_{2} \in \evctxAppp{\uvarset_{2}}{\avarset_{2}}$ is derived via an application of rule $\ruleUseUselessExp$, the statement follows by \ih like in the previous item.

								\item Suppose $\pctx_{2} \in \evctxAppp{\uvarset_{2}}{\avarset_{2}}$ is derived as follows
									$$
									\infer
										[\ruleUseHerExp]
										{\appES{\pctxtwo_{2} \ctxholep{\var}}{\esub{\var}{\ctxhole}} \in \evctxAppp{\uvarsettwo_{2} \setminus \set{\var}}{\avarset_{2}}}
										{\pctxtwo_{2} \in \evctxAppp{\uvarsettwo_{2}}{\avarset_{2}}
										\quad
										\var \notin \avarset_{2}}
									$$
								with $\pctx_{2} = \appES{\pctxtwo_{2} \ctxholep{\var}}{\esub{\var}{\ctxhole}}$ and $\uvarset_{2} = \uvarsettwo_{2} \setminus \set{\var}$. However, note that it should be that $\val_{1} = \tm_{2}$. By \reflemma{Redex_in_non-normal_terms-usefulopencbneed} (Redex in non-normal terms), this is absurd.

							\end{itemize}

						\item \emph{Rule $\ruleUseGcMult$}: Let $\pctx_{1} \in \evctxUsep{\uvarset_{1}}{\avarset_{1}}$ be derived as follows:
							$$
							\infer
								[\ruleUseGcMult]
								{\appES{\pctxtwo_{1}}{\esub{\var}{\tmseven_{1}}} \in \evctxUsep{\uvarset_{1}}{\avarset_{1}}}
								{\pctxtwo_{1} \in \evctxUsep{\uvarset_{1}}{\avarset_{1}}
								\quad
								\var \notin (\uvarsettwo_{1} \cup \avarset_{1})}
							$$
						where $\pctx_{1} = \appES{\pctxtwo_{1}}{\esub{\var}{\tmseven_{1}}}$. Case analysis on the last derivation rule in $\pctx_{2}  \in \evctxAppp{\uvarset_{2}}{\avarset_{2}}$:
							\begin{itemize}
								\item The case of rule $\ruleUseAxOneExp$ is impossible.

								\item Suppose $\pctx_{2} \in \evctxAppp{\uvarset_{2}}{\avarset_{2}}$ is derived as follows
									$$
									\infer
										[\ruleUseAxTwoExp]
										{\appES{\pctxtwo_{2} \ctxholep{\var}}{\esub{\var}{\hctxappsub{2}}} \in \evctxAppp{(\uvarsettwo_{2} \setminus \set{\var}) \cup \unv{\hctxappsub{2}}}{(\avarsettwo_{2} \setminus \set{\var}) \cup \anv{\hctxappsub{2}}}}
										{\pctxtwo_{2} \in \evctxUsep{\uvarsettwo_{2}}{\avarsettwo_{2}}
										\quad
										\var \notin (\uvarsettwo_{2} \cup \avarsettwo_{2})}
									$$
								with $\pctx_{2} = \appES{\pctxtwo_{2} \ctxholep{\var}}{\esub{\var}{\hctxappsub{2}}}$, $\uvarset_{2} = (\uvarsettwo_{2} \setminus \set{\var}) \cup \unv{\hctxappsub{2}}$ and $\avarset_{2} = (\avarsettwo_{2} \setminus \set{\var}) \cup \anv{\hctxappsub{2}}$. However, by application of the \ih on $\pctxtwo_{1} \ctxholep{\tm_{1}}$ and $\pctxtwo_{2} \ctxholep{\var}$ we would have that $\tm_{1} = \var$, forcing $\tmseven_{1} = \hctxappsub{2} \ctxholep{\tm_{2}}$ to be a value---by \refdef{Reduction_places_in_usefulopencbneed} (Reduction places in \usefulopencbneednsp); absurd---by \reflemma{Redex_in_non-normal_terms-usefulopencbneed} (Redex in non-normal terms).

								\item Let $\pctx_{2} \in \evctxAppp{\uvarset_{2}}{\avarset_{2}}$ be derived as follows
									$$
									\infer
										[\ruleUseVarExp]
										{\appES{\pctxtwo_{2}}{\esub{\var}{\varthree_{2}}} \in \evctxAppp{\updp{\uvarsettwo_{2}}{\var}{\varthree_{2}}}{\updp{\avarsettwo_{2}}{\var}{\varthree_{2}}}}
										{\pctxtwo_{2} \in \evctxAppp{\uvarsettwo_{2}}{\avarsettwo_{2}}
										\quad
										}
									$$
								where $\pctx_{2} = \appES{\pctxtwo_{2}}{\esub{\var}{\varthree_{2}}}$, $\uvarset_{2} = \updp{\uvarsettwo_{2}}{\var}{\varthree_{2}}$ and $\avarset_{2} = \updp{\avarsettwo_{2}}{\var}{\varthree_{2}}$. We can then apply the \ih on $\pctxtwo_{1} \ctxholep{\tm_{1}}$ and $\pctxtwo_{2} \ctxholep{\tm_{2}}$ to obtain that $\pctxtwo_{1} = \pctxtwo_{2}$---and so $\pctx_{1} = \pctx_{2}$---and $\tm_{1} = \tm_{2}$.

								\item In the cases where $\pctx_{2} \in \evctxAppp{\uvarset_{2}}{\avarset_{2}}$ s derived via an application of rules $\ruleUseNonVarExp$, $\ruleUseUselessExp$ or $\ruleUseGcExp$, the statement follows by \ih.

								\item Finally, the case where $\pctx_{2} \in \evctxAppp{\uvarset_{2}}{\avarset_{2}}$ is derived via an application of rule $\ruleUseHerExp$ is ruled out as the case where it is derived via an application of rule $\ruleUseAxTwoExp$.

							\end{itemize}

					\end{itemize}

			\end{itemize}

		\item Let $\pctx_{1} \in \evctxAppp{\uvarset_{1}}{\avarset_{1}}$. The case where $\pctx_{2} \in \evctxUsep{\uvarset_{2}}{\avarset_{2}}$ was already covered when we took $\pctx_{1}$ to be a multiplicative evaluation context and $\pctx_{2}$ to be an exponential one. Let $\pctx_{2} \in \evctxAppp{\uvarset_{2}}{\avarset{2}}$. We proceed by induction on the derivation of $\pctx_{1} \in \evctxAppp{\uvarset_{1}}{\avarset_{1}}$. Case analysis on the last derivation rule:
					\begin{itemize}
						\item \emph{Rule $\ruleUseAxOneExp$}: Let $\pctx_{1} \in \evctxAppp{\uvarset_{1}}{\avarset_{1}}$ be derived as follows
							$$
							\infer
								[\ruleUseAxOneExp]
								{\progentry{\hctxappsub{1}}{\emptyenv} \in 
								\evctxAppp{\unv{\hctxappsub{1}}}{\anv{\hctxappsub{1}}}}
								{}
							$$
						where $\pctx_{1} = \progentry{\hctxappsub{1}}{\emptyenv}$, $\uvarset_{1} = $ and $\avarset_{1} = \anv{\hctxappsub{1}}$. Then it must be that $\pctx_{2} = \progentry{\hctxappsub{2}}{\emptyenv}$, and so \reflemma{Unique_decomposition_of_Lambda-terms-usefulopencbneed} (Unique decomposition of terms) gives us that $\hctxappsub{1} = \hctxappsub{2}$ and so $\pctx_{1} = \pctx_{2}$ and $\tm_{1} = \tm_{2}$.

						\item \emph{Rule $\ruleUseAxTwoExp$}: Let $\pctx_{1} \in \evctxAppp{\uvarset_{1}}{\avarset_{1}}$ be derived as follows
							$$
							\infer
								[\ruleUseAxTwoExp]
								{\appES{\pctxtwo_{1} \ctxholep{\var}} {\esub{\var}{\hctxappsub{1}}} \in \evctxAppp{(\uvarsettwo_{1} \setminus \set{\var}) \cup \unv{\hctxappsub{1}}}{(\avarsettwo_{1} \setminus \set{\var}) \cup \anv{\hctxappsub{1}}}}
								{\pctxtwo \in \evctxUsep{\uvarsettwo_{1}}{\avarsettwo_{1}}
								\quad
								\var \notin (\uvarsettwo_{1} \cup \avarsettwo_{1})}
							$$
						where $\pctx_{1} = \appES{\pctxtwo_{1} \ctxholep{\var}}{\esub{\var}{\hctxappsub{1}}}$, $\uvarset_{1} = (\uvarsettwo_{1} \setminus \set{\var}) \cup \unv{\hctxappsub{1}}$ and $\avarset_{1} = \anv{\hctxappsub{1}}$. By proceeding by case analysis on the last derivation rule in $\pctx_{2} \in \evctxAppp{\uvarset_{2}}{\avarset_{2}}$, we find that 
							\begin{itemize}
								\item Rule $\ruleUseAxOneExp$ is impossible.

								\item In the case where $\ruleUseAxTwoExp$ is the last derivation rule in $\pctx_{2} \in \evctxAppp{\uvarset_{2}}{\avarset_{2}}$, the statement follows by \ih and by application of \reflemma{Unique_decomposition_of_Lambda-terms-usefulopencbneed} (Unique decomposition of terms).

								\item Rules $\ruleUseVarExp$, $\ruleUseNonVarExp$ and $\ruleUseUselessExp$ are impossible, because normal terms cannot be rewritten as $\hctxappsub{1} \ctxholep{\tm_{1}}$---by \reflemma{Redex_in_non-normal_terms-usefulopencbneed} (Redex in non-normal terms).

								\item Suppose $\pctx_{2} \in \evctxAppp{\uvarset_{2}}{\avarset_{2}}$ was derived as follows
									$$
									\infer
										[\ruleUseGcExp]
										{\appES{\pctxtwo_{2}}{\esub{\var}{\tmseven_{2}}} \in \evctxAppp{\uvarset_{2}}{\avarset_{2}}}
										{\pctxtwo_{2} \in \evctxAppp{\uvarset_{2}}{\avarset_{2}}
										\quad
										\var \notin (\uvarset_{2} \cup \avarset_{2})}
									$$
								where $\pctx_{2} =\appES{\pctxtwo_{2}}{\esub{\var}{\tmseven_{2}}}$. But then application of the \ih gives that $\var = \tm_{2}$, implying that $\hctxappsub{1} \ctxholep{\tm_{1}} = \tmseven_{2}$ is a value; by application of \reflemma{Redex_in_non-normal_terms-usefulopencbneed} (Redex in non-normal terms), this is absurd.

								\item Rule $\ruleUseHerExp$ is impossible because $\ctxhole$ is not applicative---\reflemma{Unique_decomposition_of_Lambda-terms-usefulopencbneed} (Unique decomposition of terms) would otherwise give that $\hctxappsub{1} = \ctxhole$.

							\end{itemize}

						\item \emph{Rule $\ruleUseVarExp$}: Let $\pctx_{1} \in \evctxAppp{\uvarset_{1}}{\avarset_{1}}$ be derived as follows
							$$
							\infer
								[\ruleUseVarExp]
								{\appES{\pctxtwo_{1}}{\esub{\var}{\varthree_{1}}} \in \evctxAppp{\updp{\uvarsettwo_{1}}{\var}{\varthree_{1}}}{\updp{\avarsettwo_{1}}{\var}{\varthree_{1}}}}
								{\pctxtwo \in \evctxUsep{\uvarsettwo_{1}}{\avarsettwo_{1}}
								\quad
								\var \in (\uvarsettwo_{1} \cup \avarsettwo_{1})}
							$$
						where $\pctx_{1} = \appES{\pctxtwo_{1}}{\esub{\var}{\varthree_{1}}}$, $\uvarset_{1} = \updp{\uvarsettwo_{1}}{\var}{\varthree_{1}}$ and $\avarset_{1} = \updp{\avarsettwo_{1}}{\var}{\varthree_{1}}$. By proceeding by case analysis on the last derivation rule in $\pctx_{2} \in \evctxAppp{\uvarset_{2}}{\avarset_{2}}$, we find that 
							\begin{itemize}
								\item Rule $\ruleUseAxOneExp$ is impossible.

								\item Rules $\ruleUseAxTwoExp$ and $\ruleUseHerExp$ are ruled out by application of the \ih on the underlying exponential evaluation contexts.

								\item In the cases where $\pctx_{2} \in \evctxAppp{\uvarset_{2}}{\avarset_{2}}$ is derived via an application of rule $\ruleUseVarExp$ or an application of rule $\ruleUseGcExp$, the statement follows by \ih.

								\item Rules $\ruleUseNonVarExp$ and $\ruleUseUselessExp$ are impossible because $\varthree$ is not a non-variable inert term nor is it a value.

							\end{itemize}

						\item \emph{Rule $\ruleUseNonVarExp$}, \emph{Rule $\ruleUseUselessExp$}, \emph{Rule $\ruleUseGcExp$}: This cases are treated analogously to rule $\ruleUseVarExp$.

						\item \emph{Rule $\ruleUseHerExp$}: Let $\pctx_{1} \in \evctxAppp{\uvarset_{1}}{\avarset_{1}}$ be derived as follows
							$$
							\infer
								[\ruleUseHerExp]
								{\appES{\pctxtwo_{1} \ctxholep{\var}}{\esub{\var}{\ctxhole}} \in \evctxAppp{\uvarsettwo_{1} \setminus \set{\var}}{\avarset_{1}}}
								{\pctxtwo_{1} \in \evctxAppp{\uvarsettwo_{1}}{\avarset_{1}}
								\quad
								\var \notin \avarset_{1}}
							$$
						where $\pctx_{1} = \appES{\pctxtwo_{1} \ctxholep{\var}}{\esub{\var}{\ctxhole}}$ and $\uvarset_{1} = \uvarsettwo_{1} \setminus \set{\var}$. Note that $\tm_{1} = (\la{\var_{1}}{\tmtwo_{1}}) \tmthree_{1}$. By proceeding by case analysis on the last derivation rule in $\pctx_{2} \in \evctxAppp{\uvarset_{2}}{\avarset_{2}}$, we find that 
							\begin{itemize}
								\item Rule $\ruleUseAxOneExp$ is impossible.

								\item Rule $\ruleUseAxTwoExp$ is impossible because $\ctxhole$ is not applicative---\reflemma{Unique_decomposition_of_Lambda-terms-usefulopencbneed} (Unique decomposition of terms) would otherwise give that $\hctxappsub{2} = \ctxhole$.

								\item By application of the \ih, rules $\ruleUseVarExp$ (resp. $\ruleUseNonVarExp$ ; $\ruleUseUselessExp$) is excluded from the last possible derivation rule in $\pctx_{2} \in \evctxAppp{\uvarset_{2}}{\avarset_{2}}$ because variables (resp. useful inert terms ; values) cannot be rewritten as $\tm_{1} = (\la{\var_{1}}{\tmtwo_{1}}) \tmthree_{1}$.

								\item Suppose $\pctx_{2} \in \evctxAppp{\uvarset_{2}}{\avarset_{2}}$ is derived as follows
									$$
									\infer
										[\ruleUseGcExp]
										{\appES{\pctxtwo_{2}}{\esub{\var}{\tmseven_{2}}} \in \evctxAppp{\uvarset_{2}}{\avarset_{2}}}
										{\pctxtwo_{2} \in \evctxAppp{\uvarset_{2}}{\avarset_{2}}
										\quad
										\var \notin (\uvarset_{2} \cup \avarset_{2})}
									$$
								where $\pctx_{2} = $. But then application of the \ih on $\pctxtwo_{1} \ctxholep{\var}$ and $\pctxtwo_{2} \ctxholep{\tm_{2}}$ give that $\var = \tm_{2}$ and so it should be that $\tmseven_{2}$ is a value. But then $\tmseven_{2} = \tm_{1} = (\la{\var_{1}}{\tmtwo_{1}}) \tmthree_{1}$; absurd by \reflemma{Redex_in_non-normal_terms-usefulopencbneed} (Redex in non-normal terms).

								\item In the case where $\ruleUseHerExp$ is the last derivation rule in $\pctx_{2} \in \evctxAppp{\uvarset_{2}}{\avarset_{2}}$, the statement follows by \ih.\qedhere

							\end{itemize}

					\end{itemize}

	\end{itemize}

\end{proof}

\gettoappendix{prop:Determinism_of_usefulopencbneed}

\begin{proof}
We prove that if $\prog \tound \progtwo$ and $\prog \tound \progthree$ then $\progtwo = \progthree$. Let $\pctx_{1} \in (\evctxUsep{\uvarset_{1}}{\avarset_{1}} \cup \evctxAppp{\uvarset_{1}}{\avarset_{1}})$ be such that $\prog = \pctx_{1} \ctxholep{\tm_{1}} \tound \pctx_{1} \ctxholep{\tmfour_{1}} = \progtwo$, and let $\pctx_{2} \in (\evctxUsep{\uvarset_{2}}{\avarset_{2}} \cup \evctxAppp{\uvarset_{2}}{\avarset_{2}})$ be such that $\prog = \pctx_{2} \ctxholep{\tm_{2}} \tound \pctx_{2} \ctxholep{\tmfour_{2}} = \progthree$. 

First, note that if $\prog = \pctx_{1} \ctxholep{\tm_{1}} \toum \pctx_{1} \ctxholep{\tmfour_{1}} = \progtwo$, then it must be that $\tm_{1}$ is a $\beta$-redex. Similarly, if $\prog = \pctx_{1} \ctxholep{\tm_{1}} \toue \pctx_{1} \ctxholep{\tmfour_{1}} = \progtwo$, then it must be that $\tm_{1} \in \Var$, $\tm_{1} \in \dom{\pctx_{1}}$, and $\pctx(\tm_{1})$ is a value. This implies that $\tm_{1}$ is a $S$-reduction place of $\pctx_{1} \ctxholep{\tm_{1}}$ for \emph{any} $S$. Similarly, we can prove that $\tm_{2}$ is a $S$-reduction place $\pctx_{2} \ctxholep{\tm_{2}}$ for \emph{any} $S$.

Thus, if we take $S \defeq \Var$, we may apply \reflemma{Unique_decomposition_of_programs-usefulopencbneed} (Unique decomposition of programs) to obtain that $\pctx_{1} = \pctx_{2}$ and $\tm_{1} = \tm_{2}$. That is, $\progtwo = \progthree$.\qedhere

\end{proof}

\subsection{The Usefulness Criterion}
\label{subsect:The_usefulness_criterion-Proofs_of_usefulopencbneed}

For proving \refprop{Usefulness_of_exponential_steps} (Usefulness of exponential steps), we first need to characterize the shape of exponential in a very practical way, in the specific terms required for proving \refprop{Usefulness_of_exponential_steps}. That is, we need the following

\begin{lemma}[Characterization of exponential evaluation contexts]
\label{l:Characterization_of_exponential_evaluation_contexts}
\hfill

Let $\pctx \in \evctxAppp{\uvarset}{\avarset}$ be an exponential evaluation context. Then $\pctx$ has one of the following three forms:
	\begin{enumerate}
		\item $\pctx = \progentry{\hctx}{\env}$, where 
			\begin{enumerate}
				\item $\hctx = \hctxtwop{\ctxhole \, \tmtwo}$ is an applicative term context, for some term context $\hctxtwo$ and term $\tmtwo$.

				\item $\pctxtwo = \progentry{\hctxtwo}{\env} \in \evctxUsep{\uvarset}{\avarset}$. 

				\item The derivation of $\pctx \in \evctxAppp{\uvarset}{\avarset}$ has no applications of rule $\ruleUseHerExp$.

			\end{enumerate}

		\item $\pctx = \progentry{\tm}{\env \, \esub{\varthree}{\hctx} \, \envtwo}$, where 
			\begin{enumerate}
				\item $\hctx = \hctxtwop{\ctxhole \, \tmtwo}$ is an applicative term context, for some term context $\hctxtwo$ and term $\tmtwo$.

				\item $\pctxtwo = \progentry{\tm}{\env \, \esub{\varthree}{\hctxtwo} \, \envtwo} \in \evctxUsep{\uvarset}{\avarset}$.

				\item The derivation of $\pctx \in \evctxAppp{\uvarset}{\avarset}$ has no applications of rule $\ruleUseHerExp$.

			\end{enumerate}

		\item $\pctx = \progentry{\tm}{\env \, \esub{\varthree}{\ctxhole} \, \envtwo}$, where
			\begin{enumerate}
				\item There exists exponential evaluation context $\pctxtwo \in \evctxAppp{\uvarsettwo}{\avarsettwo}$ such that $\progentry{\tm}{\env} = \pctxtwop{\varthree}$ and such that for every value $\val$, $\pctxthree \defeq \appES{\pctxtwo}{\esub{\varthree}{\val} \, \envtwo} \in \evctxAppp{\uvarset}{\avarset}$.

				\item The number of applications of rule $\ruleUseHerExp$ in the derivation of $\pctx \in \evctxAppp{\uvarset}{\avarset}$ is the number of applications of rule $\ruleUseHerExp$ in the derivation of $\pctxtwo \in \evctxAppp{\uvarsettwo}{\avarsettwo}$ \emph{plus 1}. The same relation holds between the derivations of $\pctx \in \evctxAppp{\uvarset}{\avarset}$ and of $\pctxthree \in \evctxAppp{\uvarset}{\avarset}$.

			\end{enumerate}

	\end{enumerate}

\end{lemma}

\begin{proof}
\hfill

By induction on $\pctx \in \evctxAppp{\uvarset}{\avarset}$. Cases:
	\begin{itemize}
		\item Base rule $\ruleUseAxOneExp$: then we are in the first case.

		\item Base rule $\ruleUseAxTwoExp$: then we are in the second case.

		\item Inductive rules $\ruleUseNonVarExp$, $\ruleUseVarExp$, $\ruleUseGcExp$, and $\ruleUseUselessExp$: they simply preserve the case given by the \ih.

		\item Inductive rule $\ruleUseHerExp$: then $\pctx = \appES{\pctx_1\ctxholep{\var}}{\esub{\var}{\ctxhole}}$ and it is derived as follows:
		$$			\infer
					[\ruleUseHerExp]
					{\appES{\pctx_1\ctxholep{\var}}{\esub{\var}{\ctxhole}} \in \evctxAppp{\uvarset}{\avarset}}
					{\pctx_1 \in \evctxAppp{\uvarset}{\avarset}			
					\quad
					\var \notin (\uvarset \cup \avarset)}		$$
		By \ih, $\pctx_1$ has one of the 3 shapes in the statement. If it is the first or the second then we end up in the third case (details similar to the next case, that is the more interesting one). If it is the third case, then $\pctx_1 = \progentry\tm{\env_1\esub\varthree\ctxhole\envtwo_1}$ and $\progentry\tm{\env_1} = \pctxtwo_1\ctxholep\varthree$ for some $\pctxtwo_1 \in \evctxAppp{\uvarsettwo}{\avarsettwo}$ such that $\pctxthree_1\defeq \appES{\pctxtwo_1}{(\esub\varthree\val\envtwo_1)} \in \evctxAppp{\uvarset}{\avarset}$ for every value $\val$. Now note that by taking $\pctxtwo \defeq \pctx_1$, $\env \defeq \env_1\esub\varthree\var\envtwo_1$, and $\envtwo \defeq \emptyenv$ we do satisfy the statement because we have 
			$$
			\infer
				[\ruleUseGcExp]
				{\appES{\pctxtwo}{\esub{\var}{\val}} \in \evctxAppp{\uvarset}{\avarset}}
				{\pctxtwo \in \evctxAppp{\uvarset}{\avarset}
				\quad
				\var \notin (\uvarset \cup \avarset)}
			$$
		as required, and that the number of $\ruleUseHerExp$ in the derivation of $\pctx \in \evctxAppp{\uvarset}{\avarset}$ is 1 plus the number in the derivation of $\pctxtwo \in \evctxAppp{\uvarset}{\avarset}$.\qedhere

	\end{itemize}

\end{proof}


\gettoappendix{prop:Usefulness_of_exponential_steps}

\begin{proof}
\hypertarget{targ-pr:Usefulness_of_exponential_steps}{}
\emph{(Click \hyperlink{targ-st:Usefulness_of_exponential_steps}{here} to go back to main chapter.)}
\hfill

Let $\prog= \pctxp{\var} \toue \pctxp{\rename\val} = \progtwo$ with $\pctx \in \evctxAppp{\uvarset}{\avarset}$ and $\pctx(\var) = \val$. The proof is by induction on the derivation of $\pctx \in \evctxAppp{\uvarset}{\avarset}$. Cases:
\begin{itemize}
	\item Rules $\ruleUseAxOneExp$,$\ruleUseAxTwoExp$, and $\ruleUseHerExp$: impossible, because if $\pctx = \progentry{\hctx}{\emptyenv}$, $\pctx =\appES{\pctxtwop{\var}}{\esub{\var}{\hctx}}$, or $\pctx = \appES{\pctxtwop{\var}}{\esub{\var}{\ctxhole}}$ then $\pctx(\var) = \bot$, against hypothesis.

	\item Rule $\ruleUseNonVarExp$: then $\pctx = \appES{\pctxtwo}{\esub{\vartwo}{\tm}}$ and it is derived as follows:
		$$			
		\infer
			[\ruleUseNonVarExp]
			{\appES{\pctxtwo}{\esub{\vartwo}{\tm}} \in \evctxAppp{(\uvarsettwo {\setminus} \set{\vartwo}) \cup \unv{\tm}}{(\avarsettwo {\setminus} \set{\vartwo}) \cup \anv{\tm}}}
			{\pctxtwo \in \evctxAppp{\uvarsettwo}{\avarsettwo}
			\quad
			\vartwo \in (\uvarsettwo \cup \avarsettwo)
			\quad
			\ufinertpr{\tm}}
		$$
	with $\uvarset = (\uvarsettwo {\setminus} \set{\vartwo}) \cup \unv{\tm}$ and $\avarset = (\avarsettwo {\setminus} \set{\vartwo}) \cup \anv{\tm}$. Note that $\var\neq \vartwo$ because $\pctx(\var)$ is not an inert term. Let $\progtwo' \defeq \pctxtwop\var$. By \ih, there exists $\derivtwo:\progtwo' \toue^k \toum \progthree'$ as in the statement. Let the evaluation context of each $\toue$-step in $\derivtwo$ be $\pctxtwo_{\usefulExpSym,1},..., \pctxtwo_{\usefulExpSym,k} \in \evctxAppp{\uvarsettwo}{\avarsettwo}$, and let the $\toum$-step in $\derivtwo$ be $\pctxtwo_{\usefulMultSym} \in \evctxUsep{\uvarsettwo}{\avarsettwo}$. We can derive the evaluation contexts of each $\toue$-step in $\deriv$, say $\pctx_{\usefulExpSym,1},..., \pctx_{\usefulExpSym,1} \in \evctxAppp{\uvarsettwo}{\avarsettwo}$, and the evaluation context of the $\toum$ in $\deriv$, $\pctx_{\usefulMultSym}$ by applying the $\ruleUseNonVarExp$ rule obtaining a derivation $\derivtwo:\progtwo = \appES{\progtwo'}{\esub\var\tm} \toue^k \toum \appES{\progthree'}{\esub\var\tm}$ as in the statement.

	\item Rule $\ruleUseVarExp$: exactly as the previous one.

	\item Rule $\ruleUseGcExp$: then $\pctx = \appES{\pctxtwo}{\esub{\vartwo}{\tm}}$ and it is derived as follows:
	$$			\infer
				[\ruleUseGcExp]
				{\appES{\pctxtwo}{\esub{\vartwo}{\tm}} \in \evctxAppp{\uvarset}{\avarset}}
				{\pctxtwo \in \evctxAppp{\uvarset}{\avarset}
				\quad
				\vartwo \notin (\uvarset \cup \avarset)}$$
	If $\var\neq\vartwo$ then it goes as in the previous case. Otherwise, $\tm = \val$ and $\progtwo = \appES{\pctxtwop{\rename\val}}{\esub{\var}{\val}}$. By \reflemma{Characterization_of_exponential_evaluation_contexts}, $\pctxtwo$ has one of the following three forms:
		\begin{enumerate}
			\item $\pctxtwo = \progentry\hctx\env$, where $\hctx$ is applicative, that is $\hctx = \hctxtwop{\ctxhole\tmtwo}$ for some term $\tmtwo$, and $\pctxthree \defeq \progentry\hctxtwo\env \in \evctxUsep{\uvarset}{\avarset}$. Then $\progtwo = \pctxtwop{\rename\val} = \pctxthreep{\rename\val \tm}$ is a multiplicative redex in a useful multiplicative context, and so $\progtwo \toum \progthree$ for some $\progthree$. As required by the statement, the lemma gives also that there are no $\ruleUseHerExp$ rules in $\pctxtwo$, and thus in $\pctx$.

			\item $\pctxtwo = \progentry\tm{\env\esub\varthree\hctx\envtwo}$, where $\hctx$ is applicative. It goes as in the previous case.
			\item $\pctxtwo = \progentry\tm{\env\esub\varthree\ctxhole\envtwo}$ and $\progentry\tm\env = \pctxthreep\varthree$ for some $\pctxthree \in \evctxAppp{\uvarsettwo}{\avarsettwo}$ having one less  $\ruleUseHerExp$ rule than $\pctxtwo$ and such that $\ctxthree_1\defeq \appES\pctxtwo{(\esub\varthree{\valtwo}\envtwo)} \in \evctxAppp{\uvarset}{\avarset}$ for every value $\valtwo$. Now, consider taking $\valtwo$ as $\rename\val$ and note that 
				$$
				\progtwo'\defeq \pctxtwop{\rename\val} = \appES{\pctxthreep\varthree}{(\esub\varthree{\rename\val}\envtwo)} \toue \appES{\pctxthreep{\rename\val}}{(\esub\varthree{\rename\val}\envtwo)} \eqdef \progtwo''
				$$ 
			By \ih, there exists $\deriv':\progtwo''\toue^{k'} \toum \progthree'$ for some $\progthree'$, where $k'$ is the number of $\ruleUseHerExp$ in $\pctxtwo$ and the evaluation contexts satisfy the statement. Then we can apply rule $\ruleUseGcExp$ to all these evaluation contexts, and to $\pctxtwo$ as well, obtaining the following derivation, that satisfies the statement.
				$$
				\deriv:\progtwo = \appES{\progtwo'}{\esub{\var}{\val}}\ \  \toue\ \  \appES{\progtwo''}{\esub{\var}{\val}}\ \  \toue^{k'} \toum\ \  \appES{\progthree'}{\esub{\var}{\val}}
				$$

		\end{enumerate}
	
		\item Rule $\ruleUseUselessExp$: exactly as the previous one.\qedhere

	\end{itemize}

\end{proof}


\subsection{Characterizing {\usefulopencbneednsp}-normal forms.}
\label{subsect:Characterizing_usefulopencbneedn-normal_forms-Proofs_of_usefulopencbneed}

The following proposition from the body of the paper:
\gettoappendix{l:unf-usef-preds}
\noindent is actually decomposed into a sequence of lemmas here in the appendix. First, we prove the disjointness property, that is, the fact that no two predicates about the shape of normal forms can be true at the same time, in \reflemma{Disjointness_of_generalized_variables_useful_abstraction_programs_and_useful_inert_programs} below. Then, the three enumerated points are proved separately, by the lemmas about the properties of generalized variables (\reflemma{Properties_of_generalized_variables}), useful inert programs (\reflemma{Properties_of_useful_inert_programs}), and useful abstraction programs (\reflemma{Properties_of_useful_abstraction_programs}) given below.

\begin{lemma}[Disjointness of generalized variables, useful abstraction programs and useful inert programs]
For every program $\prog$, \emph{at most one} of the following holds: $\genVarSetpr{\prog}$, $\ufabspr{\prog}$, or $\ufinertpr{\prog}$.
\label{l:Disjointness_of_generalized_variables_useful_abstraction_programs_and_useful_inert_programs}
\end{lemma}

\begin{proof}
Let $\prog = \progentry{\tm}{\env}$. We proceed by induction on $\size{\env}$:
	\begin{itemize}
		\item \emph{Base case}: Let $\env \defeq \emptyenv$. The statement follows by the fact that the base derivation rules for $\genVarpr{.}$, $\ufabspr{.}$ and $\ufinertpr{.}$ can only be applied to pairwise syntactically distinct terms.

		\item \emph{Inductive case}: Let $\env \defeq \envtwo \esub{\vartwo}{\tmtwo}$. By \ih, $\progentry{\tm}{\envtwo}$ can satisfy \emph{at most} one of the three predicates. The proof proceeds by assuming that $\genVarentrypr{\tm}{\envtwo \esub{\vartwo}{\tmtwo}}$ (resp. $\ufabsentrypr{\tm}{\envtwo \esub{\vartwo}{\tmtwo}}$ ; $\ufinertpr{\tm}{\envtwo \esub{\vartwo}{\tmtwo}}$), and then proving that the other predicates cannot hold for $\progentry{\tm}{\envtwo \esub{\vartwo}{\tmtwo}}$. We avoid the repetitive work here, since every case follows from the \ih and the fact that each derivation rule for each of the predicates takes as premise the specific syntactical category of $\tmtwo$.\qedhere

	\end{itemize}

\end{proof}


\begin{lemma}[Properties of useful inert terms]
\hypertarget{targ-st:Properties_of_useful_inert_terms}{}
\hfill

Let $\itmplus$ be a non-variable inert term. Then $\sizend{\itmplus} \geq 1$ and $\anv{\itmplus} \neq \emptyset$.

\label{l:Properties_of_useful_inert_terms}
\end{lemma}

\begin{proof}
The fact that $\anv{\itmplus} \neq \emptyset$ can be easily proven by induction on the structure of $\itmplus$, while $\sizend{\itmplus} \geq 1$ simply follows by the fact that $\itmplus$ is an application term.
\end{proof}


\begin{lemma}[Properties of generalized variables]
Let $\genVarpr{\var}{\prog}$. Then $\sizend{\prog} = 0$, $\nv{\prog} = \unv{\prog} = \set{\var}$ and $\anv{\prog} = \emptyset$. Moreover, $\unf\prog = \var$.

\label{l:Properties_of_generalized_variables}
\end{lemma}


\begin{proof}
We proceed by induction on the derivation of $\genVarpr{\var}{\prog}$:
	\begin{itemize}
		\item If $\prog = \progentry{\var}{\emptyenv}$ then the statement clearly follows.
		
		\item Let $\genVarpr{\var}{\prog}$ be derived as follows
			$$
			\infer
				[\ruleGenVarHer]
				{\genVarpr{\var}{\appES{\progtwo}{\esub{\vartwo}{\var}}}}
				{\genVarpr{\vartwo}{\progtwo}}
			$$
		By \ih on $\progtwo$, $\nv{\progtwo} = \unv{\progtwo} = \set{\vartwo}$, $\anv{\progtwo} = \emptyset$ and $\sizend{\progtwo} = 0$\erase{ and $\unf{\progtwo} = \vartwo$}. The statement follows easily. In particular, note that $\unf{\prog} = \unf{\progtwo} \isub{\vartwo}{\var} =_{\ih} \vartwo \isub{\vartwo}{\var} = \var$.
		
		\item Let $\genVarpr{\var}{\prog}$ be derived as follows
			$$
			\infer
				[\ruleGenVarGc]
				{\genVarpr{\var}{\appES{\progtwo}{\esub{\vartwo}{\tm}}}}
				{\genVarpr{\var}{\progtwo}
				\quad
				\vartwo \neq \var}
			$$
		By \ih on $\progtwo$, $\nv{\progtwo} = \unv{\progtwo} = \set{\var}$, $\anv{\progtwo} = \emptyset$ and $\sizend{\progtwo} = 0$\erase{ and $\unf{\progtwo} = \var$}. The statement follows easily. In particular, note that $\unf{\prog} = \unf{\progtwo} \isub{\vartwo}{\var} =_{\ih} \var \isub{\vartwo}{\var} = \var$.\qedhere
		
	\end{itemize}
	
\end{proof}

\begin{lemma}[Inert terms, substitutions, and variables]
\label{l:properties-of-inert-terms-and-subs} 
Let $\itm$ be an inert term.
\begin{enumerate}
\item \label{p:properties-of-inert-terms-and-subs-one} Let $\itm'$ be a inert term such that $\var\notin\fv{\itm'}$. If $\var\in\nv\itm$ then $\tm \defeq \itm\isub\var{\itm'}$ is an inert term such that  $\anv{\tm} = \anv{\itm,\esub\var{\itm'}}$ and  $\unv{\tm} = \unv{\itm,\esub\var{\itm'}}$.


\item \label{p:properties-of-inert-terms-and-subs-three}
If $\var\in\uselesspr\itm$ then $\tm\defeq\itm\isub\var\val$ is an inert term such that $\anv\tm = \anv\itm$ and $\unv\tm = \unv\itm\setminus\set\var$.

\item \label{p:properties-of-inert-terms-and-subs-two}
If $\var\notin\nv\itm$ then $\tm\defeq\itm\isub\var\tmtwo$ is an inert term such that $\anv\tm = \anv\itm$ and $\unv\tm = \unv\itm$.
\end{enumerate}
\end{lemma}

\begin{proof}
By induction on the structure of $\itm$.
\end{proof}

\begin{lemma}[Properties of useful inert programs]
\hypertarget{targ-st:Properties_of_useful_inert_programs}{}
Let $\ufinertpr{\prog}$. Then $\sizend{\prog} \geq 1$ and $\anv{\prog} \neq \emptyset$. Moreover, $\anv{\prog} = \anv{\unf\prog}$, $\unv{\prog} = \unv{\unf\prog}$, and $\unf\prog$ is a non-variable inert term.

\label{l:Properties_of_useful_inert_programs}
\end{lemma}

\gettoappendix{l:Properties_of_useful_inert_programs}

\begin{proof}
We proceed by induction on the derivation of $\ufinertpr{\prog}$:
	\begin{itemize}
		\item If $\ufinertpr{\prog}$ is derived as
			$$
			\infer
				[\ufRuleInertProm]
				{\ufinertentrypr{\itmplus}{\emptyenv}}
				{}
			$$
		with $\prog = \progentry{\itmplus}{\emptyenv}$, then the statement follows by application of \reflemma{Properties_of_useful_inert_terms} (Properties of useful inert terms) on $\itmplus$. Moreover, $\unf{\progentry{\itmplus}{\emptyenv}} = \itmplus$, $\anv{\progentry{\itmplus}{\emptyenv}} = \anv{\itmplus}$, and $\unv{\progentry{\itmplus}{\emptyenv}} = \unv{\itmplus}$.
		
		\item Let $\ufinertpr{\prog}$ be derived as
			$$
			\infer
				[\ufRuleInertGenVar]
				{\ufinertpr{\appES{\progtwo}{\esub{\var}{\itmplus}}}}
				{\genVarpr{\var}{\progtwo}}
			$$
		with $\prog = \appES{\progtwo}{\esub{\var}{\itmplus}}$. By \reflemma{Properties_of_generalized_variables} (Properties of generalized variables) on $\progtwo$, $\nv{\progtwo} = \unv{\progtwo} = \set{\var}$, $\anv{\progtwo} = \emptyset$ and $\sizend{\progtwo} = 0$\erase{, and $\unfoldpr{\prog} = \var$}. By \reflemma{Properties_of_useful_inert_terms} (Properties of useful inert terms) on $\itmplus$, $\anv{\itmplus} \neq \emptyset$, $\sizend{\itmplus} \geq 1$, and $\itmplus$ is a non-variable inert term.
		
		Therefore, the statement holds for $\prog$ by noting that $\anv{\prog} = (\anv{\progtwo} \setminus \set{\var}) \cup \anv{\tm} \geq \anv{\tm} \neq \emptyset$, $\sizend{\prog} = \sizend{\progtwo} + \sizend{\tm} \geq \sizend{\tm} \geq 1$\erase{, and that---by \reflemma{Properties_of_generalized_variables} (Properties of generalized variables)---$\unfoldpr{\prog} = (\unfoldpr{\progtwo}) \isub{\var}{\itmplus} = \var \isub{\var}{\itmplus} = \itmplus$}. Moreover, $\unf\progtwo = \var$ by \reflemma{Properties_of_generalized_variables}, which gives $\unf{\appES{\progtwo}{\esub{\var}{\itmplus}}} = \unf\progtwo\isub\var{\itmplus} = \var \isub\var{\itmplus} = \itmplus$. By \reflemma{Properties_of_generalized_variables}, $\unv{\progtwo} = \set{\var}$ and $\anv{\progtwo} = \emptyset$, from which it follows that $\unv{\appES{\progtwo}{\esub{\var}{\itmplus}}} = \unv{\itmplus} $, that is, $\unv\prog =  \unv{\unf\prog}$, and similarly for applied variables, as required.
		
		\item Let $\ufinertpr{\prog}$ be derived as
			$$
			\infer
				[\ufRuleInertInert]
				{\ufinertpr{\appES{\progtwo}{\esub{\var}{\itm}}}}
				{\ufinertpr{\progtwo}
				\quad
				\var \in \nv{\progtwo}}
			$$
		with $\prog = \appES{\progtwo}{\esub{\var}{\itm}}$. By \ih on $\progtwo$, $\anv{\progtwo} \neq \emptyset$ and $\sizend{\progtwo} \geq 1$\erase{ and $\unfoldpr{\progtwo} = \itmplus$, for some non-variable inert term $\itmplus$}. Note that then $\sizend{\prog} \geq \sizend{\progtwo} \geq 1$. 
				
		Note that if $\itm \in \Var$, then the statement follows easily from the \ih Let $\itm$ be a non-variable inert term. Recall that $\var \in \nv{\progtwo}$. Two cases:
		
		\begin{itemize}
			\item $\var\in \anv\progtwo$ and $\itm=\vartwo$. Then $\anv\prog = (\anv\progtwo\setminus \set\var) \cup\set\vartwo \neq \emptyset$.
			\item $\var\notin \anv\progtwo$ or $\itm\neq\vartwo$. Then $\anv\prog = (\anv\progtwo\setminus \set\var) \cup \anv\itm$. Since then $\itm$ is a useful inert term, by \reflemma{Properties_of_useful_inert_terms} (Properties of useful inert terms) $\anv\itm \neq \emptyset$, and so $\anv\prog\neq\emptyset$.
		\end{itemize}
				
				Moreover, by \ih, $\unf\progtwo$ is of the form $\itmplustwo$. Then $\unf{\appES{\progtwo}{\esub{\var}{\itm}}} = \unf\progtwo\isub\var\itm = \itmplustwo\isub\var\itm$, which by \reflemmap{properties-of-inert-terms-and-subs}{one} is an inert term, and it is not a variable because $\itmplustwo$ is not.

By \ih, we have $\anv{\itmplustwo} = \anv\progtwo$ and $\unv{\itmplustwo} = \unv\progtwo$. It follows that $\anv{\progentry{\itmplustwo}{\esub{\var}{\itm}}} = \anv{\appES{\progtwo}{\esub{\var}{\itm}}}$, and similarly for unapplied variables. Then $\anv{\unf\prog} = \anv{\itmplustwo\isub\var\itm} =_{\reflemmaeqp{properties-of-inert-terms-and-subs}{one}} \anv{\progentry{\itmplustwo}{\esub{\var}{\itm}}} = \anv{\appES{\progtwo}{\esub{\var}{\itm}}} = \anv\prog$, and similarly for unapplied variables.

		\item Let $\ufinertpr{\prog}$ be derived as
			$$
			\infer
				[\ufRuleInertUseless]
				{\ufinertpr{\appES{\progtwo}{\esub{\var}{\val}}}}
				{\ufinertpr{\progtwo}
				\quad
				\var \in \uselesspr{\progtwo}}
			$$
		with $\prog = \appES{\progtwo}{\esub{\var}{\val}}$. Note that $\var \in \nv{\progtwo}$---by \reflemmap{Unapplied_applied_and_needed_variables}{two} (Unapplied, applied and needed variables). By \ih on $\progtwo$, $\anv{\progtwo} \neq \emptyset$, $\sizend{\progtwo} \geq 1$\erase{, and $\unfoldpr{\progtwo} = \itmplus$, for some non-variable inert term $\itmplus$}.
		
		Therefore, we have that $\anv{\prog} = (\anv{\progtwo} \setminus \set{\var}) \cup \anv{\val} = \anv{\progtwo} \neq \emptyset$, $\sizend{\prog} = \sizend{\progtwo} + \sizend{\val} \geq \sizend{\progtwo} \geq 1$\erase{ and $\unfoldpr{\prog} = (\unfold{\progtwo}) \isub{\var}{\val} = \itmplus \isub{\var}{\val}$ is a non-variable inert term, by \reflemma{preservation-of-inertness-through-implicit-substitutions} (Preservation of inertness through implicit substitutions) and the fact that $\var \in \uselesspr{\unfoldpr{\progtwo}}$---by \reflemma{preservation-of-unapplied-and-applied-variables-through-unfolding} (Preservation of unapplied and applied variables through unfolding)}.
		
		Moreover, by \ih $\unf\progtwo = \itmplus$ and $\var\in\uselesspr\itmplus$. By \reflemmap{properties-of-inert-terms-and-subs}{three}, $\unf{\appES{\prog}{\esub{\var}{\val}}} = \unf\prog\isub\var\val =_{\ih} \itmplus\isub\var\val$ is an inert term, which is not a variable because $\itmplus$ is not. For applied variables, since $\var \in \uselesspr\progtwo = \uselesspr{\itmplus}$, we have $\var \notin \anv\progtwo = \anv{\itmplus}$, and so $\anv{\appES{\progtwo}{\esub{\var}{\val}}} = \anv\progtwo =_{\ih} \anv{\itmplus} =_{\reflemmaeqp{properties-of-inert-terms-and-subs}{three}} \anv{\itmplus\isub\var\val} = \anv{\unf\prog}$. For unapplied variables, we have $\var \in \unv\progtwo = \unv{\itmplus}$, and so $\unv{\appES{\progtwo}{\esub{\var}{\val}}} = \unv\progtwo \setminus\set\var =_{\ih} \unv{\itmplus}\setminus\set\var =_{\reflemmaeqp{properties-of-inert-terms-and-subs}{three}} \unv{\itmplus\isub\var\val} = \unv{\unf\prog}$.
		
		\item Let $\ufinertpr{\prog}$ be derived as
			$$
			\infer
				[\ufRuleInertGc]
				{\ufinertpr{\appES{\progtwo}{\esub{\var}{\tm}}}}
				{\ufinertpr{\progtwo}
				\quad
				\var \notin \nv{\progtwo}}
			$$
		with $\prog = \appES{\progtwo}{\esub{\var}{\tm}}$. By \ih on $\progtwo$, $\anv{\progtwo} \neq \emptyset$ and $\sizend{\progtwo} \geq 1$\erase{ and $\unfoldpr{\progtwo} = \itmplus$, for some non-variable inert term $\itmplus$}. Thus, $\anv{\prog} = \anv{\progtwo} \neq \emptyset$, $\sizend{\prog} = \sizend{\progtwo} \geq 1$\erase{ and $\unfoldpr{\prog} = (\unfoldpr{\progtwo}) \isub{\var}{\tm} = \itmplus \isub{\var}{\tm}$---by \reflemma{preservation-of-unapplied-and-applied-variables-through-unfolding} (Preservation of unapplied and applied variables through unfolding) and \reflemma{preservation-of-unapplied-and-applied-variables-through-unfolding} (Preservation of unapplied and applied variables through unfolding)}.
		
						Moreover, by \ih, $\unf\progtwo$ is of the form $\itmplus$. Then $\unf{\appES{\progtwo}{\esub{\var}{\tm}}} = \unf\progtwo\isub\var\tm = \itmplus\isub\var\tm$, which by \reflemmap{properties-of-inert-terms-and-subs}{two} is an inert term, and it is not a variable because $\itmplus$ is not.
Since $\var\notin\nv\progtwo$, we have $\anv{\appES{\progtwo}{\esub{\var}{\tm}}} = \anv\progtwo$. By \reflemmap{properties-of-inert-terms-and-subs}{two}, we have $\anv{\itmplus\isub\var\tm} = \anv{\itmplus}$. Since by \ih $\anv\progtwo = \anv{\itmplus}$, we have $\anv{\appES{\progtwo}{\esub{\var}{\tm}}} =\anv{\itmplus\isub\var\tm}$, that is, $\anv\prog = \anv{\unf\prog}$, and similarly for unapplied variables.

\qedhere
		
	\end{itemize}

\end{proof}

\begin{lemma}[Properties of useful abstraction programs]
Let $\ufabspr{\prog}$. Then $\sizend{\prog} = 0$ and $\nv{\prog} = \unv{\prog} = \anv{\prog} = \emptyset$. Moreover, $\unf\prog$ is a value.

\label{l:Properties_of_useful_abstraction_programs}
\end{lemma}

\begin{proof}
By induction on the derivation of $\ufabspr{\prog}$:
	\begin{itemize}
		\item The statement holds trivially if $\prog = \progentry{\val}{\emptyenv}$. Moreover, $\unf{\progentry{\val}{\emptyenv}} = \val$.
		
		\item Let $\ufabspr{\prog}$ be derived as
			$$
			\infer
				[\ruleAbsGenVar]
				{\ufabspr{\appES{\progtwo}{\esub{\var}{\val}}}}
				{\genVarpr{\var}{\progtwo}}
			$$
		with $\prog = \appES{\progtwo}{\esub{\var}{\val}}$. Note that $\nv{\appES{\progtwo}{\esub{\var}{\val}}}  = \nv{\val}$---by \reflemma{Properties_of_generalized_variables} (Properties of generalized variables)---and so $\nv\prog = \emptyset$. We also have $\sizend{\appES{\progtwo}{\esub{\var}{\val}}} = \sizend{\progtwo} + \sizend{\val} =_{\reflemmaeq{Properties_of_generalized_variables}}  \sizend{\val} = 0$. Moreover, $\unf{\appES{\progtwo}{\esub{\var}{\val}}} = \un\progtwo\isub\var\val =_{\reflemmaeq{Properties_of_generalized_variables}}  \var\isub\var\val = \val$ which is a value.
				
		\item Let $\ufabspr{\prog}$ be derived as
			$$
			\infer
				[\ruleGenVarGc]
				{\ufabspr{\appES{\progtwo}{\esub{\var}{\tm}}}}
				{\ufabspr{\progtwo}}
			$$
		with $\prog = \appES{\progtwo}{\esub{\var}{\tm}}$. It follows easily by the \ih\qedhere

	\end{itemize}

\end{proof}

\begin{lemma}[Rewriting: term contexts]
\hfill

\begin{enumerate}

	\item 
	\label{p:Rewriting_term_contexts-Proofs_of_usefulopencbneed-one}
	\emph{Focusing inert terms on unapplied variables}:	Let $\itm$ be an inert term and let $\var \in \unv{\itm}$. Then there exists term context $\hctx_{\var}$ such that $\hctx_{\var} \ctxholep{\var} = \itm$, with $\var \notin \unv{\hctx_{\var}} \subset \unv{\itm}$ and $\anv{\hctx_{\var}} \subseteq \anv{\itm}$.
	
	\item 
	\label{p:Rewriting_term_contexts-Proofs_of_usefulopencbneed-two}
	\emph{Focusing term contexts on unapplied varables}: Let $\hctx$ be a term context and $\var \in \unv{\hctx}$. Then for every term $\tm$ there exists a term context $\hctx_{\tm}$ such that $\hctx_{\tm} \ctxholep{\var} = \hctxp{\tm}$, with $\var \notin \unv{\hctx_{\tm}} \subset \unv{\hctx}$ and $\var \notin \anv{\hctx_{\tm}} \subseteq \anv{\hctx}$.
	
\end{enumerate}

\label{l:Rewriting_term_contexts-Proofs_of_usefulopencbneed}
\end{lemma}


\begin{proof}
\hfill
\begin{enumerate}	
	
	\item
	\emph{Focusing inert terms on unapplied variables}: By structural induction on $\itm$:
		\begin{itemize}
			\item \emph{Variable}: The statement holds by taking $\hctx \defeq \ctxhole$.
			
			\item \emph{Application}: Let $\itm = \itmtwo \ntm$. If $\var \in \itmtwo$ then the \ih gives a term context $\hctxtwo$ such that $\hctxtwop{\var} = \itm$, with $\var \notin \unv{\hctxtwo} \subset \unv{\itmtwo}$ and $\var \notin \anv{\hctxtwo} \subseteq \anv{\itmtwo}$. The statement thus holds by taking $\hctx \defeq \hctxtwo \ntm$. 
			
			If instead $\var \notin \itmtwo$, then $\var \in \ntm$. This means that $\ntm$ is itself an inert term, allowing us to apply the \ih to get a term context $\hctxtwo$ such that $\hctxtwop{\var} = \ntm$, with $\var \notin \unv{\hctxtwo} \subset \unv{\itm}$ and $\var \notin \anv{\hctxtwo} \subseteq \anv{\itm}$. The statement thus holds by taking $\hctx \defeq \itmtwo \hctxtwo$.
			
		\end{itemize}
	
	\item 
	\emph{Focusing term contexts on unapplied variables}: Let $\tm$ be a term. We proceed by structural induction on $\hctx$:
		\begin{itemize}
			\item \emph{Empty context}: This case is impossible.
			
			\item \emph{Application left}: Let $\hctx \defeq \hctxtwo \tmtwo$. Then $\var \in \unv{\hctxtwo}$ and so application of the \ih gives a term context $\hctxtwo_{\tm}$ such that $\hctxtwo_{\tm} \ctxholep{\var} = \hctxtwop{\tm}$, with $\var \notin \unv{\hctxtwo_{\tm}} \subset \unv{\hctxtwo}$ and $\var \notin \anv{\hctxtwo_{\tm}} \subseteq \anv{\hctxtwo}$. The statement then holds by taking $\hctx_{\tm} \defeq \hctxtwo_{\tm} \tmtwo$.
			
			\item \emph{Application right}: Let $\hctx \defeq \itm \hctxtwo$. Case analysis on whether $\var \in \nv{\itm}$:
				\begin{itemize}
					\item Let $\var \in \nv{\itm}$. Then \reflemmap{Rewriting_term_contexts-Proofs_of_usefulopencbneed}{one} (Focusing inert terms on unapplied variables) gives a term context $\hctxthree_{\var}$ such that $\hctxthree_{\var} \ctxholep{\var} = \itm$, with $\var \notin \unv{\hctxthree_{\var}} \subset \unv{\itm}$ and $\var \notin \anv{\hctxthree_{\var}} \subseteq \anv{\itm}$. The statement thus holds by taking $\hctx_{\tm} \defeq \hctxthree_{\var} \hctxtwop{\tm}$, noting that $\hctx_{\tm} \ctxholep{\var} = \hctxthree_{\var} \ctxholep{\var} \hctxtwop{\tm} = \itm \hctxtwop{\tm} = \hctxp{\tm}$, $\var \notin \unv{\hctxthree_{\var}} \subset \unv{\hctx_{\tm}}$, and $\var \notin \anv{\hctxthree_{\var}} \subseteq \anv{\hctx_{\tm}}$.
					
					\item Let $\var \notin \nv{\itm}$. Then it must be that $\var \in \nv{\hctxtwo}$, and so we can apply the \ih to get a term context $\hctxtwo_{\tm}$ such that $\hctxtwo_{\tm} \ctxholep{\var} = \hctxtwop{\tm}$, with $\var \notin \unv{\hctxtwo_{\tm}} \subset \unv{\hctxtwo}$ and $\var \notin \anv{\hctxtwo_{\tm}} \subseteq \anv{\hctxtwo}$. The statement thus holds by taking $\hctx_{\tm} \defeq \itm \hctxtwo_{\tm}$, noting that $\hctx_{\tm} \ctxholep{\var} = \itm \hctxtwo_{\tm} \ctxholep{\var} = \itm \hctxtwop{\tm} = \hctxp{\tm}$, $\var \notin \unv{\hctx_{\tm}} = \unv{\itm} \cup \unv{\hctxtwo_{\tm}} \subset \unv{\itm} \cup \unv{\hctxtwo} = \unv{\hctx}$, and $\var \notin \anv{\hctx_{\tm}} = \anv{\itm} \cup \anv{\hctxtwo_{\tm}} \subseteq \anv{\itm} \cup \anv{\hctxtwo} = \anv{\hctx}$.\qedhere
					
				\end{itemize}
				
		\end{itemize}
	
\end{enumerate}

\end{proof}

\begin{lemma}[Rewriting: applicative term contexts]
\hfill

\begin{enumerate}
	\item 
	\label{p:Rewriting_applicative_term_contexts-one}
	\emph{Applicative term contexts give applied variables}: Let $\hctxapp$ be an applicative term context and $\var \in \Var$. Then $\var \in \anv{\hctxappp{\var}}$.
	
	\item
	\label{p:Rewriting_applicative_term_contexts-two}
	\emph{Focusing useful inert terms on applied variables}: Let $\itmplus$ be a useful inert term and $\var \in \anv{\itmplus}$. Then there exists an applicative term context $\hctxappsub{\var}$ such that $\hctxappsub{\var} \ctxholep{\var} = \itmplus$, with $\unv{\hctxappsub{\var}} \subseteq \unv{\itmplus}$ and $\var \notin \anv{\hctxappsub{\var}} \subset \anv{\itmplus}$.
		
	\item
	\label{p:Rewriting_applicative_term_contexts-three}
	\emph{Focusing term contexts on applied variables}: Let $\var \in \anv{\hctx}$. Then for every term $\tm$ there exists an applicative term context $\hctxappsub{\tm}$ such that $\hctxappsub{\tm} \ctxholep{\var} = \hctxp{\tm}$, with $\unv{\hctxappsub{\tm}} \subseteq \unv{\hctx}$ and $\var \notin \anv{\hctxappsub{\tm}} \subset \anv{\hctx}$.
	
\end{enumerate}

\label{l:Rewriting_applicative_term_contexts}
\end{lemma}


\begin{proof}
\hfill
\begin{enumerate}
	\item 
	\emph{Applicative term contexts give applied variables}: By structural induction on $\hctxapp$:
		\begin{itemize}
			\item \emph{Base case}: If $\hctxapp = \ctxhole \tm$ then $\var \in  (\set{\var} \cup \anv{\tm}) = \anv{\var \tm} = \anv{\hctxappp{\var}}$.
			
			\item \emph{Application left}: Let $\hctxapp \defeq \hctxapptwo \tm$. By \ih, $\var \in \anv{\hctxapptwop{\var}}$ and so $\var \in \anv{\hctxapptwop{\var}} \cup \anv{\tm} = \anv{\hctxp{\var}}$.
			
			\item \emph{Application right}: Let $\hctxapp \defeq \itm \hctxapptwo$. By \ih, $\var \in \anv{\hctxapptwop{\var}}$ and so $\var \in \anv{\itm} \cup \anv{\hctxapptwop{\var}} = \anv{\hctxp{\var}}$.
			
		\end{itemize}
	
	\item 
	\emph{Focusing useful inert terms on applied variables}: By structural induction on $\itmplus \defeq \itmtwo \ntm$. Note that if $\itmtwo = \var$ then the statement holds by taking $\hctxapp \defeq \ctxhole \ntm$. Let $\itmtwo$ be a useful inert term. We proceed by case analysis on whether $\var \in \anv{\itmplus}$:
		\begin{itemize}
			\item If $\var \in \anv{\itmtwo}$, then $\itmtwo$ is a useful inert term. By application of the \ih, we obtain an applicative term context $\hctxapptwosub{\var}$ such that $\itmtwo = \hctxapptwosub{\var} \ctxholep{\var}$, with $\unv{\hctxapptwosub{\var}} \subseteq \unv{\itmtwo}$ and $\var \notin \anv{\hctxapptwosub{\var}} \subset \anv{\itmtwo}$; thus, the statement holds by taking $\hctxapp \defeq \hctxapptwo \ntm$.
			
			\item Let $\var \notin \anv{\itmtwo}$. This means that $\var \in \anv{\ntm}$, which in turn means that $\ntm$ is a useful inert term. By \ih, there exists an applicative term context $\hctxapptwo$ such that $\ntm = \hctxapptwosub{\var} \ctxholep{\var}$, with $\unv{\hctxapptwosub{\var}} \subseteq \unv{\ntm}$ and $\var \notin \anv{\hctxapptwosub{\var}} \subseteq \anv{\ntm}$; thus, the statement holds by taking $\hctxappsub{\var} \defeq \itmtwo \hctxapptwosub{\var}$, since $\hctxappsub{\var} \ctxholep{\var} = \itmtwo \hctxapptwosub{\var} \ctxholep{\var} = \itmtwo \ntm = \itmplus$, $\unv{\hctxappsub{\var}} = \unv{\itmtwo} \cup \unv{\hctxapptwosub{\var}} \subseteq \unv{\itmtwo} \cup \unv{\ntm} = \unv{\itmplus}$, $\var \notin \anv{\itmtwo} \cup \anv{\hctxapptwosub{\var}} = \anv{\hctxappsub{\var}} \subset \anv{\itmtwo} \cup \anv{\ntm} = \anv{\itmplus}$.
			
		\end{itemize}
	
	\item 
	\emph{Focusing term contexts on applied variables}:
	Let $\tm$ be a term. We proceed by structural induction on $\hctx$:
		\begin{itemize}
			\item \emph{Context hole}: Impossible.
			
			\item \emph{Application left}: Let $\hctx \defeq \hctxtwo \tmtwo$. Since then $\var \in \anv{\hctx} = \anv{\hctxtwo}$, we can apply the \ih and get an applicative term context $\hctxapptwosub{\tm}$ such that $\hctxapptwosub{\tm} \ctxholep{\var} = \hctxtwop{\tm}$, with $\unv{\hctxapptwosub{\tm}} \subseteq \unv{\hctxtwo}$ and $\var \notin \anv{\hctxapptwosub{\tm}} \subset \anv{\hctxtwo}$. The statement holds by taking $\hctxapp \defeq \hctxapptwo \tmtwo$, and noting that $\hctxappsub{\tm} \ctxholep{\var} = \hctxapptwosub{\tm} \ctxholep{\var} \tm = \hctxtwop{\tm} \tmtwo = \hctxp{\tm}$, $\unv{\hctxappsub{\tm}} = \unv{\hctxapptwosub{\tm}} \subseteq \unv{\hctxtwo} = \unv{\hctx}$, and $\var \notin \anv{\hctxapptwosub{\tm}} = \anv{\hctxappsub{\tm}} \subset \anv{\hctxtwo} = \anv{\hctx}$.
			
			\item \emph{Application right}: Let $\hctx \defeq \itm \hctxtwo$. We proceed by case analysis on whether $\var \in \anv{\itm}$:
				\begin{itemize}
					\item If $\var \in \anv{\itm}$, then $\itm$ is a useful inert term. By application of \reflemmap{Rewriting_applicative_term_contexts}{two}  (Focusing useful inert terms on applied variables) on $\itm$, we get an applicative term context $\hctxapptwosub{\var}$ such that $\hctxapptwosub{\var} \ctxholep{\var} = \itm$, with $\unv{\hctxapptwosub{\var}} \subseteq \unv{\itm}$ and $\var \notin \anv{\hctxapptwosub{\var}} \subset \anv{\itm}$. The statement holds by taking $\hctxappsub{\tm} \defeq \hctxapptwosub{\var} \hctxtwop{\tm}$, since then $\hctxappsub{\tm} \ctxholep{\var} = (\hctxapptwop{\var}) \hctxtwop{\tm} = \itm \hctxtwop{\tm} = \hctxp{\tm}$, $\unv{\hctxappsub{\tm}} = \unv{\hctxapptwosub{\var}} \subseteq \unv{\itm} \subseteq \unv{\itm} \cup \unv{\hctxtwo} = \unv{\hctx}$ and $\var \notin \anv{\hctxapptwosub{\var}} = \anv{\hctxappsub{\tm}} \subseteq \anv{\itm} \subset \anv{\itm} \cup \anv{\hctxtwo} = \anv{\hctx}$.
					
					\item Let $\var \notin \anv{\itm}$. Then it must be that $\var \in \anv{\hctxtwo}$ and so the \ih gives an applicative term context $\hctxapptwosub{\tm}$ such that $\hctxapptwosub{\tm} \ctxholep{\var} = \hctxtwop{\tm}$, with $\unv{\hctxapptwosub{\tm}} \subseteq \unv{\hctxtwo} $ and $\var \notin \anv{\hctxapptwosub{\tm}} \subset \anv{\hctxtwo}$. The statement holds by taking $\hctxappsub{\tm} \defeq \itm \hctxapptwosub{\var}$, since then $\hctxappsub{\tm} \ctxholep{\var} = \itm \hctxapptwosub{\var} \ctxholep{\var} = \itm \hctxtwop{\tm} = \hctxp{\tm}$, $\unv{\hctxappsub{\tm}} = \unv{\itm} \cup \unv{\hctxapptwosub{\var}} \subseteq \unv{\itm} \cup \unv{\hctxtwo} = \unv{\hctx}$ and $\var \notin \anv{\itm} \cup \anv{\hctxapptwosub{\var}} = \anv{\hctxappsub{\tm}} \subset \anv{\itm} \cup \anv{\hctxtwo} = \anv{\hctx}$.\qedhere
					
				\end{itemize}
			
		\end{itemize}
	
\end{enumerate}

\end{proof}

\begin{lemma}[Rewriting evaluation contexts: base cases]
\hfill

\begin{enumerate}
	\item
	\label{p:Rewriting_evaluation_contexts-base_cases-one}
	\emph{Multiplicative evaluation contexts give needed variables}: Let $\pctx \in \evctxUsep{\uvarset}{\avarset}$ and $\var \notin \dom{\pctx}$. Then $\var \in \nv{\pctxp{\var}}$.

	\item
	\label{p:Rewriting_evaluation_contexts-base_cases-two}
	\emph{Exponential evaluation contexts give applied variables}: Let $\pctx \in \evctxAppp{\uvarset}{\avarset}$ and $\var \notin \dom{\pctx}$. Then $\var \in \anv{\pctxp{\var}}$.

	\item
	\label{p:Rewriting_evaluation_contexts-base_cases-three}
	\emph{Focusing multiplicative evaluation contexts on unapplied variables}: Let $\pctx \in \evctxUsep{\uvarset}{\avarset}$ and $\var \in \uvarset$. Then for every term $\tm$ there exists multiplicative evaluation context $\pctx_{\tm} \in \evctxUsep{\uvarset_{\tm}}{\avarset_{\tm}}$ such that $\pctx_{\tm} \ctxholep{\var} = \pctxp{\tm}$, $\var \notin \uvarset_{\tm} \subset \uvarset$, and $\var \notin \avarset_{\tm} \subseteq \avarset$.

\end{enumerate}

\label{l:Rewriting_evaluation_contexts-base_cases}
\end{lemma}


\begin{proof}
\hfill
\begin{enumerate}



	\item Easily provable by induction on $\pctx \in \evctxAppp{\uvarset}{\avarset}$. The only cases that are not provable by direct application of the \ih are the following:
		\begin{itemize}
			\item If $\pctx = \progentry{\hctxapp}{\emptyenv} \in \evctxAppp{\unv{\hctxapp}}{\anv{\hctxapp}}$, then the statement holds by application of \reflemmap{Rewriting_applicative_term_contexts}{one} (Applicative term contexts give applied variables).

			\item Let $\pctx \in \evctxAppp{\uvarset}{\avarset}$ be derived as follows
				$$
				\infer
					[\ruleUseAxTwoExp]
					{\appES{\pctxtwop{\vartwo}}{\esub{\vartwo}{\hctxapp}} \in \evctxAppp{\uvarsettwo \cup \unv{\hctxapp}}{\avarsettwo \cup \anv{\hctxapp}}}
					{\pctxtwo \in \evctxUsep{\uvarsettwo}{\avarsettwo}
					\quad
					\vartwo \notin (\uvarsettwo \cup \avarsettwo)}
				$$
			where $\pctx = \appES{\pctxtwop{\vartwo}}{\esub{\vartwo}{\hctxapp}}$, $\uvarset = \uvarsettwo \cup \unv{\hctxapp}$ and $\avarset = \avarsettwo \cup \anv{\hctxapp}$. First, note that $\vartwo \in \anv{\pctxtwop{\vartwo}}$---by \ih---and so $\vartwo \in \nv{\pctxtwop{\vartwo}}$---by \reflemmap{Unapplied_applied_and_needed_variables}{two} (Unapplied, applied and needed variables). Moreover, note that $\var \in \anv{\hctxappp{\var}}$---by \reflemmap{Rewriting_applicative_term_contexts}{one} (Applicative term contexts give applied variables). Therefore, by the definition of applied variables, $\var \in \anv{\pctxp{\var}} = (\anv{\pctxtwop{\vartwo}} \setminus \set{\vartwo}) \cup \anv{\hctxappp{\var}}$.

			\item Let $\pctx \in \evctxAppp{\uvarset}{\avarset}$ be derived as follows
				$$
				\infer
					[\ruleUseHerExp]
					{\appES{\pctxtwop{\vartwo}}{\esub{\vartwo}{\ctxhole}} \in \evctxAppp{\uvarsettwo \setminus \set{\vartwo}}{\avarset}}
					{\pctxtwo \in \evctxAppp{\uvarsettwo}{\avarset}
					\quad
					\var \notin \avarset}
				$$
			where $\pctx = \appES{\pctxtwop{\vartwo}}{\esub{\vartwo}{\ctxhole}}$ and $\uvarset = \uvarsettwo \setminus \set{\vartwo}$. By \ih, $\vartwo \in \anv{\pctxtwop{\vartwo}}$, and so $\vartwo \in \nv{\pctxtwop{\vartwo}}$---by \reflemmap{Unapplied_applied_and_needed_variables}{two} (Unapplied, applied and needed variables). Therefore, by the definition of applied variables, $\var \in \anv{\pctxp{\var}} = (\anv{\pctxtwop{\vartwo}} \setminus \set{\vartwo}) \cup \set{\var}$.

		\end{itemize}

	\item Let $\var \in \uvarset$ and $\tm$ be a term. We proceed by case analysis on the last derivation rule in $\pctx \in \evctxUsep{\uvarset}{\avarset}$:
		\begin{itemize}
			\item Let $\pctx \in \evctxUsep{\uvarset}{\avarset}$ be derived as follows
				$$
				\infer	
					[\ruleUseAxMult]
					{\progentry{\hctx}{\emptyenv} \in \evctxUsep{\unv{\hctx}}{\anv{\hctx}}}
					{}
				$$
			with $\pctx = \progentry{\hctx}{\emptyenv}$, $\uvarset = \unv{\hctx}$ and $\avarset = \anv{\hctx}$. Since $\var \in \unv{\hctx} \subseteq \nv{\hctx}$---by \reflemma{Term_contexts_Unapplied_applied_and_needed_variables} (Term contexts: Unapplied, applied and needed variables)---then by \reflemmap{Rewriting_term_contexts-Proofs_of_usefulopencbneed}{two} (Focusing term contexts on unapplied variables) there exists $\hctx_{\tm}$ such that $\hctx_{\tm} \ctxholep{\var} = \hctxp{\tm}$, with $\var \notin \unv{\hctx_{\tm}} \subset \unv{\hctx}$ and $\var \notin \anv{\hctx_{\tm}} \subseteq \anv{\hctx}$. The statement then follows by taking $\pctx_{\tm} = \progentry{\hctx_{\tm}}{\emptyenv}$.
		
			\item Let $\pctx \in \evctxUsep{\uvarset}{\avarset}$ be derived as follows
				$$
				\infer
					[\ruleUseVarMult]
						{\appES{\pctxtwo}{\esub{\vartwo}{\varthree}} \in \evctxUsep{\updp{\uvarsettwo}{\vartwo}{\varthree}}{\updp{\avarsettwo}{\varthree}{\varfour}}}
						{\pctxtwo \in \evctxUsep{\uvarsettwo}{\avarsettwo}
						\quad
						\vartwo \in (\uvarsettwo \cup \avarsettwo)}
				$$
			with $\pctx = \appES{\pctxtwo}{\esub{\vartwo}{\varthree}}$, $\uvarset = \updp{\uvarsettwo}{\vartwo}{\varthree}$ and $\avarset = \updp{\avarsettwo}{\vartwo}{\varthree}$. The statement follows by \ih on $\pctxtwo$, yielding a $\pctxtwo_{\tm} \in \evctxUsep{\uvarset_{\tm}}{\avarset_{\tm}}$ and finally deriving $\pctx_{\tm}$ by an application of $\ruleUseVarMult$ (resp. $\ruleUseGcMult$) if $\vartwo \in (\uvarset_{\tm} \cup \avarset_{\tm})$ (resp. if $\vartwo \notin (\uvarset_{\tm} \cup \avarset_{\tm})$).
			
			\item Let $\pctx \in \evctxUsep{\uvarset}{\avarset}$ be derived as follows
				$$
				\infer
					[\ruleUseNonVarMult]
					{\appES{\pctxtwo}{\esub{\vartwo}{\itmplus}} \in \evctxUsep{(\uvarsettwo \setminus \set{\vartwo}) \cup \unv{\itmplus}}{(\avarsettwo \setminus \set{\vartwo}) \cup \anv{\itmplus}}}
					{\pctxtwo \in \evctxUsep{\uvarsettwo}{\avarsettwo}
					\quad
					\vartwo \in (\uvarsettwo \cup \avarsettwo)}
				$$
			with $\pctx = \appES{\pctxtwo}{\esub{\vartwo}{\itmplus}}$, $\uvarset = (\uvarsettwo \setminus \set{\vartwo}) \cup \unv{\itmplus}$ and $\avarset = (\avarsettwo \setminus \set{\vartwo}) \cup \anv{\itmplus}$. The statement follows by \ih on $\pctxtwo$, yielding a $\pctxtwo_{\tm} \in \evctxUsep{\uvarset_{\tm}}{\avarset_{\tm}}$ and finally deriving $\pctx_{\tm}$ by an application of $\ruleUseNonVarMult$ (resp. $\ruleUseGcMult$) if $\vartwo \in (\uvarset_{\tm} \cup \avarset_{\tm})$ (resp. if $\vartwo \notin (\uvarset_{\tm} \cup \avarset_{\tm})$).
					
			\item Let $\pctx \in \evctxUsep{\uvarset}{\avarset}$ be derived as follows
				$$
				\infer
					[\ruleUseGcMult]
					{\appES{\pctxtwo}{\esub{\vartwo}{\tmthree}} \in \evctxUsep{\uvarset}{\avarset}}
					{\pctxtwo \in \evctxUsep{\uvarsettwo}{\avarsettwo}
					\quad
					\vartwo \notin (\uvarsettwo \cup \avarsettwo)}
				$$
			with $\pctx = \appES{\pctxtwo}{\esub{\vartwo}{\tmthree}}$. The statement follows by \ih on $\pctxtwo$ and finally deriving $\pctx_{\tm}$ by an application of $\ruleUseGcMult$.
						
			\item Let $\pctx \in \evctxUsep{\uvarset}{\avarset}$ be derived as follows
				$$
				\infer
					[\ruleUseUselessMult]
					{\appES{\pctxtwo}{\esub{\vartwo}{\val}} \in \evctxUsep{\uvarsettwo \setminus \set{\vartwo}}{\avarsettwo}}	
					{\pctxtwo \in \evctxUsep{\uvarsettwo}{\avarsettwo}
					\quad
					\vartwo \in (\uvarsettwo \setminus \avarsettwo)}
				$$
			with $\pctx = \appES{\pctxtwo}{\esub{\vartwo}{\val}}$, $\uvarset = \uvarsettwo \setminus \set{\vartwo}$ and $\avarset = \avarsettwo$. The statement follows by \ih on $\pctxtwo$, yielding a $\pctxtwo_{\tm} \in \evctxUsep{\uvarset_{\tm}}{\avarset_{\tm}}$ and finally deriving $\pctx_{\tm}$ by an application of $\ruleUseUselessMult$ (resp. $\ruleUseGcMult$) if $\vartwo \in (\uvarset_{\tm} \setminus \avarset_{\tm})$ (resp. if $\vartwo \notin (\uvarset_{\tm} \setminus \avarset_{\tm})$).
			
			\item Let $\pctx \in \evctxUsep{\uvarset}{\avarset}$ be derived as follows
				$$
				\infer
					[\ruleUseHerMult]
					{\appES{\pctxtwop{\varthree}}{\esub{\vartwo}{\hctx}} \in \evctxUsep{\uvarsettwo  \cup \unv{\hctx}}{\avarsettwo \cup \anv{\hctx}}}
					{\pctxtwo \in \evctxUsep{\uvarsettwo}{\avarsettwo}
					\quad
					\vartwo \notin (\uvarsettwo \cup \avarsettwo)}
				$$
			with $\pctx = \appES{\pctxtwop{\varthree}}{\esub{\vartwo}{\hctx}}$, $\uvarset = \uvarsettwo \cup \unv{\hctx}$ and $\avarset = \avarsettwo \cup \anv{\hctx}$. The statement follows by \reflemmap{Rewriting_term_contexts-Proofs_of_usefulopencbneed}{two} (Focusing term contexts on unapplied variables) on $\hctx$, yielding a term context $\hctx_{\tm}$ and finally deriving $\pctx_{\tm}$ by an application of $\ruleUseHerMult$ combining $\pctxtwo$ and $\hctx_{\tm}$.\qedhere
					
		\end{itemize}

\end{enumerate}

\end{proof}

\gettoappendix{prop:exp-ctx-are-mult}

\begin{proof}
By induction on the derivation of $\pctx \in \evctxAppp{\uvarset}{\avarset}$:
	\begin{itemize}
		\item Let $\pctx \in \evctxAppp{\uvarset}{\avarset}$ be derived as
			$$
			\infer	
				[\ruleUseAxOneExp]
				{\progentry{\hctxapp}{\emptyenv} \in \evctxAppp{\unv{\hctxapp}}{\anv{\hctxapp}}}
				{}
			$$
		with $\pctx = \progentry{\hctxapp}{\emptyenv}$, $\uvarset = \unv{\hctxapp}$ and $\avarset = \anv{\hctxapp}$. The statement simply follows by the fact that $\hctxapp$ is a term context.
		
		\item Let $\pctx \in \evctxAppp{\uvarset}{\avarset}$ be derived as
			$$
			\infer
				[\ruleUseAxTwoExp]
				{\appES{\pctxtwop{\var}}{\esub{\var}{\hctxapp}} \in \evctxAppp{\uvarsetthree \cup \unv{\hctxapp}}{\avarsetthree \cup \anv{\hctxapp}}}
				{\pctxtwo \in \evctxUsep{\uvarsetthree}{\avarsetthree}
				\quad
				\var \notin (\uvarsetthree \cup \avarsetthree)}
			$$
		where $\pctx = \appES{\pctxtwop{\var}}{\esub{\var}{\hctxapp}}$, $\uvarset = \uvarsetthree \cup \unv{\hctxapp}$ and $\avarset = \avarsetthree \cup \anv{\hctxapp}$. Since $\hctxapp$ is a term context, we get that
			$$
			\infer
				[\ruleUseHerMult]
				{\appES{\pctxtwop{\var}}{\esub{\var}{\hctxapp}} \in \evctxUsep{\uvarsetthree \cup \unv{\hctxapp}}{\avarsetthree \cup \anv{\hctxapp}}}
				{\pctxtwo \in \evctxUsep{\uvarsetthree}{\avarsetthree}
				\quad
				\var \notin (\uvarsetthree \cup \avarsetthree)}
			$$
		
		\item Let $\pctx \in \evctxAppp{\uvarset}{\avarset}$ be derived as
			$$
			\infer
				[\ruleUseVarExp]
				{\appES{\pctxtwo}{\esub{\var}{\vartwo}} \in \evctxAppp{\updp{\uvarsetthree}{\var}{\vartwo}}{\updp{\avarsetthree}{\var}{\vartwo}}}
				{\pctxtwo \in \evctxAppp{\uvarsetthree}{\avarsetthree}
				\quad
				\var \in (\uvarsetthree \cup \avarsetthree)}
			$$
		where $\pctx = \appES{\pctxtwo}{\esub{\var}{\vartwo}}$, $\uvarset = \updp{\uvarsetthree}{\var}{\vartwo}$ and $\avarset = \updp{\avarsetthree}{\var}{\vartwo}$. By \ih, $\pctxtwo \in \evctxUsep{\uvarsetseven}{\avarsetseven}$ for some $\uvarsetseven \subseteq \uvarsetthree$ and $\avarsetseven \subseteq \avarsetthree$. Case analysis on whether $\var \in (\uvarsetseven \cup \avarsetseven)$:
			\begin{itemize}
				\item Let $\var \in (\uvarsetseven \cup \avarsetseven)$. Then  we can derive $\appES{\pctxtwo}{\esub{\var}{\vartwo}} \in \evctxUsep{\updp{\uvarsetseven}{\var}{\vartwo}}{\updp{\avarsetseven}{\var}{\vartwo}}$ via rule $\ruleUseVarMult$.
				
				\item Let $\var \notin (\uvarsetseven \cup \avarsetseven)$. Then  we can derive $\appES{\pctxtwo}{\esub{\var}{\vartwo}} \in \evctxUsep{\uvarsetseven}{\avarsetseven}$ via rule $\ruleUseGcMult$.
				
			\end{itemize}
			
		\item Let $\pctx \in \evctxAppp{\uvarset}{\avarset}$ be derived as
			$$
			\infer
				[\ruleUseNonVarExp]
				{\appES{\pctxtwo}{\esub{\var}{\itmplus}} \in \evctxAppp{(\uvarsetthree \setminus \set{\var}) \cup \unv{\itmplus}}{(\avarsetthree \setminus \set{\var}) \cup \anv{\itmplus}}}
				{\pctxtwo \in \evctxAppp{\uvarsetthree}{\avarsetthree}
				\quad
				\var \in (\uvarsetthree \cup \avarsetthree)}
			$$
		where $\pctx = \appES{\pctxtwo}{\esub{\var}{\itmplus}}$, $\uvarset = (\uvarsetthree \setminus \set{\var}) \cup \unv{\itmplus}$ and $\avarset = (\avarsetthree \setminus \set{\var}) \cup \anv{\itmplus}$. By \ih, $\pctxtwo \in \evctxUsep{\uvarsetseven}{\avarsetseven}$ for some $\uvarsetseven \subseteq \uvarsetthree$ and $\avarsetseven \subseteq \avarsetthree$. Case analysis on whether $\var \in (\uvarsetseven \cup \avarsetseven)$:
			\begin{itemize}
				\item Let $\var \in (\uvarsetseven \cup \avarsetseven)$. Then  we can derive $\appES{\pctxtwo}{\esub{\var}{\itmplus}} \in \evctxUsep{(\uvarsetsix \setminus \set{\var}) \cup \unv{\itmplus}}{(\avarsetsix \setminus \set{\var}) \cup \anv{\itmplus}}$ via rule $\ruleUseNonVarMult$.
				
				\item Let $\var \notin (\uvarsetseven \cup \avarsetseven)$. Then  we can derive $\appES{\pctxtwo}{\esub{\var}{\itmplus}} \in \evctxUsep{\uvarsetseven}{\avarsetseven}$ via rule $\ruleUseGcMult$. Note that $\uvarsetseven = \uvarsetseven \setminus \set{\var} \subseteq (\uvarsetthree \setminus \set{\var}) \subseteq (\uvarsetthree \setminus \set{\var}) \cup \unv{\itmplus} = \uvarset$ and, similarly,  $\avarsetseven \subseteq \avarset$.
				
			\end{itemize}
			
		\item Let $\pctx \in \evctxAppp{\uvarset}{\avarset}$ be derived as
			$$
			\infer
				[\ruleUseGcExp]
				{\appES{\pctxtwo}{\esub{\var}{\tm}} \in \evctxAppp{\uvarset}{\avarset}}
				{\pctxtwo \in \evctxAppp{\uvarset}{\avarset}
				\quad
				\var \notin (\uvarset \cup \avarset)}
			$$
		where $\pctx = \appES{\pctxtwo}{\esub{\var}{\tm}}$. By \ih, $\pctxtwo \in \evctxAppp{\uvarsetfive}{\avarsetfive}$ for some $\uvarsetfive \subseteq \uvarset$ and $\avarsetfive \subseteq \avarset $, and we can derive $\appES{\pctxtwo}{\esub{\var}{\tm}} \in \evctxUsep{\uvarsetfive}{\avarsetfive}$ via rule $\ruleUseGcMult$.
			
		\item Let $\pctx \in \evctxAppp{\uvarset}{\avarset}$ be derived as
			$$
			\infer
				[\ruleUseUselessExp]
				{\appES{\pctxtwo}{\esub{\var}{\val}} \in \evctxAppp{\uvarsetthree \setminus \set{\var}}{\avarsetthree}}
				{\pctxtwo \in \evctxAppp{\uvarsetthree}{\avarsetthree}
				\quad
				\var \in (\uvarsetthree \setminus \avarsetthree)}
			$$
		where $\pctxtwo = \appES{\pctx}{\esub{\var}{\val}}$, $\uvarset = \uvarsetthree \setminus \set{\var}$ and $\avarset = \avarsetthree$. By \ih, $\pctxtwo \in \evctxUsep{\uvarsetseven}{\avarsetseven}$ for some $\uvarsetseven \subseteq \uvarsetthree$ and $\avarsetseven \subseteq \avarsetthree$ . Case analysis on whether $\var \in (\uvarsetseven \setminus \avarsetseven)$:
			\begin{itemize}
				\item Let $\var \in (\uvarsetseven \setminus \avarsetseven)$. Then we can derive $\appES{\pctxtwo}{\esub{\var}{\tm}} \in \evctxUsep{\uvarsetseven \setminus \set{\var}}{\avarsetseven}$ via rule $\ruleUseUselessMult$.
				
				\item Let $\var \notin (\uvarsetseven \setminus \avarsetseven)$. Then we can derive $\appES{\pctxtwo}{\esub{\var}{\tm}} \in \evctxUsep{\uvarsetseven}{\avarsetseven}$ via rule $\ruleUseGcMult$.
				
			\end{itemize}
			
		\item Let $\pctx \in \evctxAppp{\uvarset}{\avarset}$ be derived as
			$$
			\infer
				[\ruleUseHerExp]
				{\appES{\pctxp{\var}}{\esub{\var}{\ctxhole}} \in \evctxAppp{\uvarsetthree \setminus \set{\var}}{\avarsetthree}}
				{\pctx \in \evctxAppp{\uvarsetthree}{\avarsetthree}
				\quad
				\var \notin \avarsetthree}
			$$
		where $\pctx = \appES{\pctxp{\var}}{\esub{\var}{\ctxhole}}$, $\uvarset = \uvarsetthree \setminus \set{\var}$ and $\avarset = \avarsetthree$. By \ih, $\pctx \in \evctxUsep{\uvarsetseven}{\avarsetseven}$ for some $\uvarsetseven \subseteq \uvarsetthree$ and $\avarsetseven \subseteq \avarsetthree$. Case analysis on whether $\var \in \uvarsetseven$:
			\begin{itemize}
				\item Let $\var \in \uvarsetseven$. By \reflemmap{Rewriting_evaluation_contexts-base_cases}{three} (Focusing multiplicative evaluation contexts on unapplied variables), there exists $\pctx_{\var} \in \evctxUsep{\uvarseteleven}{\avarseteleven}$ such that $\pctx_{\var} \ctxholep{\var} = \pctxp{\var}$, with $\var \notin \uvarseteleven \subset \uvarsetseven$ and $\var \notin \avarseteleven \subseteq \avarsetseven$. Hence, we can derive 
					$$
					\infer
						[\ruleUseHerMult]
						{\appES{\pctx_{\var} \ctxholep{\var}}{\esub{\var}{\ctxhole}} \in \evctxUsep{\uvarseteleven}{\avarseteleven}}
						{\pctx_{\var} \in \evctxUsep{\uvarseteleven}{\avarseteleven}
						\quad
						\var \notin (\uvarseteleven \cup \avarseteleven)}
					$$
				noting in particular that $\appES{\pctx_{\var} \ctxholep{\var}}{\esub{\var}{\ctxhole}} = \appES{\pctxp{\var}}{\esub{\var}{\ctxhole}}$.
				
				\item Let $\var \notin \uvarsetseven$. Then we can derive $\appES{\pctxp{\var}}{\esub{\var}{\ctxhole}} \in \evctxUsep{\uvarsetseven}{\avarsetseven}$ via rule $\ruleUseHerMult$.\qedhere
				
			\end{itemize}

	\end{itemize}
\end{proof}
	
\begin{lemma}[Rewriting evaluation contexts]
\hfill

\begin{enumerate}
	\item
	\label{p:Rewriting_evaluation_contexts-one}
	\emph{Focusing multiplicative evaluation contexts on applied variables}: Let $\pctx \in \evctxUsep{\uvarset}{\avarset}$ and $\var \in \avarset$. Then for every normal term $\ntm$, there exists exponential evaluation context $\pctx_{\ntm} \in \evctxAppp{\uvarset_{\ntm}}{\avarset_{\ntm}}$ such that $\pctx_{\ntm} \ctxholep{\var} = \pctxp{\ntm}$, with $\uvarset_{\ntm} \subseteq \uvarset$ and $\var \notin \avarset_{\ntm} \subset \avarset$.

	\item
	\label{p:Rewriting_evaluation_contexts-two}
	\emph{Focusing exponential evaluation contexts on unapplied variables}: Let $\pctx \in \evctxAppp{\uvarset}{\avarset}$ and $\var \in \uvarset$. Then for every normal term $\ntm$, there exists exponential evaluation context $\pctx_{\ntm} \in \evctxUsep{\uvarset_{\ntm}}{\avarset_{\ntm}}$ such that $\pctx_{\ntm} \ctxholep{\var} = \pctxp{\ntm}$, with $\var \notin \uvarset_{\ntm} \subset \uvarset$ and $\avarset_{\ntm} \subseteq \avarset$.

	\item
	\label{p:Rewriting_evaluation_contexts-three}
	\emph{Focusing exponential evaluation contexts on applied variables}: Let $\pctx \in \evctxAppp{\uvarset}{\avarset}$ and $\var \in \avarset$. Then for every normal term $\ntm$, there exists exponential evaluation context $\pctx_{\ntm} \in \evctxAppp{\uvarset_{\ntm}}{\avarset_{\ntm}}$ such that $\pctx_{\ntm} \ctxholep{\var} = \pctxp{\ntm}$, with $\uvarset_{\ntm} \subseteq \uvarset$ and $\var \notin \avarset_{\ntm} \subset \avarset$.

\end{enumerate}

\label{l:Rewriting_evaluation_contexts}
\end{lemma}


\begin{proof}

All three statements are proven by mutual induction on the derivation of $\pctx \in \evctxUsep{\uvarset}{\avarset}$ or $\pctx \in \evctxAppp{\uvarset}{\avarset}$:
\begin{enumerate}	
	\item \emph{Focusing multiplicative evaluation contexts on applied variables}:
	
	Let $\var \in \avarset$ and $\tm$ be normal. We proceed by case analysis on the last derivation rule in $\pctx \in \evctxUsep{\uvarset}{\avarset}$:
		\begin{itemize}
			\item Let $\pctx \in \evctxUsep{\uvarset}{\avarset}$ be derived as follows
				$$
				\infer
					[\ruleUseAxMult]
					{\progentry{\hctx}{\emptyenv} \in \evctxUsep{\unv{\hctx}}{\anv{\hctx}}}
					{}
				$$
			with $\pctx = \progentry{\hctx}{\emptyenv}$, $\uvarset = \unv{\hctx}$ and $\avarset = \anv{\hctx}$. By application of \reflemmap{Rewriting_applicative_term_contexts}{three} (Focusing term contexts on applied variables) on $\hctx$, we get an applicative term context $\hctxapptwosub{\tm}$ such that $\hctxapptwosub{\tm} \ctxholep{\var} = \hctxp{\tm}$, with $\unv{\hctxapptwosub{\tm}} \subseteq \unv{\hctx}$ and $\var \notin \anv{\hctxapptwosub{\tm}} \subset \anv{\hctx}$. The statement then follows by taking $\pctx_{\tm} \defeq \progentry{\hctxapptwosub{\tm}}{\emptyenv}$.
	
			\item Let $\pctx \in \evctxUsep{\uvarset}{\avarset}$ be derived as follows
				$$
				\infer
					[\ruleUseVarMult]
					{\appES{\pctxtwo}{\esub{\vartwo}{\varthree}} \in \evctxUsep{\updp{\uvarsettwo}{\vartwo}{\varthree}}{\updp{\avarsettwo}{\vartwo}{\varthree}}}
					{\pctxtwo \in \evctxUsep{\uvarsettwo}{\avarsettwo}
					\quad
					\vartwo \in (\uvarsettwo \cup \avarsettwo)}
				$$
			with $\pctx = \appES{\pctxtwo}{\esub{\vartwo}{\varthree}}$, $\uvarset = \updp{\uvarsettwo}{\vartwo}{\varthree}$ and $\avarset = \updp{\avarsettwo}{\vartwo}{\varthree}$. We proceed by case analysis on whether $\vartwo \in \avarsettwo$:
				\begin{itemize}
					\item Let $\vartwo \notin \avarsettwo$. Since then $\avarsettwo = \updp{\avarsettwo}{\vartwo}{\varthree} \ni \var$, application of the \ih (Focusing multiplicative evaluation contexts on applied variables) gives us $\pctxtwo_{\tm} \in \evctxAppp{\uvarsettwo_{\tm}}{\avarsettwo_{\tm}}$ such that $\pctxtwo_{\tm} \ctxholep{\var} = \pctxtwop{\tm}$, with $\uvarsettwo_{\tm} \subseteq \uvarsettwo$ and $\var \notin \avarsettwo_{\tm} \subset \avarsettwo$. We can then derive $\pctx_{\tm} \defeq \appES{\pctxtwo_{\tm}}{\esub{\vartwo}{\varthree}} \in \evctxAppp{\updp{\uvarsettwo_{\tm}}{\vartwo}{\varthree}}{\updp{\avarsettwo_{\tm}}{\vartwo}{\varthree}}$ (resp., $\pctx_{\tm} \defeq \appES{\pctxtwo_{\tm}}{\esub{\vartwo}{\varthree}} \in \evctxAppp{\uvarsettwo_{\tm}}{\avarsettwo_{\tm}}$) via rule $\ruleUseVarExp$ (resp., rule $\ruleUseGcExp$) if $\vartwo \in (\uvarsettwo_{\tm} \cup \avarsettwo_{\tm})$ (resp., if $\vartwo \notin (\uvarsettwo_{\tm} \cup \avarsettwo_{\tm})$).
					
					\item Let $\vartwo \in \avarsettwo$. Note that then then $\updp{\avarsettwo}{\vartwo}{\varthree} = (\avarsettwo \setminus \set{\vartwo}) \cup \set{\varthree}$. We do case analysis on whether $\var \in \avarsettwo$:
						\begin{itemize}
							\item If $\var \in \avarsettwo$, then we can prove the statement similarly to how we proved it for when $\vartwo \notin \avarsettwo$.
							
							\item If $\var \notin \avarsettwo$, then it must be that $\var = \varthree$. By \ih (Focusing multiplicative evaluation contexts on applied variables) with respect to $\vartwo \in \avarsettwo$, there exists $\pctxtwo_{\tm} \in \evctxAppp{\uvarsettwo_{\tm}}{\avarsettwo_{\tm}}$ such that $\pctxtwo_{\tm} \ctxholep{\vartwo} = \pctxtwop{\tm}$, with $\uvarsettwo_{\tm} \subseteq \uvarsettwo$ and $\vartwo \notin \avarsettwo_{\tm} \subset \avarsettwo$. We can then derive $\pctx_{\tm} \defeq \appES{\pctxtwo_{\tm} \ctxholep{\vartwo}}{\esub{\vartwo}{\ctxhole}} \in \evctxAppp{\uvarsettwo_{\tm} \setminus \set{\vartwo}}{\avarsettwo_{\tm}}$ via rule $\ruleUseHerExp$. 
							
						\end{itemize}
					
				\end{itemize}
	
			\item Let $\pctx \in \evctxUsep{\uvarset}{\avarset}$ be derived as follows
				$$
				\infer
					[\ruleUseNonVarMult]
					{\appES{\pctxtwo}{\esub{\vartwo}{\itmplus}} \in \evctxUsep{(\uvarsettwo \setminus \set{\vartwo}) \cup \unv{\itmplus}}{(\avarsettwo \setminus \set{\vartwo}) \cup \anv{\itmplus}}}
					{\pctxtwo \in \evctxUsep{\uvarsettwo}{\avarsettwo}
					\quad
					\vartwo \in (\uvarsettwo \cup \avarsettwo)}
				$$
			with $\pctx = \appES{\pctxtwo}{\esub{\vartwo}{\itmplus}}$, $\uvarset = (\uvarsettwo \setminus \set{\vartwo}) \cup \unv{\itmplus}$ and $\avarset = (\avarsettwo \setminus \set{\vartwo}) \cup \anv{\itmplus}$. Case analysis on whether $\var \in \avarsettwo$:
				\begin{itemize}
					\item Let $\var \in \avarsettwo$. By \ih (Focusing multiplicative evaluation contexts on applied variables), there exists $\pctxtwo_{\tm} \in \evctxAppp{\uvarsettwo_{\tm}}{\avarsettwo_{\tm}}$ such that $\pctxtwo_{\tm} \ctxholep{\var} = \pctxtwop{\tm}$, with $\uvarsettwo_{\tm} \subseteq \uvarsettwo$ and $\var \notin \avarsettwo_{\tm} \subset \avarsettwo$. 
					
					First, note that if $\vartwo \notin (\uvarsettwo_{\tm} \cup \avarsettwo_{\tm})$, then we can derive $\pctx_{\tm} \defeq \appES{\pctxtwo_{\tm}}{\esub{\vartwo}{\itmplus}} \in \evctxAppp{\uvarsettwo_{\tm}}{\avarsettwo_{\tm}}$ via rule $\ruleUseGcExp$. 

					Second, if $\vartwo \in (\uvarsettwo_{\tm} \cup \avarsettwo_{\tm})$ and $\var \notin \anv{\itmplus}$, then we can derive $\pctx_{\tm} \defeq \appES{\pctxtwo_{\tm}}{\esub{\vartwo}{\itmplus}} \in \evctxAppp{(\uvarsettwo_{\tm} \setminus \set{\vartwo}) \cup \unv{\itmplus}}{(\avarsettwo_{\tm} \setminus \set{\vartwo}) \cup \anv{\itmplus}}$ via rule $\ruleUseNonVarExp$. 
							
					Finally, let us consider the case where $\vartwo \in (\uvarsettwo_{\tm} \cup \avarsettwo_{\tm})$ and $\var \in \anv{\itmplus}$. Notice that the latter implies---by \reflemmap{Rewriting_applicative_term_contexts}{two} (Focusing useful inert terms on applied variables)---the existence of an applicative term context $\hctxappsub{\var}$ such that $\hctxappsub{\var} \ctxholep{\var} = \itmplus$, with $\unv{\hctxappsub{\var}} \subseteq \unv{\itmplus}$ and $\var \notin \anv{\hctxappsub{\var}} \subset \anv{\itmplus}$. We now proceed by case analysis on how exactly $\vartwo \in (\uvarsettwo_{\tm} \cup \avarsettwo_{\tm})$:
						\begin{itemize}
							\item Let $\vartwo \in \uvarsettwo_{\tm}$. By \ih (Focusing exponential evaluation contexts on unapplied variables), there exists $\pctxthree_{\var} \in \evctxUsep{\uvarsetthree_{\var}}{\avarsetthree_{\var}}$ such that $\pctxthree_{\var} \ctxholep{\vartwo} = \pctxtwo_{\tm} \ctxholep{\var} = \pctxtwop{\tm}$, with $\vartwo \notin \uvarsetthree_{\var} \subset \uvarsettwo_{\tm} \subseteq \uvarsettwo$ and $\avarsetthree_{\var} \subseteq \avarsettwo_{\tm} \subset \avarsettwo$. Case analysis on whether $\vartwo \in \avarsetthree_{\var}$:
								\begin{itemize}	
									\item If $\vartwo \notin \avarsetthree_{\var}$, then we can apply rule $\ruleUseAxTwoExp$ to derive
										$$
										\pctx_{\tm} \defeq \appES{\pctxthree_{\var} \ctxholep{\vartwo}}{\esub{\vartwo}{\hctxappsub{\var}}} \in \evctxAppp{\uvarsetthree_{\var} \cup \unv{\hctxappsub{\var}}}{\avarsetthree_{\var} \cup \anv{\hctxappsub{\var}}}
										$$

									\item Let $\vartwo \in \avarsetthree_{\var}$. By \ih (Focusing multiplicative evaluation contexts on applied variables), there exists $\pctxfour_{\vartwo} \in \evctxAppp{\uvarsetfour_{\vartwo}}{\avarsetfour_{\vartwo}}$ such that $\pctxfour_{\vartwo} \ctxholep{\vartwo} = \pctxthree_{\var} \ctxholep{\vartwo} = \pctxtwo_{\tm} \ctxholep{\var} = \pctxtwop{\tm}$, with $\uvarsetfour_{\vartwo} \subseteq \uvarsetthree_{\var} \subset \uvarsettwo_{\tm} \subseteq \uvarsettwo$ and $\vartwo \notin \avarsetfour_{\vartwo} \subset \avarsetthree_{\var} \subseteq \avarsettwo_{\tm} \subset \avarsettwo$. In addition, \refprop{exp-ctx-are-mult} (Exponential evaluation contexts are multiplicative) gives that $\pctxfour_{\vartwo} \in \evctxUsep{\uvarseteight_{\vartwo}}{\avarseteight_{\vartwo}}$ for some $\uvarseteight_{\vartwo} \subseteq \uvarsetfour_{\vartwo}$ and $\avarseteight_{\vartwo} \subseteq \avarsetfour_{\vartwo}$. We can then derive $\pctx_{\tm} \defeq \appES{\pctxfour_{\vartwo} \ctxholep{\vartwo}}{\esub{\vartwo}{\hctxappsub{\var}}} \in \evctxAppp{\uvarseteight_{\vartwo} \cup \unv{\hctxappsub{\var}}}{\avarseteight_{\vartwo} \cup \anv{\hctxappsub{\var}}}$ via rule $\ruleUseHerExp$.

								\end{itemize}

							\item Let $\vartwo \notin \uvarsettwo_{\tm}$. Then $\vartwo \in \avarsettwo_{\tm}$, and by \ih (Focusing exponential evaluation contexts on applied variables) there exists $\pctxthree_{\var} \in \evctxAppp{\uvarsetthree_{\var}}{\avarsetthree_{\var}}$ such that $\pctxthree_{\var} \ctxholep{\vartwo} = \pctxtwo_{\tm} \ctxholep{\var} = \pctxtwop{\tm}$, with $\uvarsetthree_{\var} \subseteq \uvarsettwo_{\tm} \subseteq \uvarsettwo$ and $\vartwo \notin \avarsetthree_{\var} \subset \avarsettwo_{\tm} \subset \avarsettwo$. In addition, \refprop{exp-ctx-are-mult} (Exponential evaluation contexts are multiplicative) gives that $\pctxthree_{\var} \in \evctxUsep{\uvarsetseven_{\var}}{\avarsetseven_{\var}}$ for some $\uvarsetseven_{\var} \subseteq \uvarsetthree_{\var}$ and $\avarsetseven_{\var} \subseteq \avarsetthree_{\var}$. We can then derive $\pctx_{\tm} \defeq \appES{\pctxthree_{\var}}{\esub{\vartwo}{\hctxappsub{\var}}} \in \evctxAppp{\uvarsetseven_{\var} \cup \unv{\hctxappsub{\var}}}{\avarsetseven_{\var} \cup \anv{\hctxappsub{\var}}}$ via rule $\ruleUseAxTwoExp$.
							
						\end{itemize}
					
					\item Let $\var \notin \avarsettwo$. Then it must be that $\var \in \anv{\itmplus}$, which implies---by \reflemmap{Rewriting_applicative_term_contexts}{two} (Focusing useful inert terms on applied variables)---the existence of an applicative term context $\hctxappsub{\var}$ such that $\hctxappsub{\var} \ctxholep{\var} = \itmplus$, with $\unv{\hctxappsub{\var}} \subseteq \unv{\itmplus}$ and $\var \notin \anv{\hctxappsub{\var}} \subset \anv{\itmplus}$.
					
					We proceed by case analysis on whether $\vartwo \in \uvarsettwo$:
						\begin{itemize}
							\item If $\vartwo \in \uvarsettwo$, then \ih (Focusing multiplicative evaluation contexts on applied varibales) gives the existence of $\pctxtwo_{\tm} \in \evctxUsep{\uvarsettwo_{\tm}}{\avarsettwo_{\tm}}$ such that $\pctxtwo_{\tm} \ctxholep{\vartwo} = \pctxtwop{\tm}$ such that $\var \notin \uvarsettwo_{\tm} \subset \uvarsettwo$ and $\avarsettwo_{\tm} \subseteq \avarsettwo$. Case analyss on whether $\vartwo \in \avarsettwo_{\tm}$:
								\begin{itemize}
									\item If $\vartwo \notin \avarsettwo_{\tm}$, then application of the \ih (Focusing multiplicative evaluation contexts on applied variables) gives $\pctxthree_{\vartwo} \in \evctxAppp{\uvarsetthree_{\vartwo}}{\avarsetthree_{\vartwo}}$ such that $\pctxthree_{\vartwo} \ctxholep{\vartwo} = \pctxtwo_{\tm} \ctxholep{\vartwo} = \pctxtwop{\tm}$, with $\uvarsetthree_{\vartwo} \subseteq \uvarsettwo_{\tm} \subset \uvarsettwo$ and $\vartwo \notin \avarsetthree_{\vartwo} \subset \avarsettwo_{\tm} \subseteq \avarsettwo$. In addition, $\pctxthree_{\vartwo} \in \evctxUsep{\uvarsetseven_{\vartwo}}{\avarsetseven_{\vartwo}}$ for some $\uvarsetseven_{\vartwo} \subseteq \uvarsetthree_{\vartwo}$ and $\avarsetseven_{\vartwo} \subseteq \avarsetthree_{\vartwo}$. We can then derive $\pctx_{\tm} \defeq \appES{\pctxthree_{\vartwo} \ctxholep{\vartwo}}{\esub{\vartwo}{\hctxappsub{\var}}} \in \evctxAppp{\uvarsetseven_{\vartwo} \cup \unv{\hctxappsub{\var}}}{\avarsetseven_{\vartwo} \cup \anv{\hctxappsub{\var}}}$ via rule $\ruleUseAxTwoExp$. 
 
									\item If $\vartwo \in \avarsettwo_{\tm}$, then we can derive $\pctx_{\tm} \defeq \appES{\pctxtwo_{\tm} \ctxholep{\vartwo}}{\esub{\vartwo}{\hctxappsub{\var}}} \in \evctxAppp{\uvarsettwo_{\tm} \cup \unv{\hctxappsub{\var}}}{\avarsettwo_{\tm} \cup \anv{\hctxappsub{\var}}}$ via rule $\ruleUseAxTwoExp$.

								\end{itemize}
														
							\item Let $\vartwo \notin \uvarsettwo$. Then it must be that $\vartwo \in \avarsettwo$, and so applying the \ih gives $\pctxthree_{\tm} \in \evctxAppp{\uvarsetthree_{\tm}}{\avarsetthree_{\tm}}$ such that $\pctxthree_{\tm} \ctxholep{\vartwo} = \pctxtwop{\tm}$, with $\uvarsetthree_{\tm} \subseteq \uvarsettwo$ and $\vartwo \notin \avarsetthree_{\tm} \subset \avarsettwo$. Moreover, \refprop{exp-ctx-are-mult} (Exponential evaluation contexts are multiplicative) proves that $\pctxthree_{\tm} \in \evctxUsep{\uvarsetfour}{\avarsetfour}$ such that $\uvarsetfour \subseteq \uvarsetthree_{\tm}$ and $\avarsetfour \subseteq \avarsetthree_{\tm}$. Thus, we can derive $\pctx_{\tm} \defeq \appES{\pctxthree_{\tm} \ctxholep{\vartwo}}{\esub{\vartwo}{\hctxappsub{\var}}} \in \evctxAppp{\uvarsetfour \cup \unv{\hctxappsub{\var}}}{\avarsetfour \cup \anv{\hctxappsub{\var}}}$ via rule $\ruleUseAxTwoExp$.
							
						\end{itemize}
					
				\end{itemize}
	
			\item Let $\pctx \in \evctxUsep{\uvarset}{\avarset}$ be derived as follows
				$$
				\infer
					[\ruleUseGcMult]
					{\appES{\pctxtwo}{\esub{\vartwo}{\tmtwo}} \in \evctxUsep{\uvarset}{\avarset}}
					{\pctxtwo \in \evctxUsep{\uvarset}{\avarset}
					\quad
					\vartwo \notin (\uvarset \cup \avarset)}
				$$
			with $\pctx = \appES{\pctxtwo}{\esub{\vartwo}{\tmtwo}}$. By \ih (Focusing multiplicative evaluation contexts on applied variables), there exists $\pctxtwo_{\tm} \in \evctxAppp{\uvarset_{\tm}}{\avarset_{\tm}}$ such that $\pctxtwo_{\tm} \ctxholep{\var} = \pctxtwop{\tm}$, with $\uvarset_{\tm} \subseteq \uvarset$ and $\var \notin \avarset_{\tm} \subset \avarset$. We can then derive $\pctx_{\tm} \defeq \appES{\pctxtwo_{\tm}}{\esub{\vartwo}{\tmtwo}} \in \evctxAppp{\uvarset_{\tm}}{\avarset_{\tm}}$ via rule $\ruleUseGcExp$.
		
			\item Let $\pctx \in \evctxUsep{\uvarset}{\avarset}$ be derived as follows
				$$
				\infer
					[\ruleUseUselessMult]
					{\appES{\pctxtwo}{\esub{\vartwo}{\val}} \in \evctxUsep{\uvarsettwo \setminus \set{\vartwo}}{\avarsettwo}}
					{\pctxtwo \in \evctxUsep{\uvarsettwo}{\avarsettwo}
					\quad
					\vartwo \in (\uvarsettwo \setminus \avarsettwo)}
				$$
			with $\pctx = \appES{\pctxtwo}{\esub{\vartwo}{\val}}$, $\uvarset = \uvarsettwo \setminus \set{\vartwo}$ and $\avarset = \avarsettwo$. By \ih (Focusing multiplicative evaluation contexts on applied variables), there exists $\pctxtwo_{\tm} \in \evctxAppp{\uvarsettwo_{\tm}}{\avarsettwo_{\tm}}$ such that $\pctxtwo_{\tm} \ctxholep{\var} = \pctxtwop{\tm}$, with $\uvarsettwo_{\tm} \subseteq \uvarsettwo$ and $\var \notin \avarsettwo_{\tm} \subset \avarsettwo$.  We can then derive $\pctx_{\tm} \defeq \appES{\pctxtwo_{\tm}}{\esub{\vartwo}{\tm}} \in \evctxAppp{\uvarsettwo_{\tm} \setminus \set{\vartwo}}{\avarsettwo_{\tm}}$ (resp. $\pctx_{\tm} \defeq \appES{\pctxtwo_{\tm}}{\esub{\vartwo}{\tm}} \in \evctxAppp{\uvarsettwo_{\tm}}{\avarsettwo_{\tm}}$) via rule $\ruleUseUselessExp$ (resp. rule $\ruleUseGcExp$) if $\vartwo \in \uvarsettwo_{\tm}$ (resp. if $\vartwo \notin \uvarsettwo_{\tm}$).
	
			\item Let $\pctx \in \evctxUsep{\uvarset}{\avarset}$ be derived as follows
				$$
				\infer
					[\ruleUseHerMult]
					{\appES{\pctxtwop{\vartwo}}{\esub{\vartwo}{\ctxhole}} \in \evctxUsep{\uvarset}{\avarset}}
					{\pctxtwo \in \evctxUsep{\uvarset}{\avarset}
					\quad
					\vartwo \notin (\uvarset \cup \avarset)}
				$$
			with $\pctx = \appES{\pctxtwop{\vartwo}}{\esub{\vartwo}{\ctxhole}}$. By \ih (Focusing multiplicative evaluation contexts on applied variables), there exists $\pctxtwo_{\vartwo} \in \evctxAppp{\uvarset_{\vartwo}}{\avarset_{\vartwo}}$ such that $\pctxtwo_{\vartwo} \ctxholep{\var} = \pctxtwop{\vartwo}$, with $\uvarset_{\vartwo} \subseteq \uvarset$ and $\var \notin \avarset_{\vartwo} \subset \avarset$.  We can then derive $\pctx_{\tm} \defeq \appES{\pctxtwo_{\vartwo}}{\esub{\vartwo}{\tm}} \in \evctxAppp{\uvarset_{\vartwo}}{\avarset_{\tm}}$ via rule $\ruleUseGcExp$.
			
		\end{itemize}

	\item \emph{Focusing exponential evaluation contexts on unapplied variables}: Let $\var \in \uvarset$ and $\tm$ be normal. We proceed by case analysis on the last derivation rule in $\pctx \in \evctxAppp{\uvarset}{\avarset}$:
		\begin{itemize}
			\item Let $\pctx \in \evctxAppp{\uvarset}{\avarset}$ be derived as follows
				$$
				\infer
					[\ruleUseAxOneExp]
					{\progentry{\hctxapp}{\emptyenv} \in \evctxAppp{\unv{\hctxapp}}{\anv{\hctxapp}}}
					{}
				$$
			with $\pctx = \progentry{\hctxapp}{\emptyenv}$, $\uvarset = \unv{\hctxapp}$ and $\avarset = \anv{\hctxapp}$. The statement is given by applying \reflemmap{Rewriting_term_contexts-Proofs_of_usefulopencbneed}{two} (Focusing terms contexts on unapplied variables) on $\hctxapp$, yielding a term context $\hctx_{\tm}$ such that $\hctx_{\tm} \ctxholep{\var} = \hctxappp{\tm}$, with $\var \notin \unv{\hctx_{\tm}} \subset \unv{\hctxapp}$ and $\anv{\hctx_{\tm}} \subseteq \anv{\hctxapp}$, and finally taking $\pctx_{\tm} \defeq \progentry{\hctx_{\tm}}{\emptyenv}$.
	
			\item Let $\pctx \in \evctxAppp{\uvarset}{\avarset}$ be derived as follows
				$$
				\infer
					[\ruleUseAxTwoExp]
					{\appES{\pctxtwop{\vartwo}}{\esub{\vartwo}{\hctxapp}} \in \evctxAppp{\uvarsettwo \cup \unv{\hctxapp}}{\avarsettwo \cup \anv{\hctxapp}}}
					{\pctxtwo \in \evctxUsep{\uvarsettwo}{\avarsettwo}
					\quad
					\vartwo \notin (\uvarsettwo \cup \avarsettwo)}
				$$
			with $\pctx = \appES{\pctxtwop{\vartwo}}{\esub{\vartwo}{\hctxapp}}$, $\uvarset = \uvarsettwo \cup \unv{\hctxapp}$ and $\avarset = \avarsettwo \cup \anv{\hctxapp}$. Case analysis on whether $\var \in \uvarsettwo$:
				\begin{itemize}
					\item Let $\var \in \uvarsettwo$. By \reflemmap{Rewriting_evaluation_contexts-base_cases}{three} (Focusing multiplicative evaluation contexts on unapplied variables), there exists $\pctxtwo_{\vartwo} \in \evctxUsep{\uvarsettwo_{\vartwo}}{\avarsettwo_{\vartwo}}$ such that $\pctxtwo_{\vartwo} \ctxholep{\var} = \pctxtwop{\vartwo}$, with $\var \notin \uvarsettwo_{\vartwo} \subset \uvarsettwo$ and $\avarsettwo_{\vartwo} \subseteq \avarsettwo$. We can then derive $\pctx_{\tm} \in \evctxUsep{\uvarset_{\tm}}{\avarset_{\tm}}$ as follows 
						$$
						\infer
							[\ruleUseGcMult]
							{\appES{\pctxtwo_{\vartwo}}{\esub{\vartwo}{\hctxappp{\tm}}} \in \evctxUsep{\uvarsettwo_{\vartwo}}{\avarsettwo_{\vartwo}}}
							{\pctxtwo_{\vartwo} \in \evctxUsep{\uvarsettwo_{\vartwo}}{\avarsettwo_{\vartwo}}
							\quad
							\vartwo \notin (\uvarsettwo_{\vartwo} \cup \avarsettwo_{\vartwo})}
						$$
					
					\item Let $\var \notin \uvarsettwo$. Then $\var \in \unv{\hctxapp}$. By \reflemmap{Rewriting_term_contexts-Proofs_of_usefulopencbneed}{two} (Focusing term contexts on unapplied variables), there exists term context $\hctx_{\tm}$ such that $\hctx_{\tm} \ctxholep{\var} = \hctxappp{\tm}$, with $\var \notin \unv{\hctx_{\tm}} \subset \unv{\hctxapp}$ and $\anv{\hctx_{\tm}} \subseteq \anv{\hctxapp}$. We can then derive $\pctx_{\tm} \in \evctxUsep{\uvarset_{\tm}}{\avarset_{\tm}}$ as follows 
						$$
						\infer
							[\ruleUseAxTwoExp]
							{\appES{\pctxtwop{\vartwo}}{\esub{\vartwo}{\hctx_{\tm}}} \in \evctxAppp{\uvarsettwo \cup \unv{\hctx_{\tm}}}{\avarsettwo \cup \anv{\hctx_{\tm}}}}
							{\pctxtwo \in \evctxUsep{\uvarsettwo}{\avarsettwo}
							\quad
							\vartwo \notin (\uvarsettwo \cup \avarsettwo)}
						$$
				
				\end{itemize}
	
			\item Let $\pctx \in \evctxAppp{\uvarset}{\avarset}$ be derived as follows
				$$
				\infer
					[\ruleUseVarExp]
					{\appES{\pctxtwo}{\esub{\vartwo}{\varthree}} \in \evctxAppp{\updp{\uvarsettwo}{\vartwo}{\varthree}}{\updp{\avarsettwo}{\vartwo}{\varthree}}}
					{\pctxtwo \in \evctxAppp{\uvarsettwo}{\avarsettwo}
					\quad
					\vartwo \in (\uvarsettwo \cup \avarsettwo)}
				$$
			with $\pctx = \appES{\pctxtwo}{\esub{\vartwo}{\varthree}}$, $\uvarset = \updp{\uvarsettwo}{\vartwo}{\varthree}$ and $\avarset = \updp{\avarsettwo}{\vartwo}{\varthree}$. We proceed by case analysis on whether $\vartwo \in \uvarsettwo$:
				\begin{itemize}
					\item Let $\vartwo \notin \uvarsettwo$. Then $\uvarsettwo = \updp{\uvarsettwo}{\vartwo}{\varthree} \ni \var$ and so application of the \ih (Focusing exponential evaluation contexts on unapplied variables) yields multiplicative evaluation context $\pctxtwo_{\tm} \in \evctxUsep{\uvarsettwo_{\tm}}{\avarsettwo_{\tm}}$ such that $\pctxtwo_{\tm} \ctxholep{\var} = \pctxtwop{\tm}$, with $\var \notin \uvarsettwo_{\tm} \subset \uvarsettwo$ and $\avarsettwo_{\tm} \subseteq \avarsettwo$. Case analysis on whether $\vartwo \in \avarsettwo_{\tm}$:
						\begin{itemize}
							\item Let $\vartwo \in \avarsettwo_{\tm}$. Then we can derive $\pctx_{\tm} \defeq \appES{\pctxtwo_{\tm}}{\esub{\vartwo}{\varthree}} \in \evctxUsep{\updp{\uvarsettwo_{\tm}}{\vartwo}{\varthree}}{\updp{\avarsettwo_{\tm}}{\vartwo}{\varthree}}$ via rule $\ruleUseVarExp$.

							\item Let $\vartwo \notin \avarsettwo_{\tm}$. Then we can derive $\pctx_{\tm} \defeq \appES{\pctxtwo_{\tm}}{\esub{\vartwo}{\varthree}} \in \evctxUsep{\uvarsettwo_{\tm}}{\avarsettwo_{\tm}}$ via rule $\ruleUseVarExp$.

						\end{itemize}

					\item Let $\vartwo \in \uvarsettwo$. Then $\uvarset = (\uvarsettwo \setminus \set{\vartwo}) \cup \set{\varthree}$. Case analysis on whether $\var \in (\uvarsettwo \setminus \set{\vartwo})$ and on whether $\var = \varthree$:
						\begin{itemize}
							\item Let $\var \neq \varthree$. Then $\var \in (\uvarsettwo \setminus \set{\vartwo})$. By \ih (Focusing exponential evaluation contexts on unapplied variables), there exists $\pctxtwo_{\tm} \in \evctxUsep{\uvarsettwo_{\tm}}{\avarsettwo_{\tm}}$ such that $\pctxtwo_{\tm} \ctxholep{\var} = \pctxtwop{\tm}$, with $\var \notin \uvarsettwo_{\tm} \subset \uvarsettwo$ and $\avarsettwo_{\tm} \subseteq \avarsettwo$. 

							Finally, if $\vartwo \in (\uvarsettwo_{\tm} \cup \avarsettwo_{\tm})$, then we can derive $\pctx_{\tm} \defeq \appES{\pctxtwo_{\tm}}{\esub{\vartwo}{\varthree}} \in \evctxUsep{\updp{\uvarsettwo_{\tm}}{\vartwo}{\varthree}}{\updp{\avarsettwo_{\tm}}{\vartwo}{\varthree}}$ via rule $\ruleUseVarMult$. If instead $\vartwo \notin (\uvarsettwo_{\tm} \cup \avarsettwo_{\tm})$, then we can derive $\pctx_{\tm} \defeq \appES{\pctxtwo_{\tm}}{\esub{\vartwo}{\varthree}} \in \evctxUsep{\uvarsettwo_{\tm}}{\avarsettwo_{\tm}}$.

							\item Let $\var = \varthree$ and $\var \notin (\uvarsettwo \setminus \set{\vartwo})$. By \ih (Focusing exponential evaluation contexts on unapplied variables), there exists $\pctxtwo_{\tm} \in \evctxUsep{\uvarsettwo_{\tm}}{\avarsettwo_{\tm}}$ such that $\pctxtwo_{\tm} \ctxholep{\vartwo} = \pctxtwop{\tm}$, with $\vartwo \notin \uvarsettwo_{\tm} \subset \uvarsettwo$ and $\avarsettwo_{\tm} \subseteq \avarsettwo$. Case analysis on whether $\vartwo \in \avarsettwo_{\tm}$:
								\begin{itemize}
									\item Let $\vartwo \in \avarsettwo_{\tm}$. By \ih (Focusing multiplicative evaluation contexts on applied variables), there exists $\pctxthree_{\vartwo} \in \evctxAppp{\uvarsetthree_{\vartwo}}{\avarsetthree_{\vartwo}}$ such that $\pctxthree_{\vartwo} \ctxholep{\vartwo} = \pctxtwo_{\tm} \ctxholep{\var} = \pctxtwop{\tm}$, with $\uvarsetthree_{\vartwo} \subseteq \uvarsettwo_{\tm} \subset \uvarsettwo$ and $\vartwo \notin \avarsetthree_{\vartwo} \subset \avarsettwo_{\tm} \subseteq \avarsettwo$. Moreover, by \refprop{exp-ctx-are-mult} (Exponential evaluation contexts are multiplicative), $\pctxthree_{\vartwo} \in \evctxUsep{\uvarsetseven_{\vartwo}}{\avarsetseven_{\vartwo}}$ for some $\uvarsetseven_{\vartwo} \subseteq \uvarsetthree_{\vartwo}$ and $\avarsetseven_{\vartwo} \subseteq \avarsetthree_{\vartwo}$. We can finally derive $\pctx_{\tm} \defeq \appES{\pctxthree_{\vartwo} \ctxholep{\vartwo}}{\esub{\vartwo}{\ctxhole}} \in \evctxUsep{\uvarsetseven_{\vartwo} \setminus \set{\vartwo}}{\avarsetseven_{\vartwo}}$ via rule $\ruleUseHerMult$.

									\item Let $\vartwo \notin \avarsettwo_{\tm}$. Then we can derive $\pctx_{\tm} \defeq \appES{\pctxtwo_{\tm} \ctxholep{\vartwo}}{\esub{\vartwo}{\ctxhole}} \in \evctxUsep{\uvarsettwo_{\tm} \setminus \set{\vartwo}}{\avarsettwo_{\tm}}$ via rule $\ruleUseHerMult$.

								\end{itemize}

							\item Let $\var = \varthree$ and $\var \in (\uvarsettwo \setminus \set{\vartwo})$. By \ih (Focusing exponential evaluation contexts on unapplied variables), there exists $\pctxtwo_{\tm} \in \evctxUsep{\uvarsettwo_{\tm}}{\avarsettwo_{\tm}}$ such that $\pctxtwo_{\tm} \ctxholep{\var} = \pctxtwop{\tm}$, with $\var \notin \uvarsettwo_{\tm} \subset \uvarsettwo $ and $\avarsettwo_{\tm} \subseteq \avarsettwo$. Cases analysis on whether $\vartwo \in \avarsettwo_{\tm}$:
								\begin{itemize}
									\item Let $\vartwo \in \avarsettwo_{\tm}$. By \ih (Focusing multiplicative evaluation contexts on applied variables), there exists $\pctxthree_{\var} \in \evctxAppp{\uvarsetthree_{\var}}{\avarsetthree_{\var}}$ such that $\pctxthree_{\var} \ctxholep{\vartwo} = \pctxtwo_{\tm} \ctxholep{\var} = \pctxtwop{\tm}$, with $\uvarsetthree_{\var} \subseteq \uvarsettwo_{\tm} \subset \uvarsettwo$ and $\vartwo \notin \avarsetthree_{\var} \subset \avarsettwo_{\tm} \subseteq \avarsettwo$. Moreover, by \refprop{exp-ctx-are-mult} (Exponential evaluation contexts are multiplicative), $\pctxthree_{\var} \in \evctxUsep{\uvarsetsix_{\var}}{\avarsetsix_{\var}}$ for some $\uvarsetsix_{\var} \subseteq \uvarsetthree_{\var}$ and $\avarsetsix_{\var} \subseteq \avarsetthree_{\var}$. 

									Finally, if $\vartwo \in \uvarsetsix_{\var}$, then \reflemmap{Rewriting_evaluation_contexts-base_cases}{three} (Focusing multiplicative evaluation contexts on unapplied variables) gives multiplicative evaluation context $\pctxfour_{\vartwo} \in \evctxUsep{\uvarsetfour_{\vartwo}}{\avarsetfour_{\vartwo}}$ such that $\pctxfour_{\vartwo} \ctxholep{\vartwo} =\pctxthree_{\var} \ctxholep{\vartwo} = \pctxtwo_{\tm} \ctxholep{\var} = \pctxtwop{\tm}$, with $\vartwo \notin \uvarsetfour_{\vartwo} \subset \uvarsetthree_{\var} \subseteq \uvarsettwo_{\tm} \subset \uvarsettwo$ and $\avarsetfour_{\vartwo} \subseteq \avarsetthree_{\var} \subset \avarsettwo_{\tm} \subseteq \avarsettwo$; we can derive $\pctx_{\tm} \defeq \appES{\pctxfour_{\vartwo} \ctxholep{\vartwo}}{\esub{\vartwo}{\ctxhole}} \in \evctxUsep{\uvarsetfour_{\vartwo} \setminus \set{\vartwo}}{\avarsetfour_{\vartwo}}$ via rule $\ruleUseHerMult$. If instead $\vartwo \notin \uvarsetsix_{\var}$, then we can derive $\pctx_{\tm} \defeq \appES{\pctxthree_{\var} \ctxholep{\vartwo}}{\esub{\vartwo}{\ctxhole}} \in \evctxUsep{\uvarsetsix_{\var} \setminus \set{\vartwo}}{\avarsetsix_{\var}}$ via rule $\ruleUseHerMult$.

									\item Let $\vartwo \notin \avarsettwo_{\tm}$. If $\vartwo \notin \uvarsettwo_{\tm}$, then we can simply derive $\appES{\pctxtwo_{\tm}}{\esub{\vartwo}{\varthree}} \in \evctxUsep{\uvarsettwo_{\tm} \setminus \set{\vartwo}}{\avarsettwo_{\tm}}$ via rule $\ruleUseHerMult$. If $\vartwo \in \uvarsettwo_{\tm}$ instead, then application of \reflemmap{Rewriting_evaluation_contexts-base_cases}{three} (Focusing mutliplicative evaluation contexts on unapplied varibales) gives the existence of $\pctxthree_{\var} \in \evctxUsep{\uvarsetthree_{\var}}{\avarsetthree_{\var}}$ such that $\pctxthree_{\var} \ctxholep{\vartwo} = \pctxtwo_{\tm} \ctxholep{\var} = \pctxtwop{\tm}$, with $\vartwo \notin \uvarsetthree_{\var} \subset \uvarsettwo_{\tm} \subset \uvarsettwo$ and $\avarsetthree_{\var} \subseteq \avarsettwo_{\tm} \subseteq \avarsettwo$; we can derive $\pctx_{\tm} \defeq \appES{\pctxthree_{\var} \ctxholep{\vartwo}}{\esub{\vartwo}{\ctxhole}} \in \evctxUsep{\uvarsetthree_{\var} \setminus \set{\vartwo}}{\avarsetthree_{\var}}$ via rule $\ruleUseHerMult$.

								\end{itemize}

						\end{itemize}

				\end{itemize}
	
			\item Let $\pctx \in \evctxAppp{\uvarset}{\avarset}$ be derived as follows
				$$
				\infer
					[\ruleUseNonVarExp]
					{\appES{\pctxtwo}{\esub{\vartwo}{\itmplus}} \in \evctxAppp{(\uvarsettwo \setminus \set{\vartwo}) \cup \unv{\itmplus}}{(\avarsettwo \setminus \set{\vartwo}) \cup \anv{\itmplus}}}
					{\pctxtwo \in \evctxAppp{\uvarsettwo}{\avarsettwo}
					\quad
					\vartwo \in (\uvarsettwo \cup \avarsettwo)}
				$$
			with $\pctx = \appES{\pctxtwo}{\esub{\vartwo}{\itmplus}}$, $\uvarset = (\uvarsettwo \setminus \set{\vartwo}) \cup \unv{\itmplus}$ and $\avarset = (\avarsettwo \setminus \set{\vartwo}) \cup \anv{\itmplus}$. Case analysis on whether $\var \in \uvarsettwo$:
				\begin{itemize}
					\item Let $\var \in \uvarsettwo$. By \ih (Focusing exponential evaluation contexts on unapplied variables), there exists $\pctxtwo_{\tm} \in \evctxUsep{\uvarsettwo_{\tm}}{\avarsettwo_{\tm}}$ such that $\pctxtwo_{\tm} \ctxholep{\var} = \pctxtwop{\tm}$, with $\var \notin \uvarsettwo_{\tm} \subset \uvarsettwo$ and $\avarsettwo_{\tm} \subseteq \avarsettwo$. Case analysis on whether $\var \in \unv{\itmplus}$:
						\begin{itemize}
							\item Let $\var \notin \unv{\itmplus}$. If $\vartwo \in (\uvarsettwo_{\tm} \cup \avarsettwo_{\tm})$, then we can derive $\pctx \defeq \appES{\pctxtwo_{\tm}}{\esub{\vartwo}{\itmplus}} \in \evctxUsep{(\uvarsettwo_{\tm} \setminus \set{\vartwo}) \cup \unv{\itmplus}}{(\avarsettwo_{\tm} \setminus \set{\vartwo}) \cup \anv{\itmplus}}$ via rule $\ruleUseNonVarMult$. If $\vartwo \notin (\uvarsettwo_{\tm} \cup \avarsettwo_{\tm})$, then we can derive $\pctx_{\tm} \defeq \appES{\pctxtwo_{\tm}}{\esub{\vartwo}{\itmplus}} \in \evctxUsep{\uvarsettwo_{\tm}}{\avarsettwo_{\tm}}$ via rule $\ruleUseGcMult$.

							\item Let $\var \in \unv{\itmplus}$. By \reflemmap{Rewriting_term_contexts-Proofs_of_usefulopencbneed}{two} (Focusing inert terms on unapplied variables) gives the existence of a term context $\hctx_{\var}$ such that $\hctx_{\var} \ctxholep{\var} = \itmplus$, with $\var \notin \unv{\hctx_{\var}} \subset \unv{\itmplus}$ and $\anv{\hctx_{\var}} \subseteq \anv{\itmplus}$. Case analysis on whether $\vartwo \in \uvarsettwo_{\tm}$ and on whether $\vartwo \in \avarsettwo_{\tm}$:
								\begin{itemize}
									\item Let $\vartwo \notin (\uvarsettwo_{\tm} \cup \avarsettwo_{\tm})$. Then we can derive $\pctx_{\tm} \defeq \appES{\pctxtwo_{\tm}}{\esub{\vartwo}{\itmplus}} \in \evctxUsep{\uvarsettwo_{\tm}}{\avarsettwo_{\tm}}$ via rule $\ruleUseGcMult$.

									\item Let $\vartwo \in \uvarsettwo_{\tm}$ and $\vartwo \notin \avarsettwo_{\tm}$. By \reflemmap{Rewriting_evaluation_contexts-base_cases}{three} (Focusing multiplicative evaluation context on unapplied variables), there exists $\pctxthree_{\var} \in \evctxUsep{\uvarsetthree_{\var}}{\avarsetthree_{\var}}$ such that $\pctxthree_{\var} \ctxholep{\vartwo} = \pctxtwo_{\tm} \ctxholep{\var} = \pctxtwop{\tm}$, with $\vartwo \notin \uvarsetthree_{\var} \subset \uvarsettwo_{\tm} \subseteq \uvarsettwo$ and $\avarsetthree_{\var} \subseteq \avarsettwo_{\tm} \subseteq \avarsettwo$. We can finally derive $\pctx_{\tm} \defeq \appES{\pctxthree_{\var} \ctxholep{\vartwo}}{\esub{\vartwo}{\hctx_{\var}}} \in \evctxUsep{\uvarsetthree_{\var} \cup \unv{\hctx_{\var}}}{\avarsetthree_{\var} \cup \anv{\hctx_{\var}}}$ via rule $\ruleUseHerMult$.

									\item Let $\vartwo \in \avarsettwo_{\tm}$. By \ih (Focusing multiplicative evaluation contexts on applied variables), there exists $\pctxthree_{\var} \in \evctxAppp{\uvarsetthree_{\var}}{\avarsetthree_{\var}}$ such that $\pctxthree_{\var} \ctxholep{\vartwo} = \pctxtwo_{\tm} \ctxholep{\vartwo} = \pctxtwop{\tm}$, with $\uvarsetthree_{\var} \subseteq \uvarsettwo_{\tm} \subset \uvarsettwo$ and $\vartwo \notin \avarsetthree_{\var} \subset \avarsettwo_{\tm} \subseteq \avarsettwo$. Moreover, \refprop{exp-ctx-are-mult} (Exponential evaluation contexts are multiplicative) gives that $\pctxthree_{\var} \in \evctxUsep{\uvarsetseven_{\var}}{\avarsetseven_{\var}}$ for some $\uvarsetseven_{\var} \subseteq \uvarsetthree_{\var}$ and $\avarsetseven_{\var} \subseteq \avarsetthree_{\var}$.

									Finally, if $\vartwo \notin \uvarsetseven_{\var}$, then we can derive $\pctx_{\tm} \defeq \appES{\pctxthree_{\var} \ctxholep{\vartwo}}{\esub{\vartwo}{\hctx_{\var}}} \in \evctxUsep{\uvarsetseven_{\var} \cup \unv{\hctx_{\var}}}{\avarsetseven_{\var} \cup \anv{\hctx_{\var}}}$ via rule $\ruleUseHerMult$. If $\vartwo \in \uvarsetseven_{\var}$ instead, then \reflemmap{Rewriting_evaluation_contexts-base_cases}{three} (Focusing multiplicative evaluation contexts on unapplied variables) gives $\pctxfour_{\vartwo} \in \evctxUsep{\uvarsetfour_{\vartwo}}{\avarsetfour_{\vartwo}}$ such that $\pctxfour_{\vartwo} \ctxholep{\vartwo} = \pctxthree_{\var} \ctxholep{\var} = \pctxtwo_{\tm} \ctxholep{\var} = \pctxtwop{\tm}$, with $\vartwo \notin \uvarsetfour_{\vartwo} \subset \uvarsetseven_{\var} \subseteq \uvarsetthree_{\var} \subseteq \uvarsettwo_{\tm} \subset \uvarsettwo$ and $\avarsetfour_{\vartwo} \subseteq \avarsetseven_{\var} \subset \avarsettwo_{\tm} \subseteq \uvarsettwo$; we can then derive $\pctx_{\tm} \defeq \appES{\pctxfour_{\vartwo} \ctxholep{\vartwo}}{\esub{\vartwo}{\hctx_{\var}}} \in \evctxUsep{\uvarsetfour_{\vartwo} \cup \unv{\hctx_{\var}}}{\avarsetfour_{\vartwo} \cup \anv{\hctx_{\var}}}$ via rule $\ruleUseHerMult$.

								\end{itemize}

						\end{itemize}

					\item Let $\var \notin \uvarsettwo$. Then $\var \in \unv{\itmplus}$. Note that \reflemmap{Rewriting_term_contexts-Proofs_of_usefulopencbneed}{one} (Focusing inert terms on unapplied variables) gives the existence of term context $\hctx_{\var}$ such that $\hctx_{\var} \ctxholep{\var} = \itmplus$, with $\var \notin \unv{\hctx_{\var}} \subset \unv{\itmplus}$ and $\anv{\hctx_{\var}} \subseteq \anv{\itmplus}$. Case analysis on whether $\vartwo \in \uvarsettwo$ and on whether $\vartwo \in \avarsettwo$:
						\begin{itemize}
							\item Let $\vartwo \in \uvarsettwo$ and $\vartwo \notin \avarsettwo$. By \ih (Focusing exponential evaluation contexts on unapplied variables), there exists $\pctxtwo_{\tm} \in \evctxUsep{\uvarsettwo_{\tm}}{\avarsettwo_{\tm}}$ such that $\vartwo \notin \uvarsettwo_{\tm} \subset \uvarsettwo$ and $\avarsettwo_{\tm} \subseteq \avarsettwo$. We can then derive $\pctx_{\tm} \defeq \appES{\pctxtwo_{\tm} \ctxholep{\vartwo}}{\esub{\vartwo}{\hctx_{\var}}} \in \evctxUsep{\uvarsettwo_{\tm} \cup \unv{\hctx_{\var}}}{\avarsettwo \cup \anv{\hctx_{\var}}}	$ via rule $\ruleUseHerMult$. 

							\item Let $\vartwo \notin \uvarsettwo$ and $\vartwo \in \avarsettwo$. By \ih (Focusing exponential evaluation contexts on applied variables), there exists $\pctxtwo_{\tm} \in \evctxAppp{\uvarsettwo_{\tm}}{\avarsettwo_{\tm}}$ such that $\pctxtwo_{\tm} \ctxholep{\vartwo} = \pctxtwop{\tm}$, with $\uvarsettwo_{\tm} \subseteq \uvarsettwo$ and $\vartwo \notin \avarsettwo_{\tm} \subset \avarsettwo$. Moreover, \refprop{exp-ctx-are-mult} (Exponential evaluation contexts are multiplicative) gives that $\pctxtwo_{\tm} \in \evctxUsep{\uvarsetsix_{\tm}}{\avarsetsix_{\tm}}$ for some $\uvarsetsix_{\tm} \subseteq \uvarsettwo_{\tm}$ and $\avarsetsix_{\tm} \subseteq \avarsettwo$. We can finally derive $\pctx_{\tm} \defeq \appES{\pctxtwo_{\tm} \ctxholep{\vartwo}}{\esub{\vartwo}{\hctx_{\var}}} \in \evctxUsep{\uvarsetsix_{\tm} \cup \unv{\hctx_{\var}}}{\avarsetsix_{\tm} \cup \anv{\hctx_{\var}}}$ via rule $\ruleUseHerMult$.

							\item Let $\vartwo \in \uvarsettwo$ and $\vartwo \in \avarsettwo$. By \ih (Focusing exponential evaluation contexts on applied variables), there exists $\pctxtwo_{\tm} \in \evctxAppp{\uvarsettwo_{\tm}}{\avarsettwo_{\tm}}$ such that $\pctxtwo_{\tm} \ctxholep{\vartwo} = \pctxtwop{\tm}$, with $\uvarsettwo_{\tm} \subseteq \uvarsettwo$ and $\vartwo \notin \avarsettwo_{\tm} \subset \avarsettwo$. Case analysis on whether $\var \in \uvarsettwo_{\tm}$:
								\begin{itemize}
									\item Let $\vartwo \in \uvarsettwo_{\tm}$. By \ih (Focusing exponential evaluation contexts on unapplied variables), there exists $\pctxthree_{\vartwo} \in \evctxUsep{\uvarsetthree_{\vartwo}}{\avarsetthree_{\vartwo}}$ such that $\pctxthree_{\vartwo} \ctxholep{\vartwo} = \pctxtwo_{\tm} \ctxholep{\vartwo} = \pctxtwop{\tm}$, with $\vartwo \notin \uvarsetthree_{\vartwo} \subset \uvarsettwo_{\tm} \subseteq \uvarsettwo$ and $\avarsetthree_{\vartwo} \subseteq \avarsettwo_{\tm} \subset \avarsettwo$. We can then derive $\pctx_{\tm} \defeq \appES{\pctxthree_{\vartwo} \ctxholep{\vartwo}}{\esub{\vartwo}{\hctx_{\var}}} \in \evctxUsep{\uvarsetthree_{\vartwo} \cup \unv{\hctx_{\var}}}{\avarsetthree_{\vartwo} \cup \anv{\hctx_{\var}}}$ via rule $\ruleUseHerMult$.

									\item Let $\vartwo \notin \uvarsettwo_{\tm}$. By \refprop{exp-ctx-are-mult} (Exponential evaluation contexts are multiplicative) gives that $\pctxtwo_{\tm} \in \evctxUsep{\uvarsetsix_{\tm}}{\avarsetsix_{\tm}}$ for some $\uvarsetsix_{\tm} \subseteq \uvarsettwo_{\tm}$ and $\avarsetsix_{\tm} \subseteq \avarsettwo$. We can then derive $\pctx_{\tm} \defeq \appES{\pctxtwo_{\tm} \ctxholep{\vartwo}}{\esub{\vartwo}{\hctx_{\var}}} \in \evctxUsep{\uvarsetsix_{\tm} \cup \unv{\hctx_{\var}}}{\avarsetsix_{\tm} \cup \anv{\hctx_{\var}}}$.

								\end{itemize}

						\end{itemize}

				\end{itemize}
	
			\item Let $\pctx \in \evctxAppp{\uvarset}{\avarset}$ be derived as follows
				$$
				\infer
					[\ruleUseGcExp]
					{\appES{\pctxtwo}{\esub{\vartwo}{\tmtwo}} \in \evctxAppp{\uvarset}{\avarset }}
					{\pctxtwo \in \evctxAppp{\uvarset}{\avarset}
					\quad
					\vartwo \notin (\uvarset \cup \avarset)}
				$$
			with $\pctx = \appES{\pctxtwo}{\esub{\vartwo}{\tmtwo}}$. By \ih (Focusing exponential evaluation contexts on unapplied variables), there exists $\pctxtwo_{\tm} \in \evctxUsep{\uvarsettwo_{\tm}}{\avarsettwo_{\tm}}$ such that $\pctxtwo_{\tm} \ctxholep{\var} = \pctxtwop{\tm}$, $\var \notin \uvarset_{\tm} \subset \uvarset$ and $\avarset_{\tm} \subseteq \avarset$. We can then derive $\pctx_{\tm} \defeq \appES{\pctxtwo_{\tm}}{\esub{\vartwo}{\tmtwo}} \in \evctxAppp{\uvarset_{\tm}}{\avarset_{\tm}}$ via rule $\ruleUseGcExp$.
		
			\item Let $\pctx \in \evctxAppp{\uvarset}{\avarset}$ be derived as follows
				$$
				\infer
					[\ruleUseUselessExp]
					{\appES{\pctxtwo}{\esub{\vartwo}{\val}} \in \evctxAppp{\uvarsettwo \setminus \set{\vartwo}}{\avarsettwo}}
					{\pctxtwo \in \evctxAppp{\uvarsettwo}{\avarsettwo}
					\quad
					\vartwo \in (\uvarsettwo \setminus \avarsettwo)}
				$$
			with $\pctx = \appES{\pctxtwo}{\esub{\vartwo}{\val}}$, $\uvarset = \uvarsettwo \setminus \set{\vartwo}$ and $\avarset = \avarsettwo$. By \ih (Focusing exponential evaluation contexts on unapplied variables), there exists $\pctxtwo_{\tm} \in \evctxUsep{\uvarsettwo_{\tm}}{\avarsettwo_{\tm}}$ such that $\pctxtwo_{\tm} \ctxholep{\var} = \pctxtwop{\tm}$, with $\var \notin \uvarsettwo_{\tm} \subset \uvarsettwo$ and $\avarsettwo_{\tm} \subseteq \avarsettwo$. 

			We can then derive $\pctx_{\tm} \defeq \appES{\pctxtwo_{\tm}}{\esub{\vartwo}{\val}} \in \evctxUsep{\uvarsettwo_{\tm} \setminus \set{\vartwo}}{\avarsettwo_{\tm}}$ (resp. $\pctx_{\tm} \defeq \appES{\pctxtwo_{\tm}}{\esub{\vartwo}{\val}} \in \evctxUsep{\uvarsettwo_{\tm}}{\avarsettwo_{\tm}}$) via rule $\ruleUseUselessMult$ (resp. via rule $\ruleUseGcMult$) if $\vartwo \in \uvarsettwo_{\tm}$ (resp. if $\vartwo \notin \uvarsettwo_{\tm}$).
	
			\item Let $\pctx \in \evctxAppp{\uvarset}{\avarset}$ be derived as follows
				$$
				\infer
					[\ruleUseHerExp]
					{\appES{\pctxtwop{\vartwo}}{\esub{\vartwo}{\ctxhole}} \in \evctxAppp{\uvarsettwo \setminus \set{\vartwo}}{\avarsettwo}}
					{\pctxtwo \in \evctxAppp{\uvarsettwo}{\avarsettwo}
					\quad
					\vartwo \notin \avarsettwo}
				$$
			with $\pctx = \appES{\pctxtwop{\vartwo}}{\esub{\vartwo}{\ctxhole}}$, $\uvarset = \uvarsettwo \setminus \set{\vartwo}$ and $\avarset = \avarsettwo$. By \ih (Focusing exponential evaluation contexts on unapplied variables) on $\var$ and $\pctxtwo$, there exists $\pctxtwo_{\vartwo} \in \evctxUsep{\uvarsettwo_{\vartwo}}{\avarsettwo_{\vartwo}}$ such that $\pctxtwo_{\vartwo} \ctxholep{\var} = \pctxtwop{\vartwo}$, with $\var \notin \uvarsettwo_{\vartwo} \subset \uvarsettwo$ and $\avarsettwo_{\vartwo} \subseteq \avarsettwo$.

			If $\vartwo \notin \uvarsettwo_{\tm}$, then we can derive $\pctx_{\tm} \defeq \appES{\pctxtwo_{\vartwo}}{\esub{\vartwo}{\tm}} \in \evctxAppp{\uvarsettwo_{\vartwo}}{\avarsettwo_{\tm}}$ via rule $\ruleUseGcMult$.

			Else, we proceed by case analysis on the shape of $\tm$:
				\begin{itemize}
					\item Let $\tm \in \Var$. Then we can derive $\pctx_{\tm} \defeq \appES{\pctxtwo_{\tm}}{\esub{\vartwo}{\tm}} \in \evctxUsep{\updp{\uvarsettwo_{\tm}}{\vartwo}{\tm}}{\updp{\avarsettwo_{\tm}}{\vartwo}{\tm}}$ via rule $\ruleUseVarMult$.

					\item Let $\tm$ be a useful inert term. Then we apply rule $\ruleUseNonVarMult$ to derive 
						$$
						\pctx_{\tm} \defeq \appES{\pctxtwo_{\tm}}{\esub{\vartwo}{\tm}} \in \evctxUsep{(\uvarsettwo_{\tm} \setminus \set{\vartwo}) \cup \unv{\tm}}{(\avarsettwo_{\tm} \setminus \set{\vartwo}) \cup \anv{\tm}}
						$$

					\item Let $\tm$ be a value. Then we can derive $\pctx_{\tm} \defeq \appES{\pctxtwo_{\tm}}{\esub{\vartwo}{\tm}} \in \evctxUsep{\uvarsettwo_{\tm} \setminus \set{\vartwo})}{\avarsettwo_{\tm}}$ via rule $\ruleUseUselessMult$.

				\end{itemize}
	
		\end{itemize}

	\item \emph{Focusing exponential evaluation contexts on applied variables}: 
	
	Let $\var \in \avarset$ and $\tm$ be normal. We proceed by case analysis on the last derivation rule in $\pctx \in \evctxAppp{\uvarset}{\avarset}$:
		\begin{itemize}
			\item Let $\pctx \in \evctxAppp{\uvarset}{\avarset}$ be derived as follows
				$$
				\infer
					[\ruleUseAxOneExp]
					{\progentry{\hctxapp}{\emptyenv} \in \evctxAppp{\unv{\hctxapp}}{\anv{\hctxapp}}}
					{}
				$$
			with $\pctx = \progentry{\hctxapp}{\emptyenv}$, $\uvarset = \unv{\hctxapp}$ and $\avarset = \anv{\hctxapp}$. The statement is given by applying \reflemmap{Rewriting_applicative_term_contexts}{three} (Focusing term contexts on applied variables) on $\hctxapp$, yielding an applicative term context $\hctxapptwosub{\tm}$ such that $\hctxapptwosub{\tm} \ctxholep{\var} = \hctxappp{\tm}$, with $\unv{\hctxapptwosub{\tm}} \subseteq \unv{\hctxapp}$ and $\var \notin \anv{\hctxapptwosub{\tm}} \subset \anv{\hctxapp}$, and finally taking $\pctx_{\tm} \defeq \progentry{\hctxapptwosub{\tm}}{\emptyenv}$.
	
			\item Let $\pctx \in \evctxAppp{\uvarset}{\avarset}$ be derived as follows
				$$
				\infer
					[\ruleUseAxTwoExp]
					{\appES{\pctxtwop{\vartwo}}{\esub{\vartwo}{\hctxapp}} \in \evctxAppp{\uvarsettwo \cup \unv{\hctxapp}}{\avarsettwo \cup \anv{\hctxapp}}}
					{\pctxtwo \in \evctxUsep{\uvarsettwo}{\avarsettwo}
					\quad
					\vartwo \notin (\uvarsettwo \cup \avarsettwo)}
				$$
			with $\pctx = \appES{\pctxtwop{\vartwo}}{\esub{\vartwo}{\hctxapp}}$, $\uvarset = \uvarsettwo \cup \unv{\hctxapp}$ and $\avarset = \avarsettwo \cup \anv{\hctxapp}$. Case analyis on whether $\var \in \avarsettwo$:
				\begin{itemize}
					\item Let $\var \in \avarsettwo$. By \ih (Focusing multiplicative evaluation contexts on applied variables), there exists $\pctxtwo_{\vartwo} \in \evctxAppp{\uvarsettwo_{\vartwo}}{\avarsettwo_{\vartwo}}$ such that $\pctxtwo_{\vartwo} \ctxholep{\var} = \pctxtwop{\vartwo}$, with $\uvarsettwo_{\vartwo} \subseteq \uvarsettwo$ and $\var \notin \avarsettwo \subset \avarsettwo_{\vartwo}$. We can then derive $\pctx_{\tm} \defeq \appES{\pctxtwo_{\vartwo}}{\esub{\vartwo}{\tm}} \in \evctxAppp{\uvarsettwo_{\vartwo}}{\avarsettwo_{\vartwo}}$ via rule $\ruleUseGcExp$.
					
					\item Let $\var \notin \avarsettwo$. Then it must be that $\var \in \anv{\hctxapp}$. By \reflemmap{Rewriting_applicative_term_contexts}{three} (Focusing term contexts on applied variables), there exists an applicative term context $\hctxappsub{\tm}$ such that $\hctxappsub{\tm} \ctxholep{\var} = \hctxappp{\tm}$, with $\unv{\hctxappsub{\tm}} \subseteq \unv{\hctxapp}$ and $\var \notin \anv{\hctxappsub{\tm}} \subset \anv{\hctxapp}$. We can then derive $\pctx_{\tm} \defeq \appES{\pctxtwop{\vartwo}}{\esub{\vartwo}{\hctxappsub{\tm}}} \in \evctxAppp{\uvarsettwo \cup \unv{\hctxappsub{\tm}}}{\avarsettwo_{\vartwo} \cup \anv{\hctxappsub{\tm}}}$ via rule $\ruleUseAxTwoExp$.
					
				\end{itemize}
	
			\item Let $\pctx \in \evctxAppp{\uvarset}{\avarset}$ be derived as follows
				$$
				\infer
					[\ruleUseVarExp]
					{\appES{\pctxtwo}{\esub{\vartwo}{\varthree}} \in \evctxAppp{\updp{\uvarsettwo}{\vartwo}{\varthree}}{\updp{\avarsettwo}{\vartwo}{\varthree}}}
					{\pctxtwo \in \evctxAppp{\uvarsettwo}{\avarsettwo}
					\quad
					\vartwo \in (\uvarsettwo \cup \avarsettwo)}
				$$
			with $\pctx = \appES{\pctxtwo}{\esub{\vartwo}{\varthree}}$, $\uvarset = \updp{\uvarsettwo}{\vartwo}{\varthree}$ and $\avarset = \updp{\avarsettwo}{\vartwo}{\varthree}$. We proceed by case analysis on whether $\vartwo \in \avarsettwo$:
				\begin{itemize}
					\item Let $\vartwo \notin \avarsettwo$. Since then $\avarsettwo = \updp{\avarsettwo}{\vartwo}{\varthree} \ni \var$, then we can apply the \ih (Focusing exponential evaluation contexts on applied variables) to obtain a $\pctxtwo_{\tm} \in \evctxAppp{\avarsettwo_{\tm}}{\uvarsettwo_{\tm}}$ such that $\pctxtwo_{\tm} \ctxholep{\var} = \pctxtwop{\tm}$, with $\uvarsettwo_{\tm} \subseteq \uvarsettwo$ and $\var \notin \avarsettwo_{\tm} \subset \avarsettwo$. Then we can derive $\pctx_{\tm} \defeq \appES{\pctxtwo_{\tm}}{\esub{\vartwo}{\varthree}} \in \evctxAppp{\updp{\uvarsettwo}{\vartwo}{\varthree}}{\updp{\avarsettwo}{\vartwo}{\varthree}}$ (resp., $\pctx_{\tm} \defeq \appES{\pctxtwo_{\tm}}{\esub{\vartwo}{\varthree}} \in \evctxAppp{\uvarsettwo}{\avarsettwo}$) via rule $\ruleUseVarExp$ (resp., $\ruleUseGcExp$) if $\vartwo \in (\avarsettwo_{\tm} \cup \uvarsettwo_{\tm})$ (resp., $\vartwo \notin (\avarsettwo_{\tm} \cup \uvarsettwo_{\tm})$).
					
					\item Let $\vartwo \in \avarsettwo$. Note that then $\updp{\avarsettwo}{\vartwo}{\varthree} = (\avarsettwo \setminus \set{\vartwo}) \cup \set{\varthree}$. We do case analysis on whether $\var \in \avarsettwo$:
						\begin{itemize}
							\item If $\var \in (\avarsettwo \setminus \set{\vartwo})$, then we can prove the statement similarly to how we proved it in the case above where $\vartwo \notin \avarsettwo$.
							
							\item If $\var \notin \avarsettwo$, then it must be that $\var = \varthree$. By \ih (Focusing exponential evaluation contexts on applied variables) with respect to $\vartwo$ and $\pctxtwo$, there exists $\pctxtwo_{\tm} \in \evctxAppp{\uvarsettwo_{\tm}}{\avarsettwo_{\tm}}$ such that $\pctxtwo_{\tm} \ctxholep{\vartwo} = \pctxtwop{\tm}$, with $\uvarsettwo_{\tm} \subseteq \uvarsettwo$ and $\vartwo \notin \avarsettwo_{\tm} \subset \avarsettwo$. Moreover, \refprop{exp-ctx-are-mult} (Exponential evaluation contexts are multiplicative) gives that $\pctxtwo_{\tm} \in \evctxUsep{\uvarsetthree}{\avarsetthree}$ for some $\uvarsetthree \subseteq \uvarsettwo_{\tm}$ and $\avarsetthree \subseteq \avarsettwo_{\tm}$. We can then derive $\pctx_{\tm} \defeq \appES{\pctxtwo_{\tm} \ctxholep{\vartwo}}{\esub{\vartwo}{\ctxhole}} \in \evctxAppp{\uvarsetthree}{\avarsetthree}$ via rule $\ruleUseAxTwoExp$.
							
						\end{itemize}
					
				\end{itemize}
	
			\item Let $\pctx \in \evctxAppp{\uvarset}{\avarset}$ be derived as follows
				$$
				\infer
					[\ruleUseNonVarExp]
					{\appES{\pctxtwo}{\esub{\vartwo}{\itmplus}} \in \evctxAppp{(\uvarsettwo \setminus \set{\vartwo}) \cup \unv{\itmplus}}{(\avarsettwo \setminus \set{\vartwo}) \cup \anv{\itmplus}}}
					{\pctxtwo \in \evctxAppp{\uvarsettwo}{\avarsettwo}
					\quad
					\vartwo \in (\uvarsettwo \cup \avarsettwo)}
				$$
			with $\pctx = \appES{\pctxtwo}{\esub{\vartwo}{\itmplus}}$, $\uvarset = (\uvarsettwo \setminus \set{\vartwo}) \cup \unv{\itmplus}$ and $\avarset = (\avarsettwo \setminus \set{\vartwo}) \cup \anv{\itmplus}$. Case analysis on whether $\var \in \avarsettwo$:
				\begin{itemize}
					\item Let $\var \in \avarsettwo$. By \ih (Focusing exponential evaluation contexts on applied variables), there exists $\pctxtwo_{\tm} \in \evctxAppp{\uvarsettwo_{\tm}}{\avarsettwo_{\tm}}$ such that $\pctxtwo_{\tm} \ctxholep{\var} = \pctxtwop{\tm}$, with $\uvarsettwo_{\tm} \subseteq \uvarsettwo$ and $\var \notin \avarsettwo_{\tm} \subset \avarsettwo$. Case analysis on whether $\var \in \anv{\itmplus}$:
						\begin{itemize}
							\item Let $\var \notin \anv{\itmplus}$. We can then derive 
								$$
								\pctx_{\tm} \defeq \appES{\pctxtwo_{\tm}}{\esub{\vartwo}{\itmplus}} \in \evctxAppp{(\uvarsettwo_{\tm} \setminus \set{\vartwo}) \cup \unv{\itmplus}}{(\avarsettwo_{\tm} \setminus \set{\vartwo}) \cup \anv{\itmplus}}
								$$
							(resp. $\pctx_{\tm} \defeq \appES{\pctxtwo_{\tm}}{\esub{\vartwo}{\itmplus}} \in \evctxAppp{\uvarsettwo_{\tm}}{\avarsettwo_{\tm}}$) via rule $\ruleUseNonVarExp$ (resp. $\ruleUseGcExp$) if $\vartwo \in (\uvarsettwo_{\tm} \cup \avarsettwo_{\tm})$ (resp. if $\vartwo \notin (\uvarsettwo_{\tm} \cup \avarsettwo_{\tm})$).

							\item Let $\var \in \anv{\itmplus}$. Note that if $\vartwo \notin (\uvarsettwo_{\tm} \cup \avarsettwo_{\tm})$ then we can derive $\pctx_{\tm} \defeq \appES{\pctxtwo_{\tm}}{\esub{\vartwo}{\itmplus}} \in \evctxAppp{\uvarsettwo_{\tm}}{\avarsettwo_{\tm}}$ via rule $\ruleUseGcExp$. Let us now consider the case of $\vartwo \in (\uvarsettwo_{\tm} \cup \avarsettwo_{\tm})$, proceeding by case analysis:
								\begin{itemize}
									\item Let $\vartwo \in \uvarsettwo_{\tm}$ and $\vartwo \notin \avarsettwo_{\tm}$. By \ih (Focusing exponential evaluation contexts on unapplied variables), there exists $\pctxthree_{\var} \in \evctxUsep{\uvarsetthree_{\var}}{\avarsetthree_{\var}}$ such that $\pctxthree_{\var} \ctxholep{\vartwo} = \pctxtwo_{\tm} \ctxholep{\var}$, with $\vartwo \notin \uvarsetthree_{\var} \subset \uvarsettwo_{\tm} \subseteq \uvarsettwo$ and $\avarsetthree_{\var} \subseteq \avarsettwo_{\tm} \subseteq \avarsettwo$. We can then derive $\pctx_{\tm} \defeq \appES{\pctxthree_{\var} \ctxholep{\vartwo}}{\esub{\vartwo}{\hctxappsub{\var}}} \in \evctxAppp{\uvarsetthree_{\var} \cup \unv{\hctxappsub{\var}}}{\avarsetthree_{\var} \cup \anv{\hctxappsub{\var}}}$ via rule $\ruleUseGcExp$.

									\item Let $\vartwo \in \avarsettwo_{\tm}$. By \ih (Focusing exponential evaluation contexts on applied variables), there exists $\pctxthree_{\var} \in \evctxAppp{\uvarsetthree_{\var}}{\avarsetthree_{\var}}$ such that $\pctxthree_{\var} \ctxholep{\vartwo} = \pctxtwo_{\tm} \ctxholep{\tm} = \pctxtwop{\tm}$, with $\uvarsetthree_{\var} \subseteq \uvarsettwo_{\tm} \subseteq \uvarsettwo$ and $\vartwo \notin \avarsetthree_{\var} \subset \avarsettwo_{\tm} \subset \avarsettwo$. In addition, \refprop{exp-ctx-are-mult} (Exponential evaluation contexts are multiplicative) gives that $\pctxthree_{\var} \in \evctxUsep{\uvarsetseven_{\var}}{\avarsetseven_{\var}}$ for some $\uvarsetseven_{\var} \subseteq \uvarsetthree_{\var}$ and $\avarsetseven_{\var} \subseteq \avarsetthree_{\var}$.

									Finally, if $\vartwo \notin \uvarsetseven_{\var}$, then we can derive $\pctx_{\tm} \defeq \appES{\pctxthree_{\var} \ctxholep{\vartwo}}{\esub{\vartwo}{\hctxappsub{\var}}}$ via rule $\ruleUseAxTwoExp$. If $\vartwo \in \uvarsetseven_{\var}$ instead, then \reflemmap{Rewriting_evaluation_contexts-base_cases}{three} (Focusing multiplicative evaluation contexts on unapplied variables) gives $\pctxfour_{\vartwo} \in \evctxUsep{\uvarsetfour_{\vartwo}}{\avarsetfour_{\vartwo}}$ such that $\pctxfour_{\vartwo} \ctxholep{\vartwo} = \pctxthree_{\var} \ctxholep{\vartwo} = \pctxtwo_{\tm} \ctxholep{\var} = \pctxtwop{\tm}$, with $\vartwo \notin \uvarsetfour_{\vartwo} \subset \uvarsetthree_{\var} \subseteq \uvarsettwo_{\tm} \subseteq \uvarsettwo$ and $\avarsetfour_{\vartwo} \subseteq \avarsetthree_{\var} \subset \avarsettwo_{\tm} \subset \avarsettwo$; we can then derive $\pctx_{\tm} \defeq \appES{\pctxfour_{\vartwo} \ctxholep{\vartwo}}{\esub{\vartwo}{\hctxappsub{\var}}} \in \evctxAppp{\uvarsetfour_{\vartwo} \cup \unv{\hctxappsub{\var}}}{\avarsetfour_{\vartwo} \cup \anv{\hctxappsub{\var}}}$ via rule $\ruleUseAxTwoExp$.

								\end{itemize}
						\end{itemize}

					\item Let $\var \notin \avarsettwo$. Then $\var \in \anv{\itmplus}$. By \reflemmap{Rewriting_applicative_term_contexts}{two} (Focusing useful inert terms on applied variables), there exists applicative term context $\hctxappsub{\var}$ such that $\hctxappsub{\var} \ctxholep{\var} = \itmplus$, with $\unv{\hctxappsub{\var}} \subseteq \unv{\itmplus}$ and $\var \notin \anv{\hctxappsub{\var}} \subset \anv{\itmplus}$. Case analysis on whether $\vartwo \in \uvarsettwo$ and on whether $\vartwo \in \avarsettwo$:
						\begin{itemize}
							\item Let $\vartwo \in \uvarsettwo$ and $\vartwo \notin \avarsettwo$. By \ih (Focusing exponential evaluation contexts on unapplied variables), there exists $\pctxtwo_{\tm} \in \evctxUsep{\uvarsettwo_{\tm}}{\avarsettwo_{\tm}}$ such that $\pctxtwo_{\tm} \ctxholep{\vartwo} = \pctxtwop{\tm}$, with $\vartwo \notin \uvarsettwo_{\tm} \subset \uvarsettwo$ and $\avarsettwo_{\tm} \subseteq \avarsettwo$. We can then derive $\pctx_{\tm} \defeq \appES{\pctxtwo_{\tm} \ctxholep{\vartwo}}{\esub{\vartwo}{\hctxappsub{\var}}} \in \evctxAppp{\uvarsettwo_{\tm} \cup \unv{\hctxappsub{\var}}}{\avarsettwo_{\tm} \cup \anv{\hctxappsub{\var}}}$ via rule $\ruleUseAxTwoExp$.

							\item Let $\vartwo \notin \uvarsettwo$ and $\vartwo \in \avarsettwo$. By \ih (Focusing exponential evaluation contexts on applied variables), there exists $\pctxtwo_{\tm} \in \evctxAppp{\uvarsettwo_{\tm}}{\avarsettwo_{\tm}}$ such that $\pctxtwo_{\tm} \ctxholep{\vartwo} = \pctxtwop{\tm}$, with $\uvarsettwo_{\tm} \subseteq \uvarsettwo$ and $\vartwo \notin \avarsettwo_{\tm} \subset \avarsettwo$. In addition, \refprop{exp-ctx-are-mult} (Exponential evaluation contexts are multiplicative) gives that $\pctxtwo_{\tm} \in \evctxUsep{\uvarsetsix_{\tm}}{\avarsetsix_{\tm}}$, for some $\uvarsetsix_{\tm} \subseteq \uvarsettwo_{\tm}$ and $\avarsetsix_{\tm} \subseteq \avarsettwo_{\tm}$. We can then derive $\pctx_{\tm} \defeq \appES{\pctxtwo_{\tm} \ctxholep{\vartwo}}{\esub{\vartwo}{\hctxappsub{\var}}} \in \evctxAppp{\uvarsetsix_{\tm} \cup \unv{\hctxappsub{\var}}}{\avarsetsix_{\tm} \cup \anv{\hctxappsub{\var}}}$ via rule $\ruleUseAxTwoExp$.

							\item Let $\vartwo \in \uvarsettwo$ and $\vartwo \in \avarsettwo$. By \ih (Focusing exponential evaluation contexts on unapplied variables), there exists $\pctxtwo_{\tm} \in \evctxUsep{\uvarsettwo_{\tm}}{\avarsettwo_{\tm}}$ such that $\pctxtwo_{\tm} \ctxholep{\vartwo} = \pctxtwop{\tm}$, with $\vartwo \notin \uvarsettwo_{\tm} \subset \uvarsettwo$ and $\avarsettwo_{\tm} \subseteq \avarsettwo$. Case analysis on whether $\vartwo \in \avarsettwo_{\tm}$:
								\begin{itemize}
									\item If $\vartwo \notin \avarsettwo_{\tm}$, then we can derived $\pctx_{\tm} \defeq \appES{\pctxtwo_{\tm} \ctxholep{\vartwo}}{\esub{\vartwo}{\hctxappsub{\var}}} \in \evctxAppp{\uvarsettwo_{\tm} \cup \unv{\hctxappsub{\var}}}{\avarsettwo_{\tm} \cup \anv{\hctxappsub{\var}}}$ via rule $\ruleUseAxTwoExp$.

									\item Let $\vartwo \in \avarsettwo_{\tm}$. By \ih (Focusing multiplicative evaluation contexts on applied variables), there exists $\pctxthree_{\vartwo} \in \evctxAppp{\uvarsetthree_{\vartwo}}{\avarsetthree_{\vartwo}}$ such that $\pctxthree_{\vartwo} \ctxholep{\vartwo} = \pctxtwo_{\tm} \ctxholep{\vartwo} = \pctxtwop{\tm}$, with $\uvarsetthree_{\vartwo} \subseteq \uvarsettwo_{\tm} \subset \uvarsettwo$ and $\vartwo \notin \avarsetthree_{\vartwo} \subset \avarsettwo_{\tm} \subseteq \avarsettwo$. In addition, \refprop{exp-ctx-are-mult} (Exponential evaluation contexts are multiplicative) gives that $\pctxthree_{\vartwo} \in \evctxUsep{\uvarsetseven_{\vartwo}}{\avarsetseven_{\vartwo}}$, for some $\uvarsetseven_{\vartwo} \subseteq \uvarsettwo_{\tm}$ and $\avarsetseven_{\vartwo} \subseteq \avarsettwo_{\tm}$. We can then derive $\pctx \defeq \appES{\pctxthree_{\vartwo} \ctxholep{\vartwo}}{\esub{\vartwo}{\hctxappsub{\var}}} \in \evctxAppp{\uvarsetseven_{\vartwo} \cup \unv{\hctxappsub{\var}}}{\avarsetseven_{\vartwo} \cup \anv{\hctxappsub{\var}}}$ via rule $\ruleUseAxTwoExp$.

								\end{itemize}
						\end{itemize}

				\end{itemize}
	
			\item Let $\pctx \in \evctxAppp{\uvarset}{\avarset}$ be derived as follows
				$$
				\infer
					[\ruleUseGcExp]
					{\appES{\pctxtwo}{\esub{\vartwo}{\tmtwo}} \in \evctxAppp{\uvarset}{\avarset }}
					{\pctxtwo \in \evctxAppp{\uvarset}{\avarset}
					\quad
					\vartwo \notin (\uvarset \cup \avarset)}
				$$
			with $\pctx = \appES{\pctxtwo}{\esub{\vartwo}{\tmtwo}}$. By \ih (Focusing exponential evaluation contexts on applied variables), there exists $\pctxtwo_{\tm} \in \evctxAppp{\uvarset_{\tm}}{\avarset_{\tm}}$ such that $\pctxtwo_{\tm} \ctxholep{\var} = \pctxtwop{\tm}$. We can then derive $\pctx_{\tm} \defeq \appES{\pctxtwo_{\tm}}{\esub{\vartwo}{\tmtwo}} \in \evctxAppp{\uvarset_{\tm}}{\avarset_{\tm}}$ via rule $\ruleUseGcExp$.
		
			\item Let $\pctx \in \evctxAppp{\uvarset}{\avarset}$ be derived as follows
				$$
				\infer
					[\ruleUseUselessExp]
					{\appES{\pctxtwo}{\esub{\vartwo}{\val}} \in \evctxAppp{\uvarsettwo \setminus \set{\vartwo}}{\avarsettwo}}
					{\pctxtwo \in \evctxAppp{\uvarsettwo}{\avarsettwo}
					\quad
					\vartwo \in (\uvarsettwo \setminus \avarsettwo)}
				$$
			with $\pctx = \appES{\pctxtwo}{\esub{\vartwo}{\val}}$, $\uvarset = \uvarsettwo \setminus \set{\vartwo}$ and $\avarset = \avarsettwo$. By \ih (Focusing exponential evaluation contexts on applied variables), there exists $\pctxtwo_{\tm} \in \evctxAppp{\uvarsettwo_{\tm}}{\avarsettwo_{\tm}}$ such that $\pctxtwo_{\tm} \ctxholep{\var} = \pctxtwop{\tm}$, with $\uvarsettwo_{\tm} \subseteq \uvarsettwo$ and $\var \notin \avarsettwo_{\tm} \subset \avarsettwo$.  We can then derive $\pctx_{\tm} \defeq \appES{\pctxtwo_{\tm}}{\esub{\vartwo}{\val}} \in \evctxAppp{\uvarsettwo_{\tm} \setminus \set{\vartwo}}{\avarsettwo_{\tm}}$ (resp. $\pctx_{\tm} \defeq \appES{\pctxtwo_{\tm}}{\esub{\vartwo}{\val}} \in \evctxAppp{\uvarsettwo_{\tm}}{\avarsettwo_{\tm}}$) via rule $\ruleUseUselessExp$ (resp. $\ruleUseGcExp$) if $\vartwo \in \uvarsettwo_{\tm}$ (resp. if $\vartwo \notin \uvarsettwo_{\tm}$).
	
			\item Let $\pctx \in \evctxAppp{\uvarset}{\avarset}$ be derived as follows
				$$
				\infer
					[\ruleUseHerExp]
					{\appES{\pctxtwop{\vartwo}}{\esub{\vartwo}{\ctxhole}} \in \evctxAppp{\uvarsettwo \setminus \set{\vartwo}}{\avarsettwo}}
					{\pctxtwo \in \evctxAppp{\uvarsettwo}{\avarsettwo}
					\quad
					\vartwo \notin \avarsettwo}
				$$
			with $\pctx = \appES{\pctxtwop{\vartwo}}{\esub{\vartwo}{\ctxhole}}$, $\uvarset = \uvarsettwo \setminus \set{\vartwo}$ and $\avarset = \avarsettwo$. By \ih (Focusing exponential evaluation contexts on applied variables) on $\var$ and $\pctxtwo$, there exists $\pctxtwo_{\vartwo} \in \evctxAppp{\uvarsettwo_{\vartwo}}{\avarsettwo_{\vartwo}}$ such that $\pctxtwo_{\vartwo} \ctxholep{\var} = \pctxtwop{\vartwo}$, with $\uvarsettwo_{\vartwo} \subseteq \uvarsettwo$ and $\var \notin \avarsettwo_{\vartwo} \subset \avarsettwo$. Case analysis on whether $\vartwo \in \uvarsettwo_{\vartwo}$:
				\begin{itemize}
					\item Let $\vartwo \in \uvarsettwo_{\vartwo}$. Case analysis on the shape of $\tm$:
						\begin{itemize}
							\item Let $\tm \in \Var$. We can then use $\ruleUseVarExp$ to derive 
								$$
								\pctx \defeq \appES{\pctxtwo_{\vartwo}}{\esub{\vartwo}{\tm}} \in \evctxAppp{\updp{\uvarsettwo_{\vartwo}}{\vartwo}{\tm}}{\updp{\avarsettwo_{\vartwo}}{\vartwo}{\tm}}
								$$

							\item Let $\tm$ be a useful inert term. We can then apply rule $\ruleUseNonVarExp$ to derive 
								$$
								\pctx \defeq \appES{\pctxtwo_{\vartwo}}{\esub{\vartwo}{\tm}} \in \evctxAppp{(\uvarsettwo_{\vartwo} \setminus \set{\vartwo}) \cup \unv{\tm}}{(\avarsettwo_{\vartwo} \setminus \set{\vartwo}) \cup \anv{\tm}}
								$$
 
							\item Let $\tm$ be a value. We can then derive $\pctx \defeq \appES{\pctxtwo_{\vartwo}}{\esub{\vartwo}{\tm}} \in \evctxAppp{\uvarsettwo_{\vartwo} \setminus \set{\vartwo}}{\avarsettwo_{\vartwo}}$ via rule $\ruleUseUselessExp$.

						\end{itemize}

					\item Let $\vartwo \notin \uvarsettwo_{\vartwo}$. We can then derive $\pctx_{\tm} \defeq \appES{\pctxtwo_{\vartwo}}{\esub{\vartwo}{\tm}} \in \evctxAppp{\uvarsettwo_{\vartwo}}{\avarsettwo_{\vartwo}}$ via rule $\ruleUseGcExp$.\qedhere

				\end{itemize}
	
		\end{itemize}
	
\end{enumerate}

\end{proof}

\begin{lemma}[Focusing for \usefulopencbneednsp-normal forms]
\hfill

Let $\prog$ be a program in $\tound$-normal form.

\begin{enumerate}
	\item
	\label{p:Focusing_for_usefulopencbneed-normal_forms-one}
	\emph{Focusing \usefulopencbneednsp-normal forms on unapplied variables}: Let $\var \in \unv{\prog}$. Then there exists $\pctx \in \evctxUsep{\uvarset}{\avarset}$ such that $\prog = \pctxp{\var}$, with $\var \notin \uvarset \subset \unv{\prog}$ and $\avarset \subseteq \unv{\prog}$.

	\item
	\label{p:Focusing_for_usefulopencbneed-normal_forms-two}
	\emph{Focusing \usefulopencbneednsp-normal forms on applied variables}: Let $\var \in \anv{\prog}$. Then there exists $\pctx \in \evctxAppp{\uvarset}{\avarset}$ such that $\prog = \pctxp{\var}$, with $\uvarset \subseteq \unv{\prog}$ and $\var \notin \avarset \subset \unv{\prog}$.

\end{enumerate}

\label{l:Focusing_for_usefulopencbneed-normal_forms}
\end{lemma}


\begin{proof}
\hfill

Both statements are proven simultaneously by mutual induction:

\begin{enumerate}
	\item
	\emph{Focusing $\tound$-normal forms on unapplied variables}: Let $\prog = \progentry{\tm}{\env}$ and $\var \in \unv{\prog}$. We proceed by induction on $\size{\env}$:
		\begin{itemize}
			\item Let $\prog = \progentry{\tm}{\emptyenv}$. Then $\unv{\tm}$. Suppose $\tm$ is not a normal term. Then there would exist a term context $\hctx$ such that $\tm = \hctxp{(\la{\varthree}{\tmthree}) \tmfour}$---by \reflemma{Redex_in_non-normal_terms-usefulopencbneed} (Redex in non-normal terms). But then we would be able to derive $\progentry{\hctx}{\emptyenv} \in \evctxUsep{\unv{\hctx}}{\anv{\hctx}}$, getting that $\prog = \progentry{\hctx}{\emptyenv} \ctxholep{(\la{\varthree}{\tmthree}) \tmfour} \toum \progentry{\hctx}{\emptyenv} \ctxholeentryp{\tmthree}{\esub{\varthree}{\tmfour}}$, which is absurd. Therefore, $\tm$ is a normal term. Moreover, if $\tm$ is a value then $\unv{\tm} = \emptyset$; absurd as well.

			Hence, $\tm$ is an inert term. By \reflemmap{Rewriting_term_contexts-Proofs_of_usefulopencbneed}{one} (Focusing inert terms on unapplied variables), there exists term context $\hctx_{\var}$ such that $\hctx_{\var} \ctxholep{\var} = \tm$, with $\var \notin \unv{\hctx_{\var}} \subset \unv{\tm}$ and $\anv{\hctx_{\var}} \subseteq \anv{\tm}$. We can then derive $\pctx \defeq \progentry{\hctx_{\var}}{\emptyenv}$ via rule $\ruleUseAxMult$.

			\item Let $\prog = \appES{\progentry{\tm}{\envtwo}}{\esub{\vartwo}{\tmtwo}}$. Case analysis on whether $\var \in \unventryp{\tm}{\envtwo}$:
				\begin{itemize}
					\item Let $\var \in \unventryp{\tm}{\envtwo}$. By \ih (Focusing $\tound$-normal forms on unapplied variables), there exists $\pctxtwo \in \evctxUsep{\uvarsettwo}{\avarsettwo}$ such that $\pctxtwop{\var} = \progentry{\tm}{\envtwo}$, with $\var \notin \uvarsettwo \subset \unventryp{\tm}{\envtwo}$ and $\avarsettwo \subseteq \anventryp{\tm}{\envtwo}$. 

					Note that if $\vartwo \notin (\uvarsettwo \cup \avarsettwo)$, then we can derive $\pctx \defeq \appES{\pctxtwo}{\esub{\vartwo}{\tmtwo}} \in \evctxUsep{\uvarsettwo}{\avarsettwo}$ via rule $\ruleUseGcMult$.

					Let us now consider the case where $\vartwo \in (\uvarsettwo \cup \avarsettwo)$, and proceed by case analysis on whether $\vartwo \in \uvarsettwo$.
						\begin{itemize}
							\item Let $\vartwo \in \uvarsettwo$. By \reflemmap{Rewriting_evaluation_contexts-base_cases}{three} (Focusing multiplicative evaluation contexts on unapplied variables), there exists $\pctxthree_{\var} \in \evctxUsep{\uvarsetthree_{\var}}{\avarsetthree_{\var}}$ such that $\pctxthree_{\var} \ctxholep{\vartwo} = \pctxtwop{\var}$, with $\vartwo \notin \uvarsetthree_{\var} \subset \unventryp{\tm}{\envtwo}$ and $\avarsetthree_{\var} \subseteq \anventryp{\tm}{\envtwo}$.

							Now, suppose $\tmtwo$ were not a normal term. By \reflemma{Redex_in_non-normal_terms-usefulopencbneed} (Redex in non-normal terms), there would exist a term context $\hctx$  such that $\tmtwo = \hctxp{(\la{\varthree}{\tmthree}) \tmfour}$. Note that if $\tmtwo \notin \avarsetthree_{\var}$, then we would be able to derive $\appES{\pctxthree_{\var} \ctxholep{\vartwo}}{\esub{\vartwo}{\hctx}}$ via rule $\ruleUseGcMult$ and get that
								$$
								\begin{array}{rcl}
									\prog & = & \appES{\progentry{\tm}{\envtwo}}{\esub{\vartwo}{\hctx}} \\
									& = & (\appES{\pctxthree_{\var} \ctxholep{\vartwo}}{\esub{\vartwo}{\hctx}}) \ctxholep{(\la{\varthree}{\tmthree}) \tmfour} \\
									& \toum & (\appES{\pctxthree_{\var} \ctxholep{\vartwo}}{\esub{\vartwo}{\hctx}}) \ctxholeentryp{\tmthree}{\esub{\varthree}{\tmfour}} \\
								\end{array}
								$$
							which is absurd. If $\tmtwo \in \avarsetthree_{\var}$ instead, then there would exist $\pctxfour_{\vartwo} \in \evctxAppp{\uvarsetfour_{\vartwo}}{\avarsetfour_{\vartwo}}$ such that $\pctxfour_{\vartwo} \ctxholep{\vartwo} = \pctxthree_{\var} \ctxholep{\vartwo}$, with $\uvarsetfour_{\vartwo} \subseteq \uvarsetthree_{\var}$ and $\vartwo \notin \avarsetfour_{\vartwo} \subset \avarsetthree_{\var}$---by \reflemmap{Rewriting_evaluation_contexts}{one} (Focusing multiplicative evaluation contexts on applied variables). In addition, \refprop{exp-ctx-are-mult} (Exponential evaluation contexts are multiplicative) would give that $\pctxfour_{\vartwo} \in \evctxUsep{\uvarseteight_{\vartwo}}{\avarseteight_{\vartwo}}$, for some $\uvarseteight_{\vartwo} \subseteq \uvarsetfour_{\vartwo}$ and $\avarseteight_{\vartwo} \subseteq \avarsetfour_{\vartwo}$. But then we would be able to derive $\appES{\pctxfour_{\vartwo} }{\esub{\vartwo}{\hctx}} \in \evctxUsep{\uvarseteight_{\vartwo} \cup \unv{\hctx}}{\avarseteight_{\vartwo} \cup \anv{\hctx}}$ and get that 
								$$
								\begin{array}{rcl}
									\prog & = & \appES{\progentry{\tm}{\envtwo}}{\esub{\vartwo}{\tmtwo}} \\
									& = & (\appES{\pctxthree_{\var} \ctxholep{\vartwo}}{\esub{\vartwo}{\hctx}}) \ctxholep{(\la{\varthree}{\tmthree}) \tmfour} \\
									& \toum & (\appES{\pctxthree_{\var} \ctxholep{\vartwo}}{\esub{\vartwo}{\hctx}}) \ctxholeentryp{\tmthree}{\esub{\varthree}{\tmfour}} \\
								\end{array}
								$$
							which is also absurd.

							Therefore, $\tmtwo$ must be a normal term. We proceed by case analysis on the shape of $\tmtwo$ as a normal term, and on whether $\var \in \unv{\tmtwo}$:
								\begin{itemize}
									\item Let $\var = \tmtwo \in \Var$. If $\vartwo \notin \avarsetthree_{\var}$, then we can derive $\pctx \defeq \appES{\pctxthree_{\var} \ctxholep{\vartwo}}{\esub{\vartwo}{\ctxhole}} \in \evctxUsep{\uvarsetthree_{\var}}{\avarsetthree_{\var}}$ via rule $\ruleUseHerMult$. If $\vartwo \in \avarsetthree_{\var}$ instead, then \reflemmap{Rewriting_evaluation_contexts}{one} (Focusing multiplicative evaluation contexts on applied variables) gives the existence of $\pctxfour_{\vartwo} \in \evctxAppp{\uvarsetfour_{\vartwo}}{\avarsetfour_{\vartwo}}$ such that $\pctxfour_{\vartwo} \ctxholep{\vartwo} = \pctxthree_{\var} \ctxholep{\vartwo}$, with $\uvarsetfour_{\vartwo} \subseteq \uvarsetthree_{\var}$ and $\vartwo \notin \avarsetfour_{\vartwo} \subset \avarsetthree_{\var}$. In addition, \refprop{exp-ctx-are-mult} (Exponential evaluation contexts are multiplicative) gives that $\pctxfour_{\vartwo} \in \evctxUsep{\uvarseteight_{\vartwo}}{\avarseteight_{\vartwo}}$, for some $\uvarseteight_{\vartwo} \subseteq \uvarsetfour_{\vartwo} $ and $\avarseteight_{\vartwo} \subseteq \avarsetfour_{\vartwo}$. We can then derive $\pctx \defeq \appES{\pctxfour_{\vartwo} \ctxholep{\vartwo}}{\esub{\vartwo}{\ctxhole}} \in \evctxUsep{\uvarseteight_{\vartwo}}{\avarseteight_{\vartwo}}$ via rule $\ruleUseHerMult$.

									\item Let $\var \neq \tmtwo \in \Var$. Then we can derive 
										$$
										\pctx \defeq \appES{\pctxthree_{\var} \ctxholep{\vartwo}}{\esub{\vartwo}{\tmtwo}} \in \evctxUsep{\updp{\uvarsetthree_{\var}}{\vartwo}{\tmtwo}}{\updp{\avarsetthree_{\var}}{\vartwo}{\tmtwo}}
										$$
									via rule $\ruleUseVarMult$.

									\item Let $\tmtwo$ be an inert term, with $\var \in \unv{\tmtwo}$. By \reflemmap{Rewriting_term_contexts-Proofs_of_usefulopencbneed}{one} (Focusing inert terms on unapplied variables), there exists term context $\hctx_{\var}$ such that $\hctx_{\var} \ctxholep{\var} = \tmtwo$, with $\var \notin \unv{\hctx_{\var}} \subset \unv{\tmtwo}$ and $\anv{\hctx_{\var}} \subseteq \anv{\tmtwo}$.

									Now, if $\tmtwo \notin \avarsetthree_{\var}$, then we can derive 
										$$
										\pctx \defeq \appES{\pctxthree_{\var} \ctxholep{\vartwo}}{\esub{\vartwo}{\hctx_{\var}}} \in \evctxUsep{\uvarsetthree_{\var} \cup \unv{\hctx_{\var}}}{\avarsetthree_{\var} \cup \anv{\hctx_{\var}}}
										$$
									via rule $\ruleUseHerMult$. If $\vartwo \in \avarsetthree_{\var}$ instead, then \reflemmap{Rewriting_evaluation_contexts}{one} (Focusing multiplicative evaluation contexts on applied variables) gives the existence of $\pctxfour_{\vartwo} \in \evctxAppp{\uvarsetfour_{\vartwo}}{\avarsetfour_{\vartwo}}$ such that $\pctxfour_{\vartwo} \ctxholep{\vartwo} = \pctxthree_{\var} \ctxholep{\vartwo}$, with $\uvarsetfour_{\vartwo} \subseteq \uvarsetthree_{\var}$ and $\vartwo \notin \avarsetfour_{\vartwo} \subset \avarsetthree_{\var}$. In addition, \refprop{exp-ctx-are-mult} (Exponential evaluation contexts are multiplicative) gives that $\pctxfour_{\vartwo} \in \evctxUsep{\uvarseteight_{\vartwo}}{\avarseteight_{\vartwo}}$, for some $\uvarseteight_{\vartwo} \subseteq \uvarsetfour_{\vartwo} $ and $\avarseteight_{\vartwo} \subseteq \avarsetfour_{\vartwo}$. We can then derive $\pctx \defeq \appES{\pctxfour_{\vartwo} \ctxholep{\vartwo}}{\esub{\vartwo}{\hctx_{\var}}} \in \evctxUsep{\uvarseteight_{\vartwo} \cup \unv{\hctx_{\var}}}{\avarseteight_{\vartwo} \cup \anv{\hctx_{\var}}}$ via rule $\ruleUseHerMult$.
									
									\item Let $\tmtwo$ be an inert term, with $\var \notin \unv{\tmtwo}$. We can then derive $\pctx \defeq \appES{\pctxtwo}{\esub{\vartwo}{\tmtwo}} \in \evctxUsep{(\uvarsettwo \setminus \set{\vartwo}) \cup \unv{\tmtwo}}{(\avarsettwo \setminus \set{\vartwo}) \cup \anv{\tmtwo}}$ via rule $\ruleUseNonVarMult$.

									\item Let $\tmtwo$ be a value. Suppose that $\vartwo \in \avarsetthree_{\var}$. By \reflemmap{Rewriting_evaluation_contexts}{one} (Focusing multiplicative evaluation contexts on applied variables), there would exist $\pctxfour_{\vartwo} \in \evctxAppp{\uvarsetfour_{\vartwo}}{\avarsetfour_{\vartwo}}$ such that $\pctxfour_{\vartwo} \ctxholep{\vartwo} = \pctxthree_{\var} \ctxholep{\vartwo}$, with $\uvarsetfour_{\vartwo} \subseteq \uvarsetthree_{\var}$ and $\vartwo \notin \avarsetfour_{\vartwo} \subset \avarsetthree_{\var}$. But then we would be able to derive $\appES{\pctxfour_{\vartwo} \ctxholep{\vartwo}}{\esub{\vartwo}{\tmtwo}} \in \evctxAppp{\uvarsetfour_{\vartwo}}{\avarsetfour_{\vartwo}}$ via rule $\ruleUseGcMult$ and get that
										$$
										\pctx = \appES{\progentry{\tm}{\envtwo}}{\esub{\vartwo}{\tmtwo}} = (\appES{\pctxfour_{\vartwo}}{\esub{\vartwo}{\tmtwo}}) \ctxholep{\vartwo} \toue (\appES{\pctxfour_{\vartwo}}{\esub{\vartwo}{\tmtwo}}) \ctxholep{\rename{\tmtwo}}
										$$
									which is absurd.

									Therefore, $\vartwo \notin \avarsetthree_{\var} \subseteq \avarsettwo$. Since $\vartwo \in \uvarsettwo$, then we can then derive $\pctx \defeq \appES{\pctxtwo}{\esub{\vartwo}{\tmtwo}} \in \evctxUsep{\uvarsettwo}{\avarsettwo}$ via rule $\ruleUseUselessMult$.

								\end{itemize}

							\item Let $\vartwo \notin \uvarsettwo$. Then $\vartwo \in \avarsettwo$. Following an analogous reasoning to the one above---where $\vartwo \in \uvarsettwo$---we can conclude that $\tmtwo$ must be a normal term. Similarly, it must be that $\tmtwo$ is not a value. The statement follows by proceeding by case analysis on the shape of $\tmtwo$, and on whether $\var \in \unv{\tmtwo}$, analogously to what we did in the case above---where $\vartwo \in \uvarsettwo$.

						\end{itemize}

					\item Let $\var \notin \unventryp{\tm}{\envtwo}$. Then $\var \in \unv{\tmtwo}$, and---by the definition of unapplied variables---it must be that either $\vartwo \in \unventryp{\tm}{\envtwo}$ or ($\vartwo \in \nventryp{\tm}{\envtwo}$ and $\tmtwo \notin \Var$)---note that these are not mutually exclusive proposition.

					Moreover, following a similar reasoning to the case above---where $\var \in \unventryp{\tm}{\envtwo}$---we can see that $\tmtwo$ must be an inert term. By \reflemmap{Rewriting_term_contexts-Proofs_of_usefulopencbneed}{one} (Focusing inert terms on unapplied variables), there exists term context $\hctx_{\var}$ such that $\hctx_{\var} \ctxholep{\var} = \tmtwo$, with $\var \notin \unv{\hctx_{\var}} \subset \unv{\tmtwo}$ and $\anv{\hctx_{\var}} \subseteq \anv{\tmtwo}$.

					Case analysis on whether $\vartwo \notin \unventryp{\tm}{\envtwo}$.
						\begin{itemize}
							\item Let $\vartwo \in \unventryp{\tm}{\envtwo}$. By \ih (Focusing $\tound$-normal forms on unapplied varibles), there exists $\pctxtwo \in \evctxUsep{\uvarsettwo}{\avarsettwo}$ such that $\pctxtwop{\vartwo} = \progentry{\tm}{\envtwo}$, with $\vartwo \notin \uvarsettwo \subset \unventryp{\tm}{\envtwo}$ and $\avarsettwo \subseteq \anventryp{\tm}{\envtwo}$. 

							Now, if $\vartwo \notin \avarsettwo$, then we can derive $\pctx \defeq \appES{\pctxtwop{\vartwo}}{\esub{\vartwo}{\hctx_{\var}}} \in \evctxUsep{\uvarsettwo \cup \unv{\hctx_{\var}}}{\avarsettwo \cup \anv{\hctx_{\var}}}$ via rule $\ruleUseHerMult$. If $\vartwo \in \avarsettwo$ instead, then \reflemmap{Rewriting_evaluation_contexts}{one} (Focusing multiplicative evaluation contexts on applied variables), there exists $\pctxthree_{\vartwo} \in \evctxAppp{\uvarsetthree_{\vartwo}}{\avarsetthree_{\vartwo}}$ such that $\pctxthree_{\vartwo} \ctxholep{\vartwo} = \pctxtwop{\vartwo}$, with $\uvarsetthree_{\vartwo} \subseteq \uvarsettwo$ and $\avarsetthree_{\vartwo} \subseteq \avarsettwo$. In addition, \refprop{exp-ctx-are-mult} (Exponential evaluation contexts are multiplicative) gives that $\pctxthree_{\vartwo} \in \evctxUsep{\uvarsetseven_{\vartwo}}{\avarsetseven_{\vartwo}}$, for some $\uvarsetseven_{\vartwo} \subseteq \uvarsetthree_{\vartwo}$ and $\avarsetseven_{\vartwo} \subseteq \avarsetthree_{\vartwo}$. We can finally derive $\pctx \defeq \appES{\pctxthree_{\vartwo} \ctxholep{\vartwo}}{\esub{\vartwo}{\hctx_{\var}}} \in \evctxUsep{\uvarsetseven_{\vartwo} \cup \unv{\hctx_{\var}}}{\avarsetseven_{\vartwo} \cup \anv{\hctx_{\var}}}$ via rule $\ruleUseHerMult$.
							
							\item Let $\vartwo \notin \unventryp{\tm}{\envtwo}$. Then $\tmtwo \notin \Var$ and $\vartwo \in \nventryp{\tm}{\envtwo}$, the latter implying that $\vartwo \in \anventryp{\tm}{\envtwo}$---by \reflemmap{Unapplied_applied_and_needed_variables}{two} (Unapplied, applied and needed variables). By \ih (Focusing $\tound$-normal forms on applied variables), there exists $\pctxtwo \in \evctxAppp{\uvarsettwo}{\avarsettwo}$ such that $\pctxtwop{\vartwo} = \progentry{\tm}{\envtwo}$, with $\uvarsettwo \subseteq \unventryp{\tm}{\envtwo}$ and $\vartwo \notin \avarsettwo \subset \anventryp{\tm}{\envtwo}$. In addition, \refprop{exp-ctx-are-mult} (Exponential evaluation contexts are multiplicative) gives that $\pctxtwo \in \evctxUsep{\uvarsetsix}{\avarsetsix}$, for some $\uvarsetsix \subseteq \uvarsettwo$ and $\avarsetsix \subseteq \avarsettwo$. Note that $\vartwo \notin \unventryp{\tm}{\envtwo} \supseteq \uvarsettwo \supseteq \uvarsetsix$. Hence, we can derive $\pctx \defeq \appES{\pctxtwop{\vartwo}}{\esub{\vartwo}{\hctx_{\var}}} \in \evctxUsep{\uvarsetsix \cup \unv{\hctx_{\var}}}{\avarsetsix \cup \anv{\hctx_{\var}}}$ via rule $\ruleUseHerMult$.
							
						\end{itemize}

				\end{itemize}

		\end{itemize}

	\item
	\emph{Focusing $\tound$-normal forms on applied variables}: Let $\prog = \progentry{\tm}{\env}$ and $\var \in \anv{\prog}$. We proceed by induction on $\size{\env}$:
		\begin{itemize}
			\item Let $\prog = \progentry{\tm}{\emptyenv}$. Note that if $\tm \in \Var$ or $\tm$ is a value then $\anv{\pctx} = \emptyset$---absurd. Thus, if $\tm$ is not a useful inert term, then it is not a normal term either and so \reflemma{Redex_in_non-normal_terms-usefulopencbneed} (Redex in non-normal terms) gives a term context $\hctx$ such that $\tm = \hctxp{(\la{\varthree}{\tmthree}) \tmfour}$. Since all this implies that $\prog = \progentry{\hctx}{\emptyenv} \ctxholep{(\la{\varthree}{\tmthree}) \tmfour} \toum \progentry{\hctx}{\emptyenv} \ctxholeentryp{\tmthree}{\esub{\varthree}{\tmfour}}$---which is absurd---we can conclude that $\tm$ is a useful inert term. Therefore, by \reflemmap{Rewriting_applicative_term_contexts}{two} (Focusing useful inert terms on applied variables) there exists an applicative term context $\hctxappsub{\var}$ such that $\tm = \hctx_{\var} \ctxholep{\var}$, with $\unv{\hctxappsub{\var}} \subseteq \unv{\tm}$ and $\var \notin \anv{\hctxappsub{\var}} \subset \anv{\tm}$. The statement then follows by taking $\pctx \defeq \progentry{\hctxappsub{\var}}{\emptyenv} \in \evctxUsep{\unv{\hctxappsub{\var}}}{\anv{\hctxappsub{\var}}}$.
				
			\item Let $\prog = \appES{\progentry{\tm}{\envtwo}}{\esub{\vartwo}{\tmtwo}}$. Case analysis on whether $\var \in \anventryp{\tm}{\envtwo}$:
				\begin{itemize}
					\item Let $\var \in \anventryp{\tm}{\envtwo}$. By \ih (Focusing $\tound$-normal forms on applied variables), there exists $\pctxtwo \in \evctxAppp{\uvarsettwo}{\avarsettwo}$ such that $\pctxtwop{\tm} = \progentry{\tm}{\envtwo}$, with $\uvarsettwo \subseteq \unventryp{\tm}{\envtwo}$ and $\var \notin \avarsettwo \subset \anventryp{\tm}{\envtwo}$.

					Note that if $\vartwo \notin (\uvarsettwo \cup \avarsettwo)$, then we can derive $\pctx \defeq \appES{\pctxtwo}{\esub{\vartwo}{\tmtwo}}$ via rule $\ruleUseGcExp$. Let us assume now that $\vartwo \in (\uvarsettwo \cup \avarsettwo)$, and proceed by case analysis on whether $\vartwo \in \avarsettwo$:
						\begin{itemize}
							\item Let $\vartwo \in \avarsettwo$. By \reflemmap{Rewriting_evaluation_contexts}{three} (Focusing exponential evaluation contexts on applied variables), there exists $\pctxthree_{\var} \in \evctxAppp{\uvarsetthree_{\var}}{\avarsetthree_{\var}}$ such that $\pctxthree_{\var} \ctxholep{\vartwo} = \pctxtwop{\var}$, with $\uvarsetthree_{\var} \subseteq \uvarsettwo$ and $\vartwo \notin \avarsetthree_{\var} \subset \avarsettwo$.

							Suppose now that $\tmtwo$ were not a normal term. By \reflemma{Redex_in_non-normal_terms-usefulopencbneed} (Redex in non-normal terms), there would exist a term context $\hctx$  such that $\tmtwo = \hctxp{(\la{\varthree}{\tmthree}) \tmfour}$. Case analysis on whether $\vartwo \in \uvarsetthree_{\var}$:
								\begin{itemize}
									\item Let $\vartwo \notin \uvarsetthree_{\var}$. Then note that $\pctxthree_{\var} \in \evctxUsep{\uvarsetseven_{\var}}{\avarsetseven_{\var}}$, for some $\uvarsetseven_{\var} \subseteq \uvarsetthree_{\var}$ and $\avarsetseven_{\var} \subseteq \avarsetthree_{\var}$---by \refprop{exp-ctx-are-mult} (Exponential evaluation contexts are multiplicative---and so we would be able to derive $\appES{\pctxthree_{\var} \ctxholep{\vartwo}}{\esub{\vartwo}{\hctx}} \in \evctxUsep{\uvarsetseven_{\var} \cup \unv{\hctx}}{\avarsetseven_{\var} \cup \anv{\hctx}}$ via rule $\ruleUseAxTwoExp$ and get that
										$$
										\begin{array}{rcl}
											\prog & = & \appES{\progentry{\tm}{\envtwo}}{\esub{\vartwo}{\tmtwo}} \\
											& = & (\appES{\pctxthree_{\var} \ctxholep{\vartwo}}{\esub{\vartwo}{\hctx}}) \ctxholep{(\la{\varthree}{\tmthree}) \tmfour} \\
											& \toum & (\appES{\pctxthree_{\var} \ctxholep{\vartwo}}{\esub{\vartwo}{\hctx}}) \ctxholeentryp{\tmthree}{\esub{\varthree}{\tmfour}} \\
										\end{array}
										$$
									which is absurd.

									\item Let $\vartwo \in \uvarsetthree_{\var}$. By \reflemmap{Rewriting_evaluation_contexts}{two} (Focusing exponential evaluation contexts on unapplied variables), there exists $\pctxfour_{\vartwo} \in \evctxUsep{\uvarsetfour_{\vartwo}}{\avarsetfour_{\var}}$ such that $\pctxfour_{\vartwo} \ctxholep{\vartwo} = \pctxthree_{\var} \ctxholep{\vartwo}$, with $\vartwo \notin \uvarsetfour_{\vartwo} \subset \uvarsetthree_{\var}$ and $\avarsetfour_{\vartwo} \subseteq \avarsetthree_{\var}$. But then we would be able to derive $\appES{\pctxfour_{\vartwo} \ctxholep{\vartwo}}{\esub{\vartwo}{\hctx}} \in \evctxUsep{\uvarsetfour_{\vartwo} \cup \unv{\hctx}}{\avarsetfour_{\vartwo} \cup \anv{\hctx}}$ via rule $\ruleUseAxTwoExp$ and get that 
										$$
										\begin{array}{rcl}
											\prog & = & \appES{\progentry{\tm}{\envtwo}}{\esub{\vartwo}{\tmtwo}} \\
											& = & (\appES{\pctxfour_{\vartwo} \ctxholep{\vartwo}}{\esub{\vartwo}{\hctx}}) \ctxholep{(\la{\varthree}{\tmthree}) \tmfour} \\
											& \toum & (\appES{\pctxfour_{\vartwo} \ctxholep{\vartwo}}{\esub{\vartwo}{\hctx}}) \ctxholeentryp{\tmthree}{\esub{\varthree}{\tmfour}} \\
										\end{array}
										$$
									which is absurd.

								\end{itemize}

							Therefore, $\tmtwo$ must be a normal term. Moreover, note that if $\tmtwo $ is a value, then we would be able to derive $\appES{\pctxthree_{\var}}{\esub{\vartwo}{\tmtwo}} \in \evctxAppp{\uvarsetthree_{\var} \setminus \set{\vartwo}}{\avarsettwo}$ via rule $\ruleUseUselessExp$ and get that 
								$$
								\begin{array}{rcl}
											\prog & = & \appES{\progentry{\tm}{\envtwo}}{\esub{\vartwo}{\tmtwo}} \\
											& = & (\appES{\pctxthree_{\var} \ctxholep{\vartwo}}{\esub{\vartwo}{\hctx}}) \ctxholep{\vartwo} \\
											& \toue & (\appES{\pctxthree_{\varthree} \ctxholep{\vartwo}}{\esub{\vartwo}{\hctx}}) \ctxholep{\rename{\tmtwo}} \\
										\end{array}
								$$
							which is absurd as well. Therefore, $\tmtwo$ must be an inert term. 

							Finally, we proceed by case analysis on the shape of $\tmtwo$:
								\begin{itemize}
									\item Let $\tmtwo \in \Var$. If $\tmtwo \neq \var$, then we can derive $\pctx \defeq \appES{\pctxtwo}{\esub{\vartwo}{\tmtwo}} \in \evctxAppp{\updp{\uvarsettwo}{\vartwo}{\tmtwo}}{\updp{\avarsettwo}{\vartwo}{\tmtwo}}$ via rule $\ruleUseVarExp$. If $\tmtwo = \var$ instead, then we can derive $\pctx \defeq \appES{\pctxthree_{\var} \ctxholep{\vartwo}}{\esub{\vartwo}{\ctxhole}} \in \evctxAppp{\uvarsetthree_{\var} \setminus \set{\vartwo}}{\avarsetthree_{\var}}$ via rule $\ruleUseHerExp$.

									\item Let $\tmtwo \notin \Var$. Then $\tm$ is a useful inert term. First, note if $\var \notin \anv{\tmtwo}$, then we can simply derive $\pctx \defeq \appES{\pctxtwo}{\esub{\vartwo}{\tmtwo}} \in \evctxAppp{(\uvarsettwo \setminus \set{\vartwo}) \cup \unv{\tmtwo}}{(\avarsettwo \setminus \set{\vartwo}) \cup \anv{\tmtwo}}$ via rule $\ruleUseNonVarExp$.

									Let us consider now the case where $\var \in \anv{\tmtwo}$. Note that there exists applicative term context $\hctxappsub{\var}$ such that $\hctxappsub{\var} \ctxholep{\var} = \tmtwo$, with $\unv{\hctxappsub{\var}} \subseteq \unv{\tmtwo}$ and $\var \notin \anv{\hctxappsub{\var}} \subset \anv{\tmtwo}$---by \reflemmap{Rewriting_applicative_term_contexts}{two} (Focusing useful inert terms on applied variables). Moreover, by \refprop{exp-ctx-are-mult} (Exponential evaluation contexts are multiplicative), it happens that $\pctxthree_{\var} \in \evctxUsep{\uvarsetseven_{\var}}{\avarsetseven_{\var}}$, for some $\uvarsetseven_{\var} \subseteq \uvarsetfour_{\var}$ and $\avarsetseven_{\var} \subseteq \avarsetfour_{\var}$.

									Now, if $\vartwo \notin \uvarsetseven_{\var}$, then we can simply apply rule $\ruleUseAxTwoExp$ to derive 
										$$
										\pctx \defeq \appES{\pctxthree_{\var} \ctxholep{\vartwo}}{\esub{\vartwo}{\hctxappsub{\var}}} \in \evctxAppp{\uvarsetthree_{\var} \cup \unv{\hctxappsub{\var}}}{\avarsetthree_{\var} \cup \anv{\hctxappsub{\var}}}
										$$
									
									If $\vartwo \in \uvarsetseven_{\var}$ instead, then \reflemmap{Rewriting_evaluation_contexts-base_cases}{three} (Focusing multiplicative evaluation contexts on unapplied varibales) gives $\pctxfour_{\vartwo} \in \evctxUsep{\uvarsetfour_{\vartwo}}{\avarsetfour_{\vartwo}}$ such that $\pctxfour_{\vartwo} \ctxholep{\vartwo} = \pctxthree_{\var} \ctxholep{\var}$, with $\vartwo \notin \uvarsetfour_{\vartwo} \subset \uvarsetthree_{\var}$ and $\avarsetfour_{\vartwo} \subseteq \avarsetthree_{\var}$; we can finally derive $\pctx \defeq \appES{\pctxfour_{\vartwo} \ctxholep{\vartwo}}{\esub{\vartwo}{\hctxappsub{\var}}} \in \evctxAppp{\uvarsetfour_{\vartwo} \cup \unv{\hctxappsub{\var}}}{\avarsetfour_{\vartwo} \cup \anv{\hctxappsub{\var}}}$ via rule $\ruleUseAxTwoExp$.

								\end{itemize}

							\item Let $\vartwo \notin \avarsettwo$. Then $\vartwo \in \uvarsettwo$. By \reflemmap{Rewriting_evaluation_contexts}{two} (Focusing exponential evaluation contexts on unapplied variables), there exists $\pctxthree_{\var} \in \evctxUsep{\uvarsetthree_{\var}}{\avarsetthree_{\var}}$ such that $\pctxthree_{\var} \ctxholep{\vartwo} = \pctxtwop{\var}$, with $\vartwo \notin \uvarsetthree_{\var} \subset \uvarsettwo$ and $\avarsetthree_{\var} \subseteq \avarsettwo$.

							We can now proceed by an analogous reasoning to the case above---where $\vartwo \in \avarsettwo$---to obtain that $\tmtwo$ must be an inert term. Moreover, we can proceed by case analysis on the shape of $\tmtwo$---following an analogous reasoning to the case above as well---to derive $\pctx$.

						\end{itemize}

					\item Let $\var \notin \anventryp{\tm}{\envtwo}$. By the definition of applied variables, there are two (mutually exclusive) possibilities, namely $\var = \tmtwo \in \Var$ or $\var \in \anv{\tmtwo}$:
						\begin{itemize}
							\item Let $\var = \tmtwo \in \Var$, with $\vartwo \in \nventryp{\tm}{\envtwo}$ and $\vartwo \in \anventryp{\tm}{\envtwo}$---by the definition of unapplied variables. Hence, we can apply the \ih (Focusing $\tound$-normal forms on applied variables) to get $\pctxtwo \in \evctxAppp{\uvarsettwo}{\avarsettwo}$ such that $\pctxtwop{\vartwo} = \progentry{\tm}{\envtwo}$, with $\uvarsettwo \subseteq \unventryp{\tm}{\envtwo}$ and $\vartwo \notin \avarsettwo \subset \anventryp{\tm}{\envtwo}$. We can finally derive $\pctx \defeq \appES{\pctxtwop{\vartwo}}{\esub{\vartwo}{\ctxhole}} \in \evctxAppp{\uvarsettwo \setminus \set{\vartwo}}{\avarsettwo}$ via rule $\ruleUseHerExp$.

							\item Let $\var \in \anv{\tmtwo}$, with $\vartwo \in \nventryp{\tm}{\envtwo}$---by the definition of unapplied variables. 

							Note that $\tmtwo$ is not a variable nor a value. We can infer that $\tmtwo$ must be a useful inert term, based on what is given by \reflemma{Redex_in_non-normal_terms-usefulopencbneed} (Redex in non-normal terms) and following an analogous reasoning to the ones given above. Therefore, there exists applicative term context $\hctxappsub{\var}$ such that $\hctxappsub{\var} \ctxholep{\var} = \tmtwo$, with $\unv{\hctxappsub{\var}} \subseteq \unv{\tmtwo}$ and $\var \notin \anv{\hctxappsub{\var}} \subset \anv{\tmtwo}$---by \reflemmap{Rewriting_applicative_term_contexts}{two} (Focusing useful inert terms on applied variables).

							Moreover, note that $\vartwo \in \nventryp{\tm}{\envtwo} = \unventryp{\tm}{\envtwo} \cup \anventryp{\tm}{\envtwo}$---by \reflemma{Unapplied_applied_and_needed_variables} (Unapplied, applied and needed variables). We proceed by case analysis on whether $\vartwo \in \unventryp{\tm}{\envtwo}$:
								\begin{itemize}
									\item Let $\vartwo \in \unventryp{\tm}{\envtwo}$. By \ih (Focusing $\tound$-normal forms on unapplied variables), there exists $\pctxtwo \in \evctxUsep{\uvarsettwo}{\avarsettwo}$ such that $\pctxtwop{\vartwo} = \progentry{\tm}{\envtwo}$, with $\vartwo \notin \uvarsettwo \subset \unventryp{\tm}{\envtwo}$ and $\avarsettwo \subseteq \anventryp{\tm}{\envtwo}$. 

									Now, if $\vartwo \notin \avarsettwo$, then we can derive $\pctx \defeq \appES{\pctxtwop{\vartwo}}{\esub{\vartwo}{\hctxappsub{\var}}} \in \evctxAppp{\uvarsettwo \cup \unv{\hctxappsub{\var}}}{\avarsettwo \cup \anv{\hctxappsub{\var}}}$ via rule $\ruleUseAxTwoExp$. If $\vartwo \in \avarsettwo$ instead, then there exists $\pctxthree_{\vartwo} \in \evctxAppp{\uvarsetthree_{\vartwo}}{\avarsetthree_{\vartwo}}$ such that $\pctxthree_{\vartwo} \ctxholep{\vartwo} = \pctxtwop{\vartwo}$, with $\uvarsetthree_{\vartwo} \subseteq \uvarsettwo$ and $\vartwo \notin \avarsetthree_{\vartwo} \subset \avarsettwo$---by \reflemmap{Rewriting_evaluation_contexts}{one} (Focusing multiplicative evaluation contexts on applied variables); we can then apply \refprop{exp-ctx-are-mult} (Exponential evaluation contexts are multiplicative) to get that $\pctxthree_{\vartwo} \in \evctxUsep{\uvarsetseven_{\var}}{\avarsetseven_{\var}}$ for some $\uvarsetseven_{\var} \subseteq \uvarsetthree_{\var}$ and $\avarsetseven_{\var} \subseteq \avarsetthree_{\var}$, and finally derive $\pctx \defeq \appES{\pctxthree_{\vartwo} \ctxholep{\vartwo}}{\esub{\vartwo}{\hctxappsub{\var}}} \in \evctxAppp{\uvarsetseven_{\vartwo} \cup \unv{\hctxappsub{\var}}}{\avarsetseven_{\vartwo} \cup \anv{\hctxappsub{\var}}}$ via rule $\ruleUseAxTwoExp$.

									\item Let $\vartwo \notin \unventryp{\tm}{\envtwo}$. Then it must be that $\vartwo \in \anventryp{\tm}{\envtwo}$. By \ih (Focusing $\tound$-normal forms on applied variables), there exists $\pctxtwo \in \evctxAppp{\uvarsettwo}{\avarsettwo}$ such that $\pctxtwop{\vartwo} = \progentry{\tm}{\envtwo}$, with $\uvarsettwo \subseteq \unventryp{\tm}{\envtwo}$ and $\vartwo \notin \avarsettwo \subset \anventryp{\tm}{\envtwo}$. By \refprop{exp-ctx-are-mult} (Exponential evaluation contexts are multiplicative), $\pctxtwo \in \evctxUsep{\uvarsetsix}{\avarsetsix}$, for some $\uvarsetsix \subseteq \uvarsettwo$ and $\avarsetsix \subseteq \avarsettwo$. 

									Now, if $\vartwo \notin \uvarsetsix$, then we can derive $\pctx \defeq \appES{\pctxtwop{\vartwo}}{\esub{\vartwo}{\hctxappsub{\var}}} \in \evctxAppp{\uvarsettwo \cup \unv{\hctxappsub{\var}}}{\avarsettwo \cup \anv{\hctxappsub{\var}}}$ via rule $\ruleUseAxTwoExp$. If $\vartwo \in \uvarsetsix$ instead, then \reflemmap{Rewriting_evaluation_contexts-base_cases}{three} (Focusing multiplicative evaluation contexts on unapplied variables) gives $\pctxthree_{\vartwo} \in \evctxUsep{\uvarsetthree_{\vartwo}}{\avarsetthree_{\vartwo}}$ such that $\pctxthree_{\vartwo} \ctxholep{\vartwo} = \pctxtwop{\vartwo}$, with $\vartwo \notin \uvarsetthree_{\vartwo} \subset \uvarsetsix$ and $\avarsetthree_{\vartwo} \subseteq \avarsetsix$; we can then derive $\pctx \defeq \appES{\pctxthree_{\vartwo} \ctxholep{\vartwo}}{\esub{\vartwo}{\hctxappsub{\var}}} \in \evctxAppp{\uvarsetthree_{\vartwo} \cup \unv{\hctxappsub{\var}}}{\avarsetthree_{\vartwo} \cup \anv{\hctxappsub{\var}}}$ via rule $\ruleUseAxTwoExp$.\qedhere

								\end{itemize}

						\end{itemize}

				\end{itemize}

		\end{itemize}

\end{enumerate}

\end{proof}

\begin{lemma}[Properties of \usefulopencbneednsp-normal forms and ESs]
\hfill

\begin{enumerate}
	\item 
	\label{p:Properties_of_usefulopencbneed-normal_forms_and_ESs-one}
	\emph{Removing ESs does not create $\tound$-redexes}: if $\progentry{\tm}{\env \esub{\vartwo}{\tmtwo}}$ is in $\tound$-normal form, then $\progentry{\tm}{\env}$ is in $\tound$-normal form.
	
	\item 
	\label{p:Properties_of_usefulopencbneed-normal_forms_and_ESs-two}
	\emph{Appending ESs that do not create $\tound$-redexes}: Let $\progentry{\tm}{\env}$ be a $\tound$-normal form such that if $\vartwo \in \nventryp{\tm}{\env}$ then $\tmtwo$ is a normal term, and, moreover, if $\vartwo \in \anventryp{\tm}{\env}$ then $\tmtwo$ is an inert term. 
	
	Then $\progentry{\tm}{\env \esub{\vartwo}{\tmtwo}}$ is in $\tound$-normal form.
	
\end{enumerate}

\label{l:Properties_of_usefulopencbneed-normal_forms_and_ESs}
\end{lemma}


\begin{proof}
\hfill	

\begin{enumerate}
	\item
	\emph{Removing ESs does not create $\tound$-redexes}: We prove the contrapositive statement: 
		\begin{center}
			If $\progentry{\tm}{\env}$ is not in $\tound$-normal form

			then

			$\progentry{\tm}{\env \esub{\vartwo}{\tmtwo}}$ is not in $\tound$-normal form
		\end{center}

	Case analysis on the kind of redex in $\progentry{\tm}{\env}$:
		\begin{itemize}
			\item Let $\progentry{\tm}{\env} = \pctxp{(\la{\var}{\tmthree}) \tmfour} \toum \pctxentryp{\tmthree}{\esub{\var}{\tmfour}}$, with $\pctx \in \evctxUsep{\uvarset}{\avarset}$. Note that if $\vartwo \notin (\uvarset \cup \avarset)$, then we can derive $\appES{\pctx}{\esub{\vartwo}{\tmtwo}} \in \evctxUsep{\uvarset}{\avarset}$ via rule $\ruleUseGcMult$ and get that 
				$$
				\progentry{\tm}{\env \esub{\vartwo}{\tmtwo}} = (\appES{\pctx}{\esub{\vartwo}{\tmtwo}}) \ctxholep{(\la{\var}{\tmthree}) \tmfour} \toum (\appES{\pctx}{\esub{\vartwo}{\tmtwo}}) \ctxholeentryp{\tmthree}{\esub{\var}{\tmfour}}
				$$
			Let us now consider the case where $\vartwo \in (\uvarset \cup \avarset)$ and proceed by case analysis on the shape of $\tmtwo$:
				\begin{itemize}
					\item Let $\tmtwo \in \Var$. We can then derive $\appES{\pctx}{\esub{\vartwo}{\tmtwo}} \in \evctxUsep{\updp{\uvarset}{\vartwo}{\tmtwo}}{\updp{\avarset}{\vartwo}{\tmtwo}}$ via rule $\ruleUseVarMult$ and get that 
						$$
						\progentry{\tm}{\env \esub{\vartwo}{\tmtwo}} = (\appES{\pctx}{\esub{\vartwo}{\tmtwo}}) \ctxholep{(\la{\var}{\tmthree}) \tmfour} \toum (\appES{\pctx}{\esub{\vartwo}{\tmtwo}}) \ctxholeentryp{\tmthree}{\esub{\var}{\tmfour}}
						$$

					\item Let $\tmtwo$ be a useful inert term. We can then derive $\appES{\pctx}{\esub{\vartwo}{\tmtwo}} \in \evctxUsep{(\uvarset \setminus \set{\vartwo}) \cup \unv{\tmtwo}}{(\avarset \setminus \set{\vartwo}) \cup \anv{\tmtwo}}$ via rule $\ruleUseNonVarMult$ and get that 
						$$
						\progentry{\tm}{\env \esub{\vartwo}{\tmtwo}} = (\appES{\pctx}{\esub{\vartwo}{\tmtwo}}) \ctxholep{(\la{\var}{\tmthree}) \tmfour} \toum (\appES{\pctx}{\esub{\vartwo}{\tmtwo}}) \ctxholeentryp{\tmthree}{\esub{\var}{\tmfour}}
						$$

					\item Let $\tmtwo$ be a value. Case analysis on whether $\vartwo \in \avarset$:
						\begin{itemize}
							\item Let $\vartwo \in \avarset$. By \reflemmap{Rewriting_evaluation_contexts}{one} (Focusing multiplicative evaluation contexts on applied variables), there exists $\pctxtwo_{\vartwo} \in \evctxAppp{\uvarsettwo_{\vartwo}}{\avarsettwo_{\vartwo}}$ such that $\pctxtwo_{\vartwo} \ctxholep{\vartwo} = \pctxp{(\la{\var}{\tmthree}) \tmfour}$, with $\uvarsettwo_{\vartwo} \subseteq \uvarset$ and $\vartwo \notin \avarsettwo_{\vartwo} \subset \avarset$. Finally, case analysis on whether $\vartwo \in \uvarsettwo_{\vartwo}$
								\begin{itemize}
									\item Let $\vartwo \in \uvarsettwo_{\vartwo}$. We can then derive $\appES{\pctxtwo_{\vartwo}}{\esub{\vartwo}{\tmtwo}} \in \evctxUsep{\uvarsettwo_{\vartwo} \setminus \set{\vartwo}}{\avarsettwo}$ via rule $\ruleUseUselessExp$ and get that 
										$$
										\progentry{\tm}{\env \esub{\vartwo}{\tmtwo}} = (\appES{\pctxtwo_{\vartwo}}{\esub{\vartwo}{\tmtwo}}) \ctxholep{\vartwo} \toue (\appES{\pctxtwo_{\vartwo}}{\esub{\vartwo}{\tmtwo}}) \ctxholep{\rename{\tmtwo}}
										$$

									\item Let $\vartwo \notin \uvarsettwo_{\vartwo}$. We can then derive $\appES{\pctxtwo_{\vartwo}}{\esub{\vartwo}{\tmtwo}} \in \evctxUsep{\uvarsettwo_{\vartwo}}{\avarsettwo}$ via rule $\ruleUseGcExp$ and get that 
										$$
										\progentry{\tm}{\env \esub{\vartwo}{\tmtwo}} = (\appES{\pctxtwo_{\vartwo}}{\esub{\vartwo}{\tmtwo}}) \ctxholep{\vartwo} \toue (\appES{\pctxtwo_{\vartwo}}{\esub{\vartwo}{\tmtwo}}) \ctxholep{\rename{\tmtwo}}
										$$

								\end{itemize}

							\item Let $\vartwo \notin \avarset$. Then it must be that $\vartwo \in \uvarset$. We can then derive $\appES{\pctx}{\esub{\vartwo}{\tmtwo}} \in \evctxUsep{\uvarset \setminus \set{\vartwo}}{\avarset}$ via rule $\ruleUseUselessMult$ and get that
								$$
								\progentry{\tm}{\env \esub{\vartwo}{\tmtwo}} = (\appES{\pctx}{\esub{\vartwo}{\tmtwo}}) \ctxholep{(\la{\var}{\tmthree}) \tmfour} \toum (\appES{\pctx}{\esub{\vartwo}{\tmtwo}}) \ctxholeentryp{\tmthree}{\esub{\var}{\tmfour}}
								$$

						\end{itemize}

					\item Let $\tmtwo$ be a non-normal term. By \reflemma{Redex_in_non-normal_terms-usefulopencbneed} (Redex in non-normal terms), there exists a term context $\hctxtwo$ such that $\tmtwo = \hctxp{(\la{\varfour}{\tmsix}) \tmseven}$. Case analysis on whether $\vartwo \in \uvarset$: 
						\begin{itemize}
							\item Let $\vartwo \in \uvarset$. By \reflemmap{Rewriting_evaluation_contexts-base_cases}{three} (Focusing multiplicative evaluation contexts on unapplied variables), there exists $\pctxtwo_{\vartwo} \in \evctxUsep{\uvarsettwo_{\vartwo}}{\avarsettwo_{\vartwo}}$ such that $\pctxtwo_{\vartwo} \ctxholep{\vartwo} = \pctxp{(\la{\var}{\tmthree}) \tmfour}$, with $\vartwo \notin \uvarsettwo_{\vartwo} \subset \uvarset$ and $\avarsettwo_{\vartwo} \subseteq \avarset$. 

							Now, if $\vartwo \notin \avarsettwo_{\vartwo}$, then we can derive $\appES{\pctxtwo_{\vartwo}}{\esub{\vartwo}{\hctxtwo}} \in \evctxUsep{\uvarsettwo_{\vartwo} \cup \unv{\hctxtwo}}{\avarsettwo_{\vartwo} \cup \anv{\hctxtwo}}$ via rule $\ruleUseHerMult$ and get that 
								$$
								\appES{\progentry{\tm}{\env}}{\esub{\vartwo}{\tmtwo}} = (\appES{\pctxtwo_{\vartwo} \ctxholep{\vartwo}}{\esub{\vartwo}{\hctxtwo}}) \ctxholep{(\la{\varfour}{\tmsix}) \tmseven} \toum (\appES{\pctxtwo_{\vartwo} \ctxholep{\vartwo}}{\esub{\vartwo}{\hctxtwo}}) \ctxholeentryp{\tmsix}{\esub{\varfour}{\tmseven}}
								$$

							Let $\vartwo \in \avarsettwo_{\vartwo}$ instead. By \reflemmap{Rewriting_evaluation_contexts}{one} (Focusing multiplicative evaluation contexts on applied variables), there exists $\pctxthree_{\vartwo} \in \evctxAppp{\uvarsetthree_{\vartwo}}{\avarsetthree_{\vartwo}}$ such that $\pctxthree_{\vartwo} \ctxholep{\vartwo} = \pctxtwo_{\vartwo} \ctxholep{\vartwo}$, with $\uvarsetthree_{\vartwo} \subseteq \uvarsettwo_{\vartwo}$ and $\vartwo \notin \avarsetthree_{\vartwo} \subset \avarsettwo_{\vartwo}$. In addition, \refprop{exp-ctx-are-mult} (Exponential evaluation contexts are multiplicative) gives that $\pctxthree_{\vartwo} \in \evctxAppp{\uvarsetseven_{\vartwo}}{\avarsetseven_{\vartwo}}$, for some $\uvarsetseven_{\vartwo} \subseteq \uvarsetthree_{\vartwo}$ and $\avarsetseven_{\vartwo} \subseteq \avarsetthree_{\vartwo}$. We can then derive $\appES{\pctxthree_{\vartwo} \ctxholep{\vartwo}}{\esub{\vartwo}{\hctxtwo}} \in \evctxUsep{\uvarsetseven_{\vartwo} \cup \unv{\hctxtwo}}{\avarsetseven_{\vartwo} \cup \anv{\hctxtwo}}$ such that 
								$$
								\appES{\progentry{\tm}{\env}}{\esub{\vartwo}{\tmtwo}} = (\appES{\pctxthree_{\vartwo} \ctxholep{\vartwo}}{\esub{\vartwo}{\hctxtwo}}) \ctxholep{(\la{\varfour}{\tmsix}) \tmseven} \toum (\appES{\pctxthree_{\vartwo} \ctxholep{\vartwo}}{\esub{\vartwo}{\hctxtwo}}) \ctxholeentryp{\tmsix}{\esub{\varfour}{\tmseven}}
								$$

							\item Let $\vartwo \notin \uvarset$. Then it must be that $\vartwo \in \avarset$. By \reflemmap{Rewriting_evaluation_contexts}{one} (Focusing multiplicative evaluation contexts on applied variables), there exists $\pctxtwo_{\vartwo} \in \evctxAppp{\uvarsettwo_{\vartwo}}{\avarsettwo_{\vartwo}}$ such that $\pctxtwo_{\vartwo} \ctxholep{\vartwo} = \pctxp{(\la{\var}{\tmthree}) \tmfour}$, with $\uvarsettwo_{\vartwo} \subseteq \uvarset$ and $\vartwo \notin \avarsettwo_{\vartwo} \subset \avarset$. In addition, \refprop{exp-ctx-are-mult} (Exponential evaluation contexts are multiplicative) gives that $\pctxthree_{\vartwo} \in \evctxAppp{\uvarsetsix_{\vartwo}}{\avarsetsix_{\vartwo}}$, for some $\uvarsetsix_{\vartwo} \subseteq \uvarsettwo_{\vartwo}$ and $\avarsetsix_{\vartwo} \subseteq \avarsettwo_{\vartwo}$. We can then derive $\appES{\pctxtwo_{\vartwo} \ctxholep{\vartwo}}{\esub{\vartwo}{\hctxtwo}} \in \evctxUsep{\uvarsetsix_{\vartwo} \cup \unv{\hctxtwo}}{\avarsetsix_{\vartwo} \cup \anv{\hctxtwo}}$ such that 
								$$
								\appES{\progentry{\tm}{\env}}{\esub{\vartwo}{\tmtwo}} = (\appES{\pctxtwo_{\vartwo} \ctxholep{\vartwo}}{\esub{\vartwo}{\hctxtwo}}) \ctxholep{(\la{\varfour}{\tmsix}) \tmseven} \toum (\appES{\pctxtwo_{\vartwo} \ctxholep{\vartwo}}{\esub{\vartwo}{\hctxtwo}}) \ctxholeentryp{\tmsix}{\esub{\varfour}{\tmseven}}
								$$

						\end{itemize}

				\end{itemize}

			\item Let $\progentry{\tm}{\env} = \pctxp{\var} \toue \pctxp{\rename{\tmthree}}$, with $\pctx \in \evctxAppp{\uvarset}{\avarset}$. Note then $\var \neq \vartwo$---because $\var \in \domprogentry{\tm}{\env}$---and that if $\vartwo \notin (\uvarset \cup \avarset)$, then we can derive $\appES{\pctx}{\esub{\vartwo}{\tmtwo}} \in \evctxAppp{\uvarset}{\avarset}$ via rule $\ruleUseGcExp$ and get that 
				$$
				\progentry{\tm}{\env \esub{\vartwo}{\tmtwo}} = (\appES{\pctx}{\esub{\vartwo}{\tmtwo}}) \ctxholep{\var} \toum (\appES{\pctx}{\esub{\vartwo}{\tmtwo}}) \ctxholep{\rename{\tmthree}}
				$$
			Let us now consider the case where $\vartwo \in (\uvarset \cup \avarset)$ and proceed by case analysis on the shape of $\tmtwo$:
				\begin{itemize}
					\item Let $\tmtwo \in \Var$. We can then derive $\appES{\pctx}{\esub{\vartwo}{\tmtwo}} \in \evctxAppp{\updp{\uvarset}{\vartwo}{\tmtwo}}{\updp{\avarset}{\vartwo}{\tmtwo}}$ via rule $\ruleUseVarExp$ and get that 
						$$
						\progentry{\tm}{\env \esub{\vartwo}{\tmtwo}} = (\appES{\pctx}{\esub{\vartwo}{\tmtwo}}) \ctxholep{\var} \toum (\appES{\pctx}{\esub{\vartwo}{\tmtwo}}) \ctxholep{\rename{\tmthree}}
						$$

					\item Let $\tmtwo$ be a useful inert term. We can then derive $\appES{\pctx}{\esub{\vartwo}{\tmtwo}} \in \evctxAppp{(\uvarset \setminus \set{\vartwo}) \cup \unv{\tmtwo}}{(\avarset \setminus \set{\vartwo}) \cup \anv{\tmtwo}}$ via rule $\ruleUseNonVarExp$ and get that 
						$$
						\progentry{\tm}{\env \esub{\vartwo}{\tmtwo}} = (\appES{\pctx}{\esub{\vartwo}{\tmtwo}}) \ctxholep{\var} \toum (\appES{\pctx}{\esub{\vartwo}{\tmtwo}}) \ctxholep{\rename{\tmthree}}
						$$

					\item Let $\tmtwo$ be a value. If $\var \notin \avarset$, we can then derive $\pctx \defeq \appES{\pctx}{\esub{\vartwo}{\tmtwo}} \in \evctxAppp{\uvarset}{\avarset}$ via rule $\ruleUseGcExp$ and get that 
						$$
						\progentry{\tm}{\env \esub{\vartwo}{\tmtwo}} = (\appES{\pctx}{\esub{\vartwo}{\tmtwo}}) \ctxholep{\var} \toum (\appES{\pctx}{\esub{\vartwo}{\tmtwo}}) \ctxholep{\rename{\tmthree}}
						$$
					Let $\var \in \avarset$ instead. By \reflemmap{Rewriting_evaluation_contexts}{three} (Focusing exponential evaluation contexts on applied variables), there exists $\pctxtwo_{\vartwo} \in \evctxAppp{\uvarsettwo_{\vartwo}}{\avarsettwo_{\vartwo}}$ such that $\pctxtwo_{\vartwo} \ctxholep{\vartwo} = \pctxp{\var}$, with $\uvarsettwo_{\vartwo} \subseteq \uvarset$ and $\vartwo \notin \avarsettwo_{\vartwo} \subset \avarset$.

					Now, if $\vartwo \notin \uvarsettwo_{\vartwo}$, then we can derive $\appES{\pctxtwo_{\vartwo}}{\esub{\vartwo}{\tmtwo}} \in \evctxAppp{\uvarsettwo_{\vartwo}}{\avarsettwo_{\vartwo}}$ via rule $\ruleUseGcExp$ and get that 	
						$$
						\progentry{\tm}{\env \esub{\vartwo}{\tmtwo}} = (\appES{\pctxtwo_{\vartwo}}{\esub{\vartwo}{\tmtwo}}) \ctxholep{\vartwo} \toum (\appES{\pctxtwo_{\vartwo}}{\esub{\vartwo}{\tmtwo}}) \ctxholep{\rename{\tmtwo}}
						$$

					Finally, if $\vartwo \in \uvarsettwo_{\vartwo}$ instead, then we can derive $\appES{\pctxtwo_{\vartwo}}{\esub{\vartwo}{\tmtwo}} \in \evctxAppp{\uvarsettwo_{\vartwo} \setminus \set{\vartwo}}{\avarsettwo_{\vartwo}}$ via rule $\ruleUseUselessExp$ and get that
						$$
						\progentry{\tm}{\env \esub{\vartwo}{\tmtwo}} = (\appES{\pctxtwo_{\vartwo}}{\esub{\vartwo}{\tmtwo}}) \ctxholep{\vartwo} \toum (\appES{\pctxtwo_{\vartwo}}{\esub{\vartwo}{\tmtwo}}) \ctxholep{\rename{\tmtwo}}
						$$

					\item Let $\tmtwo$ be a non-normal term. By \reflemma{Redex_in_non-normal_terms-usefulopencbneed} (Redex in non-normal terms), there exists a term context $\hctx$ such that $\tmtwo = \hctxp{(\la{\varthree}{\tmthree}) \tmfour}$. We proceed by case analysis on whether $\vartwo \in \uvarset$:
						\begin{itemize}
							\item Let $\vartwo \in \uvarset$. By \reflemmap{Rewriting_evaluation_contexts}{two} (Focusing exponential evaluation contexts on unapplied variables), there exists $\pctxtwo_{\vartwo} \in \evctxUsep{\uvarsettwo_{\vartwo}}{\avarsettwo_{\vartwo}}$ such that $\pctxtwo_{\vartwo} \ctxholep{\vartwo} = \pctxp{\var}$, with $\vartwo \notin \uvarsettwo_{\vartwo} \subset \uvarset$ and $\avarsettwo_{\tmtwo} \subseteq \avarset$. 

							Now, if $\vartwo \notin \avarsettwo_{\vartwo}$, then we can derive $\appES{\pctxtwo_{\vartwo} \ctxholep{\vartwo}}{\esub{\vartwo}{\hctx}} \in \evctxAppp{\pctxtwo_{\vartwo} \ctxholep{\vartwo}}{\esub{\vartwo}{\hctx}}$ via rule $\ruleUseHerMult$ and get that 
								$$
								\progentry{\tm}{\env \esub{\vartwo}{\tmtwo}} = (\appES{\pctxtwo_{\vartwo} \ctxholep{\vartwo}}{\esub{\vartwo}{\hctx}}) \ctxholep{(\la{\varthree}{\tmthree}) \tmfour} \toum (\appES{\pctxtwo_{\vartwo} \ctxholep{\vartwo}}{\esub{\vartwo}{\hctx}}) \ctxholeentryp{\tmthree}{\esub{\varthree}{\tmfour}}
								$$

							Let $\vartwo \in \avarsettwo_{\vartwo}$ instead. By \reflemmap{Rewriting_evaluation_contexts}{one} (Focusing multiplicative evaluation contexts on applied variables), there exists $\pctxthree_{\vartwo} \in \evctxAppp{\uvarsetthree_{\vartwo}}{\avarsetthree_{\vartwo}}$ such that $\pctxthree_{\vartwo} \ctxholep{\vartwo} = \pctxtwo_{\vartwo} \ctxholep{\vartwo}$, with $\uvarsetthree_{\vartwo} \subseteq \uvarsettwo_{\vartwo}$ and $\vartwo \notin \avarsetthree_{\vartwo} \subset \avarsettwo_{\vartwo}$. In addition, \refprop{exp-ctx-are-mult} (Exponential evaluation contexts are multiplicative) gives that $\pctxthree_{\vartwo} \in \evctxUsep{\uvarsetseven_{\vartwo}}{\avarsetseven_{\vartwo}}$, for some $\uvarsetseven_{\vartwo} \subseteq \uvarsetthree_{\vartwo}$ and $\avarsetseven_{\vartwo} \subseteq \avarsetthree_{\vartwo}$. We can then derive $\appES{\pctxthree_{\vartwo} \ctxholep{\vartwo}}{\esub{\vartwo}{\hctx}} \in \evctxUsep{\uvarsetseven_{\vartwo} \cup \unv{\hctx}}{\avarsetseven_{\vartwo} \cup \anv{\hctx}}$ via rule $\ruleUseHerMult$ and get that 
								$$
								\progentry{\tm}{\env \esub{\vartwo}{\tmtwo}} = (\appES{\pctxtwo_{\vartwo} \ctxholep{\vartwo}}{\esub{\vartwo}{\hctx}}) \ctxholep{(\la{\varthree}{\tmthree}) \tmfour} \toum (\appES{\pctxtwo_{\vartwo} \ctxholep{\vartwo}}{\esub{\vartwo}{\hctx}}) \ctxholeentryp{\tmthree}{\esub{\varthree}{\tmfour}}
								$$

							\item Let $\vartwo \notin \uvarset$. Then it must be that $\vartwo \in \avarset$. By \reflemmap{Rewriting_evaluation_contexts}{three} (Focusing exponential evaluatioun contexts on applied variables), there exists $\pctxtwo_{\vartwo} \in \evctxAppp{\uvarsettwo_{\vartwo}}{\avarsettwo_{\vartwo}}$ such that $\pctxtwo_{\vartwo} \ctxholep{\vartwo} = \pctxp{\var}$, with $\uvarsettwo_{\vartwo} \subseteq \uvarset$ and $\vartwo \notin \avarsettwo_{\vartwo} \subset \avarset$. In addition, \refprop{exp-ctx-are-mult} (Exponential evaluation contexts are multiplicative) gives that $\pctxtwo_{\vartwo} \in \evctxUsep{\uvarsetsix_{\vartwo}}{\avarsetsix_{\vartwo}}$, for some $\uvarsetsix_{\vartwo} \subseteq \uvarsettwo_{\vartwo}$ and $\avarsetsix_{\vartwo} \subseteq \avarsettwo_{\vartwo}$.

							Now, if $\vartwo \notin \uvarsetsix_{\vartwo}$ then we can derive $\appES{\pctxtwo_{\vartwo} \ctxholep{\vartwo}}{\esub{\vartwo}{\hctx}} \in \evctxUsep{\uvarsetsix_{\vartwo} \cup \unv{\hctx}}{\avarsetsix_{\vartwo} \cup \anv{\hctx}}$ via rule $\ruleUseHerMult$ and get that 
								$$
								\progentry{\tm}{\env \esub{\vartwo}{\tmtwo}} = (\appES{\pctxtwo_{\vartwo} \ctxholep{\vartwo}}{\esub{\vartwo}{\hctx}}) \ctxholep{(\la{\varthree}{\tmthree}) \tmfour} \toum (\appES{\pctxtwo_{\vartwo} \ctxholep{\vartwo}}{\esub{\vartwo}{\hctx}}) \ctxholeentryp{\tmthree}{\esub{\varthree}{\tmfour}}
								$$

							Finally, let $\vartwo \in \uvarsetsix_{\vartwo}$ instead. By \reflemmap{Rewriting_evaluation_contexts-base_cases}{three} (Focusing multiplicative evaluation contexts on unapplied variables), there exists $\pctxthree_{\vartwo} \in \evctxUsep{\uvarsetthree_{\vartwo}}{\avarsetthree_{\vartwo}}$ such that $\pctxthree_{\vartwo} \ctxholep{\vartwo} = \pctxtwo_{\vartwo} \ctxholep{\vartwo}$, with $\vartwo \notin \uvarsetthree_{\vartwo} \subset \uvarsetsix_{\vartwo}$ and $\avarsetthree_{\vartwo} \subseteq \avarsetsix_{\vartwo}$. In addition, \refprop{exp-ctx-are-mult} (Exponential evaluation contexts are multiplicative) gives that $\pctxtwo_{\vartwo} \in \evctxUsep{\uvarsetseven_{\vartwo}}{\avarsetseven_{\vartwo}}$, for some $\uvarsetseven_{\vartwo} \subseteq \uvarsetthree_{\vartwo}$ and $\avarsetseven_{\vartwo} \subseteq \avarsetthree_{\vartwo}$. We can derive $\appES{\pctxtwo_{\vartwo} \ctxholep{\vartwo}}{\esub{\vartwo}{\hctx}} \in \evctxUsep{\uvarsetseven_{\vartwo} \cup \unv{\hctx}}{\avarsetseven_{\vartwo} \cup \anv{\hctx}}$ via rule $\ruleUseHerMult$ and get that 
								$$
								\progentry{\tm}{\env \esub{\vartwo}{\tmtwo}} = (\appES{\pctxtwo_{\vartwo} \ctxholep{\vartwo}}{\esub{\vartwo}{\hctx}}) \ctxholep{(\la{\varthree}{\tmthree}) \tmfour} \toum (\appES{\pctxtwo_{\vartwo} \ctxholep{\vartwo}}{\esub{\vartwo}{\hctx}}) \ctxholeentryp{\tmthree}{\esub{\varthree}{\tmfour}}
								$$

						\end{itemize}

				\end{itemize}

		\end{itemize}

	\item
	\emph{Appending ESs that do not create $\tound$-redexes}: We prove the contrapositive statement:
		\begin{center}
			 Let $\tmtwo$ be such that if $\vartwo \in \nventryp{\tm}{\env}$ then $\tmtwo$ is a normal term and if $\vartwo \in \anventryp{\tm}{\env}$ then $\tmtwo$ is an inert term. Moreover, let $\progentry{\tm}{\env \esub{\vartwo}{\tmtwo}}$ not be in $\tound$-normal form
			
			then
			
			$\progentry{\tm}{\env}$ is not in $\tound$-normal form
		\end{center} 

	Case analysis on the kind of redex in $\progentry{\tm}{\env \esub{\vartwo}{\tmtwo}}$:
		\begin{itemize}
			\item Let $\progentry{\tm}{\env \esub{\vartwo}{\tmtwo}} = \pctxp{(\la{\var}{\tmthree}) \tmfour}$, with $\pctx \in \evctxUsep{\uvarset}{\avarset}$. We proceed by case analysis on the last derivation rule in $\pctx \in \evctxUsep{\uvarset}{\avarset}$:
				\begin{itemize}
					\item \emph{Rule $\ruleUseAxMult$}: The case where $\pctx = \progentry{\hctx}{\emptyenv}$ is impossible.

					\item \emph{Rule $\ruleUseVarMult$}: Let
						$$
						\infer
							[\ruleUseVarMult]
							{\appES{\pctxtwo}{\esub{\vartwo}{\tmtwo}} \in \evctxUsep{\updp{\uvarsettwo}{\vartwo}{\tmtwo}}{\updp{\avarsettwo}{\vartwo}{\tmtwo}}}
							{\pctxtwo \in \evctxUsep{\uvarsettwo}{\avarsettwo}
							\quad
							\vartwo \in (\uvarsettwo \cup \avarsettwo)}
						$$
					where $\pctx = \appES{\pctxtwo}{\esub{\vartwo}{\tmtwo}}$, $\tmtwo \in \Var$, $\uvarset = \updp{\uvarsettwo}{\vartwo}{\tmtwo}$ and $\avarset = \updp{\avarsettwo}{\vartwo}{\tmtwo}$. Then $\progentry{\tm}{\env} = \pctxtwop{(\la{\var}{\tmthree}) \tmfour} \toum \pctxtwoentryp{\tmthree}{\esub{\var}{\tmfour}}$.

					\item \emph{Rule $\ruleUseNonVarMult$}: Let
						$$
						\infer
							[\ruleUseNonVarMult]
							{\appES{\pctxtwo}{\esub{\vartwo}{\tmtwo}} \in \evctxUsep{(\uvarsettwo \setminus \set{\vartwo}) \cup \unv{\tmtwo}}{(\avarsettwo \setminus \set{\vartwo}) \cup \anv{\tmtwo}}}
							{\pctxtwo \in \evctxUsep{\uvarsettwo}{\avarsettwo}
							\quad
							\vartwo \in (\uvarsettwo \cup \avarsettwo)}
						$$
					where $\pctx = \appES{\pctxtwo}{\esub{\vartwo}{\tmtwo}}$, $\tmtwo$ is a useful inert term, $\uvarset = (\uvarsettwo \setminus \set{\vartwo}) \cup \unv{\tmtwo}$ and $\avarset = (\avarsettwo \setminus \set{\vartwo}) \cup \anv{\tmtwo}$. Then $\progentry{\tm}{\env} = \pctxtwop{(\la{\var}{\tmthree}) \tmfour} \toum \pctxtwoentryp{\tmthree}{\esub{\var}{\tmfour}}$.

					\item \emph{Rule $\ruleUseGcMult$}: Let
						$$
						\infer
							[\ruleUseGcMult]
							{\appES{\pctxtwo}{\esub{\vartwo}{\tmtwo}} \in \evctxUsep{\uvarset}{\avarset}}
							{\pctxtwo \in \evctxUsep{\uvarset}{\avarset}
							\quad
							\vartwo \notin (\uvarset \cup \avarset)}
						$$
					where $\pctx = \appES{\pctxtwo}{\esub{\vartwo}{\tmtwo}}$. Then $\progentry{\tm}{\env} = \pctxtwop{(\la{\var}{\tmthree}) \tmfour} \toum \pctxentryp{\tmthree}{\esub{\var}{\tmfour}}$.

					\item\emph{Rule $\ruleUseUselessMult$}: Let
						$$
						\infer
							[\ruleUseUselessMult]
							{\appES{\pctxtwo}{\esub{\vartwo}{\tmtwo}} \in \evctxUsep{\uvarsettwo \setminus \set{\vartwo}}{\avarsettwo}}
							{\pctxtwo \in \evctxUsep{\uvarsettwo}{\avarsettwo}
							\quad
							\vartwo \in (\uvarsettwo \setminus \avarsettwo)}
						$$
					where $\pctx = \appES{\pctxtwo}{\esub{\vartwo}{\tmtwo}}$, $\tmtwo$ is a value, $\uvarset = \uvarsettwo \setminus \set{\vartwo}$ and $\avarset = \avarsettwo$. Then $\progentry{\tm}{\env} = \pctxtwop{(\la{\var}{\tmthree}) \tmfour} \\ \toum \pctxtwoentryp{\tmthree}{\esub{\var}{\tmfour}}$.

					\item\emph{Rule $\ruleUseHerMult$}: Suppose
						$$
						\infer
							[\ruleUseHerMult]
							{\appES{\pctxtwop{\vartwo}}{\esub{\vartwo}{\hctx}} \in \evctxUsep{\uvarsettwo \cup \unv{\hctx}}{\avarsettwo \cup \anv{\hctx}}}
							{\pctxtwo \in \evctxUsep{\uvarsettwo}{\avarsettwo}
							\quad
							\vartwo \notin (\uvarsettwo \cup \avarsettwo)}
						$$
					with $\pctx = \appES{\pctxtwop{\vartwo}}{\esub{\vartwo}{\hctx}}$, $\tmtwo = \hctxp{(\la{\var}{\tmthree}) \tmfour}$, $\uvarset = \uvarsettwo \cup \unv{\hctx}$ and $\avarset = \avarsettwo \cup \anv{\hctx}$. Since $\tmtwo$ is not normal---by \reflemma{Redex_in_non-normal_terms-usefulopencbneed} (Redex in non-normal terms)---then it would have to be that $\vartwo \notin \nventryp{\tm}{\env} = \nv{\pctxtwop{\vartwo}}$, by hypothesis. But \reflemmap{Rewriting_evaluation_contexts-base_cases}{one} (Multiplicative evaluation contexts give needed variables) gives that $\vartwo \in \nv{\pctxtwop{\vartwo}}$. Hence, this case is impossible.

				\end{itemize}

			\item Let $\progentry{\tm}{\env} = \pctxp{\var}$, with $\pctx \in \evctxAppp{\uvarset}{\avarset}$. We proceed by case analysis on the last derivation rule in $\pctx \in \evctxAppp{\uvarset}{\avarset}$:
				\begin{itemize}
					\item \emph{Rule $\ruleUseAxOneExp$}: The case where $\pctx = \progentry{\hctxapp}{\emptyenv}$ is impossible.

					\item \emph{Rule $\ruleUseAxTwoExp$}: The case where $\pctx = \appES{\pctxtwop{\vartwo}}{\esub{\vartwo}{\hctxapp}}$, for some applicative term context $\hctxapp$, is impossible, as it would then be that $\var \notin \dom{\pctx}$; absurd.

					\item \emph{Rule $\ruleUseVarExp$}: Let
						$$
						\infer
							[\ruleUseVarExp]
							{\appES{\pctxtwo}{\esub{\vartwo}{\tmtwo}} \in \evctxAppp{\updp{\uvarsettwo}{\vartwo}{\tmtwo}}{\updp{\avarsettwo}{\vartwo}{\tmtwo}}}
							{\pctxtwo \in \evctxAppp{\uvarsettwo}{\avarsettwo}
							\quad
							\vartwo \in (\uvarsettwo \cup \avarsettwo)}
						$$
					where $\pctx = \appES{\pctxtwo}{\esub{\vartwo}{\tmtwo}}$, $\tmtwo \in \Var$, $\uvarset = \updp{\uvarsettwo}{\vartwo}{\tmtwo}$ and $\avarset = \updp{\avarsettwo}{\vartwo}{\tmtwo}$. Note that $\var \neq \vartwo$, because $\tmtwo$ is not a value. Hence, $\var \in \dom{\pctxtwo}$, and so we have that $\progentry{\tm}{\env} = \pctxtwop{\var} \toue \pctxtwop{\rename{\tmtwo}}$.

					\item \emph{Rule $\ruleUseNonVarExp$}: Let
						$$
						\infer
							[\ruleUseNonVarExp]
							{\appES{\pctxtwo}{\esub{\vartwo}{\tmtwo}} \in \evctxAppp{(\uvarsettwo \setminus \set{\vartwo}) \cup \unv{\tmtwo}}{(\avarsettwo \setminus \set{\vartwo}) \cup \anv{\tmtwo}}}
							{\pctxtwo \in \evctxAppp{\uvarsettwo}{\avarsettwo}
							\quad
							\vartwo \in (\uvarsettwo \cup \avarsettwo)}
						$$
					where $\pctx = \appES{\pctxtwo}{\esub{\vartwo}{\tmtwo}}$, $\tmtwo$ is a useful inert term, $\uvarset = (\uvarsettwo \setminus \set{\vartwo}) \cup \unv{\tmtwo}$ and $\avarset = (\avarsettwo \setminus \set{\vartwo}) \cup \anv{\tmtwo}$. Note that $\var \neq \vartwo$, because $\tmtwo $ is not a value. Hence, $\var \in \dom{\pctxtwo}$, and so we have that $\progentry{\tm}{\env} = \pctxtwop{\var} \toue \pctxtwop{\rename{\tmtwo}}$.

					\item \emph{Rule $\ruleUseGcExp$}: Let
						$$
						\infer
							[\ruleUseGcExp]
							{\appES{\pctxtwo}{\esub{\vartwo}{\tmtwo}} \in \evctxAppp{\uvarset}{\avarset}}
							{\pctxtwo \in \evctxAppp{\uvarset}{\avarset}
							\quad
							\vartwo \notin (\uvarset \cup \avarset)}
						$$
					where $\pctx = \appES{\pctxtwo}{\esub{\vartwo}{\tmtwo}}$. Note that if $\var \neq \vartwo$, then $\var \in \dom{\pctxtwo}$, giving that $\progentry{\tm}{\env} = \pctxtwop{\var} \toue \pctxtwop{\rename{\tmtwo}}$. Suppose $\var = \vartwo$ instead. Then it would have to be that $\tmtwo$ is a value and that $\progentry{\tm}{\env} = \pctxtwop{\vartwo}$---by definition of the \usefulopencbneed evaluation strategy. But then \reflemmap{Rewriting_evaluation_contexts-base_cases}{two} (Exponential evaluation contexts give applied variables) gives that $\vartwo \in \anv{\pctxtwop{\vartwo}}$ and so it should be that $\tmtwo$ is an inert term, by hypothesis; absurd.

					\item\emph{Rule $\ruleUseUselessExp$}: Let
						$$
						\infer
							[\ruleUseUselessExp]
							{\appES{\pctxtwo}{\esub{\vartwo}{\tmtwo}} \in \evctxAppp{\uvarsettwo \setminus \set{\vartwo}}{\avarsettwo}}
							{\pctxtwo \in \evctxAppp{\uvarsettwo}{\avarsettwo}
							\quad
							\vartwo \in (\uvarsettwo \setminus \avarsettwo)}
						$$
					where $\pctx = \appES{\pctxtwo}{\esub{\vartwo}{\tmtwo}}$, $\tmtwo $ is a value, $\uvarset = \uvarsettwo \setminus \set{\vartwo}$ and $\avarset = \avarsettwo$. Note that if $\var \neq \vartwo$, then $\var \in \dom{\pctxtwo}$, giving that $\progentry{\tm}{\env} = \pctxtwop{\var} \toue \pctxtwop{\rename{\tmtwo}}$. Suppose $\var = \vartwo$ instead. Then it would have to be that $\tmtwo $ is a value and that $\progentry{\tm}{\env} = \pctxtwop{\vartwo}$. But then \reflemmap{Rewriting_evaluation_contexts-base_cases}{two} (Exponential evaluation contexts give applied variables) gives that $\vartwo \in \anv{\pctxtwop{\vartwo}}$ and so it should be that $\tmtwo$ is an inert term, by hypothesis; absurd.

					\item \emph{Rule $\ruleUseHerExp$}: The case where $\pctx = \appES{\pctxtwop{\vartwo}}{\esub{\vartwo}{\ctxhole}}$ is impossible: it would then be that $\var \notin \dom{\pctx}$; absurd.\qedhere

				\end{itemize}

		\end{itemize}

\end{enumerate}

\end{proof}

\gettoappendix{prop:Syntactic_characterization_of_usefulopencbneed-normal_forms}

\begin{proof}
\begin{itemize}
	\item[$\Rightarrow$]: Let $\prog = \progentry{\tm}{\env}$ be in $\tound$-normal form. By induction on the length of $\env$:
		\begin{itemize}
			\item Let $\env = \emptyenv$. Note that if $\tm \in \Var$ or $\tm $ is a value then $\ufnormalentrypr{\tm}{\emptyenv}$. 
			
			Let $\tm = \tm_{1} \tm_{2}$. Case analysis on the shape of $\tm_{1}$:
				\begin{itemize}
					\item Suppose that $\tm_{1}$ is a value, say $\tm_{1} = \la{\var}{\tmtwo}$. But then we would get that
						$$
						\prog = \progentry{\ctxhole}{\emptyenv} \ctxholep{(\la{\var}{\tmtwo}) \tm_{2}} \toum \progentry{\ctxhole}{\emptyenv} \ctxholeentryp{\tmtwo}{\esub{\var}{\tm_{2}}}
						$$
					which is absurd.

					\item Let $\tm_{1}$ be an inert term. Case analysis on whether $\tm_{2}$ is a normal term:
						\begin{itemize}
							\item Suppose $\tm_{2}$ is not a normal term. By \reflemma{Redex_in_non-normal_terms-usefulopencbneed} (Redex in non-normal terms), there exists a term context $\hctx$ such that $\tm_{2} = \hctxp{(\la{\vartwo}{\tmthree}) \tmfour}$. But then $\progentry{\tm_{1} \hctx}{\emptyenv} \in \evctxUsep{\unv{\tm_{1} \hctx}}{\anv{\tm_{1} \hctx}}$, and so 
								$$
								\prog = \progentry{\tm_{1} \hctx_{1}}{\emptyenv} \ctxholep{(\la{\vartwo}{\tmthree}) \tmfour} \toum \progentry{\tm_{1} \hctx_{1}}{\emptyenv} \ctxholeentryp{\tmthree}{\esub{\vartwo}{\tmfour}}
								$$
							which is absurd.

							\item Let $\tm_{2}$ be a normal term. Then $\tm_{1} \tm_{2}$ is a useful inert term, giving that
								$$
								\infer
									[\ufRuleNormPrograms]
									{\ufnormalentrypr{\tm_{1} \tm_{2}}{\emptyenv}}
									{\infer
										[\ufRuleInertAx]
										{\ufinertentrypr{\tm_{1} \tm_{2}}{\emptyenv}}
										{}}
								$$

						\end{itemize}

					\item Suppose $\tm_{1}$ is not a normal term. By \reflemma{Redex_in_non-normal_terms-usefulopencbneed} (Redex in non-normal terms), there exists a term context $\hctx$ such that $\tm_{1} = \hctxp{(\la{\vartwo}{\tmthree}) \tmfour}$. But then $\progentry{\tm_{1} \hctx}{\emptyenv} \in \evctxUsep{\unv{\tm_{1} \hctx}}{\anv{\tm_{1} \hctx}}$, and so 
						$$
						\prog = \progentry{\hctx}{\emptyenv} \ctxholep{(\la{\vartwo}{\tmthree}) \tmfour} \toum \progentry{\hctx}{\emptyenv} \ctxholeentryp{\tmthree}{\esub{\vartwo}{\tmfour}}
						$$
					which is absurd.

				\end{itemize}
			
			\item Let $\env = \envtwo \esub{\vartwo}{\tmtwo}$---\ie, $\progentry{\tm}{\env} = \progentry{\tm}{\envtwo \esub{\vartwo}{\tmtwo}}$. By \reflemmap{Properties_of_usefulopencbneed-normal_forms_and_ESs}{one} (Removing ESs does not create $\tound$-redexes), $\progentry{\tm}{\envtwo}$ is in $\tound$-normal form. By \ih, $\ufnormalentrypr{\tm}{\envtwo}$. We proceed by case analysis on the predicate derivation of $\ufnormalentrypr{\tm}{\env}$:
				\begin{itemize}
					\item Let $\genVarentrypr{\var}{\tm}{\envtwo}$. Case analysis on the last whether $\vartwo = \var$:
						\begin{itemize}
							\item Let $\vartwo \neq \var$. We can then derive 
								$$
								\infer
									[\ufRuleNormPrograms]
									{\ufnormalentrypr{\tm}{\envtwo \esub{\vartwo}{\tmtwo}}}
									{\infer
										[\ufRuleGenVarGc]
										{\genVarentrypr{\var}{\tm}{\envtwo \esub{\vartwo}{\tmtwo}}}
										{\genVarentrypr{\var}{\tm}{\envtwo}
										\quad
										\vartwo \neq \var}}
								$$

							\item Let $\vartwo = \var$. 

							Now, if $\tmtwo \in \Var$, then we can derive 
								$$
								\infer
									[\ufRuleNormPrograms]
									{\ufnormalentrypr{\tmtwo}{\envtwo \esub{\vartwo}{\tmtwo}}}
									{\infer
										[\ufRuleGenVarHer]
										{\genVarentrypr{\tmtwo}{\tm}{\envtwo \esub{\vartwo}{\tmtwo}}}
										{\genVarentrypr{\var}{\tm}{\envtwo}}}
								$$
							If $\tmtwo$ is a useful inert term instead, then we can derive
								$$
								\infer
									[\ufRuleNormPrograms]
									{\ufnormalentrypr{\tmtwo}{\envtwo \esub{\vartwo}{\tmtwo}}}
									{\infer
										[\ufRuleInertGenVar]
										{\ufinertentrypr{\tmtwo}{\tm}{\envtwo \esub{\vartwo}{\tmtwo}}}
										{\genVarentrypr{\var}{\tm}{\envtwo}}}
								$$

							Next, if $\tmtwo $ is a value, then we can derive
								$$
								\infer
									[\ufRuleNormPrograms]
									{\ufnormalentrypr{\tmtwo}{\envtwo \esub{\vartwo}{\tmtwo}}}
									{\infer
										[\ufRuleAbsGenVar]
										{\ufinertentrypr{\tmtwo}{\tm}{\envtwo \esub{\vartwo}{\tmtwo}}}
										{\genVarentrypr{\var}{\tm}{\envtwo}}}
								$$

							Finally, suppose $\tmtwo$ is not a normal term. By \reflemma{Redex_in_non-normal_terms-usefulopencbneed} (Redex in non-normal terms), there would exist a term context $\hctx$  such that $\tmtwo = \hctxp{(\la{\vartwo}{\tmthree}) \tmfour}$. In addition, note that $\var \in \nventryp{\tm}{\envtwo}$---by \reflemma{Properties_of_generalized_variables} (Properties of generalized variables). Thus, $\var \in (\unventryp{\tm}{\envtwo} \cup \anventryp{\tm}{\envtwo})$---by \reflemmap{Unapplied_applied_and_needed_variables}{two} (Unapplied, applied and needed variables). Case analysis on whether $\var \in \unventryp{\tm}{\envtwo}$:

							If $\var \in \unventryp{\tm}{\envtwo}$, then there would exist $\pctx \in \evctxUsep{\uvarset}{\avarset}$ such that $\pctxp{\var} = \progentry{\tm}{\envtwo}$, with $\var \notin \uvarset \subset \unventryp{\tm}{\envtwo}$ and $\avarset \subseteq \anventryp{\tm}{\envtwo}$. However, if $\var \notin \avarset$, then we would be able to derive $\appES{\pctxp{\var}}{\esub{\var}{\hctx}} \in \evctxUsep{\uvarset \cup \unv{\hctx}}{\avarset \cup \anv{\hctx}}$ via rule $\ruleUseHerMult$ and get that
								$$
								\progentry{\tm}{\envtwo \esub{\vartwo}{\tmtwo}} = (\appES{\pctxp{\var}}{\esub{\var}{\hctx}}) \ctxholep{(\la{\vartwo}{\tmthree}) \tmfour} \toum (\appES{\pctxp{\var}}{\esub{\var}{\hctx}}) \ctxholeentryp{\tmthree}{\esub{\vartwo}{\tmfour}}
								$$
							which is absurd. But if $\var \in \avarset$ instead, then there would exist---by \reflemmap{Rewriting_evaluation_contexts}{one} (Focusing multiplicative evaluation contexts on applied variables)---$\pctxtwo_{\var} \in \evctxAppp{\uvarsettwo_{\var}}{\avarsettwo_{\var}}$ such that $\pctxtwo_{\var} \ctxholep{\var} = \pctxp{\var}$, with $\uvarsettwo_{\var} \subseteq \uvarset$ and $\var \notin \avarsettwo_{\var} \subset \avarset$. In addition, $\pctxtwo_{\var} \in \evctxUsep{\uvarsetsix_{\var}}{\avarsetsix_{\var}}$ for some $\uvarsetsix_{\var} \subseteq \uvarsettwo_{\var}$ and $\avarsetsix_{\var} \subseteq \avarsettwo_{\var}$. However, we would then be able to derive $\appES{\pctxtwo_{\var} \ctxholep{\var}}{\esub{\var}{\hctx}} \in \evctxUsep{\uvarsetsix_{\var} \cup \unv{\hctx}}{\avarsetsix_{\var} \cup \anv{\hctx}}$ via rule $\ruleUseHerMult$ and get that 
								$$
								\progentry{\tm}{\envtwo \esub{\vartwo}{\tmtwo}} = (\appES{\pctxp{\var}}{\esub{\var}{\hctx}}) \ctxholep{(\la{\vartwo}{\tmthree}) \tmfour} \toum (\appES{\pctxp{\var}}{\esub{\var}{\hctx}}) \ctxholeentryp{\tmthree}{\esub{\vartwo}{\tmfour}}
								$$
							which is also absurd. 

							Therefore, it would have to be that $\var \notin \unventryp{\tm}{\envtwo}$ instead, forcing that $\var \notin \anventryp{\tm}{\envtwo}$. Then, by \reflemmap{Focusing_for_usefulopencbneed-normal_forms}{one} (Focusing \usefulopencbneednsp-normal forms on applied variables), there would exist $\pctx \in \evctxAppp{\uvarset}{\avarset}$ such that $\pctxp{\var} = \progentry{\tm}{\envtwo}$, with $\uvarset \subseteq \unventryp{\tm}{\envtwo}$ and $\var \notin \avarset \subset \anventryp{\tm}{\envtwo}$. In addition, $\pctx \in \evctxUsep{\uvarsetfive}{\avarsetfive}$, for some $\uvarsetfive \subseteq \uvarset$ and $\avarsetfive \subseteq \avarset$. However, we would then be able to derive $\appES{\pctxp{\var}}{\esub{\var}{\hctx}} \in \evctxUsep{\uvarsetfive \cup \unv{\hctx}}{\avarsetfive \cup \anv{\hctx}}$ via rule $\ruleUseHerMult$ and get that 
								$$
								\progentry{\tm}{\envtwo \esub{\vartwo}{\tmtwo}} = (\appES{\pctxp{\var}}{\esub{\var}{\hctx}}) \ctxholep{(\la{\vartwo}{\tmthree}) \tmfour} \toum (\appES{\pctxp{\var}}{\esub{\var}{\hctx}}) \ctxholeentryp{\tmthree}{\esub{\vartwo}{\tmfour}}
								$$
							which is also absurd. 

							Therefore, the case where $\tmtwo$ is not a normal term is impossible.

						\end{itemize}

					\item Let $\ufabsentrypr{\tm}{\envtwo}$. Note that $\nventryp{\tm}{\envtwo} = \emptyset$---by \reflemma{Properties_of_useful_abstraction_programs} (Properties of useful abstraction programs)---and so we can derive $\ufnormalentrypr{\tm}{\envtwo \esub{\vartwo}{\tmtwo}}$ as follows:
						$$
						\infer
							[\ufRuleNormPrograms]
							{\ufnormalentrypr{\tm}{\envtwo \esub{\vartwo}{\tmtwo}}}
							{\infer
								[\ufRuleAbsGc]
								{\ufabsentrypr{\tm}{\envtwo \esub{\vartwo}{\tmtwo}}}
								{\ufabsentrypr{\tm}{\envtwo}}
							}
						$$
					
					\item Let $\ufinertentrypr{\tm}{\envtwo}$. If $\vartwo \notin \nventryp{\tm}{\envtwo}$  then we can derive $\ufnormalentrypr{\tm}{\envtwo \esub{\vartwo}{\tmtwo}}$ as follows:
						$$
						\infer
							[\ufRuleNormPrograms]
							{\ufnormalentrypr{\tm}{\envtwo \esub{\vartwo}{\tmtwo}}}
							{\infer
								[\ufRuleInertGc]
								{\ufinertentrypr{\tm}{\envtwo \esub{\vartwo}{\tmtwo}}}
								{\ufinertentrypr{\tm}{\envtwo}
								\quad
								\vartwo \notin \nventryp{\tm}{\envtwo}}
							}	
						$$
						Let $\vartwo \in \nventryp{\tm}{\envtwo} = (\unventryp{\tm}{\envtwo} \cup \anventryp{\tm}{\envtwo})$---the last equality given by \reflemmap{Unapplied_applied_and_needed_variables}{two} (Unapplied, applied and needed variables). We proceed by case analysis on the shape of $\tmtwo$:
							\begin{itemize}
								\item First, note that if $\tmtwo$ is an inert term then we can simply derive $\ufnormalentrypr{\tm}{\envtwo \esub{\vartwo}{\tmtwo}}$ as follows:
									$$
									\infer
										[\ufRuleNormPrograms]
										{\ufnormalentrypr{\tm}{\envtwo \esub{\vartwo}{\tmtwo}}}
										{\infer
											[\ufRuleInertInert]
											{\ufinertentrypr{\tm}{\envtwo \esub{\vartwo}{\tmtwo}}}
											{\ufinertentrypr{\tm}{\envtwo}
											\quad
											\vartwo \in \nventryp{\tm}{\envtwo}}
										}	
									$$
								
								\item Let $\tmtwo$ be a value. On the one hand, note that if $\vartwo \notin \anventryp{\tm}{\envtwo}$ then it must be that $\vartwo \in \unventryp{\tm}{\envtwo}$ and so we can derive $\ufnormalentrypr{\tm}{\envtwo \esub{\vartwo}{\tmtwo}}$ as follows:
									$$
									\infer
										[\ufRuleNormPrograms]
										{\ufnormalentrypr{\tm}{\envtwo \esub{\vartwo}{\tmtwo}}}
										{\infer
											[\ufRuleInertUseless]
											{\ufinertentrypr{\tm}{\envtwo \esub{\vartwo}{\tmtwo}}}
											{\ufinertentrypr{\tm}{\envtwo}
											\quad
											\vartwo \in \uselessentrypr{\tm}{\envtwo}}
										}	
									$$

								On the other hand, let us suppose that $\vartwo \in \anventryp{\tm}{\envtwo}$. Then there would exist---by \reflemmap{Focusing_for_usefulopencbneed-normal_forms}{two} (Focusing \usefulopencbneednsp-normal forms on applied variables)---$\pctx \in \evctxAppp{\uvarset}{\avarset}$ such that $\pctxp{\vartwo} = \progentry{\tm}{\envtwo}$, with $\uvarset \subseteq \unventryp{\tm}{\envtwo}$ and $\vartwo \notin \avarset \subset \anventryp{\tm}{\envtwo}$. However, this leads to a contradiction. To see why, note that if $\vartwo \notin \uvarset$ then would be able to derive
									$$
									\infer
										[\ruleUseGcExp]
										{\appES{\pctx}{\esub{\vartwo}{\tmtwo}} \in \evctxAppp{\uvarset}{\avarset}}
										{\pctx \in \evctxAppp{\uvarset}{\avarset}
										\quad
										\vartwo \notin (\uvarset \cup \avarset)}
									$$
								and if $\vartwo \in \uvarset$ then we would be able to derive
									$$
									\infer
										[\ruleUseGcExp]
										{\appES{\pctx}{\esub{\vartwo}{\tmtwo}} \in \evctxAppp{\uvarset \setminus \set{\vartwo}}{\avarset}}
										{\pctx \in \evctxAppp{\uvarset}{\avarset}
										\quad
										\vartwo \in (\uvarset \setminus \avarset)}
									$$
								having in both cases that 
									$$
									\progentry{\tm}{\envtwo} = (\appES{\pctx}{\esub{\vartwo}{\tmtwo}}) \ctxholep{\vartwo} \toue (\appES{\pctx}{\esub{\vartwo}{\tmtwo}}) \ctxholep{\rename{\tmtwo}}
									$$
								which is absurd. Therefore, the case where $\vartwo \in \anventryp{\tm}{\envtwo}$ is impossible.
								
								\item Finally, suppose $\tmtwo$ is not a normal term. By \reflemma{Redex_in_non-normal_terms-usefulopencbneed} (Redex in non-normal terms), there would exist a term context $\hctx$  such that $\tmtwo = \hctxp{(\la{\var}{\tmthree})\tmfour}$. Moreover, note that $\vartwo \in \nventryp{\tm}{\envtwo} = (\unventryp{\tm}{\envtwo} \cup \anventryp{\tm}{\envtwo})$---the equality given by \reflemmap{Unapplied_applied_and_needed_variables}{two} (Unapplied, applied and needed variables). Case analysis on whether $\vartwo \in \unventryp{\tm}{\envtwo}$:

								Let $\vartwo \in \unventryp{\tm}{\envtwo}$. By \reflemmap{Focusing_for_usefulopencbneed-normal_forms}{one} (Focusing \usefulopencbneednsp-normal forms on unapplied variables), there would exist $\pctx \in \evctxUsep{\uvarset}{\avarset}$ such that $\pctxp{\var} = \progentry{\tm}{\envtwo}$, with $\var \notin \uvarset \subset \unventryp{\tm}{\envtwo}$ and $\avarset \subset \anventryp{\tm}{\envtwo}$. But if $\var \notin \avarset$, then we would be able to derive $\appES{\pctxp{\var}}{\esub{\var}{\hctx}} \in \evctxUsep{\uvarset \cup \unv{\hctx}}{\avarset \cup \anv{\hctx}}$ via rule $\ruleUseHerMult$ and get that
									$$
									\progentry{\tm}{\envtwo \esub{\var}{\tmtwo}} = (\appES{\pctxp{\var}}{\esub{\var}{\hctx}}) \ctxholep{(\la{\var}{\tmthree})\tmfour} \toum (\appES{\pctxp{\var}}{\esub{\var}{\hctx}}) \ctxholeentryp{\tmthree}{\esub{\var}{\tmfour}}
									$$
								which is absurd. Moreover, if $\var \in \avarset$, then \reflemmap{Rewriting_evaluation_contexts}{one} (Focusing multiplicative evaluation contexts on applied variables) would give $\pctxtwo_{\var} \in \evctxAppp{\uvarsettwo_{\var}}{\avarsettwo_{\var}}$ such that $\pctxtwo_{\var} \ctxholep{\var} = \pctxp{\var}$, with $\uvarsettwo_{\var} 
								\subseteq \uvarset$ and $\var \notin \avarsettwo_{\var} \subset \avarset$; by \refprop{exp-ctx-are-mult} (Exponential evaluation contexts are multiplicative), $\pctxtwo_{\var} \in \evctxUsep{\uvarsetsix_{\var}}{\avarsetsix_{\var}}$, for some $\uvarsetsix_{\var} \subseteq \uvarsettwo_{\var}$ and $\avarsetsix_{\var} \subseteq \avarsettwo_{\var}$. However, we would then be able to derive $\appES{\pctxtwo_{\var} \ctxholep{\var}}{\esub{\var}{\hctx}} \in \evctxUsep{\uvarsetsix_{\var} \cup \unv{\hctx}}{\avarsetsix_{\var} \cup \anv{\hctx}}$ via rule $\ruleUseHerMult$ and get that 
									$$	
									\progentry{\tm}{\envtwo \esub{\var}{\tmtwo}} = (\appES{\pctxtwop{\var}}{\esub{\var}{\hctx}}) \ctxholep{(\la{\var}{\tmthree})\tmfour} \toum (\appES{\pctxtwop{\var}}{\esub{\var}{\hctx}}) \ctxholeentryp{\tmthree}{\esub{\var}{\tmfour}}
									$$
								which is absurd.
 
								Let $\vartwo \notin \unventryp{\tm}{\envtwo}$ instead, forcing that $\vartwo \in \anventryp{\tm}{\envtwo}$. It is easy to see that this case also leads to a contradiction, wrongfully proving that $\progentry{\tm}{\envtwo \esub{\var}{\tmtwo}}$ $\toum$-reduces.

								Therefore, the case where $\tmtwo$ is not a normal term is impossible.
								
							\end{itemize}
												
				\end{itemize} 
			
		\end{itemize}
	
	\item[$\Leftarrow$]: We proceed by case analysis on the derivation of $\ufnormalpr{\prog}$. We first prove the statement for $\genVarSetpr{\prog}$, and then prove it for $\ufinertpr{\prog}$ and $\ufabspr{\prog}$, both of which rely on the former:
		\begin{itemize}
			\item \emph{Generalized variables}: We proceed by induction on the derivation of $\genVarSetpr{\prog}$:
				\begin{itemize}
					\item 
					\emph{Rule $\ruleGenVarAx$}: If $\prog = \progentry{\var}{\emptyenv}$ then the statement clearly holds.
					
					\item 
					\emph{Rule $\ruleGenVarHer$}: Let $\genVarSetpr{\prog}$ be derived as
						$$
						\infer
							[\ruleGenVarHer]
							{\genVarpr{\var}{\appES{\progtwo}{\esub{\vartwo}{\var}}}}
							{\genVarpr{\vartwo}{\progtwo}}
						$$
					where $\prog = \appES{\progtwo}{\esub{\vartwo}{\var}}$. By \ih, $\progtwo$ is in $\tound$-normal form. Then,
						\begin{itemize}
							\item Suppose $\prog = \pctxp{(\la{\varthree}{\tmthree}) \tmfour} \toum \pctxentryp{\tmthree}{\esub{\var}{\tmfour}}$, for some $\pctx \in \evctxUsep{\uvarset}{\avarset}$. But then, since $\var$ cannot be rewritten as a multiplicative redex, it must be that $\progtwo = \pctxtwop{(\la{\varthree}{\tmthree}) \tmfour}$, with $\pctx = \appES{\pctxtwo}{\esub{\vartwo}{\var}}$; absurd, because $\progtwo$ is in $\toum$-normal form. Hence, $\prog$ is in $\toum$-normal form.

							\item Since neither $\var \notin \dom{\pctx}$ nor $\var$ is a value, then $\prog = \pctxp{\varthree} \toue \pctxp{\rename{\tmthree}}$, for some $\pctx \in \evctxAppp{\uvarset}{\avarset}$, only if $\pctx = \appES{\pctxtwo}{\esub{\vartwo}{\tmtwo}}$, for some $\pctxtwo \in \evctxAppp{\uvarsettwo}{\avarsettwo}$ such that $\var \in \dom{\pctxtwo}$. Absurd, since then it would be that $\progtwo = \pctxtwop{\var} \toue \pctxtwop{\rename{\tmthree}}$. Hence, $\prog$ is in $\toue$-normal form.

						\end{itemize}

					We conclude that $\prog$ is in $\tound$-normal form.
					
					\item 
					\emph{Rule $\ruleGenVarGc$}: Let $\genVarSetpr{\prog}$ be derived as
						 $$
						\infer
							[\ruleGenVarGc]
							{\genVarpr{\var}{\appES{\progtwo}{\esub{\vartwo}{\tm}}}}
							{\genVarpr{\var}{\progtwo}
							\quad
							\vartwo \neq \var}
						$$
					where $\prog = \appES{\progtwo}{\esub{\vartwo}{\tm}}$. By \ih, $\progtwo$ is in $\tound$-normal form. Then
						\begin{itemize}
							\item Suppose $\prog = \pctxp{(\la{\varthree}{\tmthree}) \tmfour} \toum \pctxentryp{\tmthree}{\esub{\varthree}{\tmfour}}$, for some $\pctx \in \evctxUsep{\uvarset}{\avarset}$. Note that if $\pctx = \appES{\pctxtwo}{\esub{\vartwo}{\tm}}$, for some $\pctxtwo \in \evctxUsep{\uvarsettwo}{\avarsettwo}$, then it would be that $\progtwo = \pctxtwop{(\la{\varthree}{\tmthree}) \tmfour} \toum \pctxtwoentryp{\tmthree}{\esub{\varthree}{\tmfour}}$; absurd. Hence, it should be that $\prog = (\appES{\pctxthreep{\vartwo}}{\esub{\vartwo}{\hctx}}) \ctxholep{(\la{\varthree}{\tmthree}) \tmfour}$ with $\progtwo = \pctxthreep{\vartwo}$ and $\tm = \hctxp{(\la{\varthree}{\tmthree}) \tmfour}$. In addition, \reflemmap{Rewriting_evaluation_contexts-base_cases}{one} (Multiplicative evaluation contexts give needed variables) would give that $\vartwo \in \nv{\pctxthreep{\vartwo}} = \nv{\progtwo}$. However, \reflemma{Properties_of_generalized_variables} (Properties of generalized variables) gives that $\nv{\progtwo} = \set{\var}$. Since $\vartwo \neq \var$, then this case is impossible, and so $\prog$ is in $\toum$-normal form. 

							\item Suppose $\prog = \pctxp{\varthree} \toue \pctxp{\rename{\tmthree}}$ for some $\pctx \in \evctxAppp{\uvarset}{\avarset}$. Note that the variable occurrence to be substituted must appear within $\progtwo$; otherwise, $\var \notin \dom{\pctx}$ which is absurd. That is, it should be that $\pctx = \appES{\pctxtwo}{\esub{\vartwo}{\tm}}$, for some $\pctxtwo \in \evctxAppp{\uvarsettwo}{\avarsettwo}$ such that $\progtwo = \pctxtwop{\varthree}$. 

							Now, let us suppose that $\vartwo = \varthree$ and that $\prog = \appES{\progtwo}{\esub{\vartwo}{\tm}} = (\appES{\pctxtwo}{\esub{\vartwo}{\tm}}) \ctxholep{\vartwo} \toue (\appES{\pctxtwo}{\esub{\vartwo}{\tm}}) \ctxholep{\rename{\tm}}$. Then \reflemmap{Rewriting_evaluation_contexts-base_cases}{two} (Exponential evaluation contexts give applied variables) would give that $\vartwo \in \anv{\pctxtwop{\vartwo}} = \anv{\progtwo}$. This is absurd, by \reflemma{Properties_of_generalized_variables} (Properties of generalized variables). 

							Hence, it should be that $\vartwo \neq \varthree \in \dom{\pctxtwo}$. However, this would in turn imply that $\progtwo = \pctxtwop{\varthree} \toue \pctxtwop{\rename{\tmthree}}$; absurd. Hence, $\prog$ is in $\toue$-normal form.

						\end{itemize}

					We conclude that $\prog$ is in $\tound$-normal form.
					
				\end{itemize}
			
			\item \emph{Useful abstraction programs}: We proceed by induction on the derivation of $\ufabspr{\prog}$:
				\begin{itemize}
					\item
					\emph{Rule $\ufRuleAbsProm$}: Let $\ufabspr{\prog}$ be derived as 
						$$
						\infer
							[\ufRuleAbsProm]
							{\ufabsentrypr{\val}{\emptyenv}}
							{}
						$$
					where $\prog = \progentry{\val}{\emptyenv}$. Note that $\env = \emptyenv$ discards the possibility that $\prog$ were not in $\toue$-normal form. Moreover, $\val$ discards the possibility that $\prog$ were not in $\toum$---by \reflemma{Redex_in_non-normal_terms-usefulopencbneed} (Redex in non-normal terms). Hence, $\prog$ is in $\tound$-normal form.

					\item 
					\emph{Rule $\ufRuleAbsGenVar$}: Let $\ufabspr{\prog}$ be derived as 
						$$
						\infer
							[\ufRuleAbsGenVar]
							{\ufabspr{\appES{\progtwo}{\esub{\var}{\val}}}}
							{\genVarpr{\var}{\progtwo}}
						$$
					where $\prog = \appES{\progtwo}{\esub{\var}{\val}}$. By \ih, $\progtwo$ is in $\tound$-normal form. Then
						\begin{itemize}
							\item Suppose $\prog = \pctxp{(\la{\varthree}{\tmthree}) \tmfour} \toum \pctxentryp{\tmthree}{\esub{\varthree}{\tmfour}}$, for some $\pctx \in \evctxUsep{\uvarset}{\avarset}$. 

							Note that if $\pctx = \appES{\pctxtwo}{\esub{\var}{\val}}$, for some $\pctxtwo \in \evctxUsep{\uvarset}{\avarset}$, then we would have that $\progtwo = \pctxtwop{(\la{\varthree}{\tmthree}) \tmfour} \toum \pctxtwoentryp{\tmthree}{\esub{\varthree}{\tmfour}}$; absurd.

							Hence, it should be that $\pctx = \appES{\pctxtwop{\var}}{\esub{\var}{\hctx}}$, with $\progtwo = \pctxtwop{\var}$ and $\val = \hctxp{(\la{\varthree}{\tmthree}) \tmfour}$. The latter is impossible, by \reflemma{Redex_in_non-normal_terms-usefulopencbneed} (Redex in non-normal terms). Hence, $\prog$ is in $\toum$-normal form.

							\item Suppose $\prog = \pctxp{\varthree} \toue \pctxp{\rename{\tmthree}}$, for some $\pctx \in \evctxAppp{\uvarset}{\avarset}$. Note that the variable occurrence to be substituted must appear within $\progtwo$; otherwise, $\var \notin \dom{\pctx}$, which is absurd. That is, it should be that $\pctx = \appES{\pctxtwo}{\esub{\vartwo}{\tm}}$, for some $\pctxtwo \in \evctxAppp{\uvarsettwo}{\avarsettwo}$ such that $\progtwo = \pctxtwop{\varthree}$. 

							Now, let us suppose that $\vartwo = \varthree$ and that $\prog = \appES{\progtwo}{\esub{\vartwo}{\tm}} = (\appES{\pctxtwo}{\esub{\vartwo}{\tm}}) \ctxholep{\vartwo} \toue (\appES{\pctxtwo}{\esub{\vartwo}{\tm}}) \ctxholep{\rename{\tm}}$. Then \reflemmap{Rewriting_evaluation_contexts-base_cases}{two} (Exponential evaluation contexts give applied variables) would give that $\vartwo \in \anv{\pctxtwop{\vartwo}} = \anv{\progtwo}$. This is absurd, by \reflemma{Properties_of_generalized_variables} (Properties of generalized variables). 

							Hence, it should be that $\vartwo \neq \varthree$ and that $\varthree \in \dom{\pctxtwo}$. However, this would in turn imply that $\progtwo = \pctxtwop{\varthree} \toue \pctxtwop{\rename{\tmthree}}$. This is absurd, and so $\prog$ is in $\toue$-normal form.

						\end{itemize}

					We conclude that $\prog$ is in $\tound$-normal form.
					
					\item 
					\emph{Rule $\ufRuleAbsGc$}: Let $\ufabspr{\prog}$ be derived as 
						$$
						\infer
							[\ufRuleAbsGc]
							{\ufabspr{\appES{\progtwo}{\esub{\var}{\tm}}}}
							{\ufabspr{\progtwo}}
						$$
					where $\prog = \appES{\progtwo}{\esub{\var}{\tm}}$. By \ih, $\progtwo$ is in $\tound$-normal form. Then
						\begin{itemize}
							\item Suppose $\prog = \pctxp{(\la{\varthree}{\tmthree}) \tmfour} \toum \pctxentryp{\tmthree}{\esub{\varthree}{\tmfour}}$, for some $\pctx \in \evctxUsep{\uvarset}{\avarset}$. 

							Note that if $\pctx = \appES{\pctxtwo}{\esub{\var}{\val}}$, for some $\pctxtwo \in \evctxUsep{\uvarset}{\avarset}$, then we would have that $\progtwo = \pctxtwop{(\la{\varthree}{\tmthree}) \tmfour} \toum \pctxtwoentryp{\tmthree}{\esub{\varthree}{\tmfour}}$; absurd.

							Hence, it should be that $\pctx = \appES{\pctxtwop{\var}}{\esub{\var}{\hctx}}$, with $\progtwo = \pctxtwop{\var}$ and $\tm = \hctxp{(\la{\varthree}{\tmthree}) \tmfour}$. However, by \reflemmap{Rewriting_evaluation_contexts-base_cases}{one} (Multiplicative evaluation contexts give needed variables), we would have that $\var \in \nv{\pctxtwop{\var}} = \nv{\progtwo}$, which is absurd by \reflemma{Properties_of_useful_abstraction_programs} (Properties of useful abstraction programs). Hence, $\prog$ is in $\toum$-normal form.

							\item Suppose $\prog = \pctxp{\varthree} \toue \pctxp{\rename{\tmthree}}$, for some $\pctx \in \evctxAppp{\uvarset}{\avarset}$. Note that the variable occurrence to be substituted must appear within $\progtwo$; otherwise, $\var \notin \dom{\pctx}$ which is absurd. That is, it should be that $\pctx = \appES{\pctxtwo}{\esub{\var}{\tm}}$, for some $\pctxtwo \in \evctxAppp{\uvarsettwo}{\avarsettwo}$ such that $\progtwo = \pctxtwop{\varthree}$.

							Now, let us suppose that $\var = \varthree$ and that $\prog = \appES{\progtwo}{\esub{\var}{\tm}} = (\appES{\pctxtwo}{\esub{\var}{\tm}}) \ctxholep{\varthree} \toue (\appES{\pctxtwo}{\esub{\var}{\tm}}) \ctxholep{\rename{\tm}}$. Then \reflemmap{Rewriting_evaluation_contexts-base_cases}{two} (Exponential evaluation contexts give applied variables) would give that $\vartwo \in \anv{\pctxtwop{\vartwo}} = \anv{\progtwo}$. This is absurd, by \reflemma{Properties_of_useful_abstraction_programs} (Properties of useful abstraction programs).

							Hence, it should be that $\var \neq \varthree \in \dom{\pctxtwo}$. However, this would in turn imply that $\progtwo = \pctxtwop{\varthree} \toue \pctxtwop{\rename{\tmthree}}$. This is absurd, and so $\prog$ is in $\toue$-normal form.

						\end{itemize}

					We conclude that $\prog$ is in $\tound$-normal form.
					
				\end{itemize}
			
			\item \emph{Useful inert programs}: We proceed by induction on the derivation of $\ufinertpr{\prog}$:
				\begin{itemize}
					\item 
					\emph{Rule $\ufRuleInertProm$}: Let $\ufinertpr{\prog}$ be derived as
						$$
						\infer
							[\ufRuleInertProm]
							{\ufinertentrypr{\itmplus}{\emptyenv}}
							{}
						$$
					where $\prog = \progentry{\itmplus}{\emptyenv}$. Note that $\env = \emptyenv$ discards the possibility that $\prog$ were not in $\toue$-normal form. Moreover, $\itmplus$ discards the possibility that $\prog$ were not in $\toum$-normal form---by  \reflemma{Redex_in_non-normal_terms-usefulopencbneed} (Redex in non-normal terms). Hence, $\prog$ is in $\tound$-normal form.
					
					\item 
					\emph{Rule $\ufRuleInertInert$}: Let $\ufinertpr{\prog}$ be derived as
						$$
						\infer
							[\ufRuleInertInert]
							{\ufinertpr{\appES{\progtwo}{\esub{\var}{\itm}}}}
							{\ufinertpr{\progtwo}
							\quad
							\var \in \nv{\itm}}
						$$
					where $\prog = \appES{\progtwo}{\esub{\var}{\itm}}$. By \ih, $\progtwo$ is in $\tound$-normal form. Then,
						\begin{itemize}
							\item Suppose $\prog = \pctxp{(\la{\varthree}{\tmthree}) \tmfour} \toum \pctxentryp{\tmthree}{\esub{\varthree}{\tmfour}}$, for some $\pctx \in \evctxUsep{\uvarset}{\avarset}$. 

							Note that if $\pctx = \appES{\pctxtwo}{\esub{\var}{\itm}}$, for some $\pctxtwo \in \evctxUsep{\uvarset}{\avarset}$, then we would have that $\progtwo = \pctxtwop{(\la{\varthree}{\tmthree}) \tmfour} \toum \pctxtwoentryp{\tmthree}{\esub{\varthree}{\tmfour}}$; absurd.

							Hence, it should be that $\pctx = \appES{\pctxtwop{\var}}{\esub{\var}{\hctx}}$, with $\progtwo = \pctxtwop{\var}$ and $\itm = \hctxp{(\la{\varthree}{\tmthree}) \tmfour}$. However, this is also absurd, by \reflemma{Redex_in_non-normal_terms-usefulopencbneed} (Redex in non-normal terms). Hence, $\prog$ is in $\tound$-normal form.

							\item Suppose $\prog = \pctxp{\varthree} \toue \pctxp{\rename{\tmthree}}$, for some $\pctx \in \evctxAppp{\uvarset}{\avarset}$. Note that the variable occurrence to be substituted must appear within $\progtwo$; otherwise, $\var \notin \dom{\pctx}$ which is absurd. That is, it should be that $\pctx = \appES{\pctxtwo}{\esub{\var}{\itm}}$, for some $\pctxtwo \in \evctxAppp{\uvarsettwo}{\avarsettwo}$ such that $\progtwo = \pctxtwop{\varthree}$. Moreover, note that since $\itm $ is not a value, then it must be that $\var \neq \varthree$ and that $\varthree \in \dom{\pctxtwo}$. However, this would imply that $\progtwo = \pctxtwop{\varthree} \toue \pctxtwop{\rename{\tmthree}}$; absurd. Hence, $\prog$ is in $\toue$-normal form.

						\end{itemize}

					We conclude that $\prog$ is in $\tound$-normal form.
					
					\item 
					\emph{Rule $\ufRuleInertGenVar$}: Let $\ufinertpr{\prog}$ be derived as
						$$
						\infer
							[\ufRuleInertGenVar]
							{\ufinertpr{\appES{\progtwo}{\esub{\var}{\itmplus}}}}
							{\genVarpr{\var}{\progtwo}}
						$$
					where $\prog = \appES{\progtwo}{\esub{\var}{\itmplus}}$. By \ih, $\progtwo$ is in $\tound$-normal form. Note that $\progtwo$ is in $\tound$-normal form---as we proved it above for every generalized variable. Then,
						\begin{itemize}
							\item Suppose $\prog = \pctxp{(\la{\varthree}{\tmthree}) \tmfour} \toum \pctxentryp{\tmthree}{\esub{\varthree}{\tmfour}}$, for some $\pctx \in \evctxUsep{\uvarset}{\avarset}$. 

							Note that if $\pctx = \appES{\pctxtwo}{\esub{\var}{\itm}}$, for some $\pctxtwo \in \evctxUsep{\uvarset}{\avarset}$, then we would have that $\progtwo = \pctxtwop{(\la{\varthree}{\tmthree}) \tmfour} \toum \pctxtwoentryp{\tmthree}{\esub{\varthree}{\tmfour}}$; absurd.

							Hence, it should be that $\pctx = \appES{\pctxtwop{\var}}{\esub{\var}{\hctx}}$, with $\progtwo = \pctxtwop{\var}$ and $\itm = \hctxp{(\la{\varthree}{\tmthree}) \tmfour}$. However, this is also absurd, by \reflemma{Redex_in_non-normal_terms-usefulopencbneed} (Redex in non-normal terms). Hence, $\prog$ is in $\tound$-normal form.

							\item Suppose $\prog = \pctxp{\varthree} \toue \pctxp{\rename{\tmthree}}$, for some $\pctx \in \evctxAppp{\uvarset}{\avarset}$. Note that the variable occurrence to be substituted must appear within $\progtwo$; otherwise, $\var \notin \dom{\pctx}$ which is absurd. That is, it should be that $\pctx = \appES{\pctxtwo}{\esub{\var}{\itm}}$, for some $\pctxtwo \in \evctxAppp{\uvarsettwo}{\avarsettwo}$ such that $\progtwo = \pctxtwop{\varthree}$ and $\varthree \in \dom{\pctxtwo}$---the latter because $\itmplus $ is not a value. However, \reflemmap{Rewriting_evaluation_contexts-base_cases}{two} (Exponential evaluation contexts give applied variables) would give that $\var \in \anv{\pctxtwop{\var}} = \anv{\progtwo}$, which is absurd---by \reflemma{Properties_of_generalized_variables} (Properties of generalized variables). Hence, $\prog$ is in $\toue$-normal form.

						\end{itemize}

					We conclude that $\prog$ is in $\tound$-normal form.
					
					\item
					\emph{Rule $\ufRuleInertUseless$}: Let $\ufinertpr{\prog}$ be derived as
						$$
						\infer
							[\ufRuleInertUseless]
							{\ufinertpr{\appES{\progtwo}{\esub{\var}{\val}}}}
							{\ufinertpr{\progtwo}
							\quad
							\var \in \unv{\progtwo}
							\quad
							\var \notin \anv{\progtwo}}
						$$
					where $\prog = \appES{\progtwo}{\esub{\var}{\val}}$. By \ih, $\progtwo$ is in $\tound$-normal form. Then,
						\begin{itemize}
							\item Suppose $\prog = \pctxp{(\la{\varthree}{\tmthree}) \tmfour} \toum \pctxentryp{\tmthree}{\esub{\varthree}{\tmfour}}$, for some $\pctx \in \evctxUsep{\uvarset}{\avarset}$. 

							Note that if $\pctx = \appES{\pctxtwo}{\esub{\var}{\val}}$, for some $\pctxtwo \in \evctxUsep{\uvarset}{\avarset}$, then we would have that $\progtwo = \pctxtwop{(\la{\varthree}{\tmthree}) \tmfour} \toum \pctxtwoentryp{\tmthree}{\esub{\varthree}{\tmfour}}$; absurd.

							Hence, it should be that $\pctx = \appES{\pctxtwop{\var}}{\esub{\var}{\hctx}}$, with $\progtwo = \pctxtwop{\var}$ and $\val = \hctxp{(\la{\varthree}{\tmthree}) \tmfour}$. However, this is also absurd, by \reflemma{Redex_in_non-normal_terms-usefulopencbneed} (Redex in non-normal terms). Hence, $\prog$ is in $\tound$-normal form.

							\item Suppose $\prog = \pctxp{\varthree} \toue \pctxp{\rename{\tmthree}}$, for some $\pctx \in \evctxAppp{\uvarset}{\avarset}$. Note that the variable occurrence to be substituted must appear within $\progtwo$; otherwise, $\var \notin \dom{\pctx}$ which is absurd. That is, it should be that $\pctx = \appES{\pctxtwo}{\esub{\var}{\val}}$, for some $\pctxtwo \in \evctxAppp{\uvarsettwo}{\avarsettwo}$ such that $\progtwo = \pctxtwop{\varthree}$. But then \reflemmap{Rewriting_evaluation_contexts-base_cases}{two} (Exponential evaluation contexts give applied variables) would give that $\var \in \anv{\pctxtwop{\var}} = \anv{\progtwo}$; absurd. Hence, $\prog$ is in $\toue$-normal form.

						\end{itemize}

					We conclude that $\prog$ is in $\tound$-normal form.
					
					\item 
					\emph{Rule $\ufRuleInertGc$}: Let $\ufinertpr{\prog}$ be derived as
						$$
						\infer
							[\ufRuleInertGc]
							{\ufinertpr{\appES{\progtwo}{\esub{\var}{\tm}}}}
							{\ufinertpr{\progtwo}
							\quad
							\var \notin \nv{\progtwo}}
						$$
					where $\prog = \appES{\progtwo}{\esub{\var}{\tm}}$. By \ih, $\progtwo$ is in $\tound$-normal form. Then,
						\begin{itemize}
							\item Suppose $\prog = \pctxp{(\la{\varthree}{\tmthree}) \tmfour} \toum \pctxentryp{\tmthree}{\esub{\varthree}{\tmfour}}$, for some $\pctx \in \evctxUsep{\uvarset}{\avarset}$. 

							Note that if $\pctx = \appES{\pctxtwo}{\esub{\var}{\tm}}$, for some $\pctxtwo \in \evctxUsep{\uvarset}{\avarset}$, then we would have that $\progtwo = \pctxtwop{(\la{\varthree}{\tmthree}) \tmfour} \toum \pctxtwoentryp{\tmthree}{\esub{\varthree}{\tmfour}}$; absurd.

							Hence, it should be that $\pctx = \appES{\pctxtwop{\var}}{\esub{\var}{\hctx}}$, with $\progtwo = \pctxtwop{\var}$ and $\tm = \hctxp{(\la{\varthree}{\tmthree}) \tmfour}$. However, \reflemmap{Rewriting_evaluation_contexts-base_cases}{one} (Multiplicative evaluation contexts give needed variables) would give that $\var \in \nv{\pctxtwop{\var}} = \nv{\progtwo}$. Hence, $\prog$ is in $\tound$-normal form.

							\item Suppose $\prog = \pctxp{\varthree} \toue \pctxp{\rename{\tmthree}}$, for some $\pctx \in \evctxAppp{\uvarset}{\avarset}$. Note that the variable occurrence to be substituted must appear within $\progtwo$; otherwise, $\var \notin \dom{\pctx}$ which is absurd. That is, it should be that $\pctx = \appES{\pctxtwo}{\esub{\var}{\val}}$, for some $\pctxtwo \in \evctxAppp{\uvarsettwo}{\avarsettwo}$ such that $\progtwo = \pctxtwop{\varthree}$. But then \reflemmap{Rewriting_evaluation_contexts-base_cases}{two} (Exponential evaluation contexts give applied variables) would give that $\var \in \anv{\pctxtwop{\var}} = \anv{\progtwo}$, with $\anv{\progtwo} \subseteq \nv{\progtwo}$---by \reflemmap{Unapplied_applied_and_needed_variables}{two} (Unapplied, applied and needed variables); absurd. This is absurd, and so $\prog$ is in $\toue$-normal form.

						\end{itemize}

					We conclude that $\prog$ is in $\tound$-normal form.\qedhere
					
				\end{itemize}
			
		\end{itemize}
	
\end{itemize}

\end{proof}




\newpage

%


\end{document}